\documentclass[12pt]{article}

\usepackage{graphicx}
\usepackage{subfig}
\usepackage{bm}
\usepackage{float}

\newcommand{\beq}{\begin{equation}}
\newcommand{\eeq}{\end{equation}}
\newcommand{\beqa}{\begin{eqnarray}}
\newcommand{\eeqa}{\end{eqnarray}}
\newcommand{\diff}{\mathrm{d}}
\newcommand{\Diff}{\mathrm{D}}

\title{Appearance of Keplerian discs orbiting Kerr superspinars}

\author{Zden\v{e}k Stuchl\'{i}k and Jan Schee\\
	\it Institute of Physics, Faculty of Philosophy and Science,\\
	\it Silesian University in Opava\\
	\it	Bezru\v{c}ovo n\'{a}m. 13, Opava, Czech Republic\\
	\small email1:zdenek.stuchlik@fpf.slu.cz\\
	\small email2:jan.schee@fpf.slu.cz
}

\date{}

\begin{document}

\maketitle

\begin{abstract}We study optical phenomena related to appearance of Keplerian accretion discs orbiting Kerr superspinars predicted by the string theory. The superspinar exterior is described by the standard Kerr naked singularity geometry breaking the black hole limit on the internal angular momentum (spin). We construct local photon escape cones for a variety of orbiting sources that enable to determine the superspinars silhouette in the case of distant observers. We show that the superspinar silhouette depends strongly on the assumed edge where the external Kerr spacetime is joined to the internal spacetime governed by the string theory and significantly differs from the black hole silhouette. The appearance of the accretion disc is strongly dependent on the value of the superspinar spin in both their shape and frequency shift profile. Apparent extension of the disc grows significantly with growing spin, while the frequency shift grows with descending spin. This behavior differs substantially from appearance of discs orbiting black holes enabling thus, at least in principle, to distinguish clearly the Kerr superspinars and black holes. In vicinity of a Kerr superspinar the non-escaped photons have to be separated to those captured by the superspinar and those being trapped in its strong gravitational field leading to self-illumination of the disc that could even influence its structure and causes self-reflection effect of radiation of the disc. The amount of trapped photons grows with descending of the superspinar spin. We thus can expect significant self-illumination effects in the field of Kerr superspinars with near-extreme spin $a \sim 1$.
\end{abstract}

\bibliographystyle{natbib}

\section{Introduction}

Kerr superspinars with mass $M$ and angular momentum $J$ violating the general relativistic bound on the spin of rotating black holes ($a=J/M^2>1$) could be primordial remnants of the high-energy phase of very early period of the evolution of the Universe when the effects of the string theory were relevant (\cite{Gim-Hor:2009:PhysLett}). The spacetime outside the superspinar, where the stringy effects are irrelevant, is assumed to be described by the standard Kerr geometry. 
It is expected that extension of the internal region is limited to $r<R<M$ covering thus the region of causality violations (naked time machine) and still allowing for the presence of the relevant astrophysical phenomena related to the Kerr naked 
singularity spacetimes. There is an expectation that the pathological naked time machine is replaced by a correctly 
behaving stringy solution \cite{Gim-Hor:2009:PhysLett}, being motivated by the resolution of the problems of 4+1 SUSY black hole solution \cite{Bre-etal:1997:PhysLettB} where the pathological time machine region is replaced by a portion 
of the Godel universe \cite{Gim-Hor:2004:,Boy-etal:2003:PhRvD}. Being assumed remnants of the early phases of the evolution of the Universe, Kerr superspinars can be considered as a modern alternative to the ideas of naked singularities (white holes, retarded cores of expansion) discussed as a model of quasars in early times of relativistic astrophysics \cite{Nov:1964:AstronZhur,Nov:1966a:SovPhys,Nov:1966b:SovPhys,Zel-Nov:1966:PhRvD,Rees:1966:Nat,Lyn-Rees:1971:MNRAS}.

It should be stressed that the assumed existence of Kerr superspinars is not in contradiction with the Penrose cosmic censorship hypothesis \cite{Pen:1969:NUOC2:} according to which the spacetime singularities generated by a gravitational collapse have to be hidden behind an event horizon.
On the other hand, the cosmic censorship hypothesis that forbids the existence of naked singularities is far from being proved and, in fact, even existence of spherically symmetric naked singularity spacetimes caused by a self-similar gravitational collepse \cite{Ori-Pir:1988:MNRAS,Ori-Piran:1990:PhysRevD} or a non-self-similar collapse \cite{Lake:1991:PhysRevD} is frequently discussed (see also \cite{Tip:1977:AnnPhys,Cla:1993:,Jos-Dwi:1993:PhysRevD,Chri:1994:AnnMath,deFel-Sim-Yun:1999:CQG}) and the related optical phenomena are studied, e.g., in  \cite{Vir-Elli:2002:PhRvD,Vir-Keet:2008:PhRvD}. The electrically charged Reissner-Nordstrom naked singularity spacetimes are also considered \cite{deFel-Yun:2001:CQG:} even with the presence of the cosmological constant \cite{Stu-Hle:2002:ACTPS2:}. Formation of naked singularities due to collapse of rotating matter is modelled too \cite{Sha-Teu:1991:AmSc}. Therefore, it is of some relevance and importance to consider even the complete Kerr naked singularity spacetimes and compare their (observationally relevant) properties to those of Kerr superspinars having the naked singularity and causality violating parts of the Kerr geometry removed. (The possibility of over-spinning black holes is discussed, e.g., in \cite{Jac-Sot:2009}.)

The Kerr superspinars should have extremely strong gravity in their vicinity and some properties substantially 
differing from those of the standard Kerr black holes \cite{deFel:1974:AA,Cun:1975:Apj, deFel:1978Natur,Stu:1980:BULAI:,Stu:1981:BULAI:,Tor-Stu:2005:AA,Sch-Stu-Jur:2005:RAGtime6and7:CrossRef,Gim-Hor:2009:PhysLett,Hio-Mae:2009:PhysRevD,Bam-Fre:2009:PhysRevD}. The differences could be related both to the accretion phenomena and optical effects and their cooperation is strongly and most clearly reflected by the optical phenomena related to appearance of the Keplerian accretion discs that will be studied in the present paper. Both optical and accretion phenomena were studied even in Kerr-de Sitter naked singularity spacetimes \cite{Asch:2004:ASTRA:, Stu:1983:BULAI:,Stu-Cal:1991:GENRG2:,Stu-Hle:2000:CLAQG:,Stu-Sla:2004:PHYSR4:,Sla-Stu:2005:CLAQG:,Stu:2005:MODPLA:}. Here we restrict our attention to the optical phenomena in Kerr naked singularity spacetimes with zero cosmological constant.

The energy efficiency of accretion in Keplerian discs around superspinars can be extremely high. In the field of Kerr naked singularities with spin $a$ very close to the minimal value of $a=1$ the efficiency overcomes substantially even the annihilation efficiency, being $\sim 157\%$ and is much larger than the efficiency $\sim 43\%$ that could be approached in the field of near-extreme Kerr black holes (\cite{Stu:1980:BULAI:}). Therefore, spectral profiles and appearance of Keplerian discs have to be strongly affected due to both the energy efficiency of the accretion process and the strong gravitational redshift of radiation coming from the deepest parts of the gravitational potential. 
Further, appearance of the Keplerian discs, their spectral continuum, and profiled spectral lines related to the innermost parts of the Keplerian discs, should reveal important differences between their character in the field of black holes and naked singularities due to the existence of bound photon orbits that appears in the vicinity of a superspinars with sufficiently small spin \cite{Stu:1980:BULAI:, Stu:1981:BULAI:} influencing thus both the optical phenomena and even the character of the innermost parts of the Keplerian disc that can extend down to the region of bound photon orbits. (Of course, no region of bound photons is possible in the field of Kerr black holes because of the presence of the event horizon.) 

Geometrically thick, toroidal accretion discs orbiting a Kerr superspinar are relevant in situations where the pressure 
gradients are important for the accretion discs structure. \footnote{Such structures were discussed (even with the cosmological constant included) in the case of both spherically symmetric Reissner-Nordstrom spacetimes \cite{Kuc-Stu-Sla:2010:} and in the rotating Kerr spacetimes \cite{Sla-Stu:2005:CLAQG:, Stu:2005:MODPLA:}. In both cases, unexpected features were found that do not appear in the black hole spacetimes.} The inner edge of the accretion disc is then closer to the superspinar surface as compared with the edge of the Keplerian disc. Here we restrict attention to the Keplerian discs only, postponing the more complex study of optical effects related to toroidal discs to future studies.

The Kerr superspinars could appear in Active Galactic Nuclei (AGN) where supermassive black holes are usually assumed, 
or in Galaxy Black Hole Candidates (GBHC) observed in some X-ray binary systems \cite{Rem-McCli:2006:ARASTRA:,Sha-Nar-McC:2008:APJ,Ste-McC-Rem:2009:APJ}. In both classes of the black hole candidates some objects are reported with the spin estimates extremely close to the limit value $a=1$ and these could serve as superspinar candidates. The high spin estimates are implied by the X-ray observations in AGN MGC-6-30-15 \cite{Tan-etal:1995:NATURE:} or the GBHC GRS1915+105 \cite{Sha-etal:2009} and are related to the spectral continuum models \cite{McC-etal:2006:ASTRJ2:,Sha-etal:2009}, profiled iron spectral lines \cite{Iwa-etal:1996:MNRAS,Rey-etal:2009:AAS}, and high-frequency quasi-periodic oscillations (QPOs) explained by the orbital resonant models applied to near-extreme black holes \cite{Tor-etal:2005AA, Stu-etal:2005:PhRvD,Stu-Sla-Tor:2007:AA:a,Stu-Sla-Tor:2007:AA:b,Sla-Stu:2008:AA}. The standard most common twin peak high-frequency QPOs observed in a variety of black hole candidate binary systems with frequency ratio $3:2$ that are in some cases quite well explained by resonant phenomena between the radial and vertical epicyclic oscillation modes in the accretion disc, giving estimates of the black hole mass and spin \cite{Tor-etal:2005AA}, could equally be explained by the resonances of the epicyclic oscillations in the field of a Kerr superpsinar with different values of the mass and spin \cite{Tor-Stu:2005:AA}. Therefore, it is quite interesting to check, if it is possible to recognize an optical effect that could, at least in principle, distinguish its demonstration in the black hole and superspinar field, or if it is necessary to combine the results of observations of different origin in order to distinguish black hole and superspinar systems. Recall that quite recently it was shown that the observed spectrum of Keplerian accretion discs cannot serve as a single indicator of the presence of a Kerr superspinar, because for a black hole with any spin $a$ an identical spectrum can be generated by properly chosen superspinar with spin in the range $5/3 < a < 8\sqrt{6}/3 \sim 6.532$ \cite{Tak-Har:2010:arXiv}.
 
Due to the accretion process, the superspinar spin can be rapidly reduced and conversion of a Kerr superspinar into 
a near-extreme black hole is possible \cite{Stu:1981:BULAI:,Cal-Nob:1979:NUOC2:}. The conversion due to counterrotating accretion discs is extremely effective and the time necessary for the conversion can be much shorter than the characteristic time of the black hole evolution \cite{Stu-Hle-Tru:2010:}. Moreover, the process of converting the Kerr superspinar into a near-extreme black hole due to corotating accretion discs is very dramatic and could be interesting with connection to the most extreme cases of Gamma Ray Bursts (GBR) since a large part of he accretion disc becomes to be dynamically unstable after the transition of the superspinar into a near-extreme black hole state \cite{Stu:1981:BULAI:}. 

Observations of a hypothetical Kerr superspinar could thus be expected at high redshift AGN and quasars. If such an extremely compact object violating the standard Kerr spin bound $a<1$ will be observed in the Universe, it could quite well be naturally interpreted in the framework of the string theory. The Kerr superspinars can thus represent one of the most relevant experimental tests of the string theory \cite{Gim-Hor:2004:}. On the other hand, one has to be very careful in making any definite conclusion based on the general-relativistic Kerr spin bound, since its breaching is also allowed in modifications of General Relativity, e.g., for braneworld Kerr black holes having quite regular event horizon when the tidal charge braneworld parameter, reflecting the braneworld black hole interaction with the external bulk space, takes negative values, and $a > 1$ \cite{Ali-Gum:2005:PHYS4:ChRBH3Brane, Ali-Tal:2009:PhRvD, Stu-Kot:2009:GReGr, Sch-Stu:2009:GReGr, Sch-Stu:2009:IJMPD}.

We shall discuss appearance of the corotating Keplerian discs orbiting Kerr superspinars and the silhouette shape of the superspinar surface. We shall focus our attention on the innermost parts of the Keplerian discs near the geodetical innermost stable circular orbit (ISCO) when the relativistic phenomena are strongest and the signatures of the spin effects related to the superspinars are most profound. We assume the superspinar boundary surface to be located at $r(\theta)=R=0.1M$ which guarantees that the surface is well under the ISCO. Further, we assume that the boundary of the superspinar has properties similar to those of the black hole horizon and its one-way membrane property, i.e., it is not radiating and eats the accreted matter with no optical response. Of course, the boundary properties has to be given by the exact solutions of the string theory that has to be joined to the standard Kerr geometry at the boundary and has to determine physical properties of the boundary. However, such solutions are not known at present time, and we follow the simplest possibility that is commonly used in most of the studies of the phenomena taking place in the field of Kerr superspinars \cite{Gim-Hor:2009:PhysLett,Bam-Fre:2009:PhysRevD,Bam-etal:2009:PhRvD}. 
In section 2 we shortly summarize properties of the photon geodetical motion in the Kerr naked singularity spacetimes. 
In section 3 we construct the local photon escape cones of a variety of relevant astrophysical observers, namely the locally non-rotating observers and the equatorial circular geodesic observers that are relevant for the Keplerian accretion discs.  In section 4 the Kerr superspinar silhouette as seen by a distant static observer is constructed in dependence on the spacetime parameters and the inclination angle of the observer. We compare the cases of superspinar with $R=0.1M$ and the Kerr naked singularity. Possibilities to determine the superspinar parameters from expected observational data (e.g. from the Galaxy central object Sgr $\mathrm{A}^*$) are briefly discussed.
In section 5 we integrate the equations of photon motion and investigate the optical phenomena relevant for the appearance of a Keplerian disc, namely the frequency shift of its radiation and deformations of the isoradial curves of direct and indirect images. The Keplerian discs appearances modified by the presence of a Kerr superspinar are confronted with those modified by the presence of a near-extreme black hole. In section 6 we briefly discuss the results and study separation of non-escaping photons into the photons captured by the superspinar surface and the trapped photons that could influence the accretion disc. In section 7 we present conclusions.

\section{Photon motion}
\subsection{Geometry and its null geodesics}
Kerr superspinars are described by the Kerr geometry that is given in the standard Boyer-Lindquist coordinates and geometric units ($c=G=1$) in the form
\beq
      \diff s^2 = -(1-\frac{2Mr}{\Sigma})\diff t^2 + \frac{\Sigma}{\Delta}\diff r^2 + \Sigma \diff \theta^2 + \frac{A}{\Sigma}\sin^2\theta \diff\varphi^2 - \frac{4 M a r \sin^2\theta}{\Sigma}\diff t \diff \varphi, 
\eeq
where 
\beq
	\Delta=r^2-2Mr+a^2,\, \Sigma=r^2+a^2\cos^2\theta, \, \mathrm{and}\, A=(r^2+a^2)^2 - \Delta a^2\sin^2\theta,
\eeq
$a$ denotes spin and $M$ mass of the spacetimes that fulfill condition $a>M$ in the superspinar case. In the following, we put $M=1$, i.e., we use dimensionless radial coordinate $r$ and spin $a>1$.
    
In order to study the optical effects we have to solve equations of motion of photons given by the null geodesics of the spacetime under consideration. The geodesic equation reads

\begin{equation}
  \frac{\Diff k^\mu}{\diff w}=0,\label{ce1}
\end{equation}
where $k^\mu=\frac{\diff x^\mu}{\diff w}$ is the wave vector tangent to the null geodesic and $w$ is the affine parameter. The normalization condition reads $g_{\mu\nu}k^\mu k^\nu = 0$. Since the components of the metric tensor do not depend on $\varphi$ and $t$ coordinates, the conjugate momenta

\begin{eqnarray}
 	k_\varphi=g_{\varphi\nu}k^\nu\equiv \Phi,\label{ce4}\\
 	k_t =g_{t\nu}k^\nu\equiv -E,\label{ce5}
\end{eqnarray}
are the integrals of motion. Carter found another integral of motion $K$ as a separation constant when solving Hamilton-Jacobi equation 

\begin{equation}
 	g^{\mu\nu}\frac{\partial S}{\partial x^\mu}\frac{\partial S}{\partial x^\nu} = 0,\label{ce6}
\end{equation}
where he assumed the action $S$ in separated form

\begin{equation}
	S=-Et+\Phi\varphi + S_r(r)+S_{\theta}(\theta).\label{ce6a}
\end{equation}

The equations of motion can be integrated and written separately in the form

\begin{eqnarray}
 	\Sigma\frac{\diff r}{\diff w}&=&\pm\sqrt{R(r)},\label{ce7}\\
 	\Sigma\frac{\diff \theta}{\diff w}&=&\pm\sqrt{W(\theta)},\label{ce8}\\
 	\Sigma\frac{\diff \varphi}{\diff w}&=&-\frac{P_W}{\sin^2\theta}+\frac{a P_R}{\Delta},\label{ce9}\\
 	\Sigma\frac{\diff t}{\diff w}&=&-a P_W + \frac{(r^2+a^2)P_R}{\Delta},\label{ce10}
\end{eqnarray}
where 

\begin{eqnarray}
  R(r)&=&P^2_R-\Delta K,\label{ce11}\\
  W(\theta)&=&K-\left(\frac{P_w}{\sin\theta}\right)^2,\label{ce12}\\
  P_R(r)&=&E(r^2+a^2)-a\Phi,\label{ce13}\\
  P_W(\theta)&=&aE\sin^2\theta - \Phi.\label{ce14}
\end{eqnarray}
It is useful to introduce integral of motion $Q$ by the formula

\begin{equation}
 Q=K-(E- a\Phi)^2.\label{ce15}
\end{equation}
Its relevance comes from the fact that in the case of astrophysically most important motion in the equatorial plane ($\Theta = \pi/2$) there is $Q=0$ for both photons and test particles with non-zero rest energy.

\subsection{Radial and latitudinal motion}
The photon motion (with fixed constants of motion $E$, $\Phi$, $Q$) is allowed in regions where $R(r;E,\Phi,Q)\ge 0$ and $W(\theta; E,\Phi,Q)\ge 0$. The conditions $R(r;E,\Phi,Q)=0$ and $W(\theta;E,\Phi,Q)=0$ determine turning points of the radial and latitudinal motion, respectively, giving boundaries of the region allowed for the motion.
Detailed analysis of the $\theta$-motion can be found in \cite{Bic-Stu:1976:BULAI:,Fel-Cal:1972:NCimB}, while the radial motion was analysed (with restrictions implied by the $\theta$-motion) in \cite{Stu:1981:BULAI:} and \cite{Stu:1981:BULAI:NullGeoKN}. Here we summarize the analysis relevant for the naked singularity spacetimes.
\par

The radial and latitudinal Carter equations read

\begin{eqnarray}
	\Sigma^2\left(\frac{\diff r}{\diff w'} \right)^2 &=& [r^2+a^2-a\lambda]^2-\Delta[\mathcal{L}-2a\lambda + a^2],\label{eq7}\\
	\Sigma^2\left(\frac{\diff \theta}{\diff w'} \right)^2 &=& \mathcal{L} + a^2  \cos^2\theta - \frac{\lambda^2}{\sin^2\theta}\label{eq8}
\end{eqnarray}
where we have introduced impact parameters

\begin{eqnarray}
	\lambda &=& \frac{\Phi}{E},\label{eq9}\\
	\mathcal{L} &=& \frac{L}{E^2} = \frac{Q+\Phi^2}{E^2} = q + \lambda^2,\label{eq10}
\end{eqnarray}
and rescaled the affine parameter by $w^\prime = E w$. We assume $a>0$ and $E>0$. (The special case of photon motion with $E=0$ is treated in \cite{Stu:1981:BULAI:}.)

\subsubsection{Latitudinal motion}
\begin{figure}[ht]
	\begin{center}
	\includegraphics[width=7.0 cm]{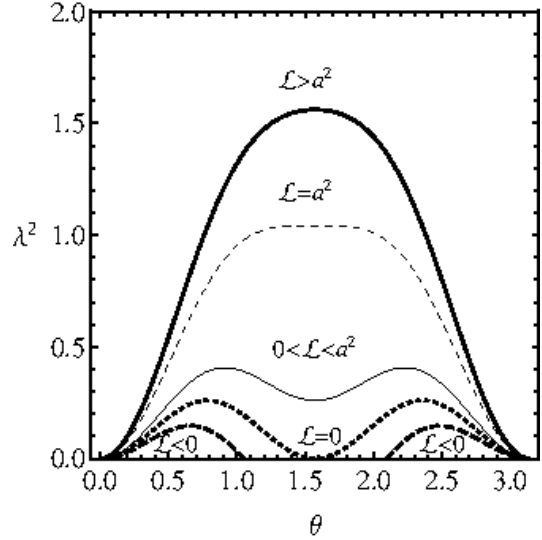}
	\caption{\label{fig1} Latitudinal motion. The curves $\lambda_t^2(\theta;\mathcal{L})$ are given for some typical values of impact parameter $\mathcal{L}$ and rotational parameter $a=1.02$ . The latitudinal motion, allowed along abscissas $\lambda^2_t = const$ below these curves, is oscillatory in general.}
	\end{center}
\end{figure}

The turning points of the latitudinal motion are determined by the condition
\beq
       \lambda^2 = \lambda^2_t \equiv \sin^{2}\theta (\mathcal{L}+a^2\cos^2\theta)
\eeq
The extrema of the function $\lambda^2_t(\theta\mathrm{;} \mathcal{L})$ are determined by
\beq
        \mathcal{L} + a^2 \cos 2\theta = 0.
\eeq
The character of the regions allowed for the latitudinal motion in dependence on the impact parameters $\lambda$ and $\mathcal{L}$ is represented by Fig.\ref{fig1} At the maxima of the function $\lambda^2_t(\theta\mathrm{;}\mathcal{L})$, there is
\beq
         \lambda^2_t = \frac{(\mathcal{L}+a^2)^2}{4a^2}.
\eeq

\begin{figure}
	\begin{center}
	\begin{tabular}{cc}
	\includegraphics[width=6.5cm]{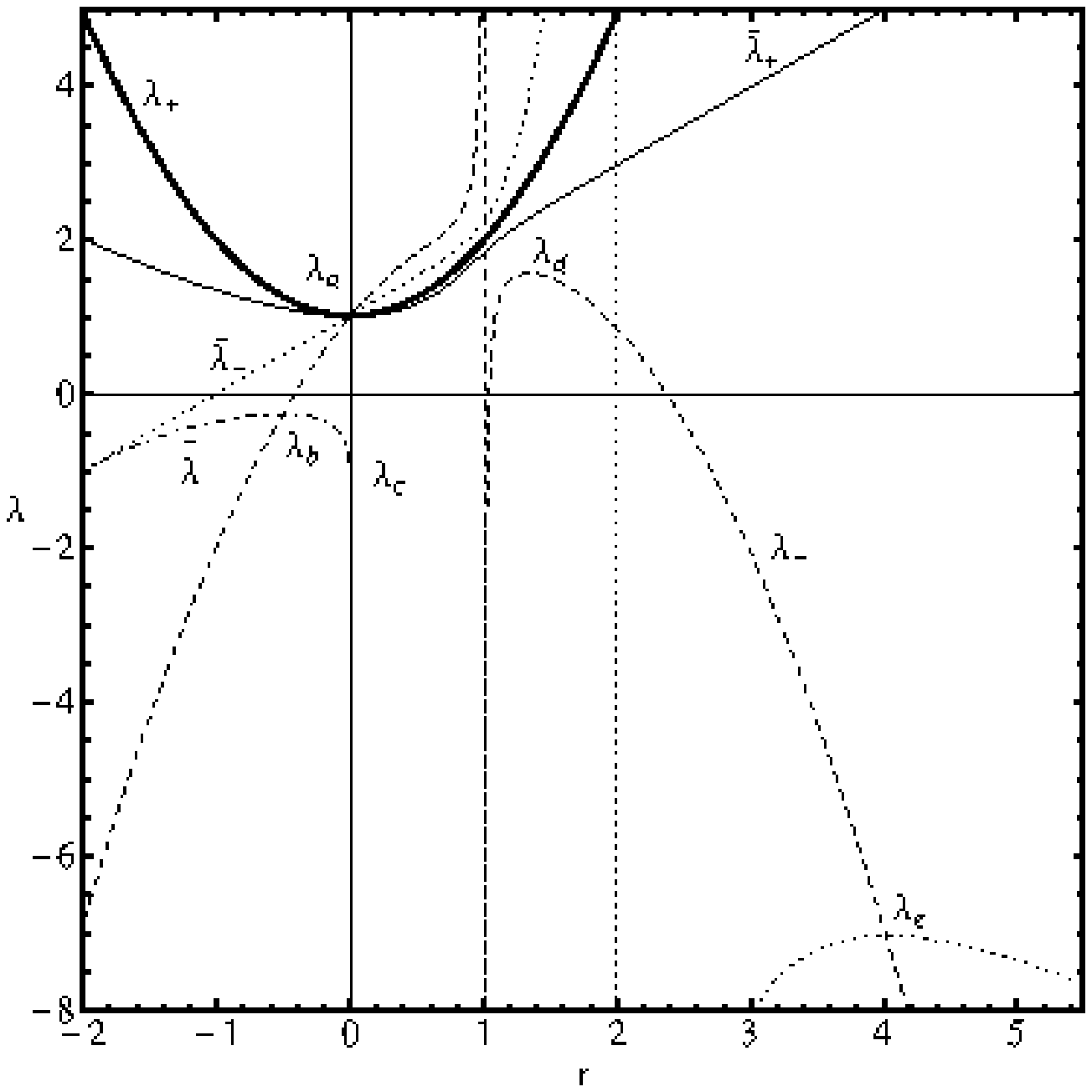}&\includegraphics[width=6.5cm]{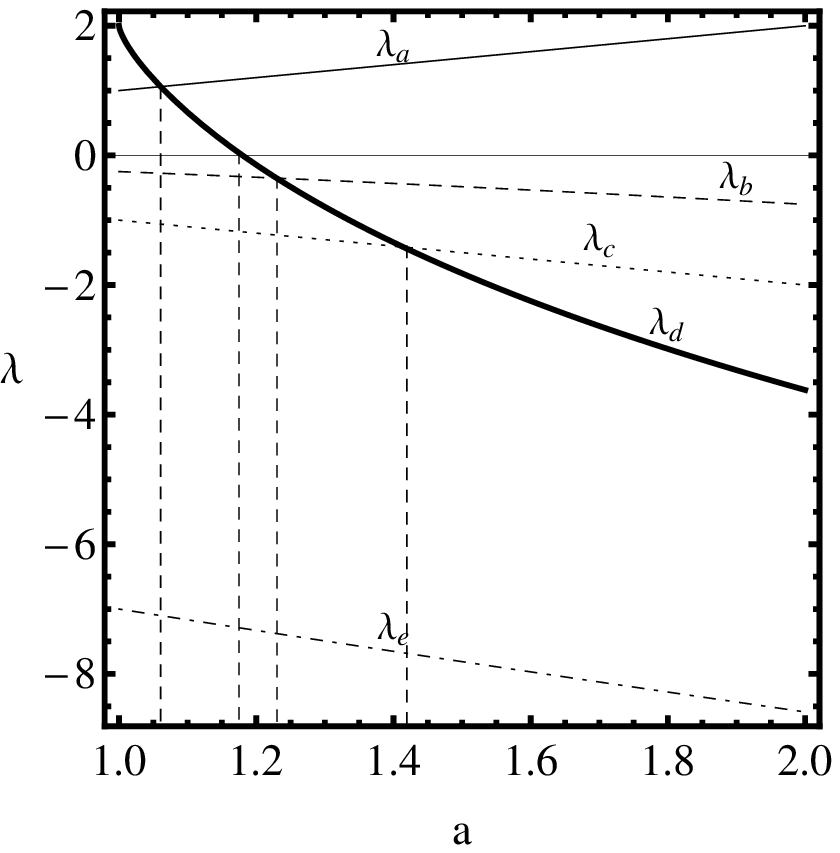}
	\end{tabular}
	\caption{\label{fig2}Left: The functions $\lambda_\pm$, $\tilde\lambda_\pm$, $\bar\lambda$ together with the characteristic  values $\lambda_a$,...,$\lambda_e$ relevant for the classification of the photon motion are plotted for a representative value of rotational parameter $a=1.02$ . Right: The characteristic values of the impact parameter$\lambda$ as functions of rotation parameter $a$.}
	\end{center}
\end{figure}

\begin{figure}[ht]
	\begin{center}
	\begin{tabular}{ccc}
		\subfloat[][]{\includegraphics[width=5.0cm]{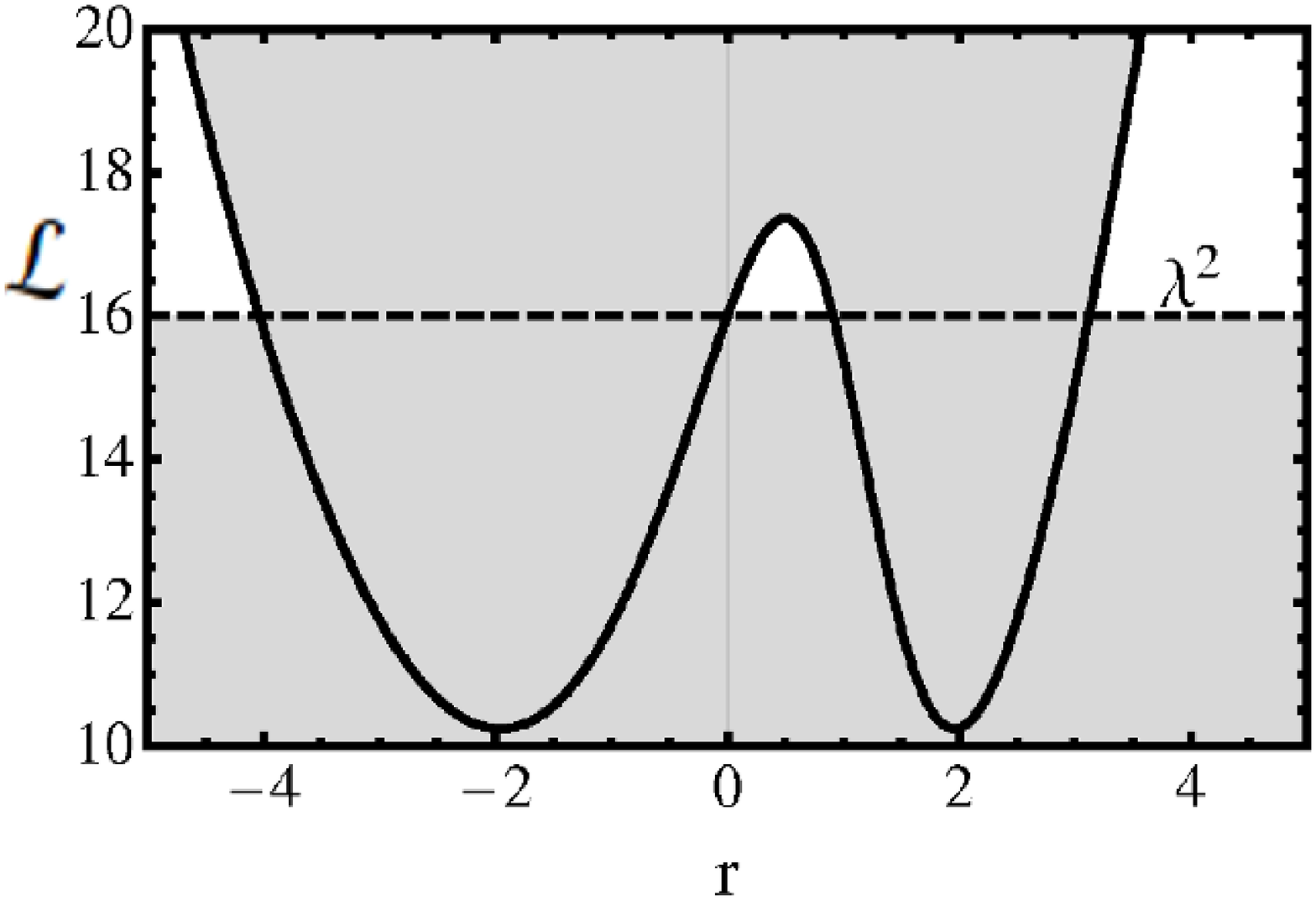}}&\subfloat[][]{\includegraphics[width=5.0cm]{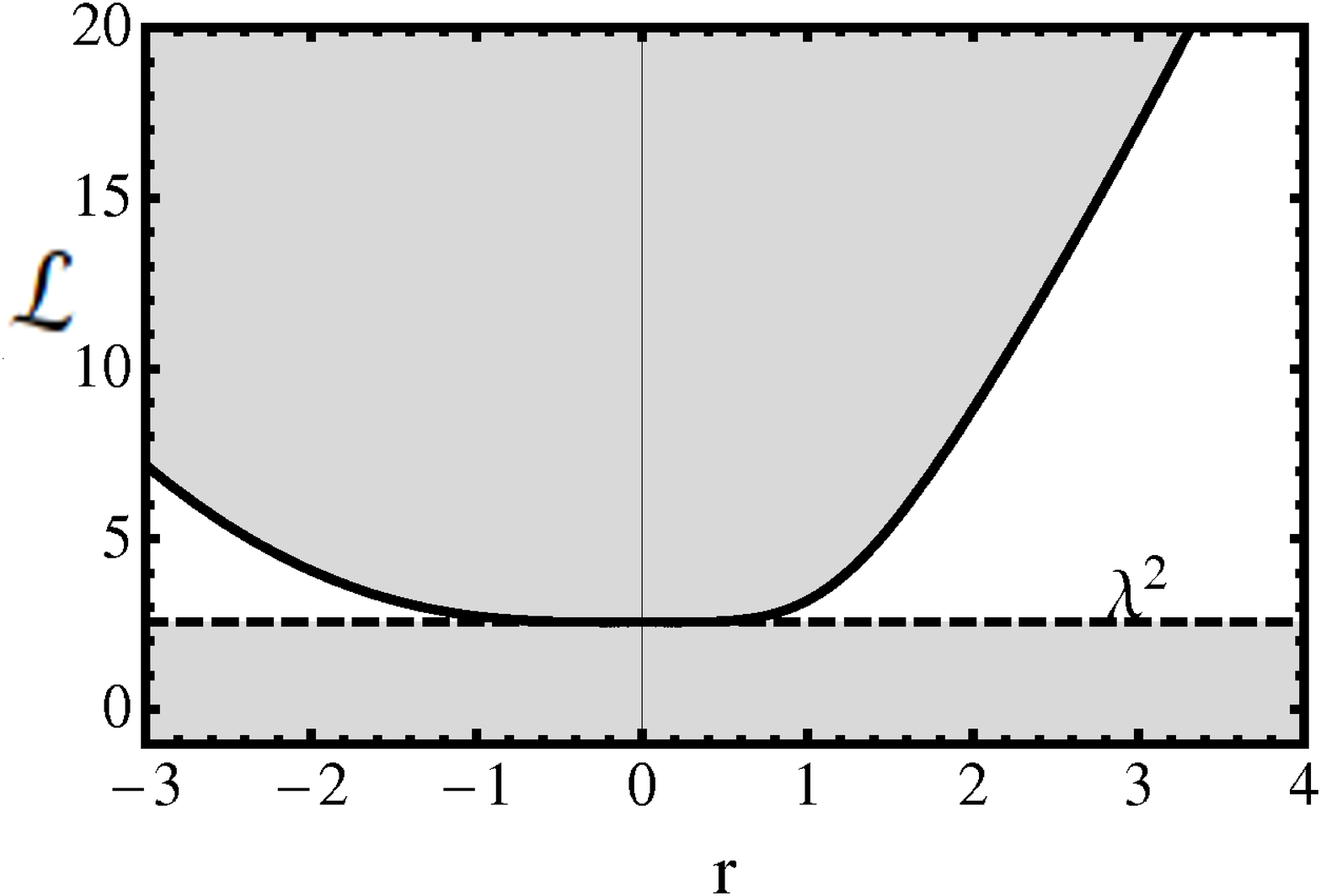}}&\subfloat[][]{\includegraphics[width=5.0cm]{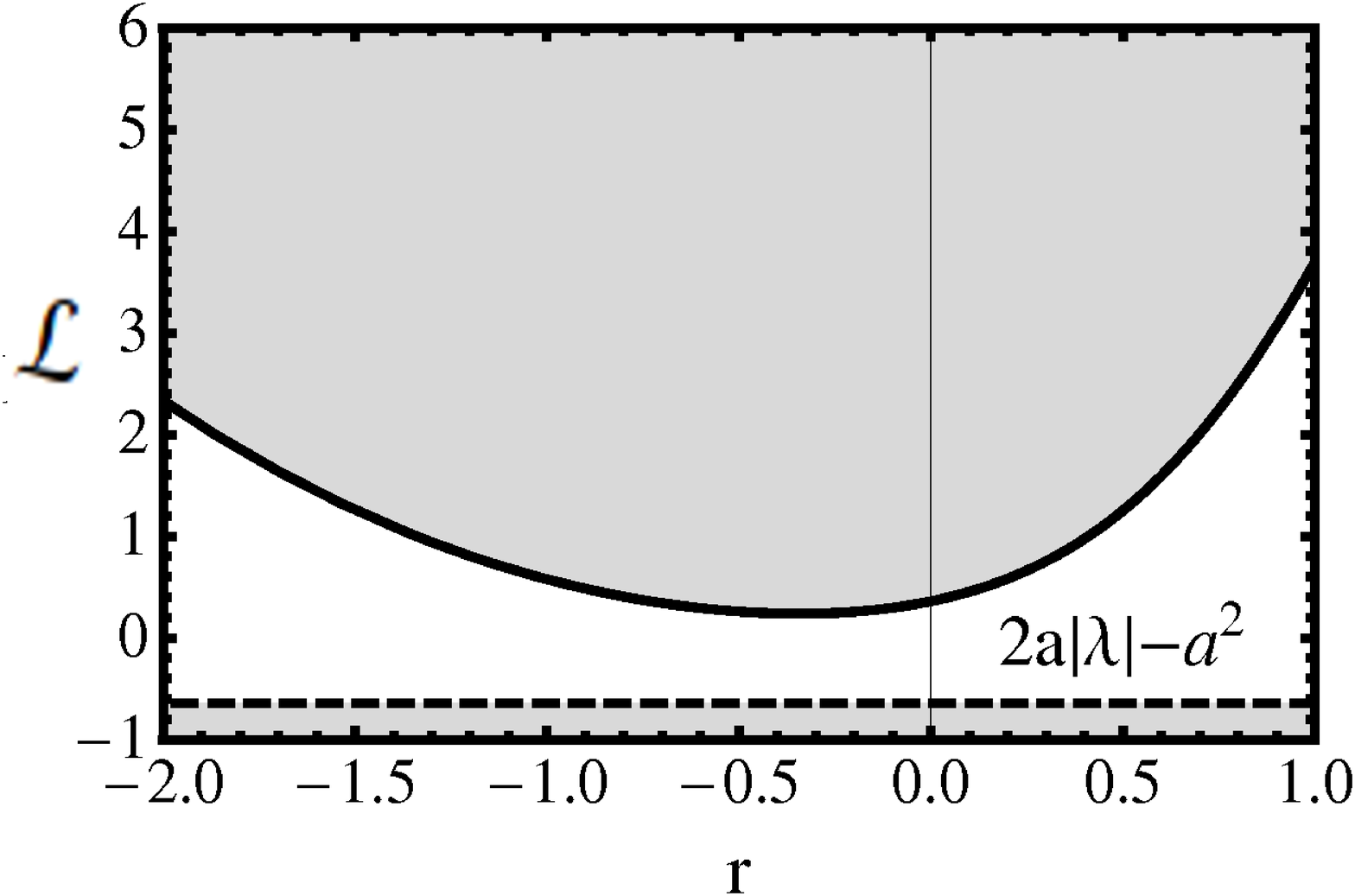}}\\
		\subfloat[][]{\includegraphics[width=5.0cm]{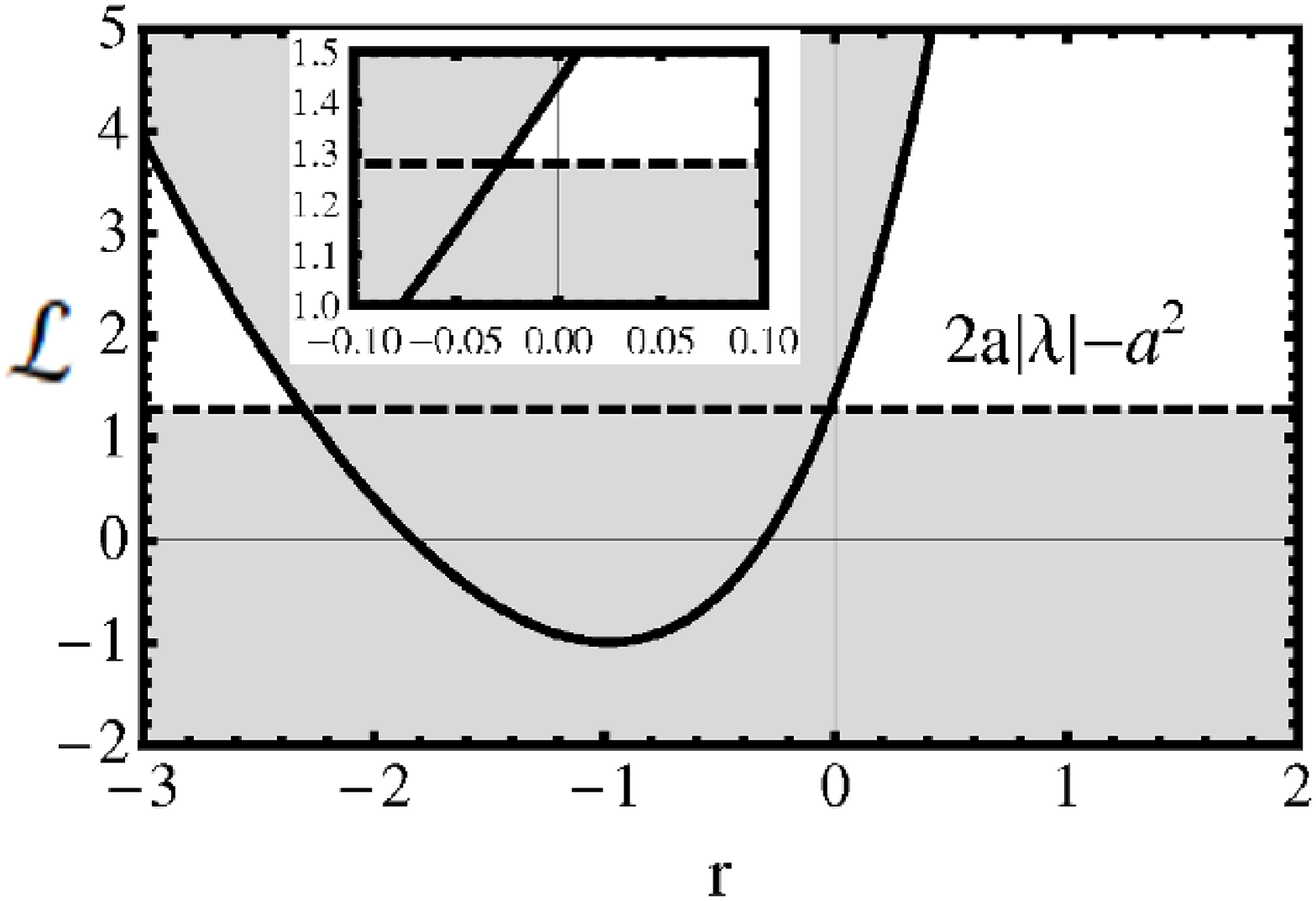}}&\subfloat[][]{\includegraphics[width=5.0cm]{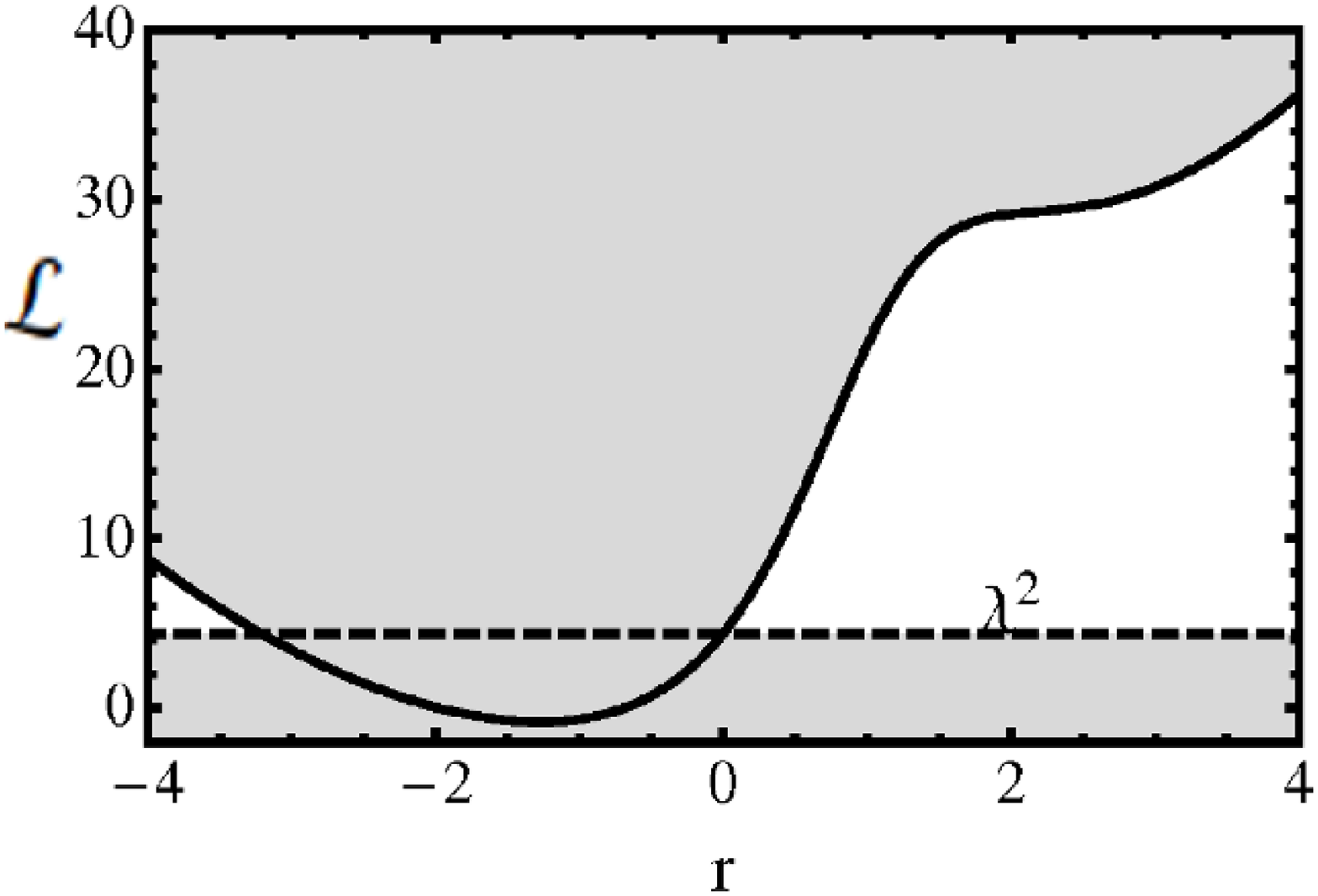}}&\subfloat[][]{\includegraphics[width=5.0cm]{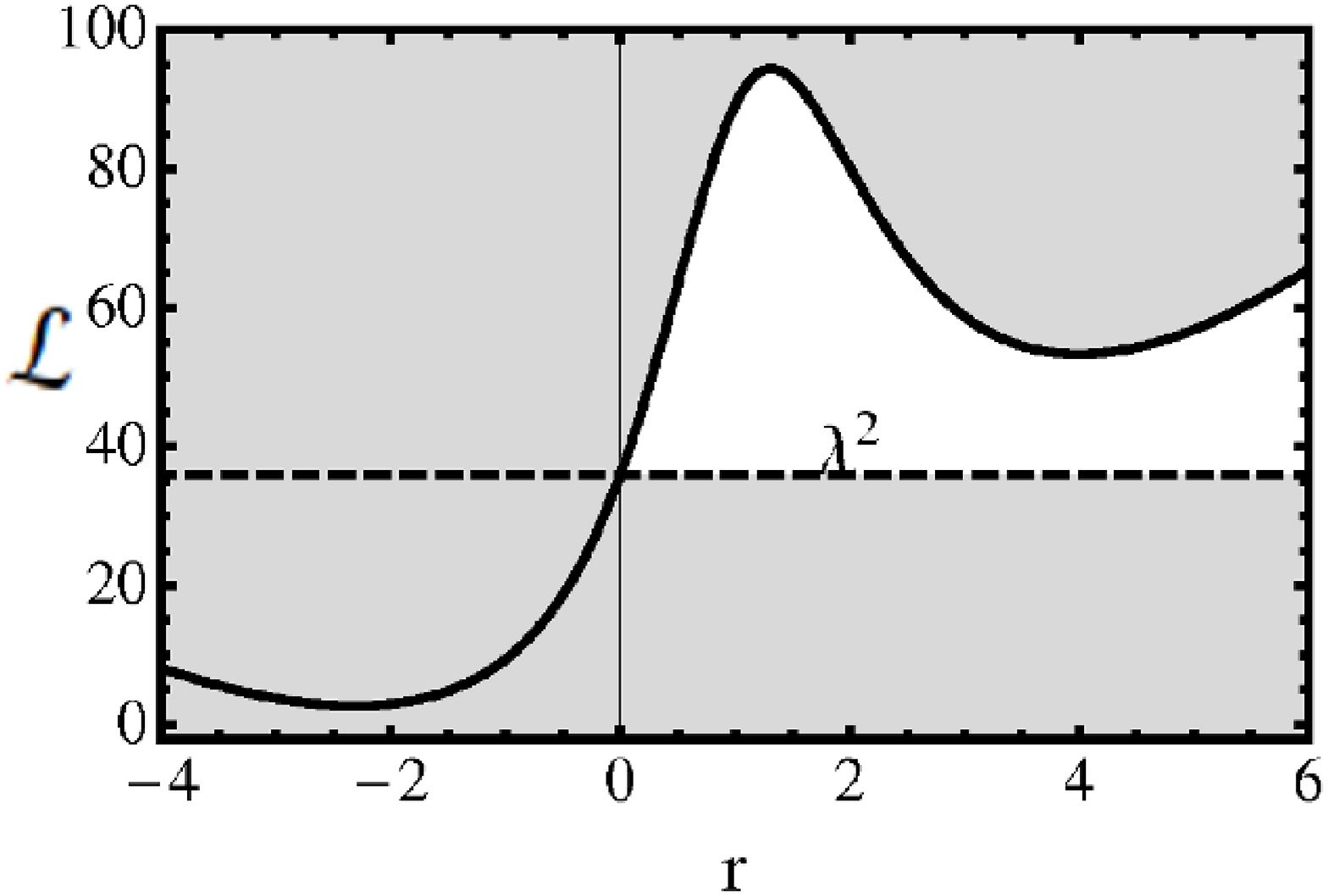}}
	\end{tabular}
	\caption{\label{fig3}The sections of constant $\lambda$-values of the surface $\mathcal{L}=\mathcal{L}(r;\lambda,a)$. The sequence is representative for Kerr naked singularities with $a>a_3=\sqrt{2}$. 
	The spin parameter is $a=1.6$ and the representative impact parameter $\lambda$ values are the following (from top left to right bottom): $\lambda=4$, $1.6$, $0.6$, $-1.2$, $-2.1$ and $-6$.
	The motion is allowed along horizontal lines, $\mathcal{L}=\mathrm{const}$, in the unshaded region.}
	\end{center}
\end{figure}

\begin{figure}[ht]
	\begin{center}
	\begin{tabular}{cc}
		\subfloat[][]{\includegraphics[width=5cm]{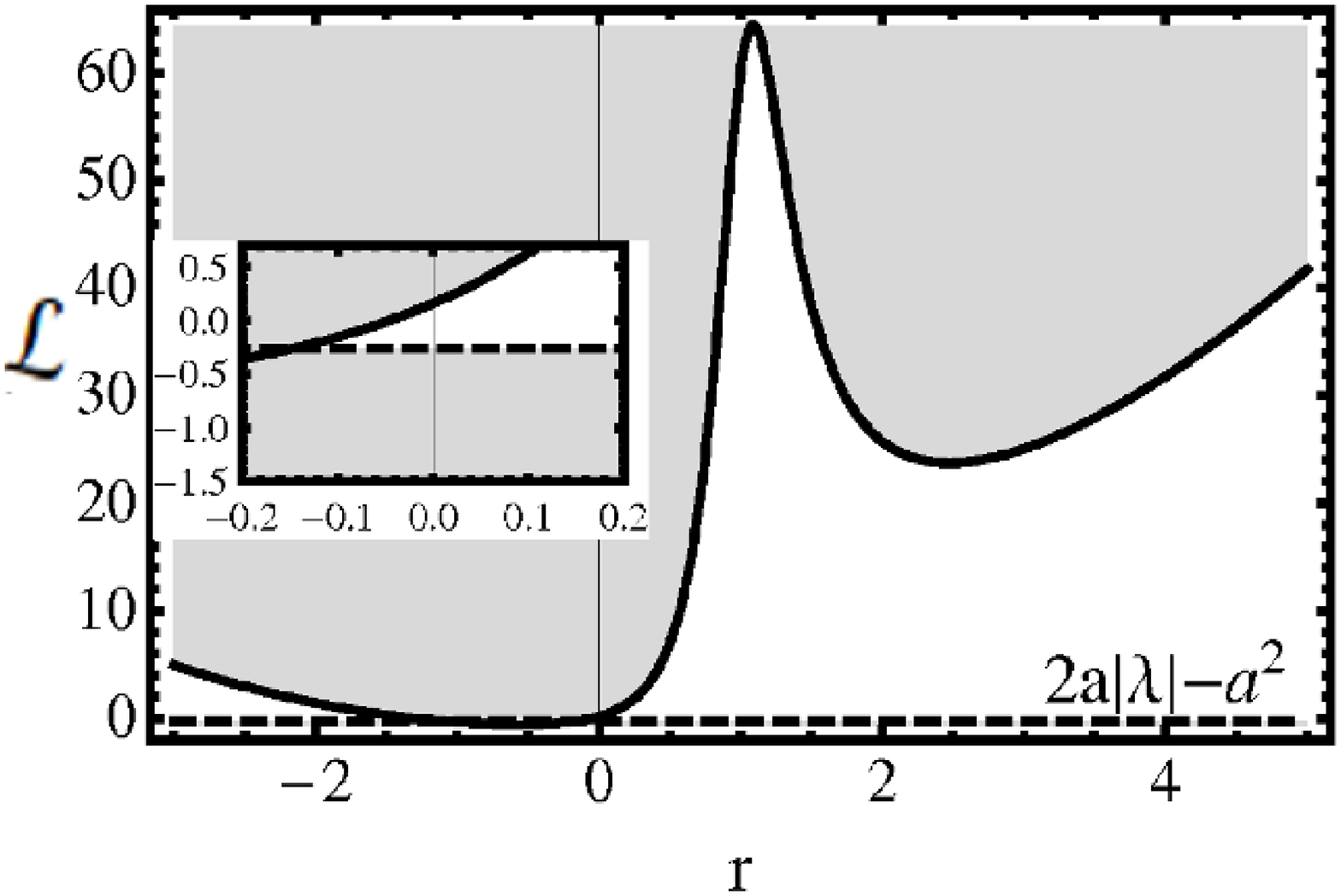}}&\subfloat[][]{\includegraphics[width=5cm]{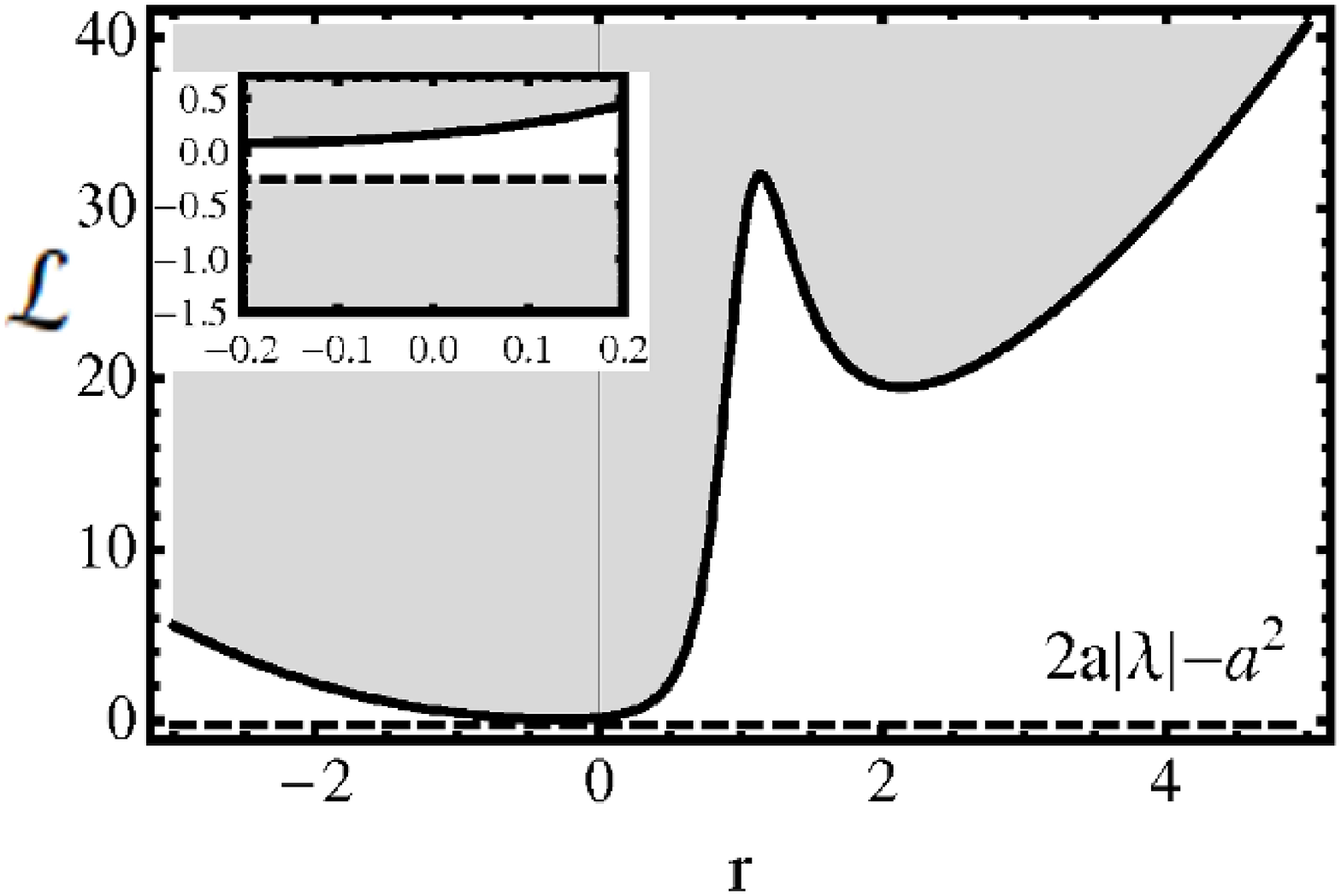}}\\
		\subfloat[][]{\includegraphics[width=5cm]{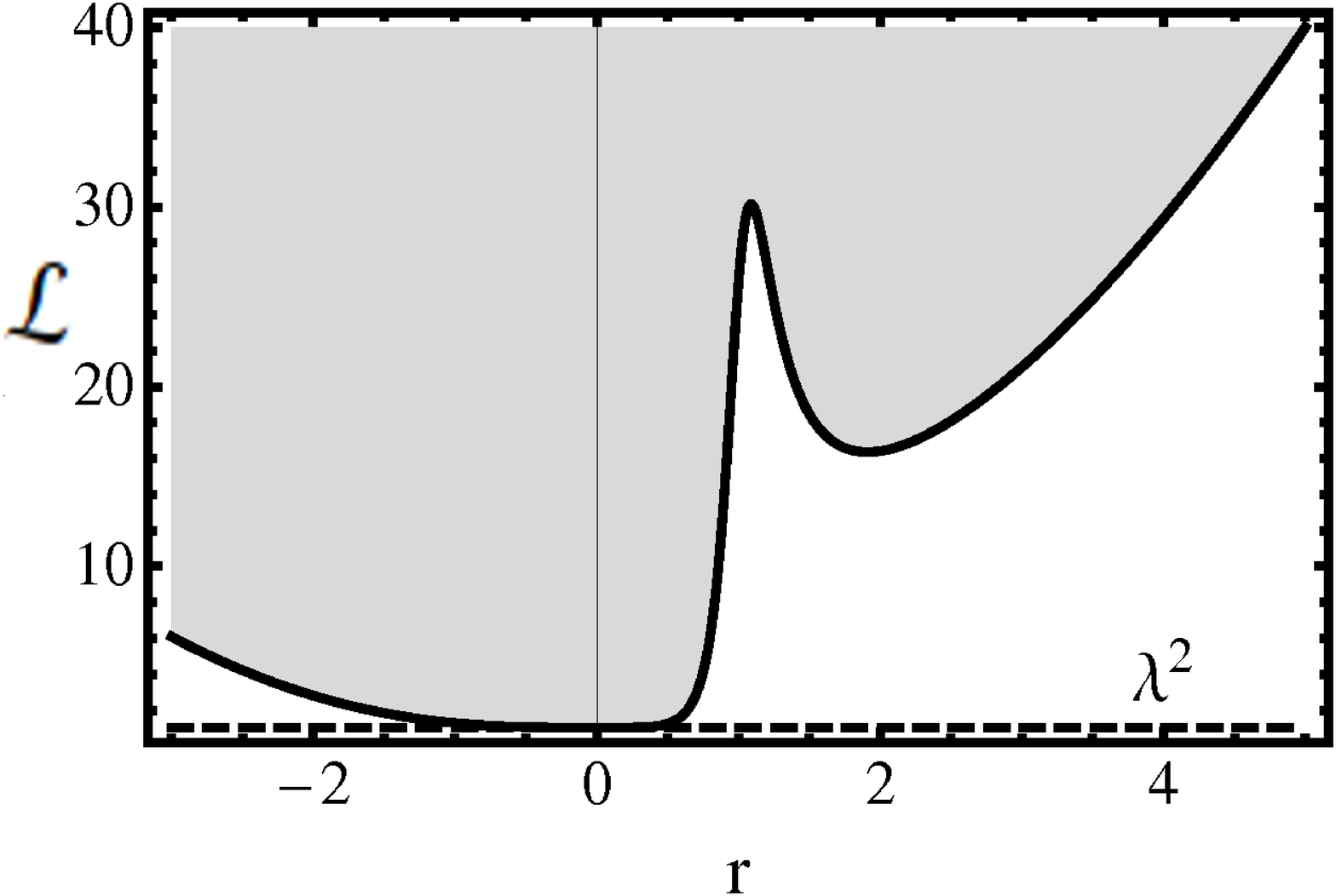}}&\subfloat[][]{\includegraphics[width=5cm]{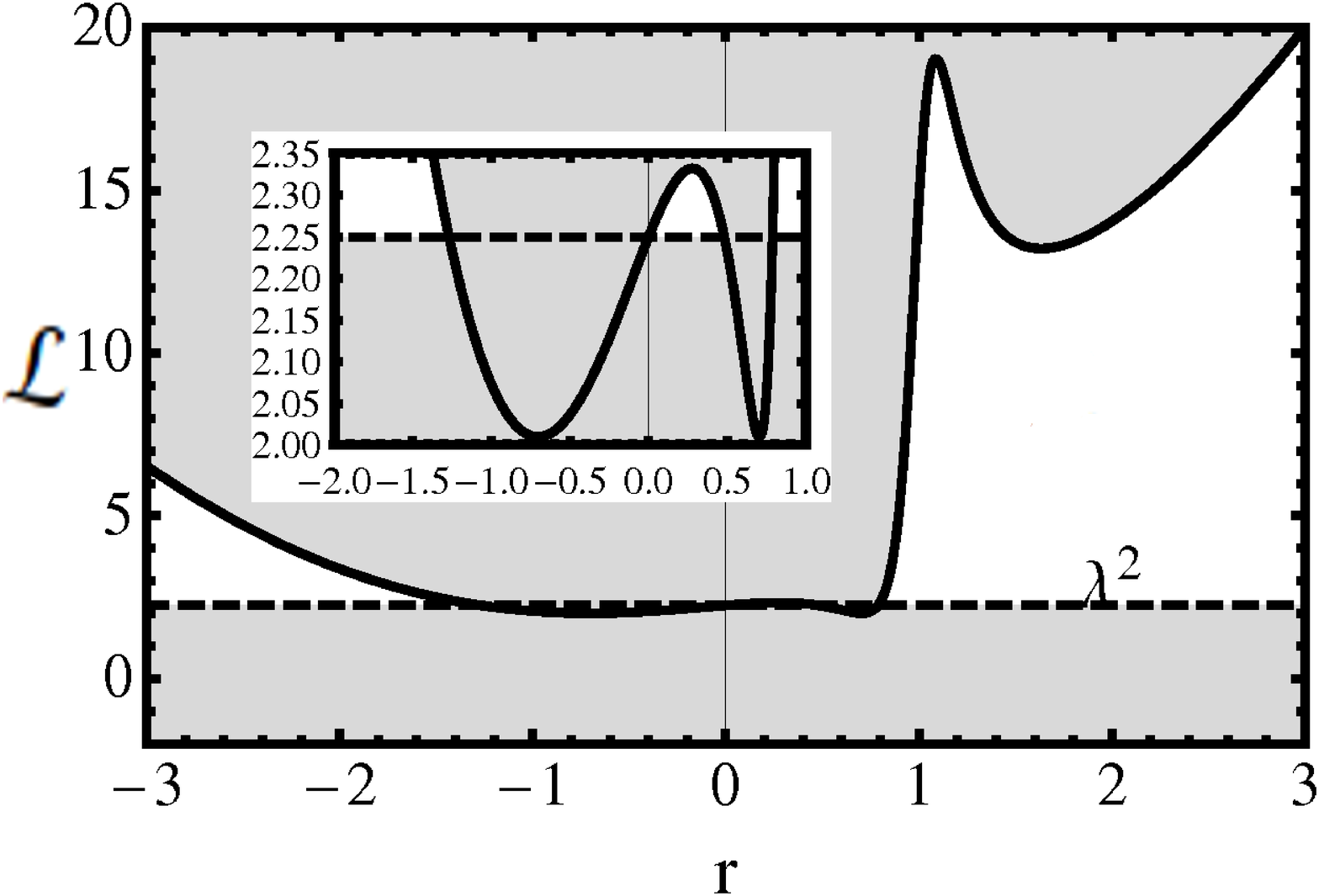}}
	\end{tabular}
	\caption{\label{fig4}The sections of constant $\lambda$-values of the surface $\mathcal{L}=\mathcal{L}(r;\lambda,a)$. The sequence for Kerr naked singularities with $a<a_3=\sqrt{2}$. 
	The doubles $(a,\lambda)$ are the following (from top left to right bottom): $(1.05, -0.4)$, $(1.05, 0.4)$, $(1.02, 1.02)$ and $(1.01, 1.5)$.
	The motion is allowed along horizontal lines, $\mathcal{L}=\mathrm{const}$, in the unshaded region.}
	\end{center}
\end{figure}

The impact parameter $\mathcal{L}$ can be negative when $q<0$ and $\lambda_t^2$ is small enough. The latitudinal motion can be divided into two subclasses, namely of the orbital motion (with $q \geq 0$) which reaches the equatorial plane and the vortical motion (with $q<0$) when the motion is limited to the region above or under the equatorial plane. Clearly, only photons of the orbital type could be radiated from the equatorial Keplerian accretion discs considered in our study.

\subsubsection{Radial motion}
The reality conditions $(\diff r/\diff w')^2 \ge 0$ and $(\diff\theta/\diff w')^2 \ge 0$ lead to the restrictions on the impact parameter $\mathcal{L}$

\begin{equation}
	\mathcal{L}_{min} \leq \mathcal{L} \leq \mathcal{L}_{max},\label{eq12}
\end{equation}
where

\begin{equation}
	\mathcal{L}_{max} \equiv \frac{(a\lambda -2r)^2}{\Delta}+ r^2+2r,\label{eq13}
\end{equation}
and
\begin{equation}
	\mathcal{L}_{min}\equiv\left\{ \begin{array}{lcr} 
				\lambda^2 & \textrm{for} & |\lambda|\geq a,\\
				2a|\lambda|-a^2 & \textrm{for} & |\lambda|\leq a. 
			   \end{array}\right.\label{eq14}
\end{equation}
The upper(lower) constraint, $\mathcal{L}_{max}$($\mathcal{L}_{min}$), comes from the radial-motion (latitudinal-motion) reality condition. The properties of the photon motion are determined by the behaviour of the surface $\mathcal{L}_{max}(r;\lambda,a,b)$, as given by (\ref{eq13}). The extrema of the surface $\mathcal{L}_{max}$ (giving spherical photon orbits) are determined by

\begin{eqnarray}
	\lambda=\lambda_+ &\equiv& \frac{r^2+a^2}{a},\label{eq15}\\
	\lambda=\lambda_- &\equiv& \frac{r^2 - a^2 - r\Delta}{a(r-1)}.\label{eq16}
\end{eqnarray}
The values of $\mathcal{L}_{max}$ at these extreme points are given by

\begin{eqnarray}
	\mathcal{L}_{max}(\lambda_{+})\equiv\mathcal{L}_+ &=& 2r^2+a^2,\label{eq17}\\
	\mathcal{L}_{max}(\lambda_{-})\equiv\mathcal{L}_- &=&\frac{2r(r^3-3r)+a^2(r+1)^2}{(r-1)^2}\label{eq18}.
\end{eqnarray}
The character of the extrema follows from the sign of $\partial^2\mathcal{L}_{max}/\partial r^2$. One finds that

\begin{eqnarray}
\frac{\partial^2 \mathcal{L}_{max}}{\partial r^2} &=& \frac{8r^2}{\Delta},\quad\textrm{for}\quad \lambda = \lambda_+,\label{eq19}\\
\frac{\partial^2 \mathcal{L}_{max}}{\partial r^2} &=&\frac{8r^2}{\Delta} - \frac{8r}{(r-1)^2},\quad\textrm{for}\quad \lambda=\lambda_-.\label{eq20}
\end{eqnarray}
Clearly, there are only minima of $\mathcal{L}_{max}$ along for $\lambda=\lambda_{+}$, corresponding to unstable
spherical orbits.

Further, we have to determine where the restrictions given by the latitudinal motion $\mathcal{L}_{min}$ meet the restrictions on the radial motion $\mathcal{L}_{max}$. We find that $\mathcal{L}_{max}=\lambda^2$ (for $|\lambda|\ge a$) is fullfilled where

\begin{equation}
 \lambda=\tilde\lambda_\pm\equiv\frac{a(-2r\pm r^2\sqrt{\Delta})}{r^2-2r},\label{eq23}
\end{equation}
while $ \quad\mathcal{L}_{max}= 2a|\lambda| - a^2$ (for $|\lambda|<a$) is fullfilled where

\begin{equation}
 \lambda=\bar\lambda \equiv \frac{1}{\Delta}[4r-r^2-a^2+2\sqrt{\Delta(-2r)}].\label{eq24}
\end{equation}
Clearly, the last function is defined only in the region of negative values of $r$ and is thus irrelevant in the case of the Kerr superspinars with surface located at $R>0$.

The extreme points of curves $\tilde\lambda_\pm$, which are also the intersection points of these curves  with $\lambda_-$, are determined by the equation

\begin{equation}
	f(r;a)\equiv r^3-6r^2+9r-4a^2=0.\label{eq25}
\end{equation}
The equation $f(r;a)=0$ determines loci of the photon equatorial circular orbits; in an implicit form the radii are given by the condition

\begin{equation}
 a^2=a^2_{ph}(r)=\frac{r(r-3)^2}{4}.
\end{equation}
In the field of Kerr superspinars with $a>1$, there exists only one (counterrotating) equatorial photon orbit at radius given by
\beq
          r_{ph} = 2 + [a + \sqrt{a^2-1}]^{2/3} + [a + \sqrt{a^2-1}]^{-2/3} .\label{rph}
\eeq
The maximum of the function $\lambda_{-}(r;a)$ is located at
\beq
          r_{m} = 1 + (a^2 -1)^{1/3}
\eeq
and the value of this maximum is 
\beq
          \lambda_{m} = a^{-1}[3 - a^2 - 3(a^2 -1)^{2/3}].
\eeq

The maxima of the curve $\bar\lambda$ (and common points with the curve $\lambda_{-}$) are located at $r$ satisfying the equation
\beq
          2r^3 - 3r^2 + a^2 = 0.
\eeq
These extremal points are irrelevant for our discussion being located at negative values of the radial coordinate, while we restrict our study to the region of $r>R=0.1$.

The Kerr superspinar spacetimes can be classified due to the properties of the photon motion as determined by the behaviour of the functions $\lambda_\pm$, $\tilde\lambda_\pm$, $\bar\lambda$. The behaviour of these functions is represented in Fig.\ref{fig2}(left) where five characteristic values of the impact parameter $\lambda$, crucial for their classification and dependent on the spin parameter $a$ of the Kerr superspinars, are introduced.  Dependence of characteristic values of $\lambda$ on the spin parameter of the spacetime is given in Fig.\ref{fig2}(right).

In the case of Kerr naked singularity spacetimes, we can distinguish four qualitatively different classes of the photon motion in dependence on the values of the spin parameter. These are given by the relative position of the characteristic values of the impact parameter $\lambda$.

\begin{itemize}
\item	{$1 < a < a_1 = 1.062$. 

The impact parameters satisfy the relation
\beq
        \lambda_{d} > \lambda_{a} > \lambda_{b} > \lambda_{c} > \lambda_{e}.
\eeq
}
\item{ $a_1 < a < a_2 = 1.23$. 

The impact parameters satisfy the relation
\beq
        \lambda_{a} > \lambda_{d} > \lambda_{b} > \lambda_{c} > \lambda_{e}.
\eeq
}

\item{ $a_2 < a < a_3 = \sqrt{2} = 1.414$. 

The impact parameters satisfy the relation
\beq
        \lambda_{a} > \lambda_{b} > \lambda_{d} > \lambda_{c} > \lambda_{e}.
\eeq
}

\item{ $a_3 < a$ 

The impact parameters satisfy the relation
\beq
        \lambda_{a} > \lambda_{b} > \lambda_{c} > \lambda_{d} > \lambda_{e}.
\eeq
}
\end{itemize}
In the case of Kerr superspinars, when we consider the photon motion in the region of $r>R=0.1$, it is enough to distinguish only two classes of the spacetimes, namely those with $1 < a < a_1$ and $a > a_1$. 

For each interval of $\lambda$ as determined by the sequence of $\lambda_a$ - $\lambda_e$ introduced in Fig.\ref{fig3}, there exists a characteristic type of behaviour of the restricting "radial" function $\mathcal{L}_{max}$ and its relation to the "latitudinal" restricting function $\mathcal{L}_{min}$, see \cite{Stu:1981:BULAI:NullGeoKN} for details. Here we give the sequence of the "effective" potential sections in Fig.\ref{fig3} for the Kerr naked singularity spacetimes with $a>a_3$. For the naked singularity spacetimes with $a<a_3$, some characteristic sections are given in Fig.\ref{fig4} - notice the case of 
$a=1.01$ and $\lambda=1.5$ when two stable spherical photon orbits exist. Further, it should be stressed the for the Kerr superspinars the effective potential sections of $\lambda = const$ are relevant only in the region of $r>R(=0.1M)$.
 
The allowed values of the impact parameter $\mathcal{L}$ lie between the limiting functions  $\mathcal{L}_{min}$ and $\mathcal{L}_{max}$. If the minimum $\mathcal{L}_{max}^{min}\equiv\mathcal{L}_{max}(r_{min},\lambda_0)$ of the limiting function $\mathcal{L}_{max}$ is less than the value of the limiting function $\mathcal{L}_{min}$, an incoming photon ($k^r < 0$) travelling from infinity will return back for all values of $\mathcal{L}_0\in[\mathcal{L}_{min};\mathcal{L}_{max}]$. If $\mathcal{L}_{max}^{min}>\mathcal{L}_{min}$, 
the incoming photon ($k^r < 0$) travelling from infinity returns back if its impact parameter $\mathcal{L}_0$  
satisfies the condition  $\mathcal{L}_{0}\ge\mathcal{L}_{max}^{min}$ and is captured by the Kerr superspinar 
if $\mathcal{L}_0<\mathcal{L}^{min}_{max}$.  
The minimum $\mathcal{L}_{max}^{min}$ determines (with the particular value of $\lambda$) an unstable photon spherical orbit, 
i.e., a sphere where photons move with $r=const$ but with varying latitude $\theta$ (and, of course, varying $\varphi$). 
When the condition  $\mathcal{L}_0 = \mathcal{L}_{min}$ is satisfied simultaneously, the spherical photon orbit is transformed to an (unstable) equatorial photon circular orbit. Photons coming from distant regions or regions close to the superpsinar surface will wind up around the photon sphere when $\mathcal{L}_0=\mathcal{L}_{max}^{min}$ for a given value of the impact parameter $\lambda$. The maxima $\mathcal{L}_{max}^{max}$ of the limiting function $\mathcal{L}_{max}$ correspond to the stable photon spherical orbits that are central to the region of photons trapped by the strong gravitational field of Kerr superspinars. 

\subsection{Keplerian discs}
We summarize shortly properties of the equatorial circular geodesic motion in the field of Kerr superspinars that are relevant for Keplerian, thin accretion discs, and in principle are important even for structure of thick or slim accretion discs. The Carter equations imply the specific energy and specific angular momentum of the circular geodesics to be given by the relations \cite{Bar-Pre-Teu:1972:ASTRJ2:,Stu:1980:BULAI:}
\beq
            \frac{E_K}{m} = \frac{r^{3/2}- 2r^{1/2} \pm a }{r^{3/4}\sqrt{r^{3/2}- 3r^{1/2} \pm 2a}}
\eeq
\beq
            \frac{\Phi_K}{m} = \pm \frac{r^2 + a^2  \mp 2a r^{1/2}}{r^{3/4}\sqrt{r^{3/2}- 3r^{1/2} \pm 2a }}.
\eeq
The angular velocity with respect to static observers at infinity $\Omega = \diff\phi/\diff t$ is given by the relation
\beq
            \Omega_K = \pm \frac{1}{r^{3/2} \pm a}
\eeq

\begin{figure}[ht]
	\begin{center}
		\begin{tabular}{cc}
			\includegraphics[width=6.5cm]{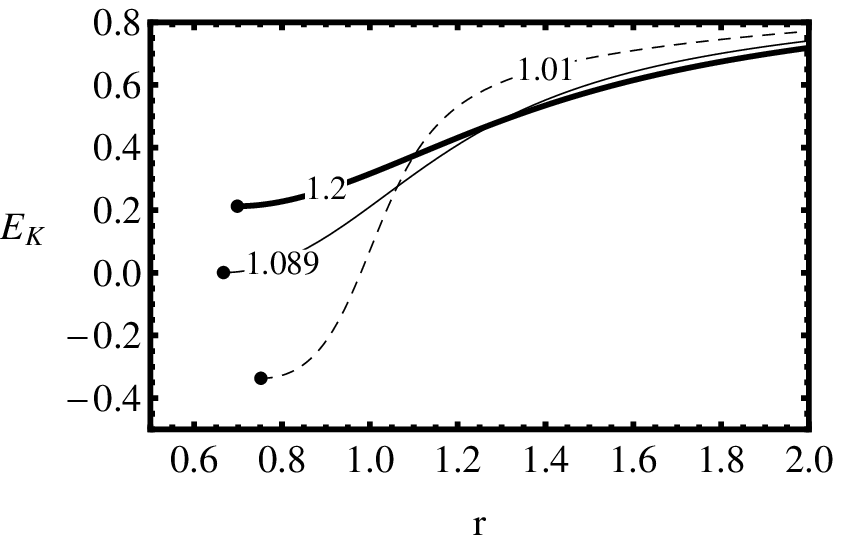}&\includegraphics[width=6.5cm]{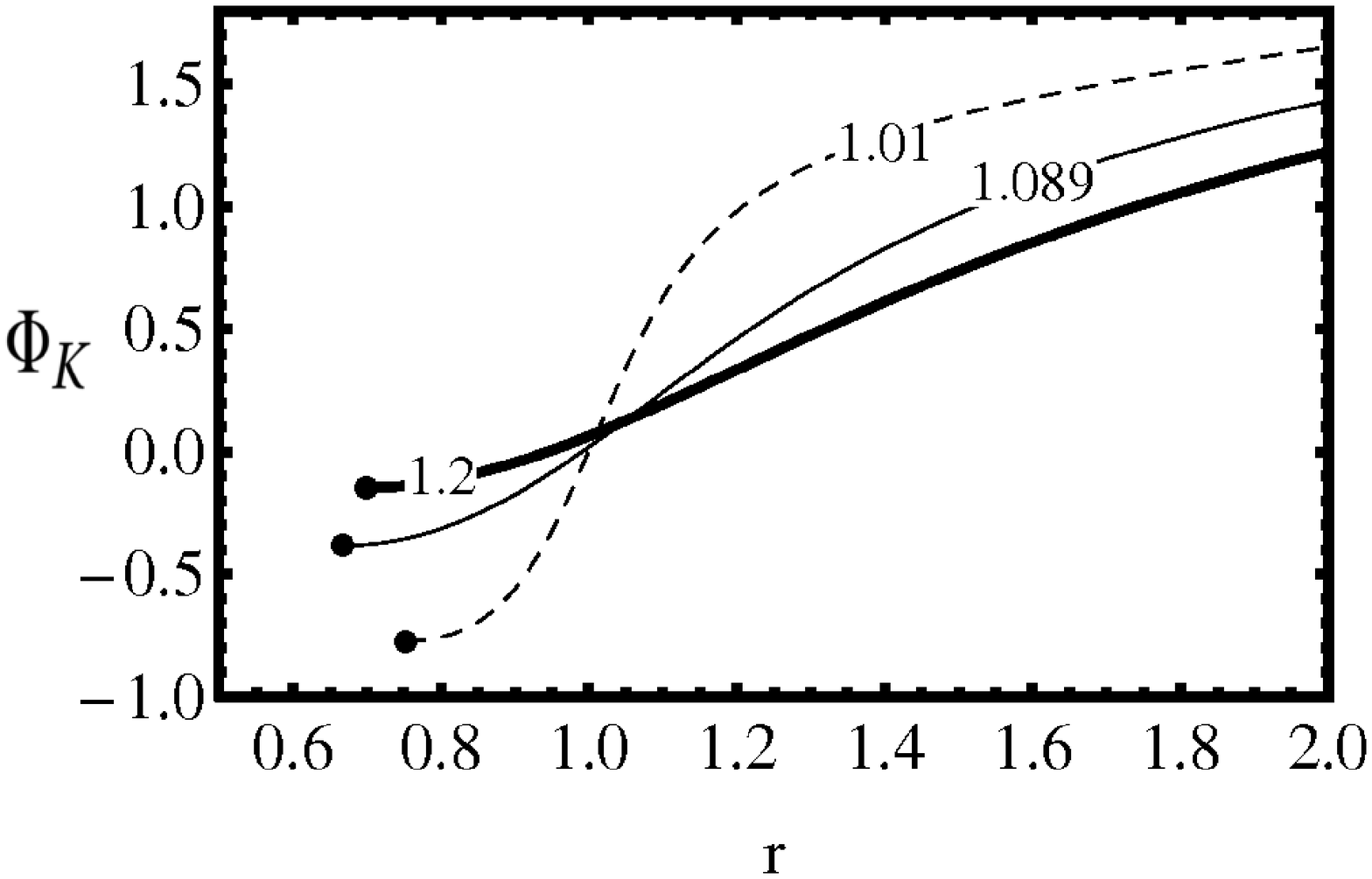}
		\end{tabular}
		\caption{\label{fig5}\emph{Left}: plots of covariant energy $E_K$ as a function of the orbit radial coordinate $r$. \emph{Right}: plots of the azimuthal angular momentum $\Phi_K$ of test particle on Keplerian orbits as a function of the orbit radial coordinate $r$. All plots are generated for three representative values of superspinar rotational parameter $a=1.01$, $1.089$ and $1.2$.}
	\end{center}
\end{figure}

\begin{figure}[ht]
	\begin{center}
		\begin{tabular}{cc}
			\includegraphics[width=6.5cm]{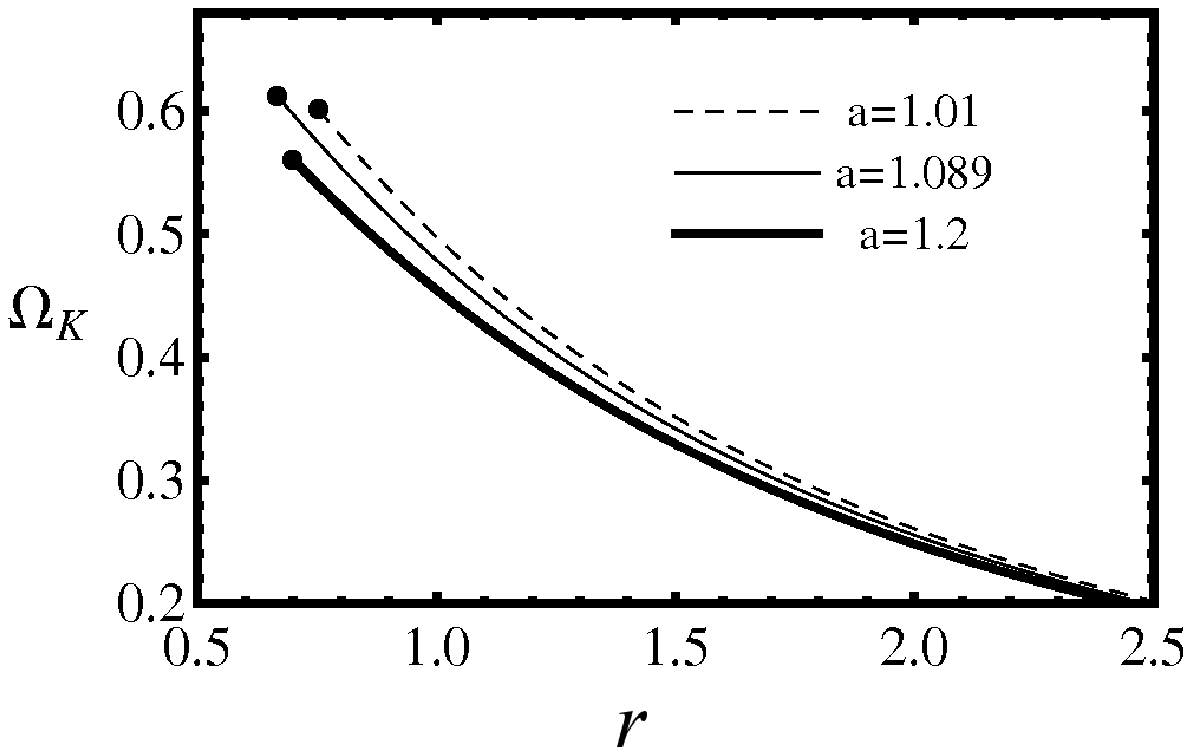}&\includegraphics[width=6.5cm]{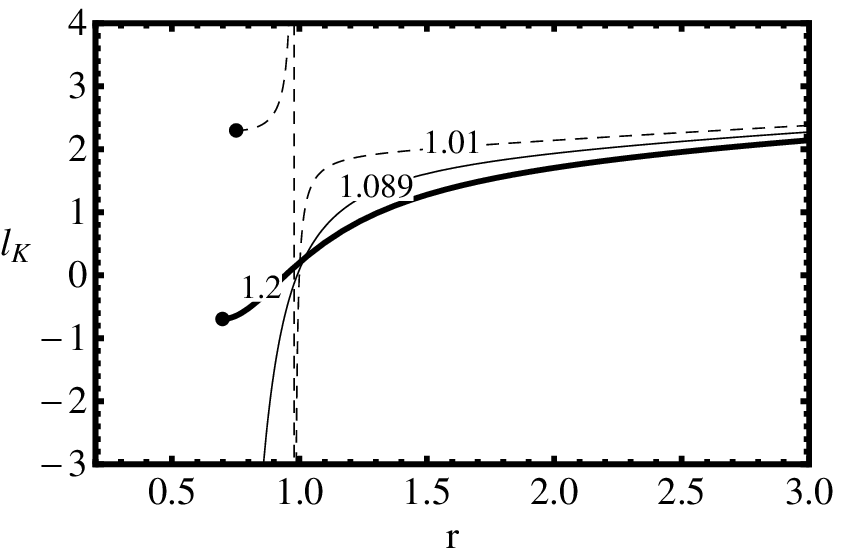}
		\end{tabular}
		\caption{\label{fig6}\emph{Left}: plots of keplerian orbital frequency $\Omega_K$ as a function of the orbit radial coordinate $r$. \emph{Right}: plots of the impact parameter $l_K$ of test particle on keplerian orbit as a function of the orbit radial coordinate $r$. All plots are generated for three representative values of superspinar rotational parameter $a=1.01$, $1.089$ and $1.2$.}
	\end{center}
\end{figure}
while the specific angular momentum related to the conservative energy parameter is given by the relation
\beq
            l_K = \frac{\Phi_K}{E_K} = \pm \frac{r^2+a^2  \mp 2 a r^{1/2}}{r^{3/2}-2r^{1/2}\pm a }.
\eeq
In theory of Keplerian discs a crucial role is devoted to the innermost stable circular orbit (ISCO) in the field of a given spacetime that is usually considered to correspond to the inner edge of Keplerian discs \cite{Nov-Tho:1973:BlaHol:}. In Kerr spacetimes the ISCO is determined by the relation \cite{Bar-Pre-Teu:1972:ASTRJ2:}
\beq
         r_{ms} = 3 + Z_2 - \sqrt{(3-Z_2)(3+Z_1+2Z_2)},\label{isco}
\eeq
where
\beq
         Z_1 = 1 + (1-a^2)^{1/3}[(1+a)^{1/3}(1-a)^{1/3}],
\eeq
\beq
         Z_2 = \sqrt{3a^2 + Z_{1}^{2}}.
\eeq
The marginally bound orbits with $E/m=1$ have radii given by
\beq
         r_{mb} = 2 + a \mp 2(1+a)^{1/2}.
\eeq

In all presented relations the upper sign corresponds to the so called 1-st family orbits that are corotating relative to distant observers ($\Omega_K > 0$) - such orbits are locally corotating ($L>0$) in regions distant from superspinars, but could be locally counterrotating ($\Phi_K<0$) in vicinity of superspinars with the spin parameter $a < a_0 = \frac{3}{4}3^{1/2} \sim 1.3$. For superspinars with spin $a < a_{cr} = \frac{3}{4} \left(\frac{3}{2}\right)^{1/2} \sim 1.089$ the 1-st family orbits with $\Phi_K<0$ could have negative energy ($E<0$), while located close enough to the superspinar boundary \cite{Stu:1980:BULAI:} see Fig.\ref{fig5}. The lower sign in all presented relations corresponds to the 2-nd family orbits that are both counterrotating relative to distant observers ($\Omega_K < 0$) and locally counterrotating ($\Phi_K<0$) everywhere for all superspinars. The Keplerian energy $E_K$ and angular momentum $\Phi_K$ radial profiles are illustrated in Fig.\ref{fig5}. for typical values of the superspinar spin. The Keplerian angular velocity $\Omega_K$ and specific angular momentum $l_K$ profiles are illustrated in Fig.\ref{fig6} for the same representative values of the superspinar spin. Notice the discontinuity of the specific angular momentum profile occuring for superspinars with spin $a < a_{cr} = \frac{3}{4} \left(\frac{3}{2}\right)^{1/2} \sim 1.089$ due to decline of energy $E_K$ to negative values in such spacetimes. Its physical implications for toroidal discs are discussed in \cite{Sla-Stu:2005:CLAQG:,Stu:2005:MODPLA:}.
 

\section{\label{sec:LEC}Light escape cones}
The optical phenomena related to accretion processes in  the field of Kerr superspinars can be efficiently studied by using the notion of light escape cones of local observers (sources) that determine which portion of radiation emitted by a source could escape to infinity and, complementary, which portion is trapped by the superspinar \cite{Sch-Stu:2009:GReGr, Sch-Stu:2009:IJMPD,Tak-Tak:2010:arXiv:,Sch-Stu-Jur:2005:RAGtime6and7:CrossRef}. Here we focus our attention to the family of observers (sources) that are of direct physical relevance to Keplerian discs, namely the circular geodetical observers. For comparison, we construct the escape cones in the locally non-rotating frames. Later we focus our attention to the corotating frames $GF_{+}$.

\subsection{Local frames of stationary observers}
We consider two families of stationary frames, namely  $LNRF$ (Locally Nonrotatig Frames) and $GF_\pm$(Circular Geodesic Frames). The $LNRF$ are of fundamental physical importance since the physical phenomena take the simplest form when expressed in such frames, because the rotational spacetime effects are maximally suppressed there \cite{Bar:1973:BlaHol:,Mis-Tho-Whe:1973:Gra:}. The $GF_\pm$ are directly related to Keplerian accretion discs in the equatorial plane of the spacetime, both corotating and counterrotating. The $GF_\pm$ are, of course, geodetical frames, while $LNRF$ are generally accelerated frames.

The radial and latitudinal 1-forms of the stationary frame tetrads are common for both families of frames and read

\begin{eqnarray}
	\omega^{(r)}&=&\left\{0,\sqrt{\Sigma/\Delta},0,0 \right\},\label{LC9}\\
	\omega^{(\theta)}&=&\left\{0,0,\sqrt{\Sigma},0 \right\}.\label{LC10}
\end{eqnarray}
$LNRF$ correspond to observers with $\Phi=0$ (zero angular momentum observers). Their time and azimuthal 1-forms read

\begin{eqnarray}
	\omega^{(t)}&=&\left\{\sqrt{\frac{\Delta\Sigma}{A}},0,0,0 \right\},\label{LC11}\\
	\omega^{(\varphi)}&=&\left\{-\Omega_{LNRF}\sqrt{\frac{A}{\Sigma}}\sin\theta,0,0,\sqrt{\frac{A}{\Sigma}}\sin\theta\right\}.\label{LC12}
\end{eqnarray}
where 

\begin{equation}
	\Omega_{LNRF}=\frac{2aMr}{A}\label{LC13}
\end{equation}
is the angular velocity of $LNRF$ as seen by observers at infinity. 
\par
The $GF_\pm$ observers move along $\varphi$-direction in the equatorial plane with velocity $V_{GF\pm}$(+...corotating, -...counterrotating) relative to the $LNRF$ and with angular velocity $\Omega$ relative to the static observers at infinity given by \cite{Bar-Pre-Teu:1972:ASTRJ2:}
\begin{equation}
\Omega_\pm=\pm\frac{1}{r^{3/2} \pm a}. \label{ang_vel_gf}
\end{equation}

The velocity $V_{GF\pm}$ is given by

\begin{equation}
	V_{GF\pm}=\pm\frac{(r^2+a^2) \mp 2ar^{1/2}}{\sqrt{\Delta}(r^{3/2} \pm a)}.\label{VGF}
\end{equation}
The standard Lorentz transformation of the $LNRF$ tetrad gives the tetrad of $GF_\pm$ in the form
\begin{eqnarray}
	\omega^{(t)}_\pm&=&\left\{ \frac{r^2-2r\pm a r^{1/2}}{Z_\pm},0,0,\mp\frac{(r^2+a^2)r^{1/2}\mp 2ar}{Z_\pm} \right\},\\
\omega^{(\varphi)}_\pm&=&\left\{\mp \frac{\sqrt{\Delta}r^{1/2}}{Z_\pm},0,0,\frac{\sqrt{\Delta(r^2\pm a r^{1/2})}}{Z_\pm}, \right\}
\end{eqnarray}
where 
\begin{equation}
	Z_\pm = r\sqrt{r^2-3r\pm2ar}.
\end{equation}
Note that the $GF_\pm$ family is restricted to the equatorial plane, while $LNRF$ are defined at any $\theta$.

\subsection{Construction of escape cones}

The analysis of the turning points of the radial motion of photons is crucial in determining the local escape cones as the boundary of the escape cones is given by directional angles related to unstable spherical photon orbits. For each direction of emission in the local frame of a source, there is a corresponding pair of values of the impact parameters $\lambda$ and $\mathcal{L}$ which can be related to the directional cosines of the photon trajectory in the local frame at the position of the source and we have to find those corresponding to the unstable spherical photon orbits. Notice that in the case of Kerr black holes the inversion of the local escape cone about the symmetry axis of the spacetime represents silhouette of the black hole as observed from the corresponding local frame \cite{Sch-Stu:2009:IJMPD,Tak-Tak:2010:CQG}, but it is not the case of the Kerr superspinar spacetimes, as we have to select photons captured by the superspinars and those trapped in their gravitational field without being captured by the superspinar surface.

Projection of a photon 4-momentum $\vec{k}$ onto the local tetrad of an observer is given by the formulae

\begin{eqnarray}
k^{(t)}&=&-k_{(t)}=1,\label{LC1}\\
k^{(r)}&=&k_{(r)}=\cos\alpha_0,\label{LC2}\\
k^{(\theta)}&=&k_{(\theta)}=\sin\alpha_0\cos\beta_0,\label{LC3}\\
k^{(\varphi)}&=&k_{(\varphi)}=\sin\alpha_0\sin\beta_0,\label{LC4}
\end{eqnarray} 
where $\alpha_0$, $\beta_0$ are directional angles of the photon in the local
frame (see \cite{Sch-Stu:2009:IJMPD}) and 
\beq
         \cos\gamma_0=\sin\alpha_0\sin\beta_0. 
\eeq
In terms of the local tetrad components of the photon 4-momentum and the related directional angles, the conserved quantities, namely, the azimutal momentum $\Phi$, energy $E$ and $K$ read

\begin{eqnarray}
	\Phi&=&k_\varphi=-\omega^{(t)}_{\phantom{(t)}\varphi}k^{(t)} + \omega^{(r)}_{\phantom{(r)}\varphi}k^{(r)}+\omega^{(\theta)}_{\phantom{(\theta)}\varphi}k^{(\theta)}+\omega^{(\varphi)}_{\phantom{(\varphi)}\varphi}k^{(\varphi)},\label{LC6}\\
	E&=&-k_t=\omega^{(t)}_{\phantom{(t)}t}k^{(t)} - \omega^{(r)}_{\phantom{(r)}t}k^{(r)}-\omega^{(\theta)}_{\phantom{(\theta)}t}k^{(\theta)}-\omega^{(t)}_{\phantom{(\varphi)}\varphi}k^{(\varphi)},\label{LC7}\\
   K&=&\frac{1}{\Delta}\left\{ [E(r^2+a^2)-a\Phi]^2-(\Sigma k^r)^2\right\}.\label{LC8}
\end{eqnarray}
The impact parameters $\lambda$ and $\mathcal{L}$ defined by relations (\ref{eq9}) and (\ref{eq10}) are thus fully determined by any double, $D$, of angles from the set $M=[\alpha_0,\beta_0,\gamma_0]$.

 In a given source frame, with fixed coordinates $r_0$, $\theta_0$, we can construct light escape cones using the following procedure:

\begin{itemize}
\item for given $D$, say $D=[\alpha_0,\beta_0]$, we calculate $\lambda=\lambda(\alpha_0,\beta_0)$,
\item $\lambda$ determines the behaviour of $\mathcal{L}_{max}=\mathcal{L}_{max}(r;\lambda)$,
\item from the analysis presented in the previous section we calculate minimum of $\mathcal{L}_{max}$, which reads $\mathcal{L}_{max}^{min}=\mathcal{L}_{max}(r_{min};\lambda)$,
\item we search for such a double $D$ which satisfies equation $\mathcal{L}_0(\alpha_0,\beta_0)=\mathcal{L}_{max}(r_{min};\lambda)$.
\end{itemize}
Here, we present in detail the construction of light escape cones in particular case of the $LNRF$. The procedure is analogous for the $GF_{+}$. Notice that for large distances from the superspinar the LNRF are almost identical to the frames of static observers. 
	\begin{figure}[h!]
		\begin{tabular}{ccc}
			\includegraphics[width=4cm]{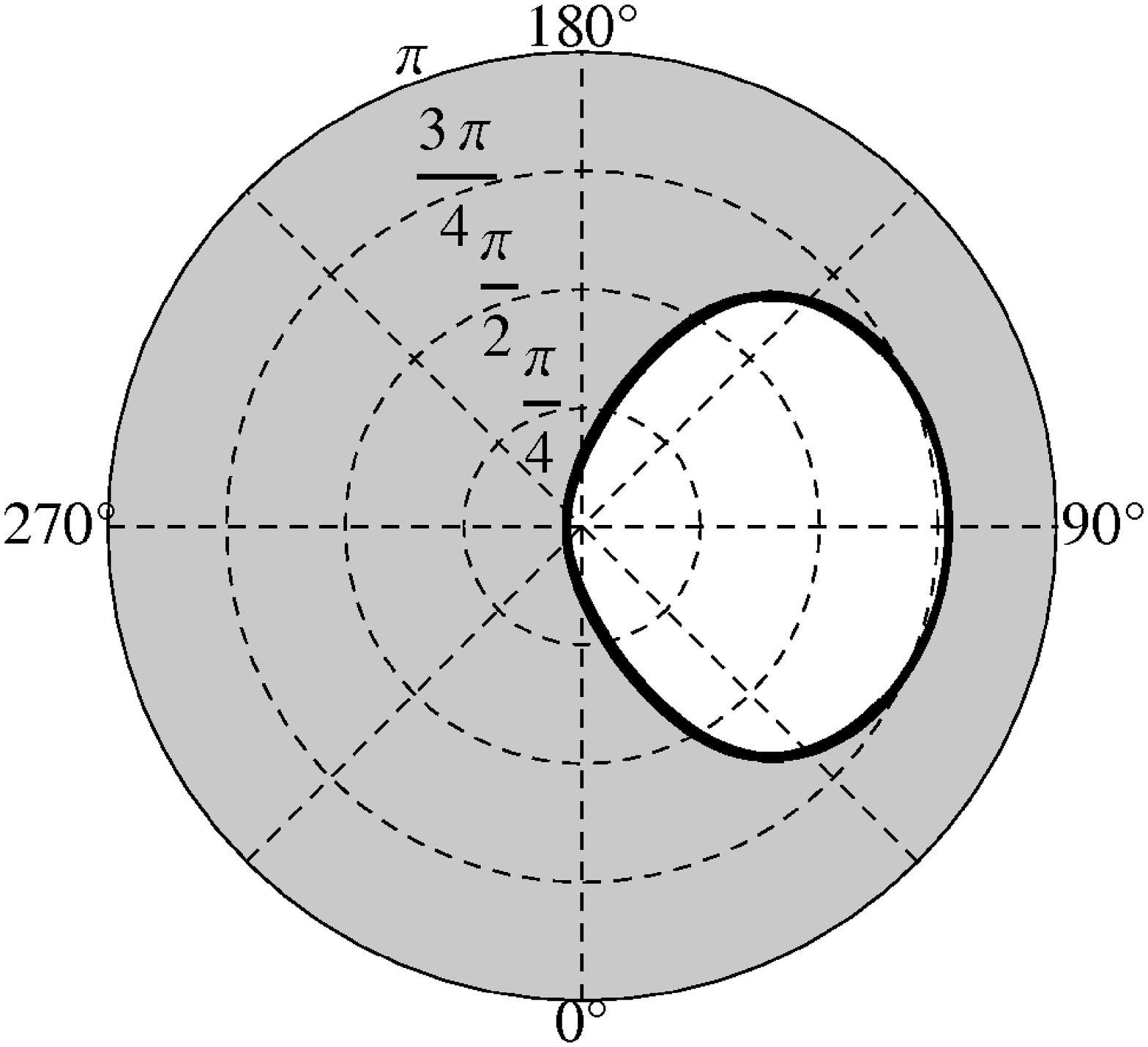}&
			\includegraphics[width=4cm]{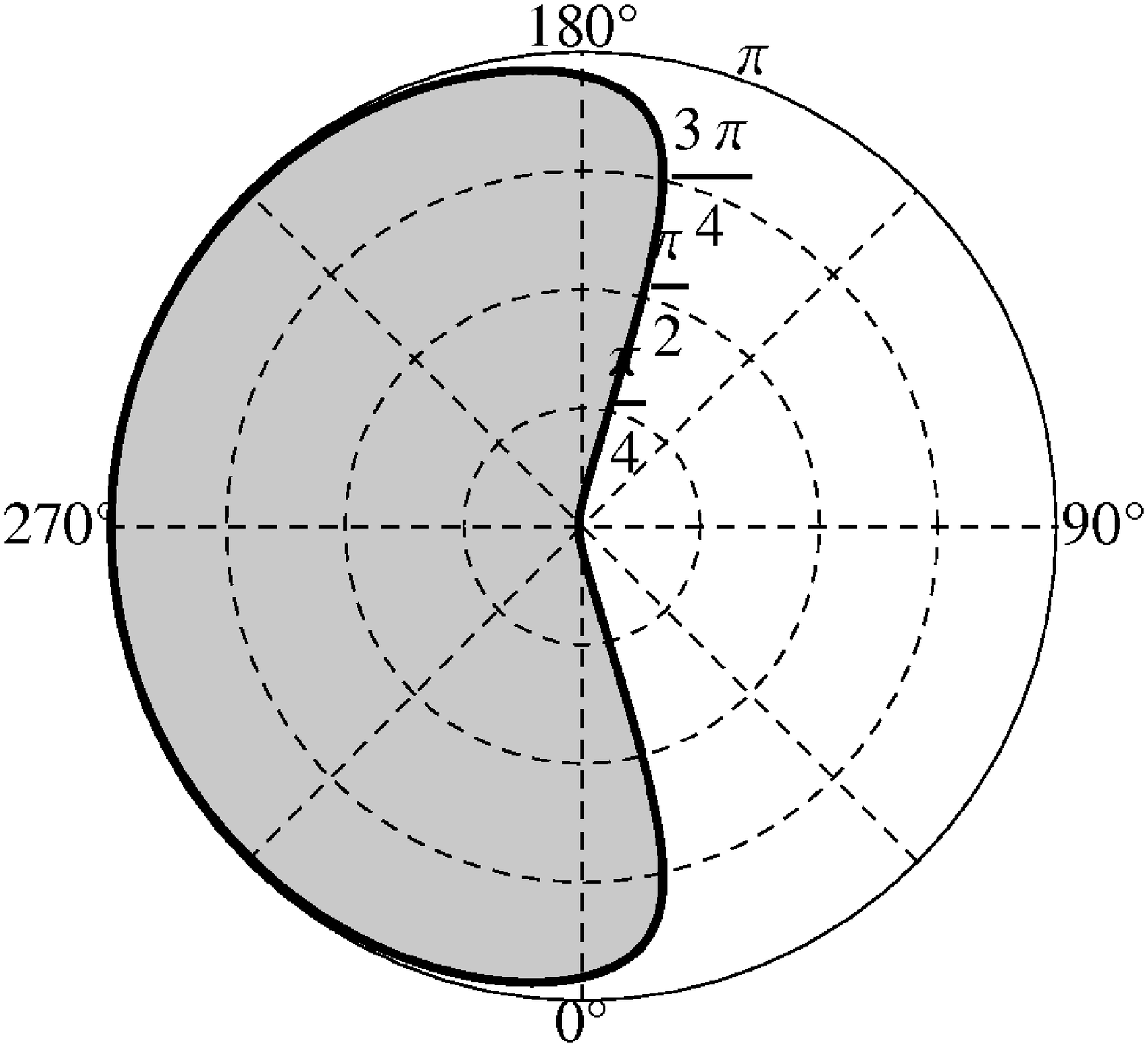}&
			\includegraphics[width=4cm]{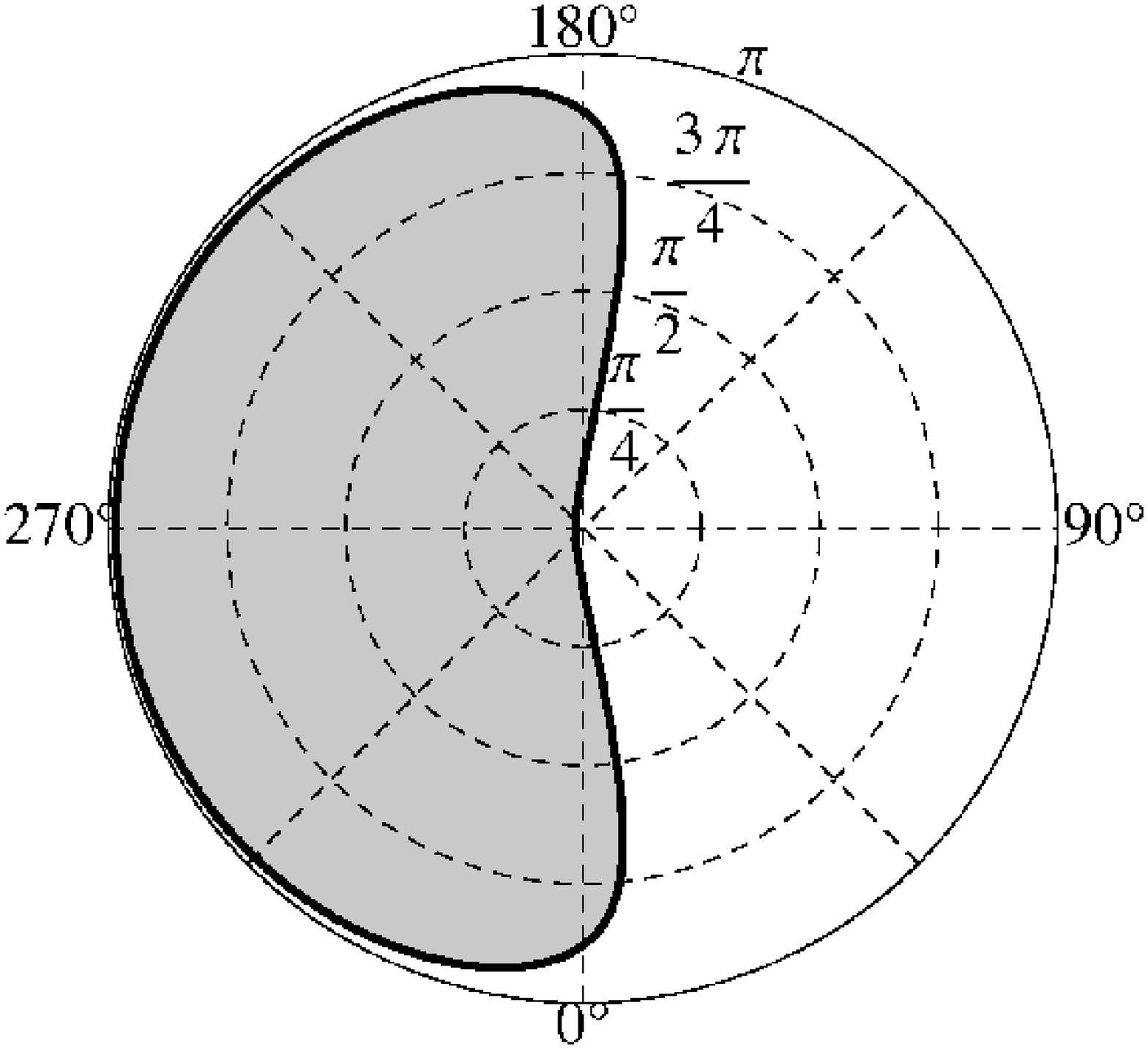}\\
			\includegraphics[width=4cm]{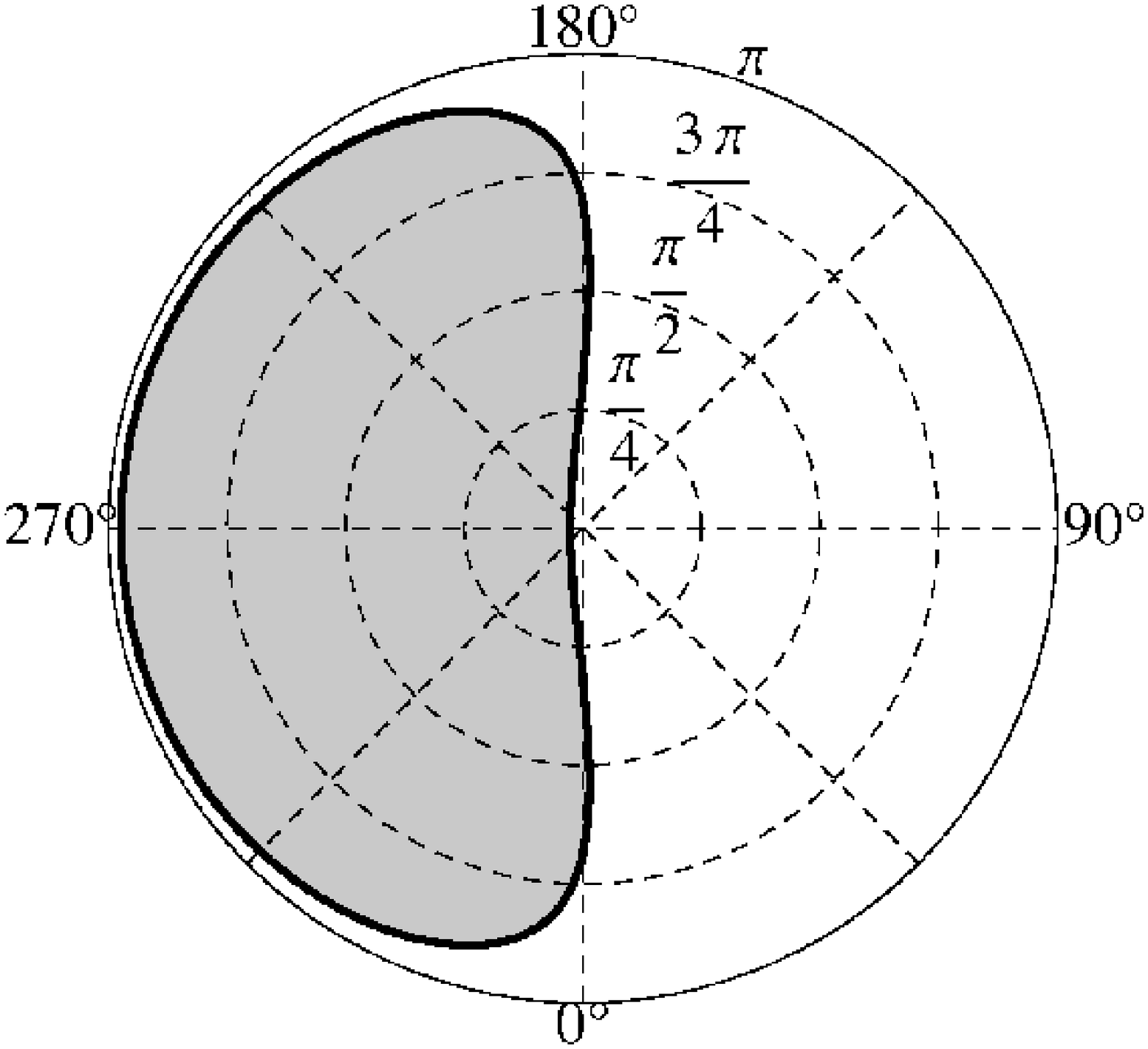}&
			\includegraphics[width=4cm]{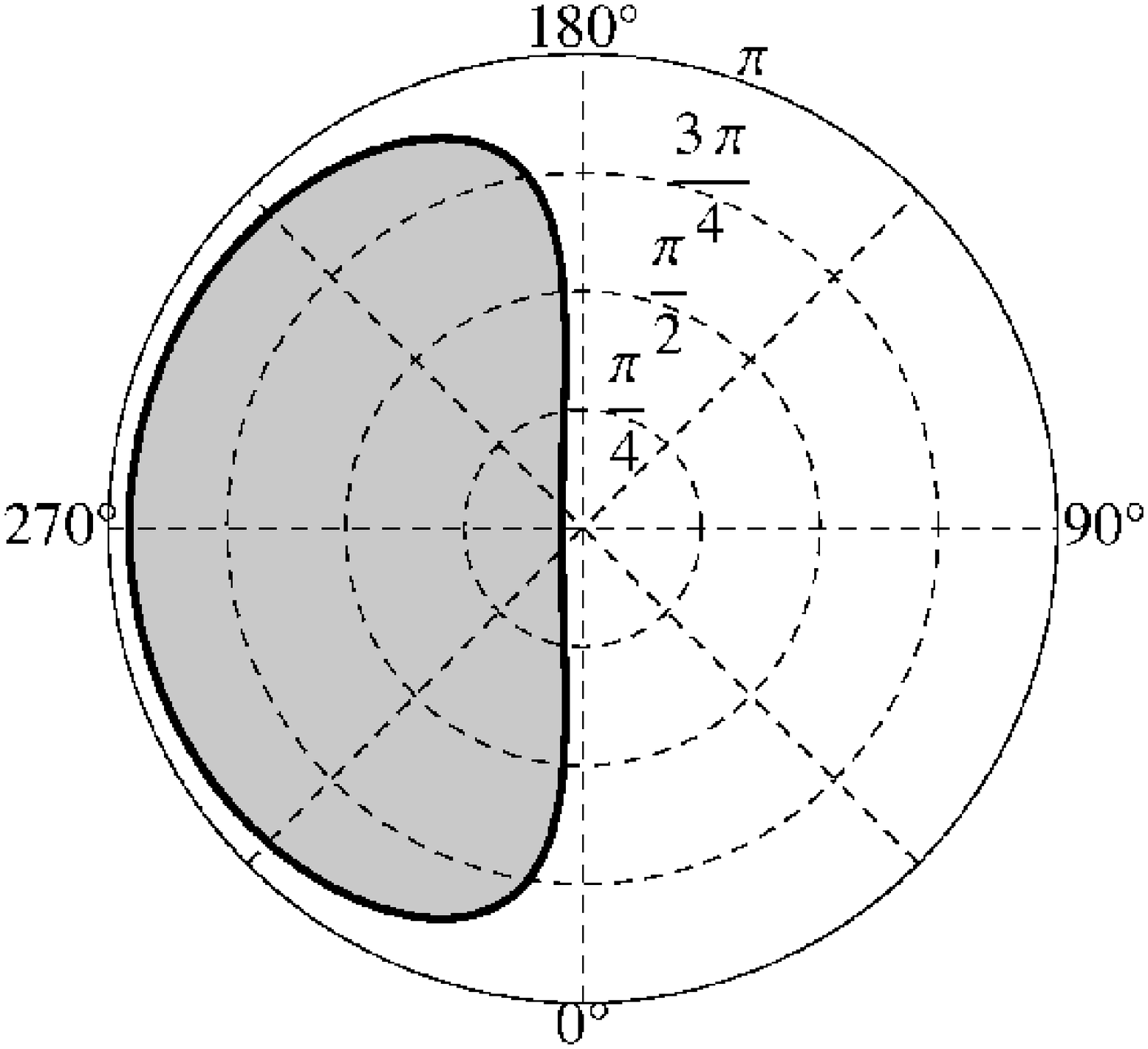}&	
			\includegraphics[width=4cm]{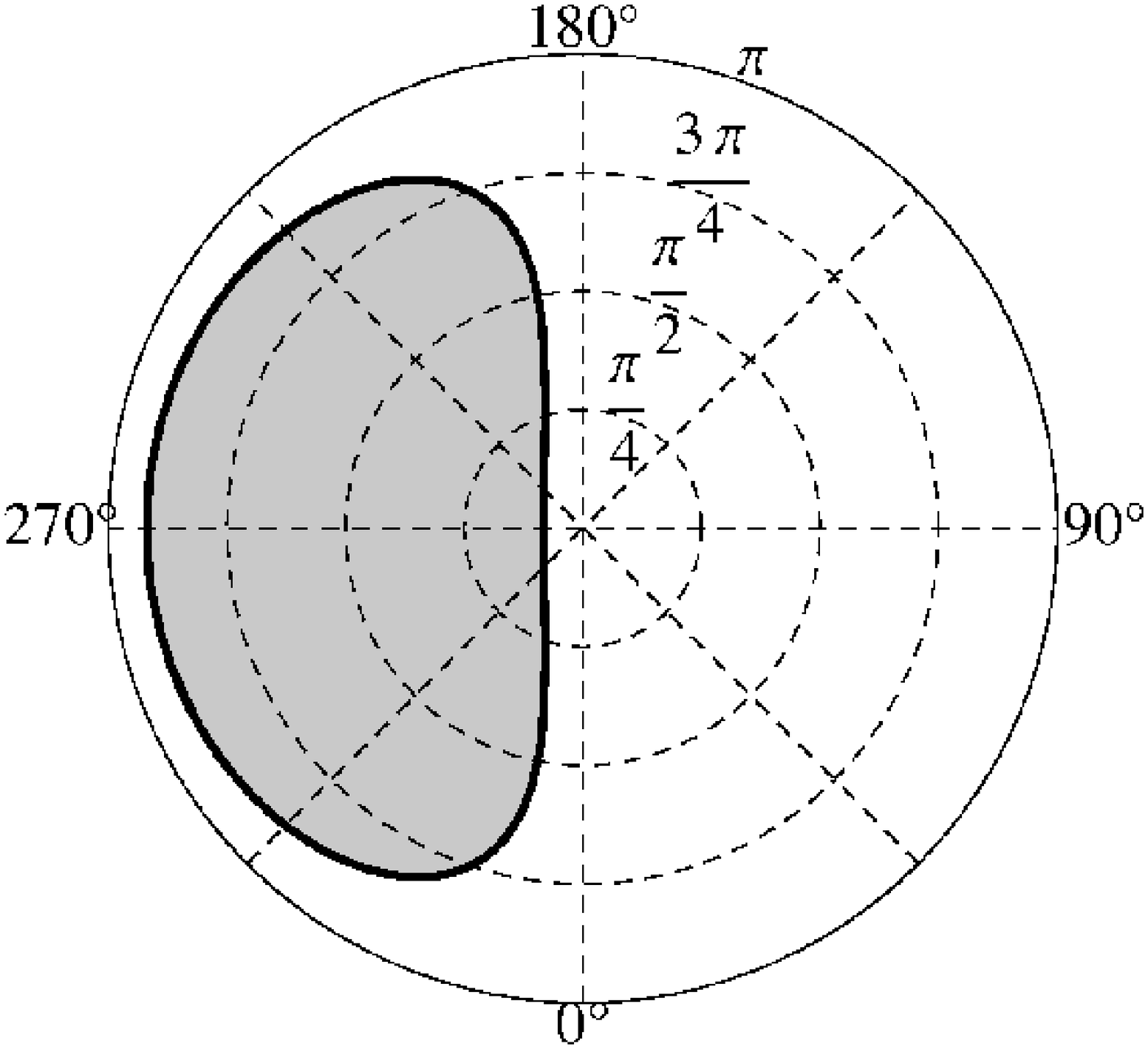}\\
			\includegraphics[width=4cm]{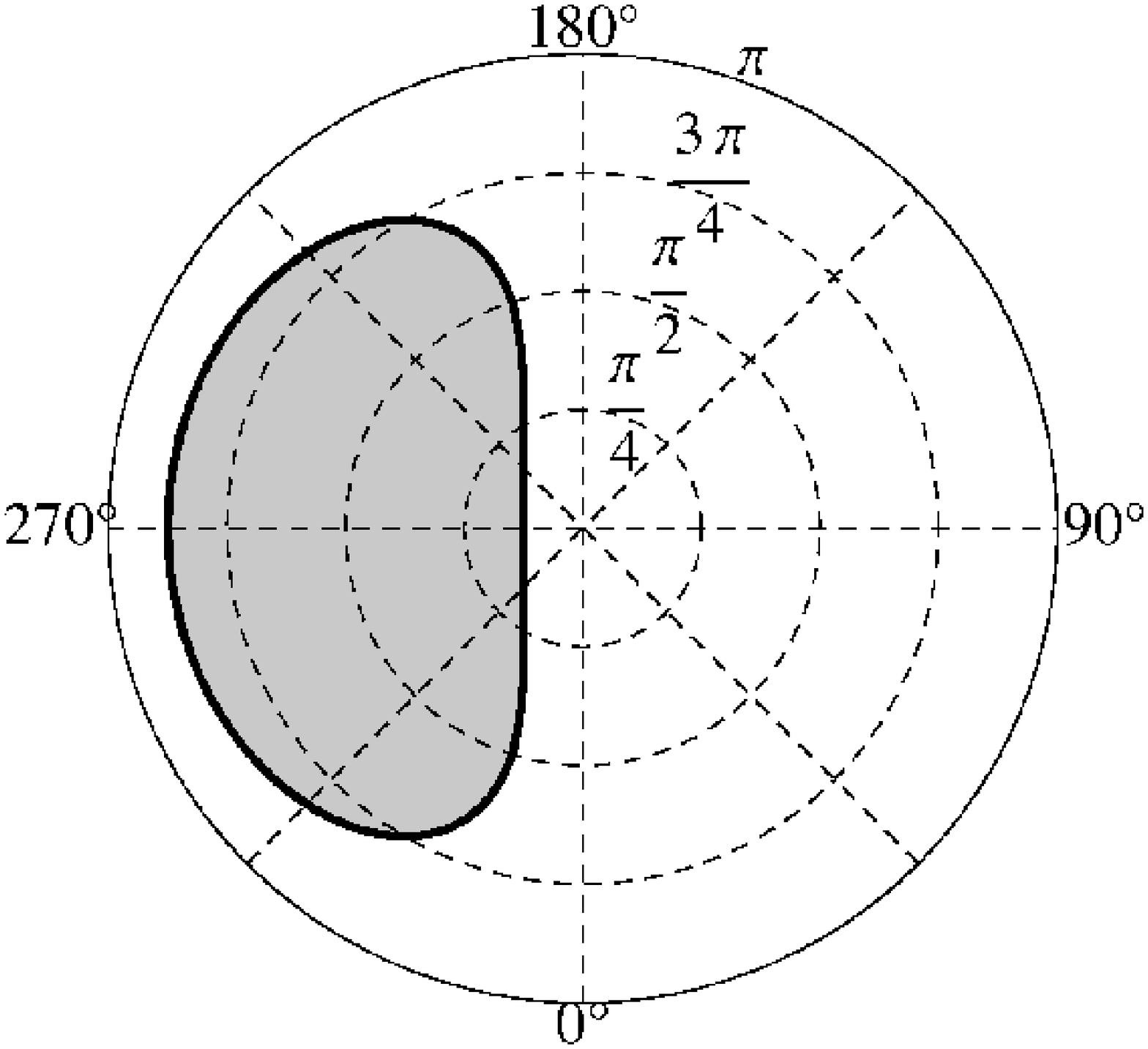}&
			\includegraphics[width=4cm]{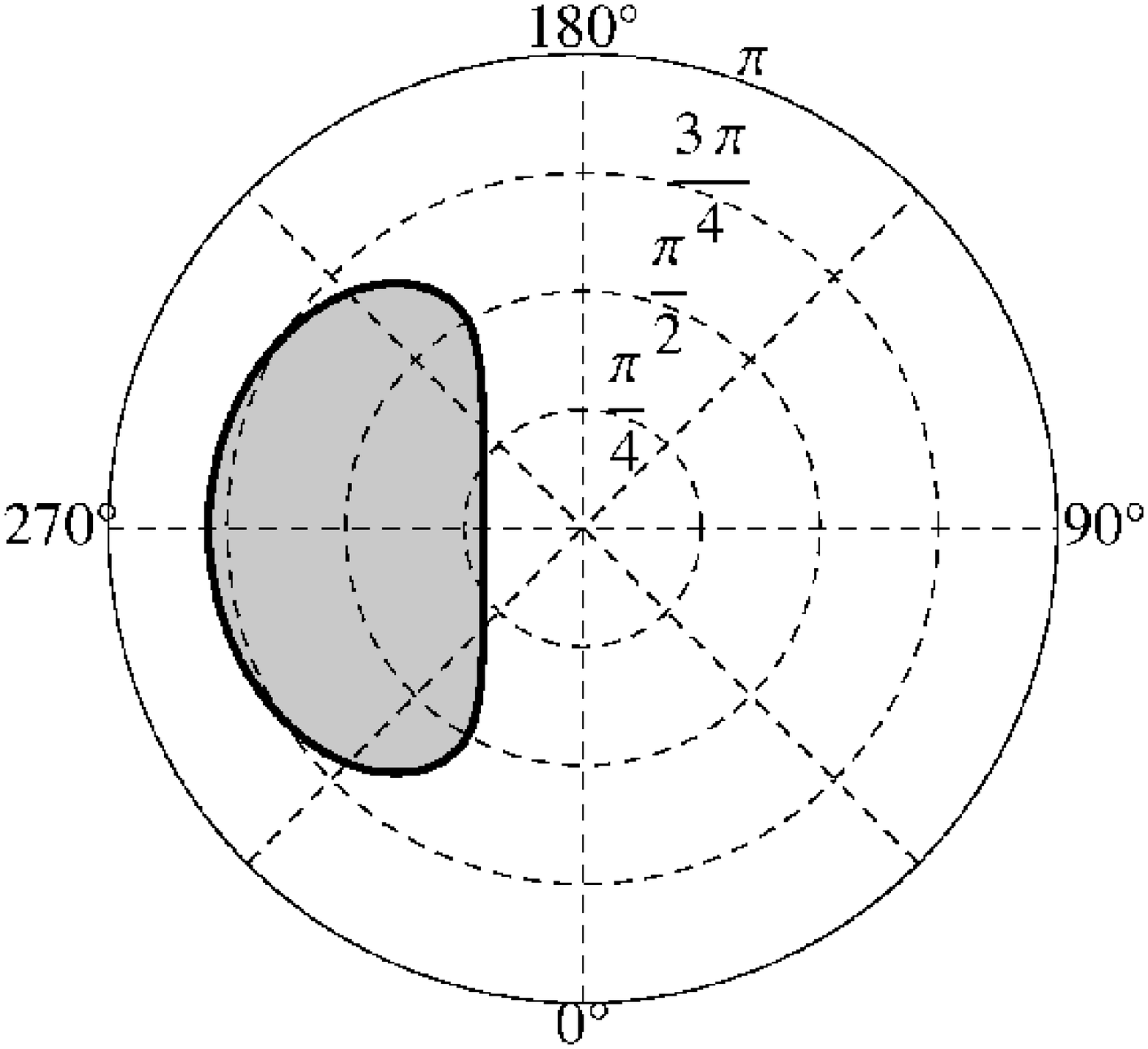}&
			\includegraphics[width=4cm]{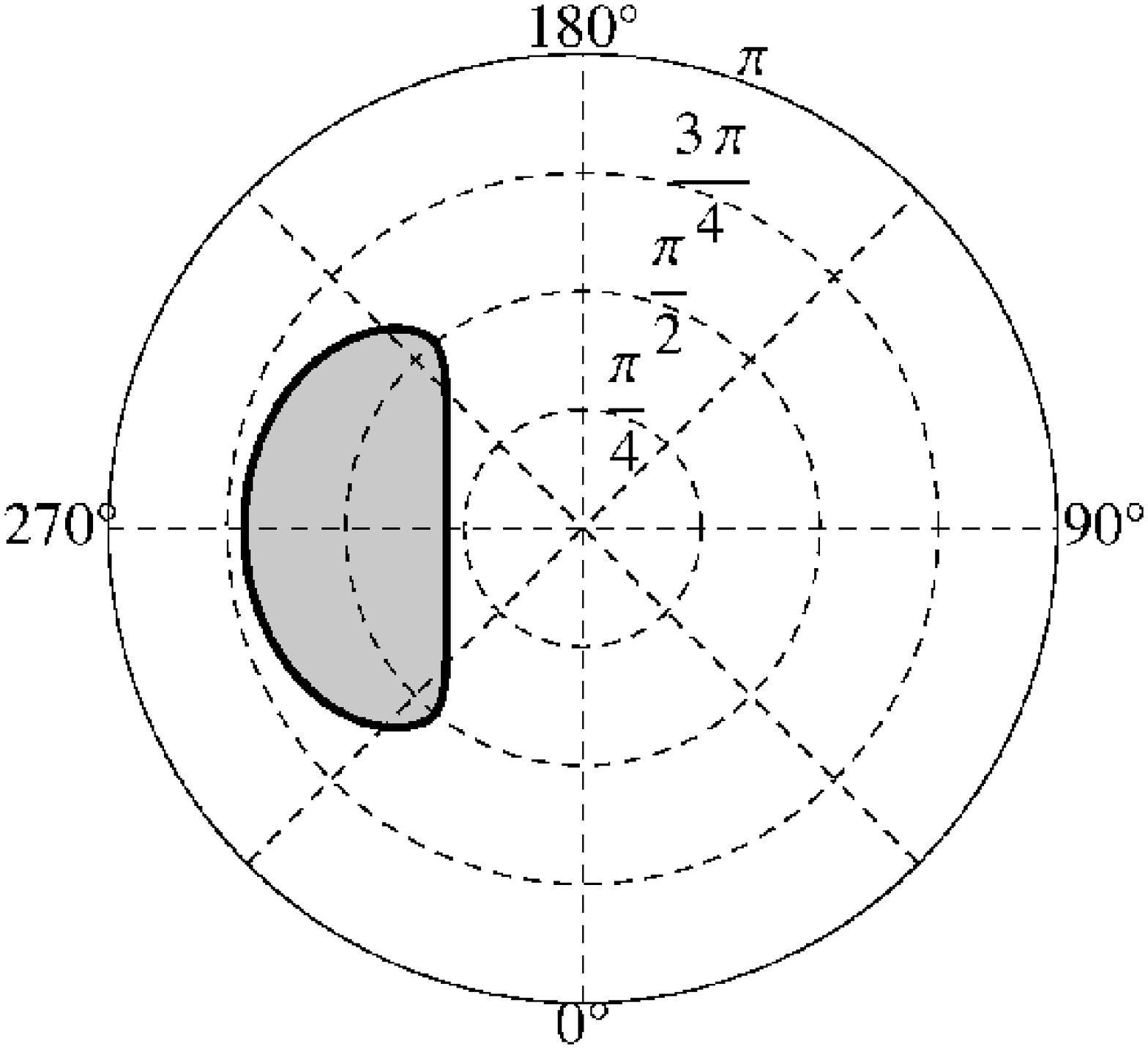}\\
			\includegraphics[width=4cm]{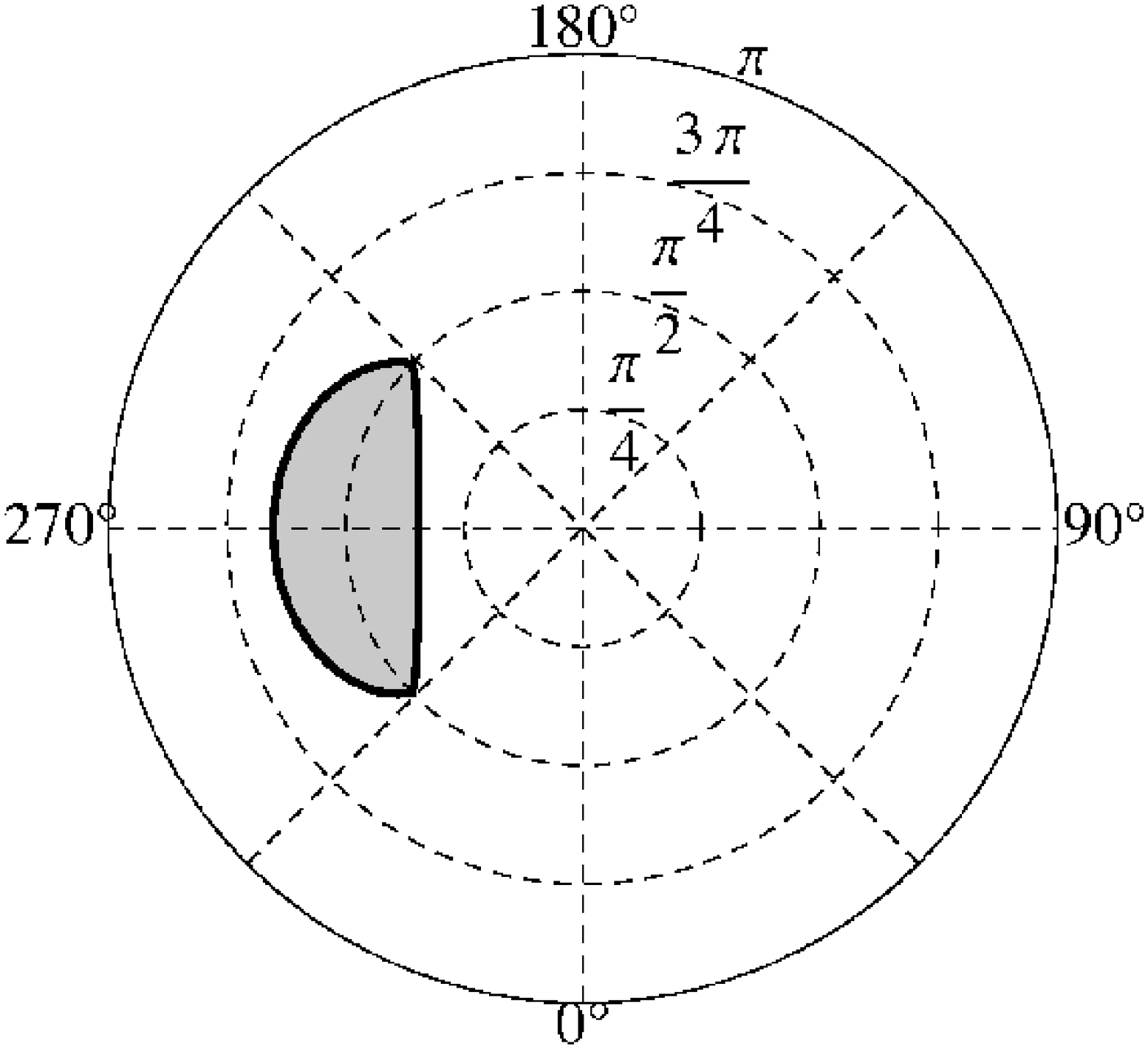}&
			\includegraphics[width=4cm]{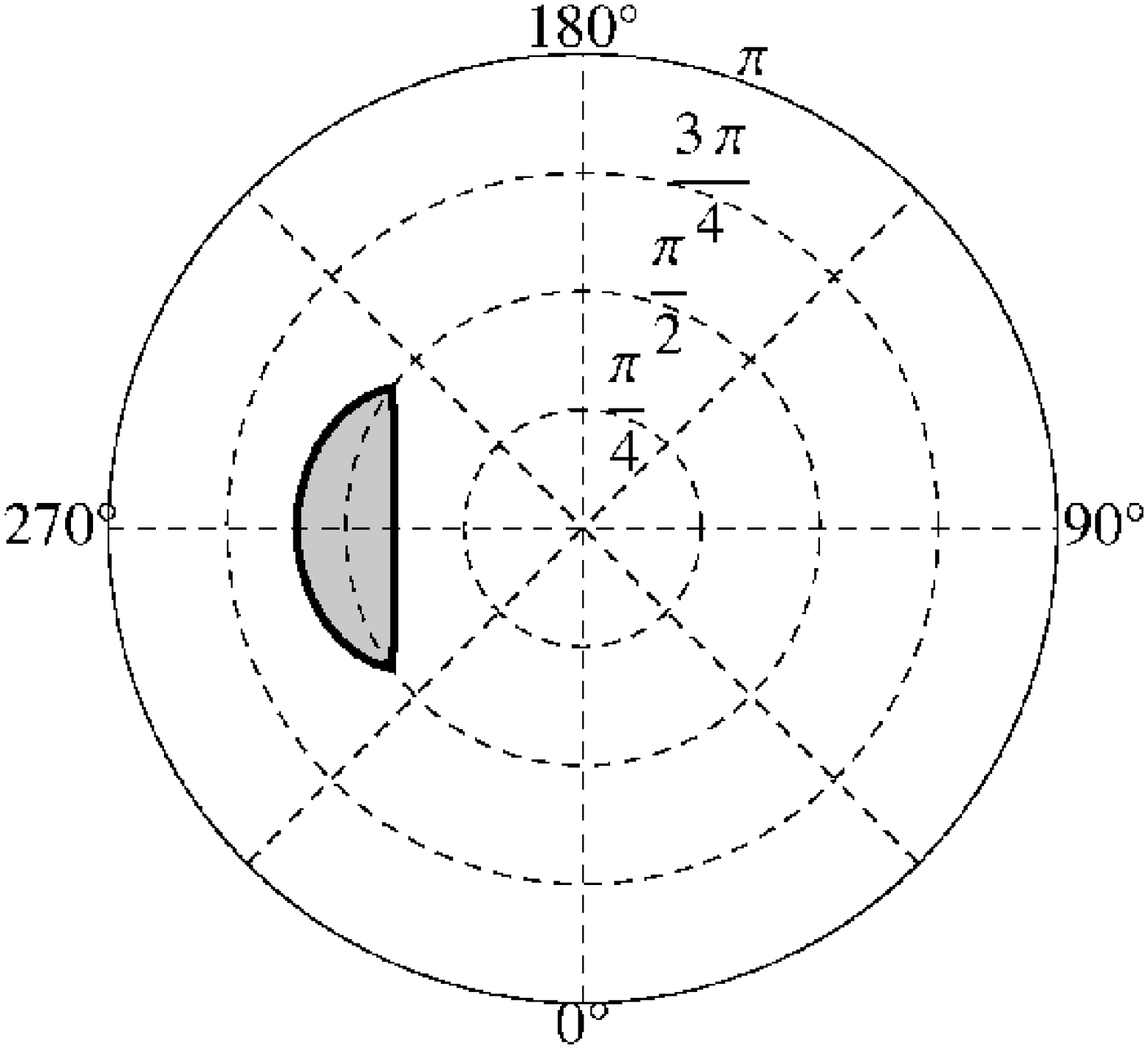}&
			\includegraphics[width=4cm]{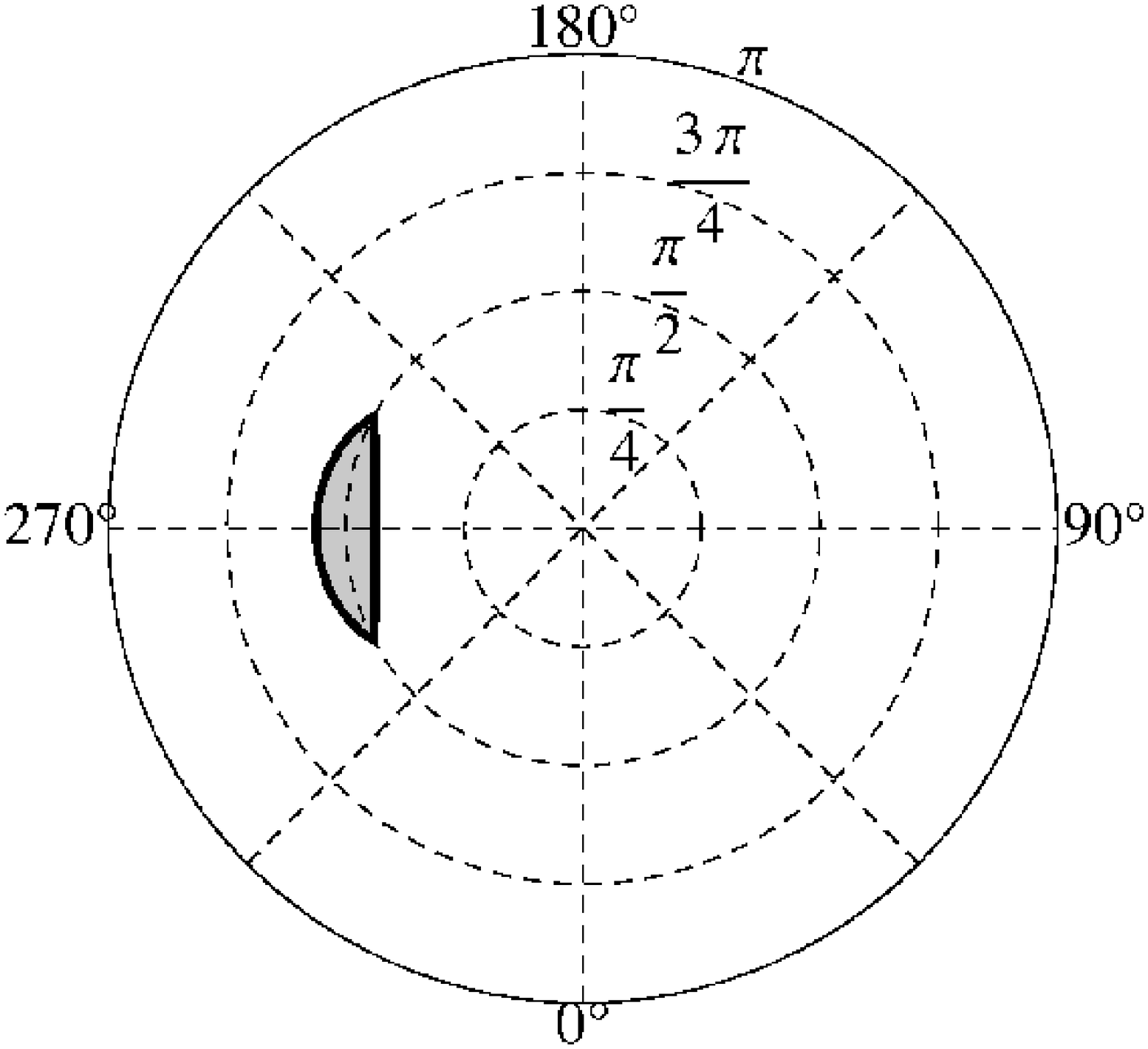}
		\end{tabular}
		\caption{\label{fig7}Escape cones of LNRF.  The LNRF observer (emitter) is at $r=r_{ms}$. The superspinar spin takes values of 1.0001,1.001,1.01,1.1, 1.5, 2.0, 3.0, 4.0, 5.0, 6.0, 7.0 (left to right, top to bottom). For comparison we take the near-extreme black hole with spin $a=0.9981$. }
	\end{figure}	
Photons radiated in the field of a Kerr superspinar by a given point-like source that are not able escape to infinity can be separated into two parts - the first one consists from the photons captured by the superspinar surface, the second one consists from those that are trapped in vicinity of the superspinar. Recall that in the field of a Kerr black hole all such photons are captured by the black hole. First we restrict our attention to the construction of the escape cones (see also \cite{Sch-Stu:2009:IJMPD}); the trapped photons will be discussed later.


	\begin{figure}[h!]
		\begin{tabular}{ccc}
			\includegraphics[width=4cm]{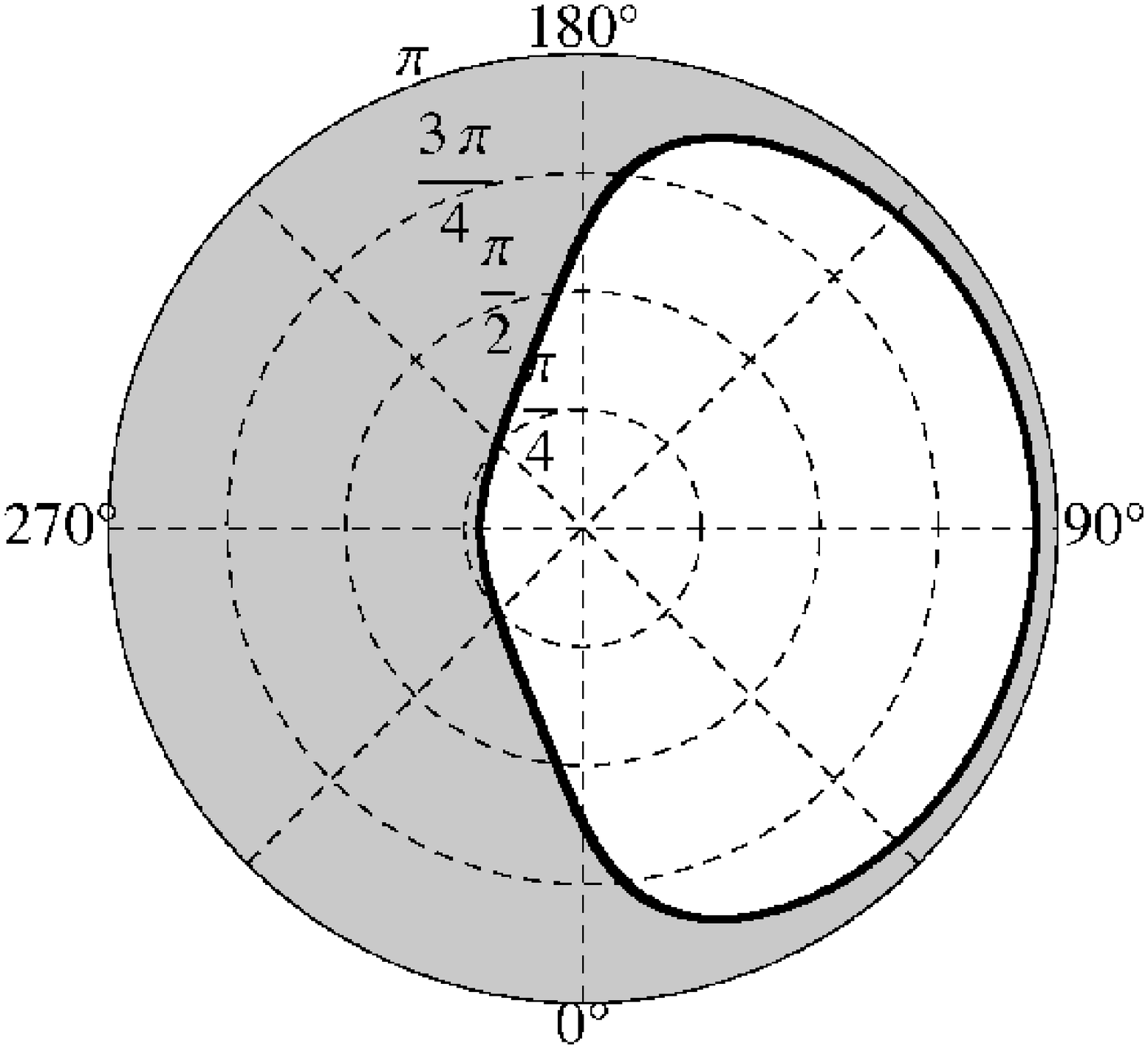}&
			\includegraphics[width=4cm]{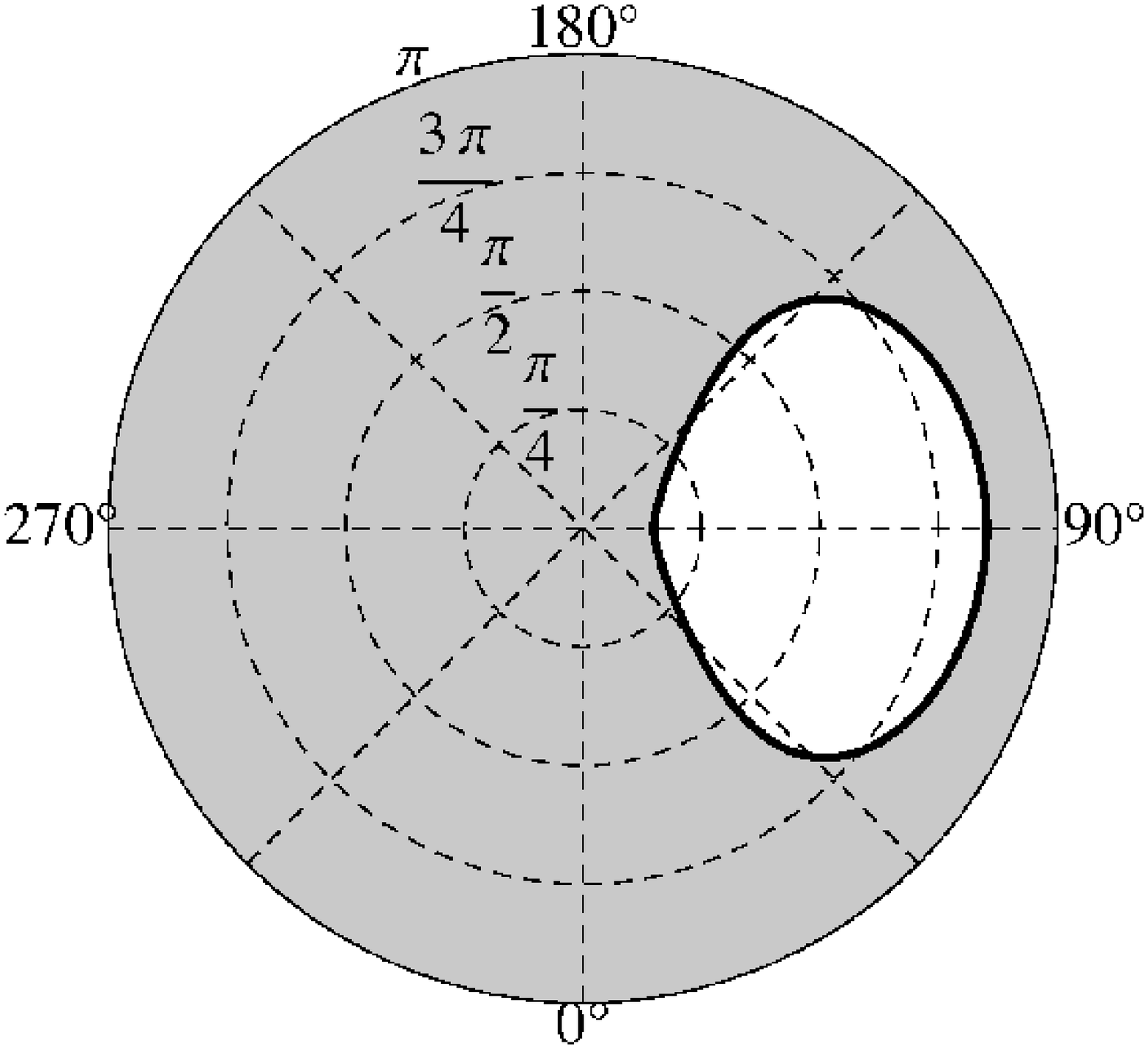}&
			\includegraphics[width=4cm]{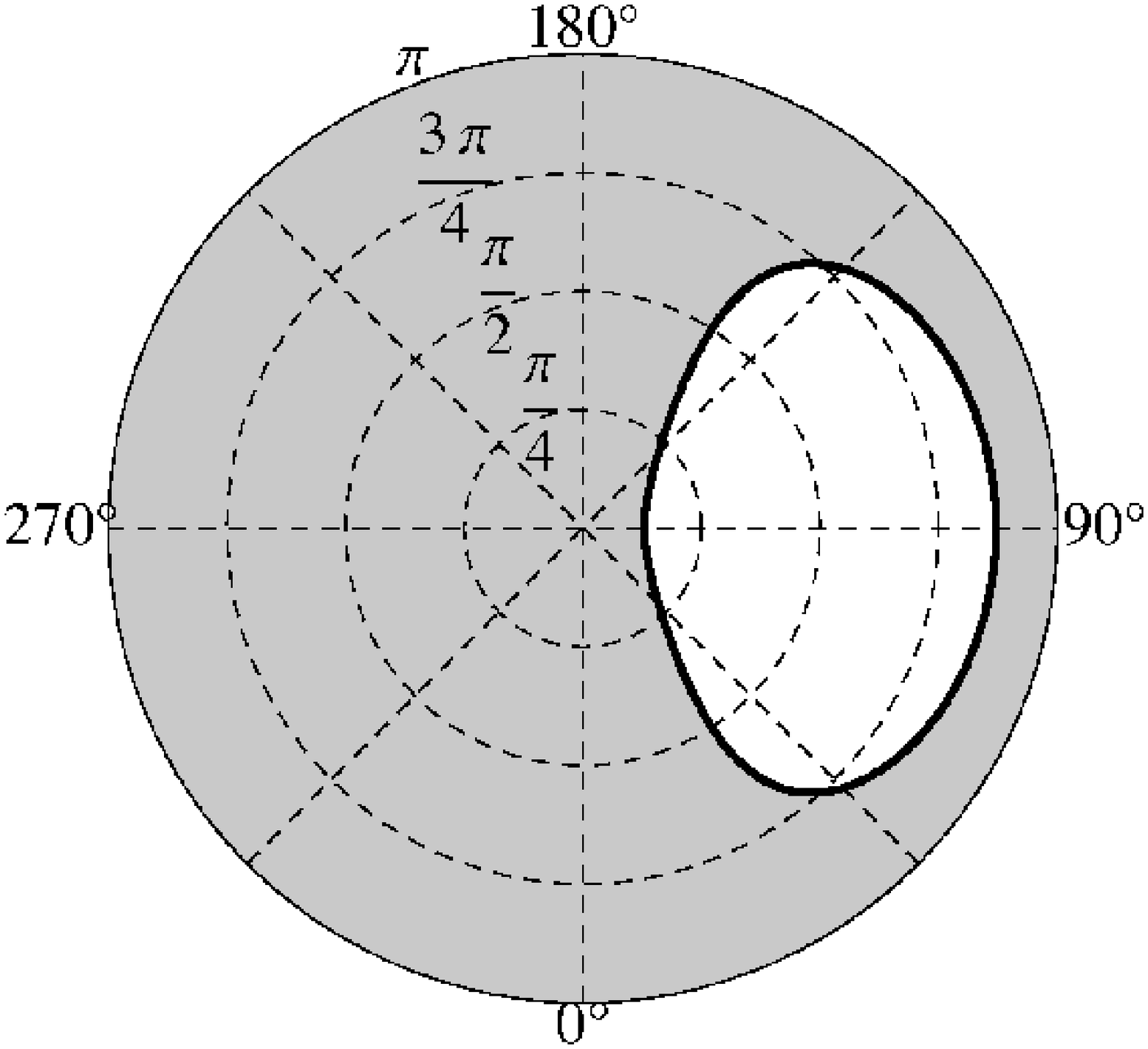}\\
			\includegraphics[width=4cm]{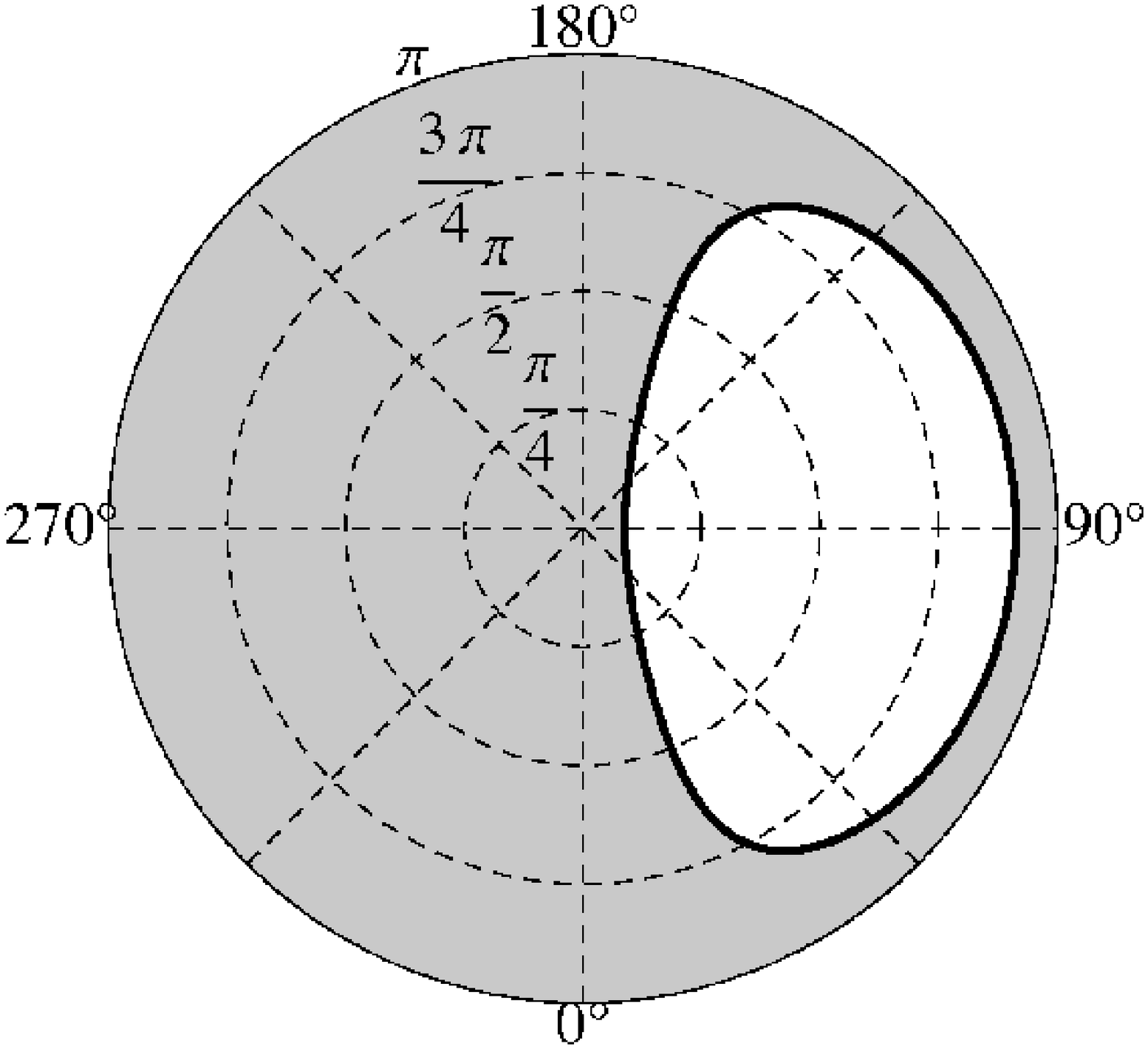}&
			\includegraphics[width=4cm]{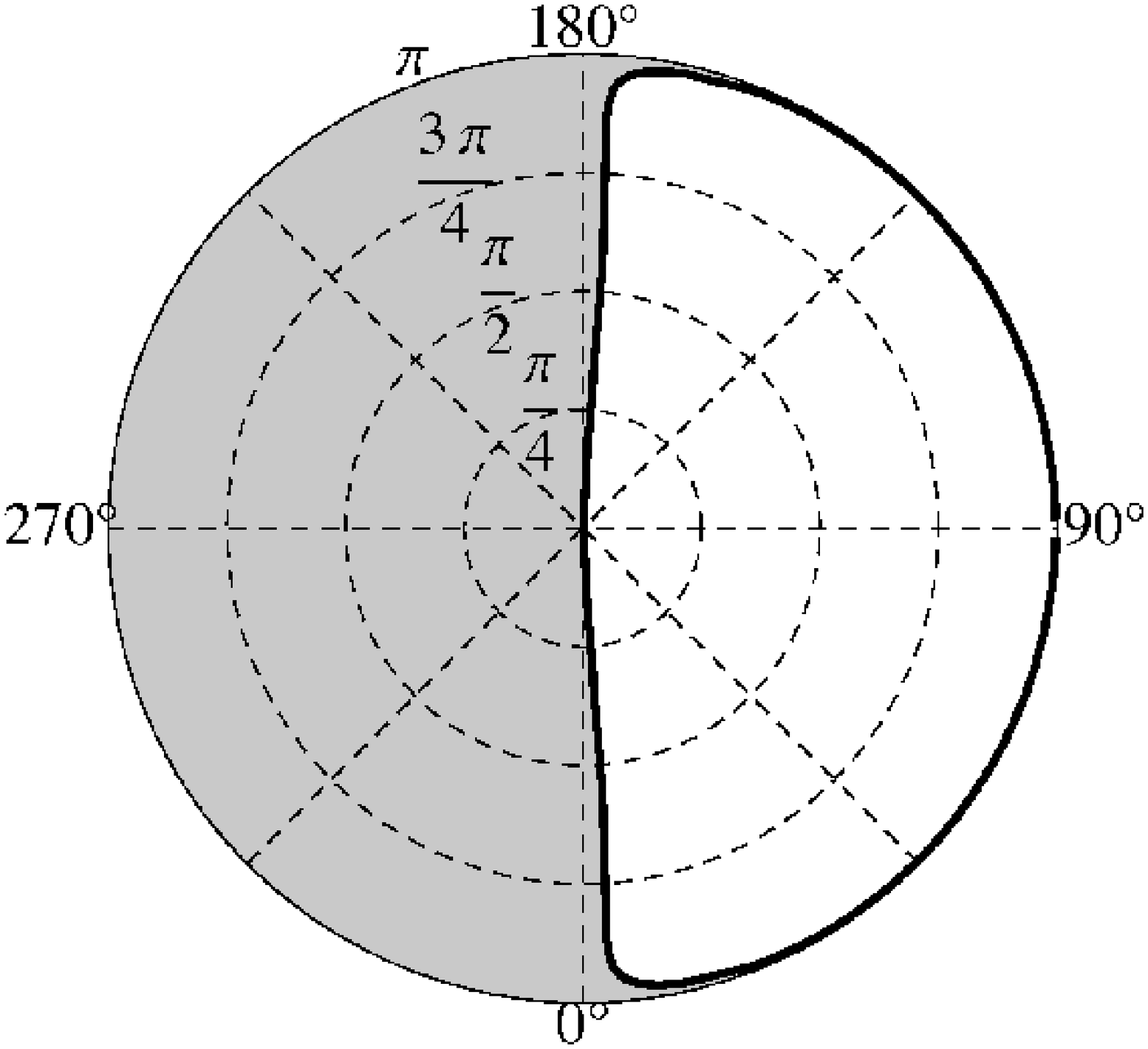}&	
			\includegraphics[width=4cm]{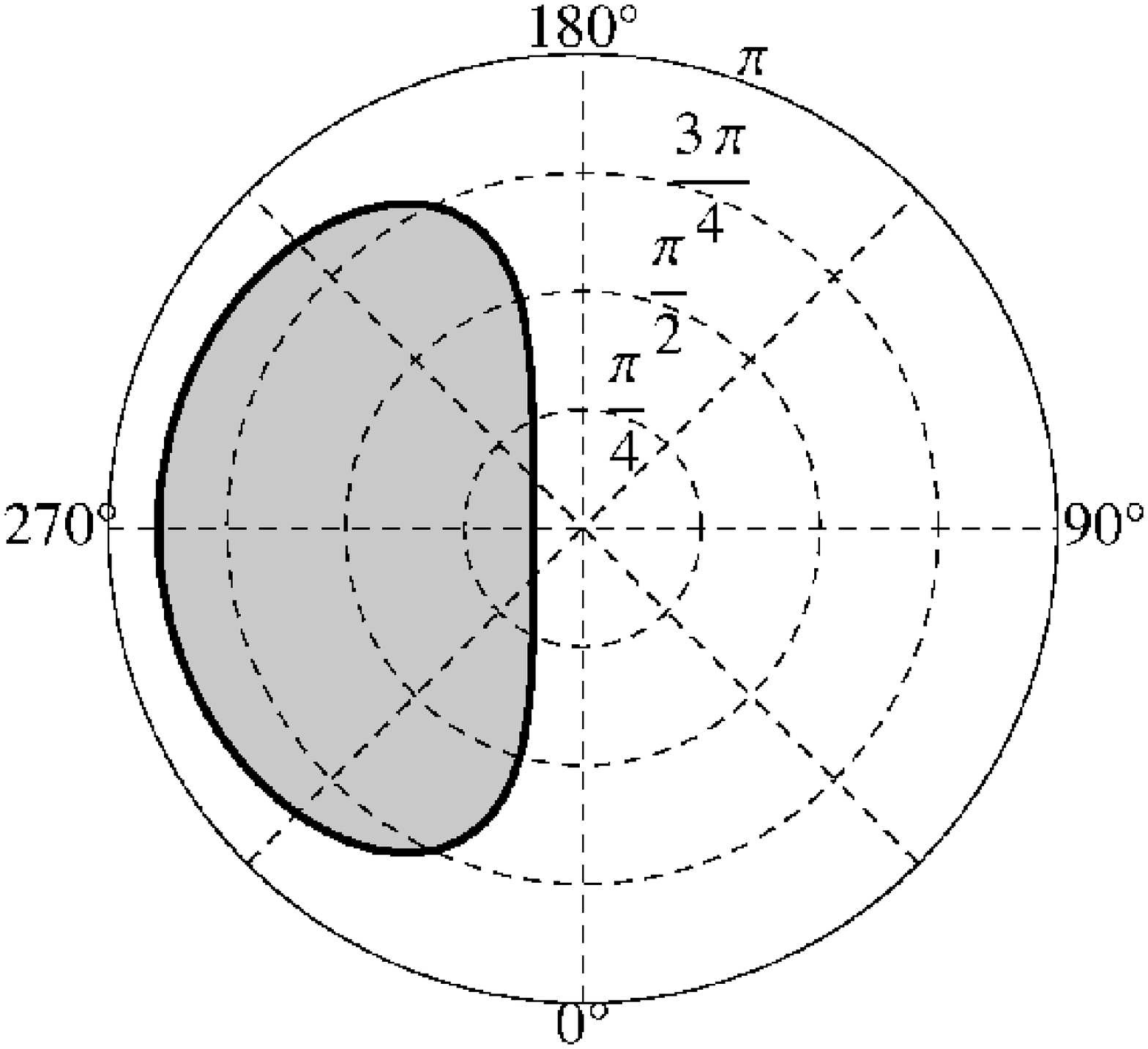}\\
			\includegraphics[width=4cm]{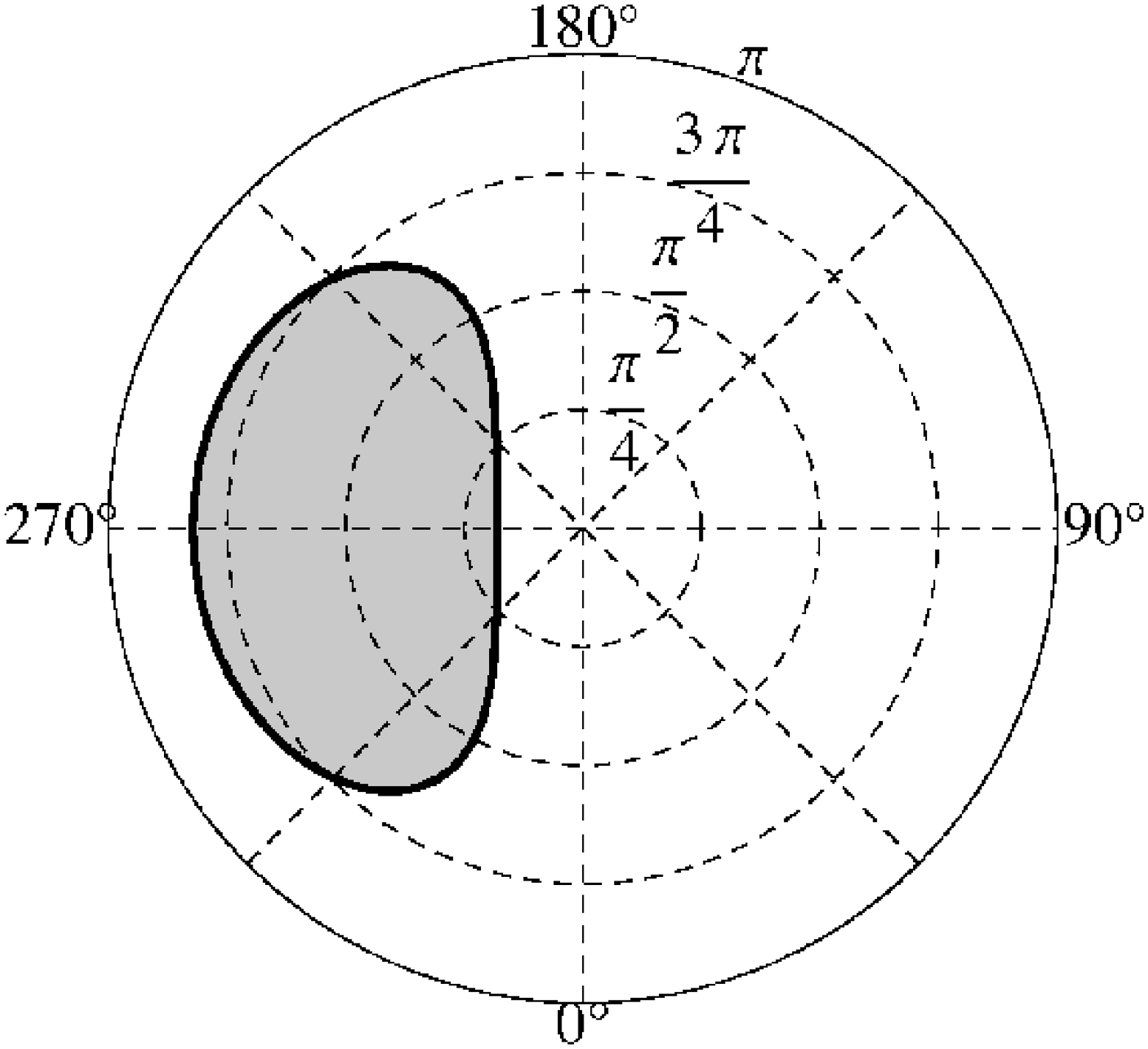}&
			\includegraphics[width=4cm]{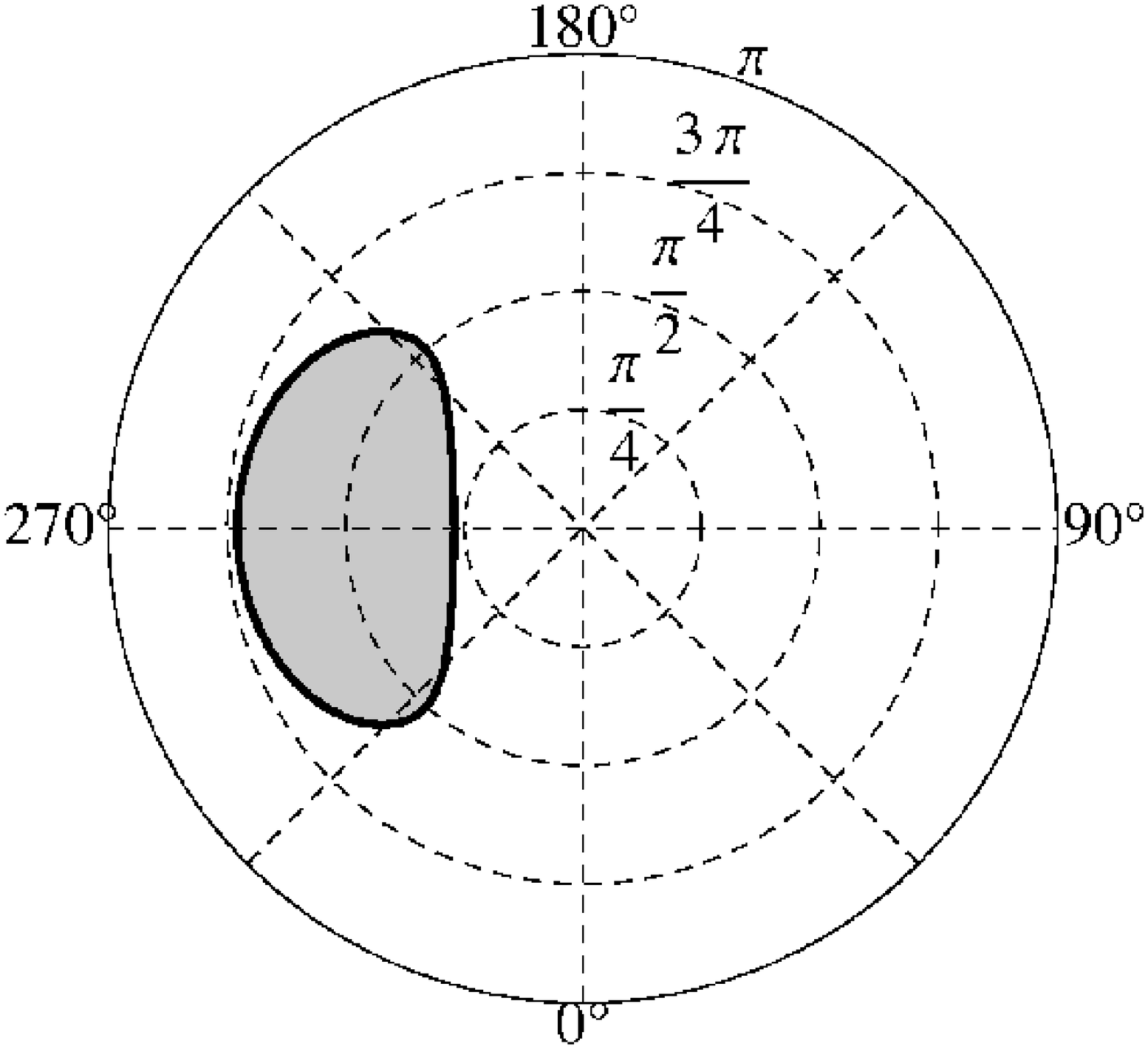}&
			\includegraphics[width=4cm]{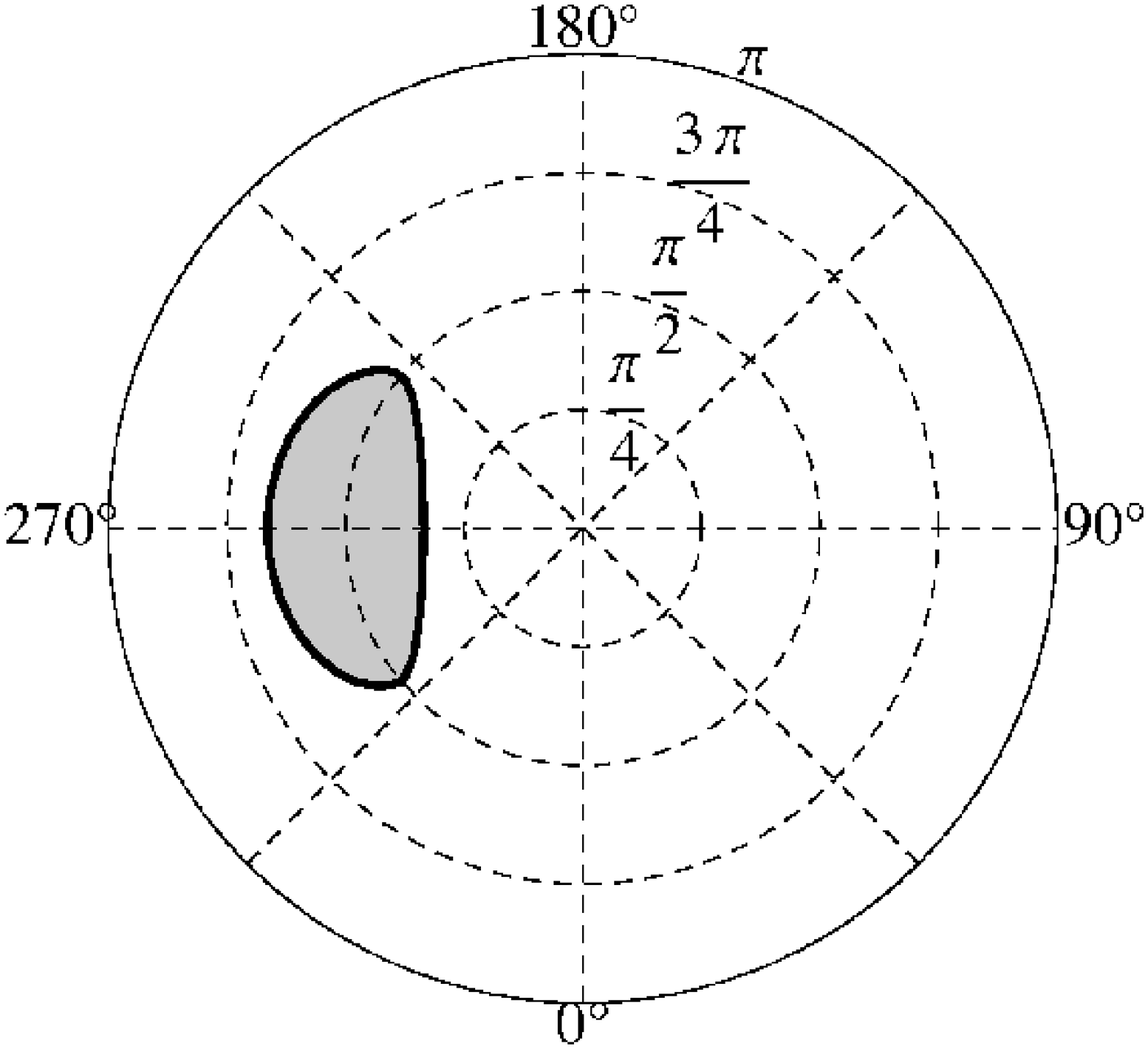}\\
			\includegraphics[width=4cm]{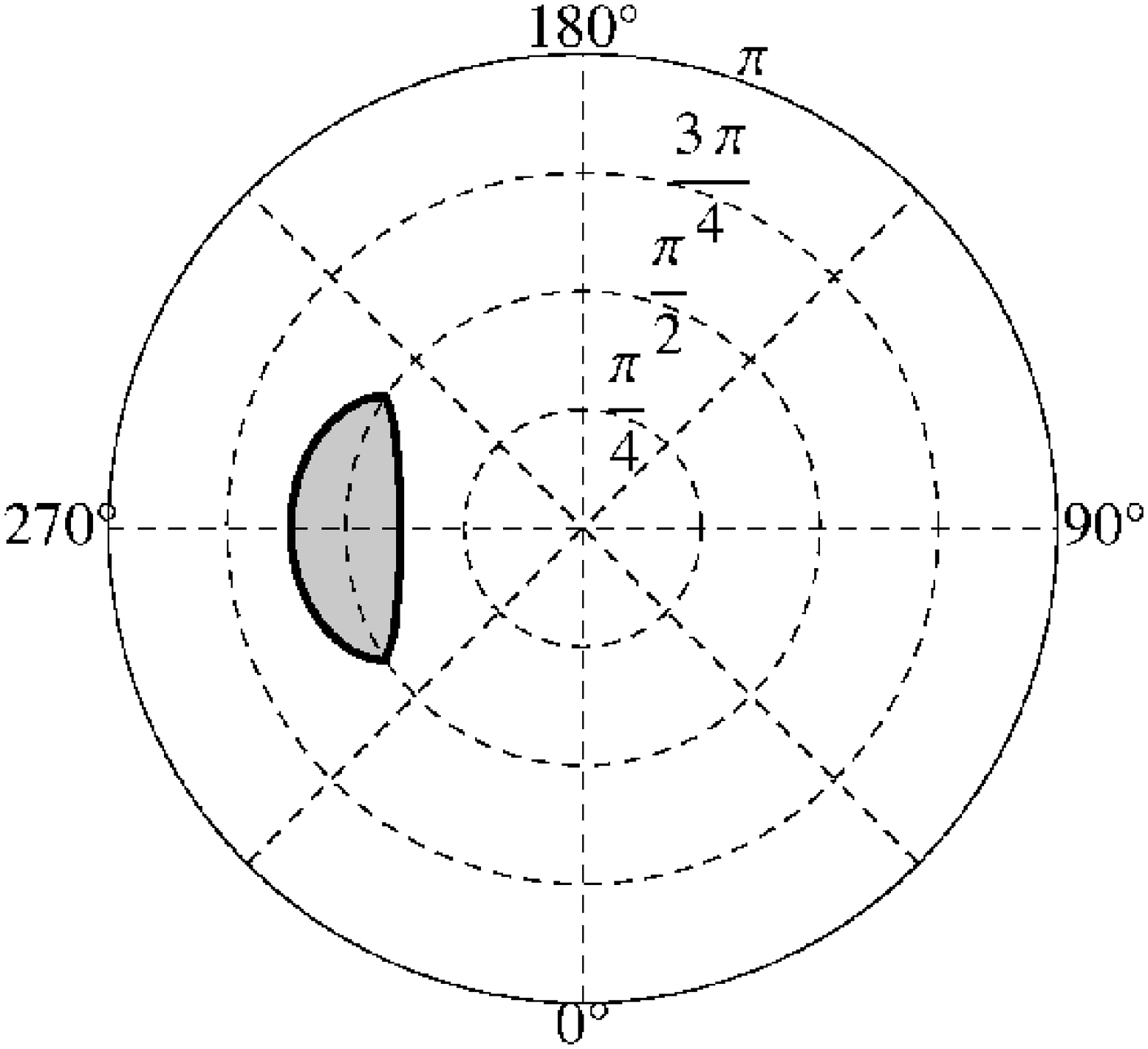}&
			\includegraphics[width=4cm]{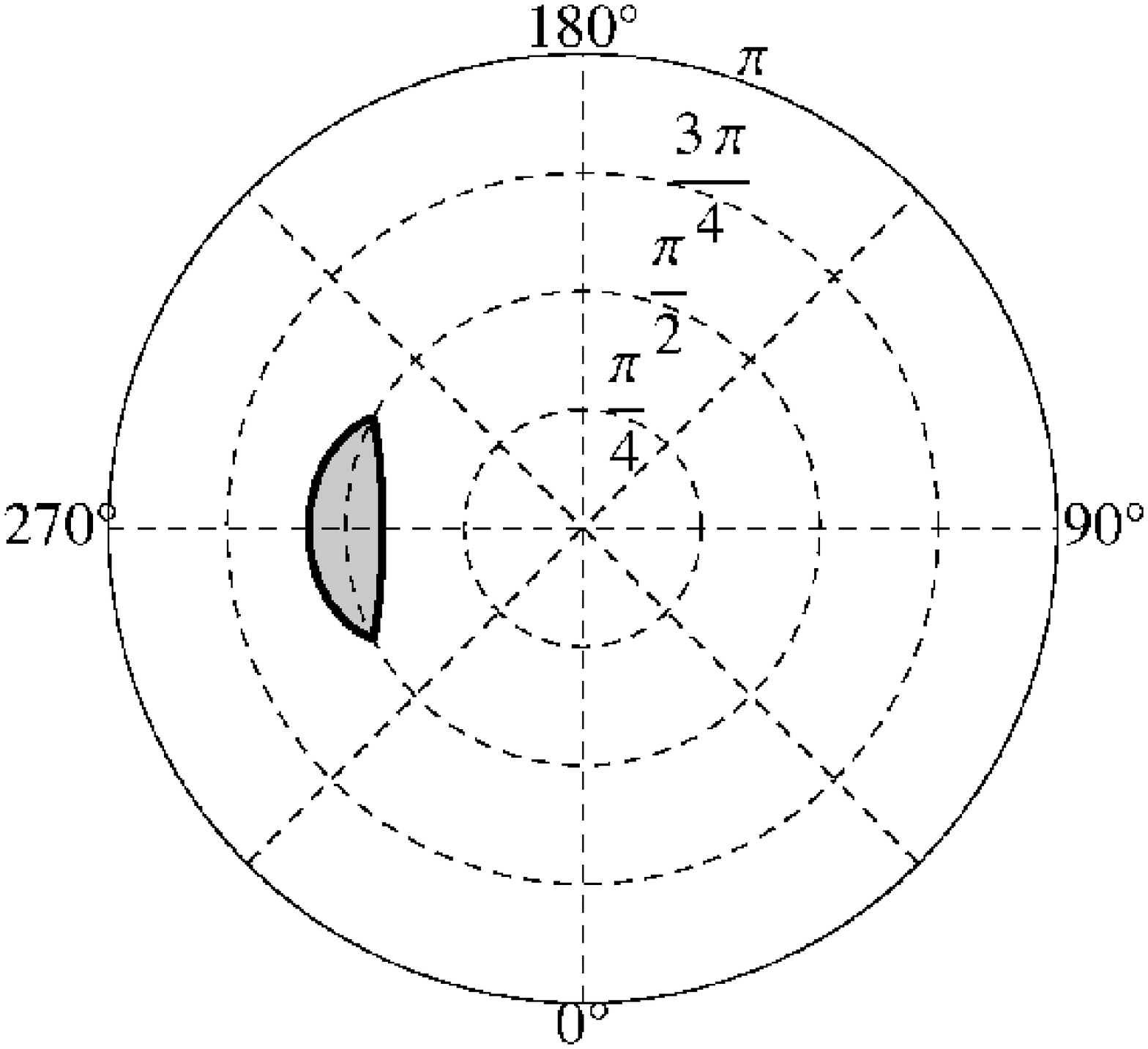}&
			\includegraphics[width=4cm]{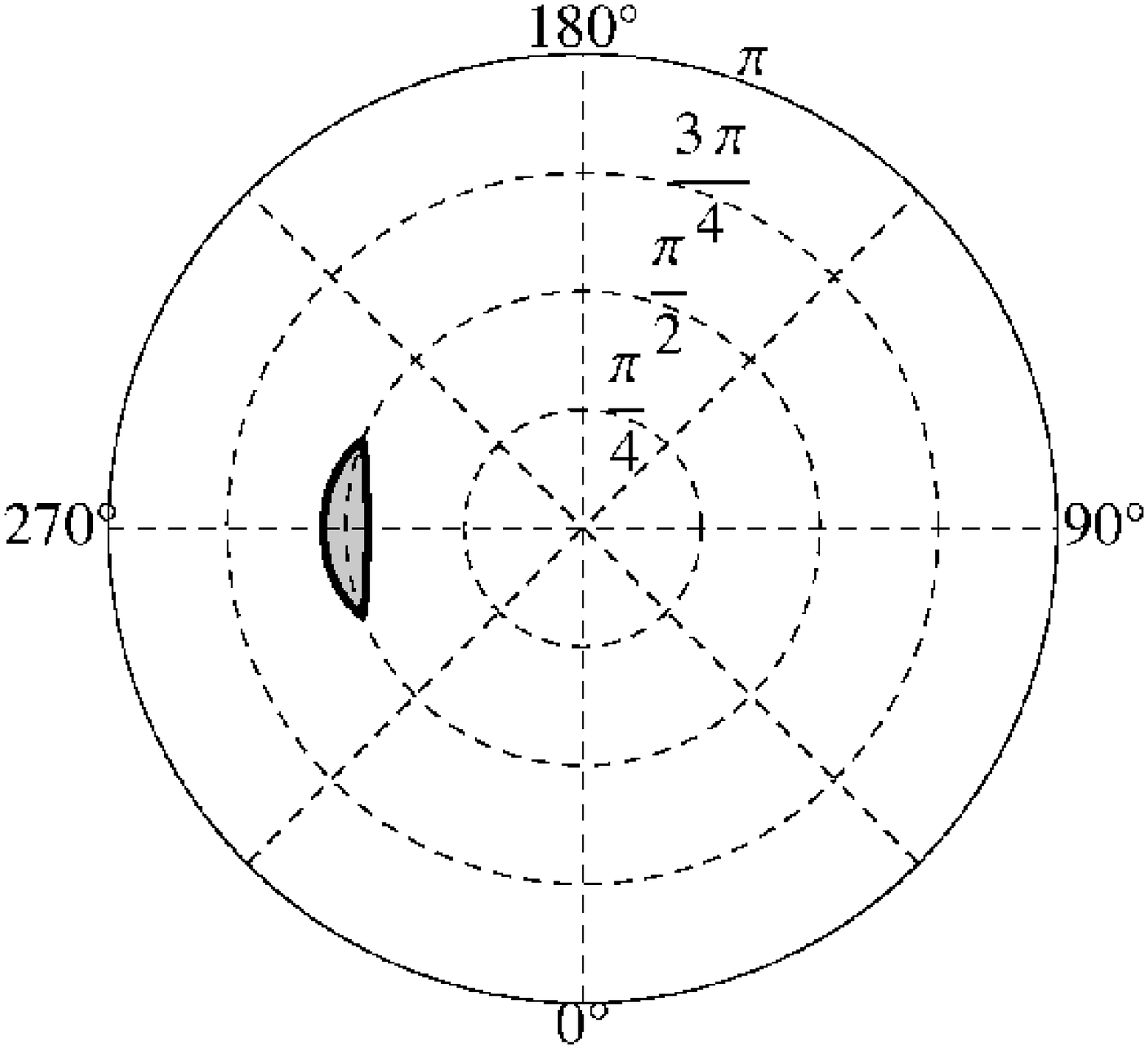}
			\end{tabular}
		\caption{\label{fig8}Escape cones of GF+ observers.  The GF+ observer (emitter) is at $r=r_{ms}$. The superspinar spin takes values of 1.0001,1.001,1.01,1.1, 1.5, 2.0, 3.0, 4.0, 5.0, 6.0, 7.0 (left to right, top to bottom). For comparison we take the near-extreme black hole with spin $a=0.9981$. }
	\end{figure}

\par
The impact parameter $\lambda$  expressed in terms of the angle $\gamma_0$, related to the $LNRF$, reads

\begin{equation}
	\lambda_0=\frac{1}{\Omega_{LNRF0}+\frac{\Sigma_0\sqrt{\Delta_0}}{A_0\sin\theta_0\cos\gamma_0}},
\end{equation}
where index '$0$' refers to the frame with coordinates $[r_0,\theta_0]$. The minimum of $\mathcal{L}_{max}$ is located at

\begin{equation}
	r_{min}=\left\{ \begin{array}{lcr}
			\sqrt{a\lambda - a^2} & \textrm{for} & \lambda\geq a\\
			1-\frac{k_1}{k_2}+\frac{k_2}{3} & \textrm{for} & \lambda<a
			\end{array}\right.\label{eq_rmin}
\end{equation} 
where 

\begin{eqnarray}
	k_1&=&a^2+a\lambda-3,\\
	k_2&=&\left\{ 27(1-a^2)+2\sqrt{3}\sqrt{27(1-a^2)^2+k_1^3}\right\}^{1/3}.
\end{eqnarray}
The relevant values of $\mathcal{L}$ lie between $\mathcal{L}_{max}$ and
$\mathcal{L}_{min}$ determined by Eqs (\ref{eq13}) and (\ref{eq14}). The
intersections of functions $\mathcal{L}_{max}=\mathcal{L}_{max}(\gamma_0)$ and
$\mathcal{L}_{min}(\gamma_0)$ give the relevant interval of angles
$\gamma\in[\gamma_{min},\gamma_{max}]$ (see figure Fig.\ref{fig9}). For each $\gamma$ from $[\gamma_{min},\gamma_{max}]$ we calculate minimal value of the photon impact parameter $\mathcal{L}$ for which the photon reaches the turning point $r_{min}$ and escapes to infinity. This minimal value is the minimum of $\mathcal{L}_{max}$ which is located at $r_{min}$, eg. $\mathcal{L}_{max}=\mathcal{L}_{max}(r_{min};\lambda_0(\gamma_0),a)$, where $r_{min}$ is given by (\ref{eq_rmin}).
Now we can calculate the value of $\alpha_0$ using equation

\begin{equation}
	\cos\alpha_0=\frac{k^{(r)}}{k^{(t)}}=\frac{\omega^{(r)}_{LNRF\mu}k^\mu}{\omega^{(t)}_{LNRF\mu}k^\mu}.
\end{equation} 
We arrive to the formula

\begin{equation}
	\cos\alpha_0=\pm\sqrt{A_0}\frac{\sqrt{(r_0^2+a^2-a\lambda_0)^2-\Delta_0(\mathcal{L}_{max}^{min}-2a\lambda_0+a^2)}}{-a(a\sin^2\theta_0-\lambda_0)\Delta_0+(r_0^2+a^2)(r_0^2+a^2-a\lambda_0)},
\end{equation}
where $A_0=A(r_0,\theta_0)$, $\Delta_0=\Delta(r_0)$  and $\mathcal{L}_{max}^{min}=\mathcal{L}_{max}(r_{min};\lambda_0,a)$. The angle $\beta_0$ can be  calculated from the formula (\ref{LC4}).
In this way we obtain angles from the arc $\beta_0\in\langle -\pi/2; \pi/2\rangle$. The remaining arc $\beta_0\in\langle \pi/2; 3\pi/2\rangle$ can be obtained by turning the arc $\beta_0\in\langle -\pi/2; \pi/2\rangle$ around the symmetry axis determined by angles $\beta_0=-\pi/2$ and $\beta_0=\pi/2$. This procedure can be done because photons released under angles $\beta_0$ and $\pi-\beta_0$ have the same constants of motion. 
Clearly, for sources under the radius corresponding to the (counterrotating) equatorial photon circular orbit, only outward directed photons with no turning point of the $r$-motion can escape. With radius of the source approaching the Kerr superspinar surface (at $R=0.1$), the escape cone shrinks, but its extension remains finite, contrary to the case of the Kerr black holes when approaching the black hole horizon, the escape cone shrinks to infinitesimal extension, with exception of the extreme black holes \cite{Bar:1973:BlaHol:}. For the $GF_{+}$, the procedure of the related light escape cone construction can be directly repeated, but with the relevant tetrad 1-form components being used in the procedure. (If the region of trapped photons in the vicinity of the superspinar surface has to be constructed, we have to consider the stable spherical photon orbits and the related impact parameters that will determine the trapped-photon  region.)

In order to reflect properly the effect of the superspinar spin $a$ on the escape cone structure, we shall give the cones for two sequences, namely for $LNRF$ and $GF_{+}$ observers located at the ISCO-radius of the spacetimes with appropriately  
chosen values of the spin.


Behaviour of the $LNRF$ escape cones in dependence on the superspinar spin $a$ is represented in Fig.\ref{fig7}.
The complementary local cones, corresponding to non-escaping photons trapped in the field of the superspinar or captured by its surface, are shaded.


 For the circular (corotating) geodesic frames $GF_{+}$ the local escape cones are presented in the Fig.\ref{fig8} for the same values of spin as in the case of LNRF. In both the $LNRF$ and $GF_{+}$ local escape cones  the latitude is kept to $\theta = \pi/2$ relevant for the Keplerian discs. For comparison, we include in the first position for both of the sequences ($LNRF$ and $GF_{+}$) the local escape cone constructed for a near-extreme Kerr black hole with the canonical value $a=0.9981$ introduced by Thorne \cite{Thorne:1974:APJ}.

The $LNRF$ local escape cones at the ISCO radius grow with the value of the spin growing in its whole considered range and for the Kerr superspinars are always larger than the one corresponding to the canonical black hole - this is the picture given for local frames (of observers or sources) that maximally restrict the rotational effects of the Kerr spacetimes \cite{Bar-Pre-Teu:1972:ASTRJ2:}. The $GF_{+}$ local escape cones at the ISCO also demonstrate growing as the superspinar spin grows, but for sufficiently small values of the spin ($a<a_{cr} \sim 1.089$) they are smaller as compared with the local escape cone of the $GF_{+}$ orbiting a canonical black hole with $a=0.9981$ while they are larger for $a>a_{cr}$.

Further, it should be noticed that at the ISCO radius of the canonical black hole the $LNRF$ local escape cone is substantially smaller in comparison with the $GF_{+}$ one, demonstrating thus the influence of the Keplerian rotation.
The rotational peculiarity of the Kerr superspinars is demonstrated by the fact that the $LNRF$ ISCO-radius escape cones are larger (smaller) in comparison with the $GF_{+}$ ISCO escape cones for $a<a_0$ ($a>a_0$). This kind of behaviour reflects the influence of the rotational state of the orbiting matter that is locally counterrotating ($\Phi_K < 0$) at ISCO for Kerr superspinars with $a<a_0 \sim 1.3$.

\section{\label{sec:Silhouette}Silhouette of Kerr superspinars}

In principle, it is of astrophysical importance to consider a Kerr superspinar (a Kerr naked singularity or a black hole \cite{Bar:1973:BlaHol:,Sch-Stu:2009:GReGr, Sch-Stu:2009:IJMPD}) being located in front of a source of illumination whose angular size is large as compared with the angular size of the superspinar. A distant observer will see a silhouette of the superspinar in the larger bright source. The superspinar silhouette is determined by photons that reach its surface and finish their travel there, contrary to the case of the rim of a black hole silhouette that corresponds to photon trajectories spiralling near the unstable spherical photon orbit around the black hole many times before they reach the observer. The spiralling photons concentrated around unstable spherical photon orbits will create an additional arc characterizing the superspinar (or a Kerr naked singularity) \cite{Hio-Mae:2009:PRD}. Of course, the shape of the silhouette and the arc enables, in principle, determination of the superspinar (or black hole) spin. But we have to be aware of the strong dependence of the silhouette shape on the observer viewing angle; clearly, the shape will be circular for observers on the superspinar rotation axis, and its deformation grows with observer approaching the equatorial plane.

Assuming that distant observers measure photon directions relative to the symmetry center of the gravitational field, the component of the angular displacement perpendicular to the symmetry axis is given by $-p^{(\varphi)}/p^{(t)}$ (for superspinar rotating anticlockwise relative to distant observers), while for angular displacement parallel to the axis it is given by $p^{(\theta)}/p^{(t)}$. These angles are proportional to $1/r_0$, therefore, it is convenient to use the impact parameters in the form independent of $r_0$ \cite{Bar:1973:BlaHol:}

\begin{equation}
 \tilde{\alpha}=-r_0\frac{p^{(\varphi)}}{p^{(t)}}=-\frac{\lambda}{\sin\theta_0},\label{silalpha}
\end{equation}

and

\begin{eqnarray}
 \tilde{\beta}&=&r_0\frac{p^{(\theta)}}{p^{(t)}}=\left[q+a^2\cos^2\theta_0-\lambda^2\cot^2\theta_0\right]^{1/2}\nonumber\\
&=&\left[\mathcal{L}+a^2\cos^2\theta-\frac{\lambda^2}{\sin^2\theta_0}\right]^{1/2}.\label{silbeta}
\end{eqnarray}
Photon trajectories reaching the observer are represented by points in the $(\tilde{\alpha}-\tilde{\beta})$ plane covering a small portion of the celestial sphere of the observer.


The shape of the superspinar silhouette (arc) is the boundary of the no-turning-point region, i.e., it is the curve $\mathcal{L}=\mathcal{L}^{min}_{max}(\lambda)$ expressed in the $(\tilde{\alpha}-\tilde{\beta})$ plane of the impact parameters. For observers in the equatorial plane $(\theta_0 = \pi/2)$, $\tilde{\alpha}=-\lambda$, $\tilde{\beta}=(\mathcal{L}-\lambda^2)^{1/2}=q^{1/2}$.

	
We consider the superspinar being observed by static distant observers that are related to the LNRF with zero angular velocity. Therefore, we use static frames at far distances, with the tetrad given by 
\beqa
 	\bf{\omega}^{(t)}&=&\left\{ \sqrt{\frac{\Delta\Sigma}{A}},0,0,0 \right\}\\
 	\bf{\omega}^{(r)}&=&\left\{ 0,\sqrt{\Sigma/\Delta},0,0 \right\}\\
 	\bf{\omega}^{(\theta)}&=&\left\{ 0,0,\sqrt{\Sigma},0 \right\}\\
 	\bf{\omega}^{(\varphi)}&=&\left\{ -\Omega_{\mathrm{LNRF}}\sqrt{\frac{A}{\Sigma}}\sin\theta,0,0,\sqrt{\frac{A}{\Sigma}} \right\}
\eeqa
The silhouette of the superspinar is quite naturally related to the trapped (escape) light cones of the static frames.

The marginal values of impact parameters $\lambda_0$ and $\mathcal{L}_0$(resp $q_0$) are obtained from the light escape cone. Using the stationary of the Kerr spacetime we ``shoot out`` virtual photons from observer (static frame at very large distance $r_0$) and we are looking for the light escape cone of this virtual source (using the results of the previous section). The trapped light cone of this virtual source is constructed from the light escape cone of the virtual source by transformations of directional angle $\alpha_0$ to $\bar{\alpha}_0=\pi - \alpha_0$ and directional angle $\beta_0$ to $\bar{\beta}_0=\beta_0$. In this way we get marginal directions for received photons from bright background behind the superspinar.  Then we can use the formulas (\ref{LC6}), (\ref{LC7}) and (\ref{LC8}) to calculate the marginal values of $\lambda_0$ and $q_0$($\mathcal{L}_0$) in order to obtain the silhouette of the Kerr superspinar in the plane $(\tilde{\alpha}-\tilde{\beta})$, i.e., the set of doubles $(\tilde{\alpha}_0,\tilde{\beta}_0)$ from equations (\ref{silalpha}) and (\ref{silbeta}). Here we plotted the silhouette directly from the trapped light cone $(\bar{\alpha}_0,\bar{\beta}_0)$ on the observer's  sky $(\bar{\alpha}_0\sin\bar{\beta}_0,\bar{\alpha}_0\cos\bar{\beta}_0)$. Note that the angle $\bar\alpha_0$ is the radial coordinate and the angle $\bar\beta_0$ is the polar coordinate in the polar graph of the silhouette.

\par 
	We shall give the silhouette of the superspinar for observers located at fixed radius $r_0=10^4$M that corresponds to the angular size of $\alpha\sim 1.4$arcsec; for higher distances the angular size falls accordingly to the $1/r_0$ dependence. We systematically compare the superspinar silhouette, when the photons defining its shape are those captured by the superspinar surface, with the silhouette of the Kerr naked singularity having the same spin when its shape is given by the photons that escape to the parallel asymptotic infinity corresponding to $r = - \infty$. Additionally, an arc corresponding to the unstable photon spherical orbits defines both the superspinars and naked singularities equivalently. (For Kerr naked singularities viewed under the inclination angle $\theta_0=90^\circ$, the silhouette shape is deformed into a line lying in the equatorial plane and terminated in the arc.) The silhouette of the superspinars and Kerr naked singularities is determined by both spin and inclination angle, while for superspinars it is further determined by the radius of its surface. In the case of Kerr naked singularities some characteristics of the silhouette could be defined properly in order to determine the spin and the inclination angle \cite{Hio-Mae:2009:PhysRevD}. Here we demonstrate that for superspinars with the additional parameter giving their radius  the situation with defining the characteristics enabling the superspinar parameters is more complicated. The rotational effect on the shape of the silhouette grows with inclination angle growing and becomes strongest when $\theta_0=\pi/2$.

\begin{figure}[h!]
	\begin{center}
	\begin{tabular}{cccc}
	\includegraphics[width=4cm]{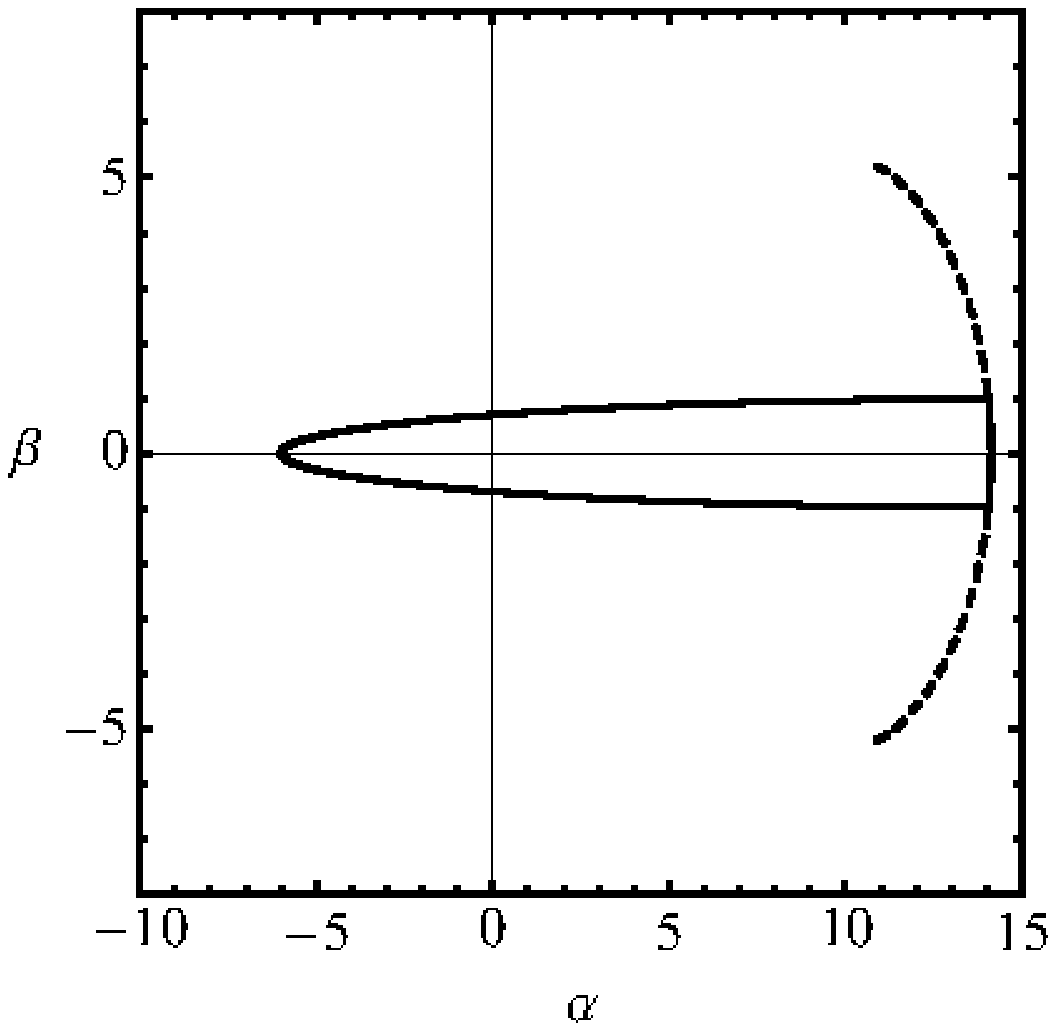}&\includegraphics[width=4cm]{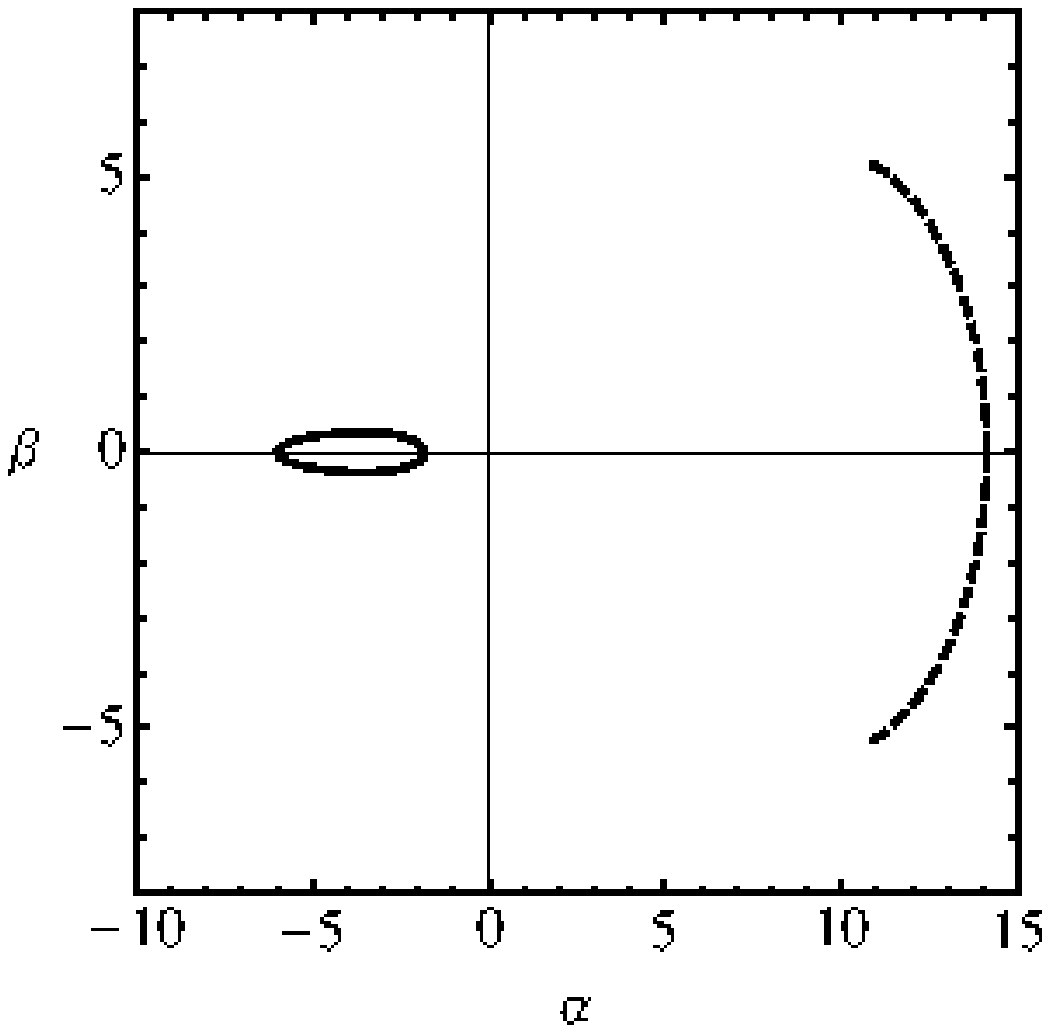}&
	\includegraphics[width=4cm]{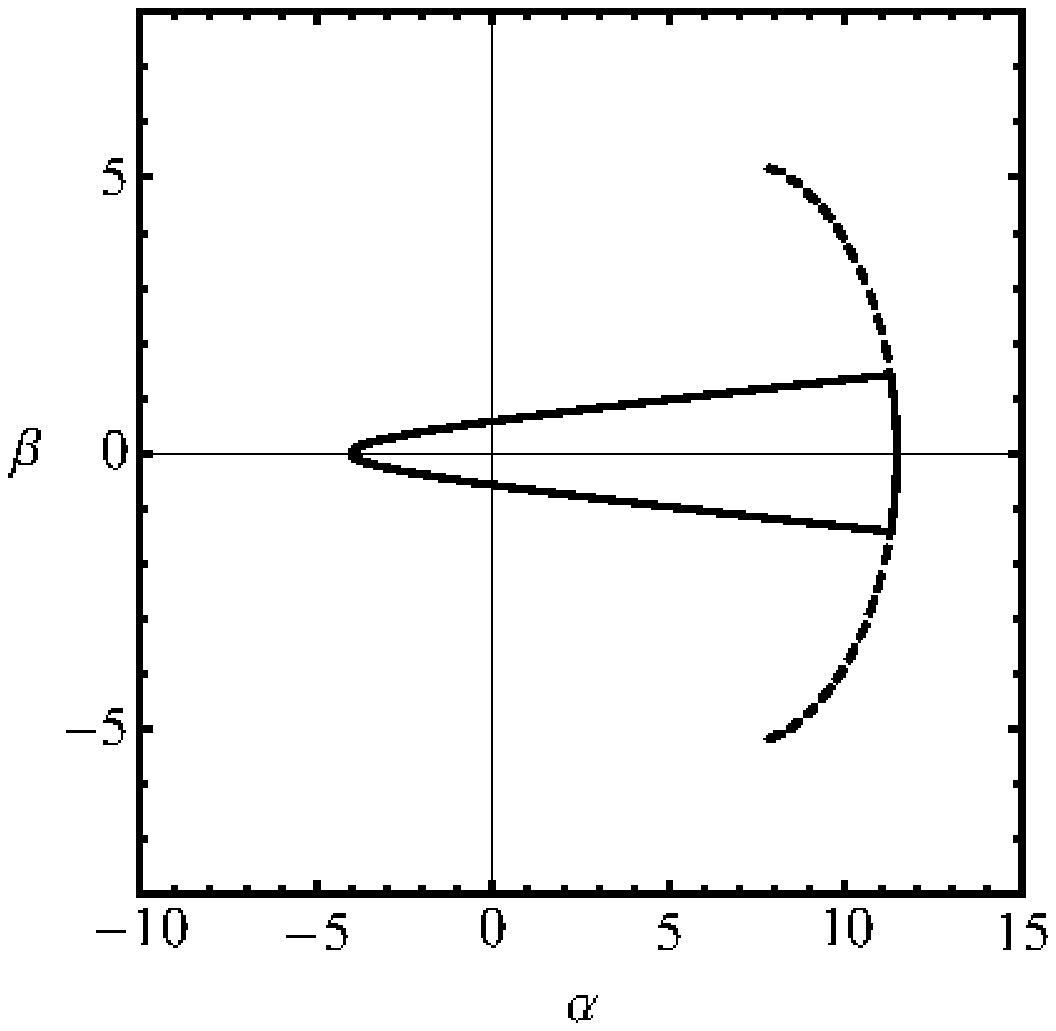}&\includegraphics[width=4cm]{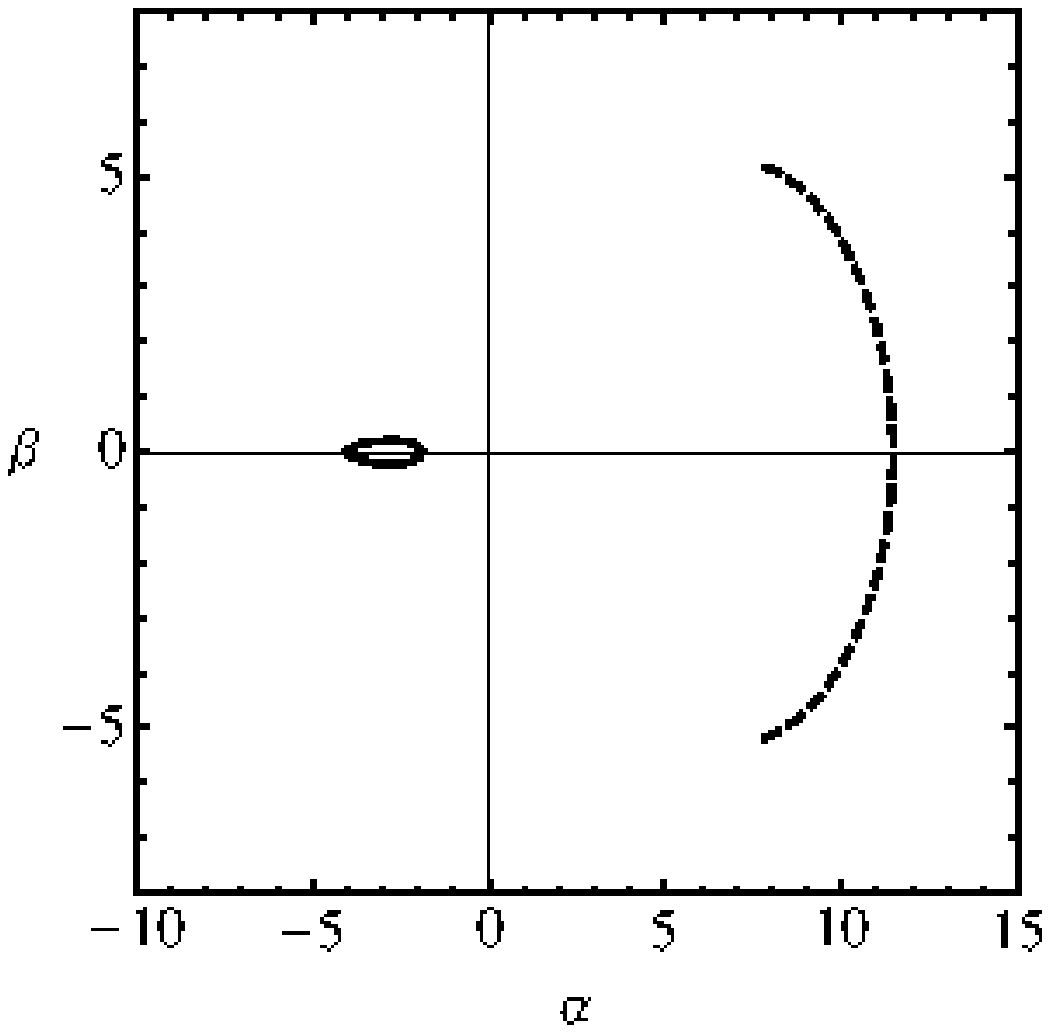}\\
	\includegraphics[width=4cm]{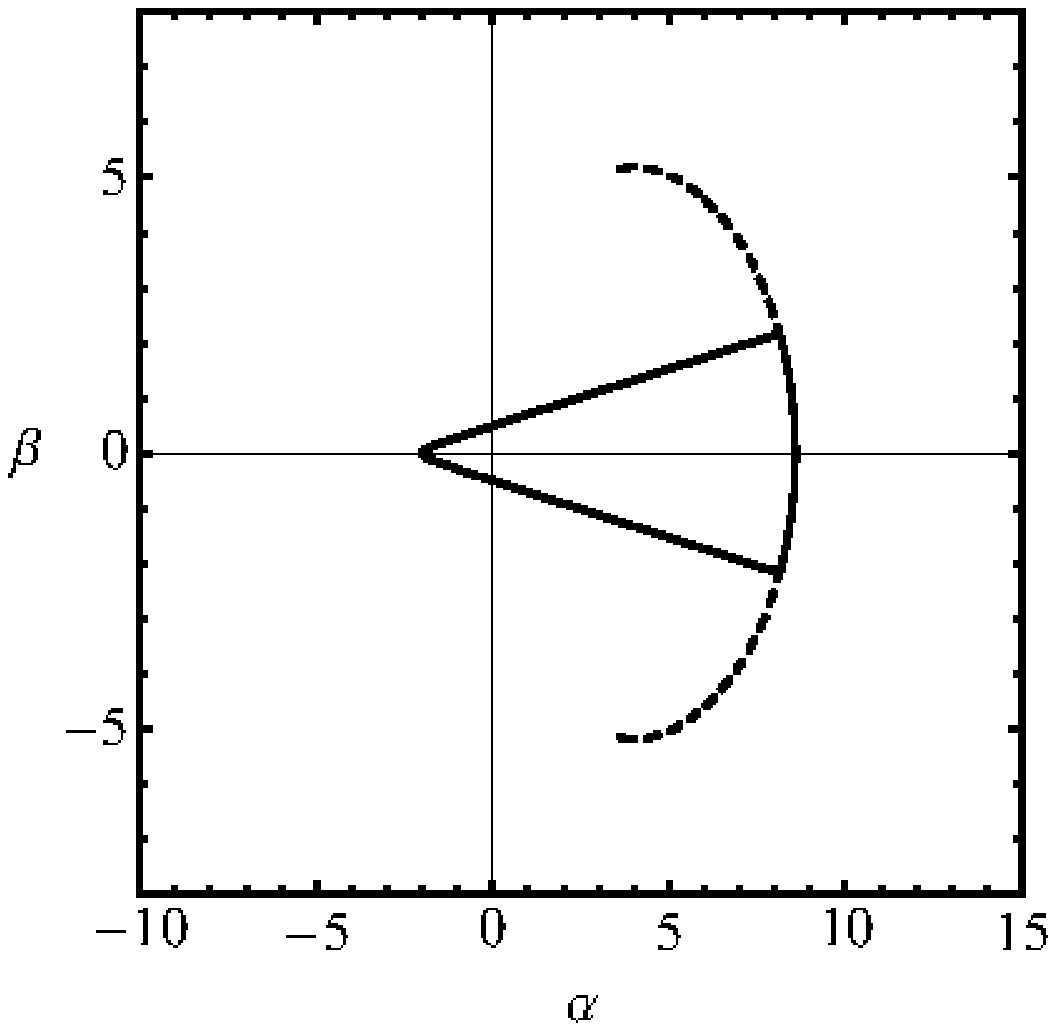}&\includegraphics[width=4cm]{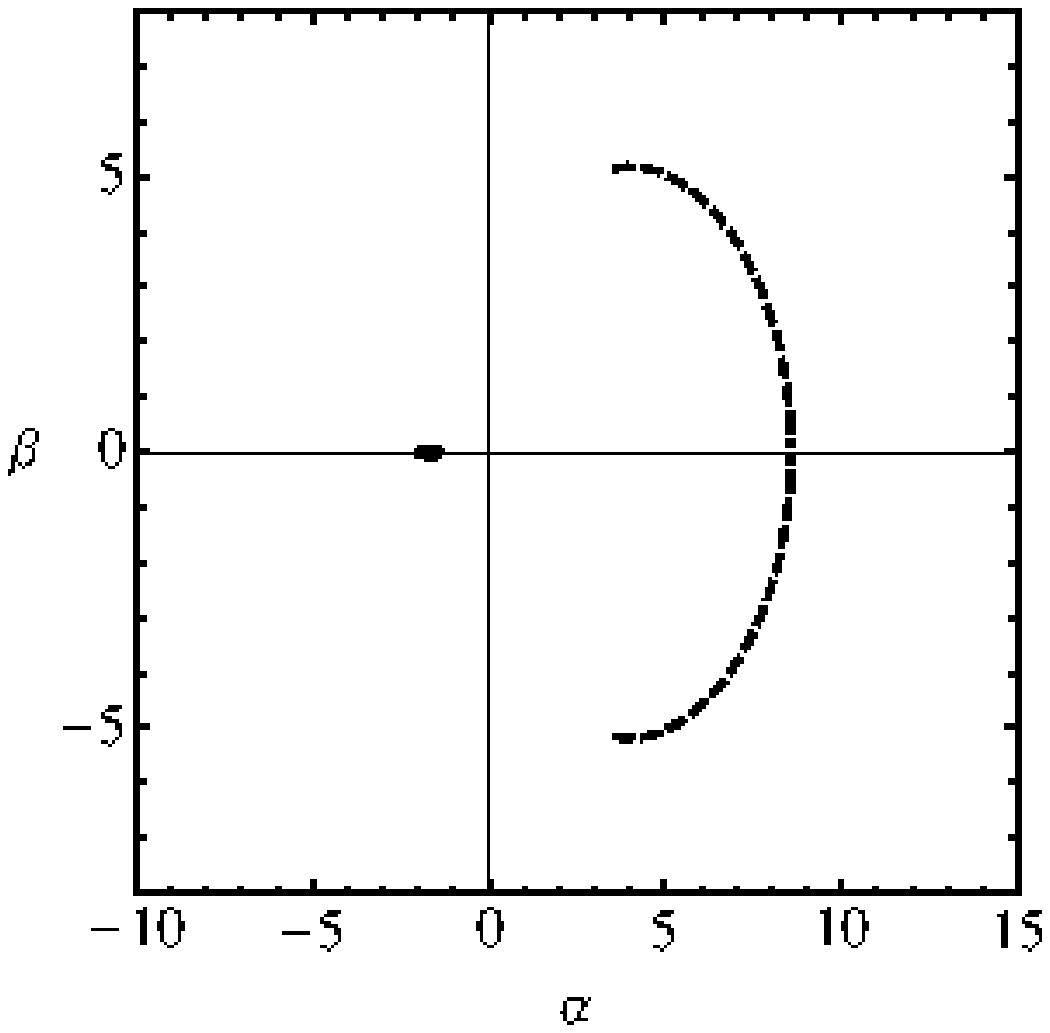}&
	\includegraphics[width=4cm]{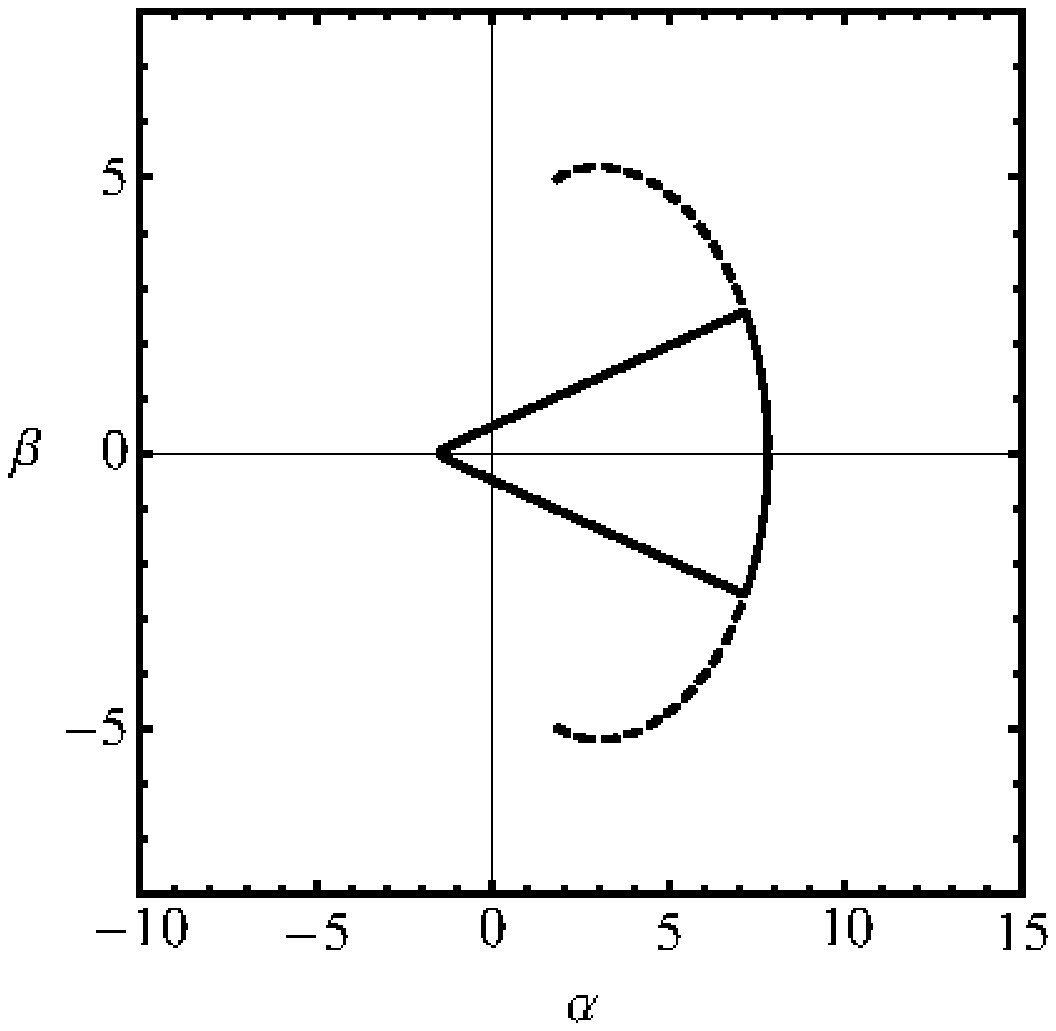}&\includegraphics[width=4cm]{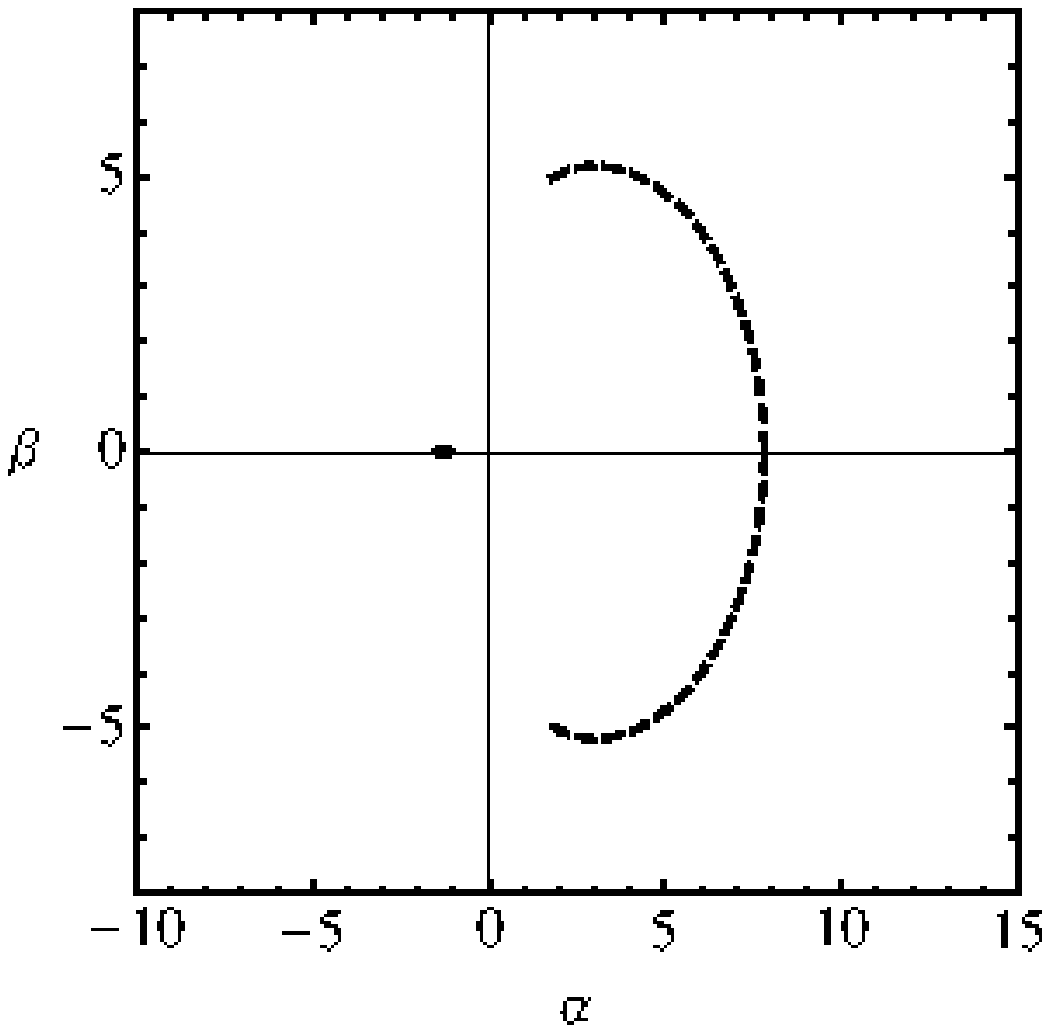}\\
	\includegraphics[width=4cm]{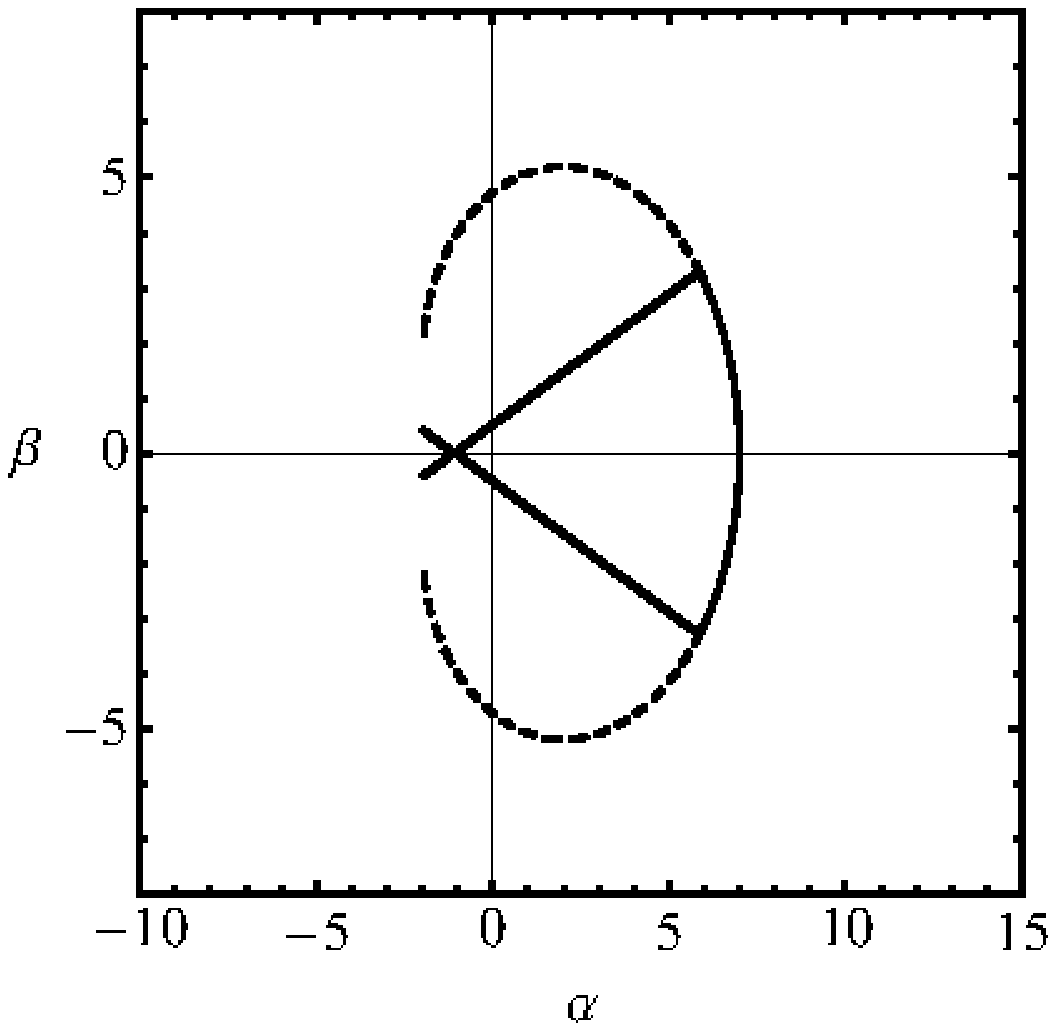}&\includegraphics[width=4cm]{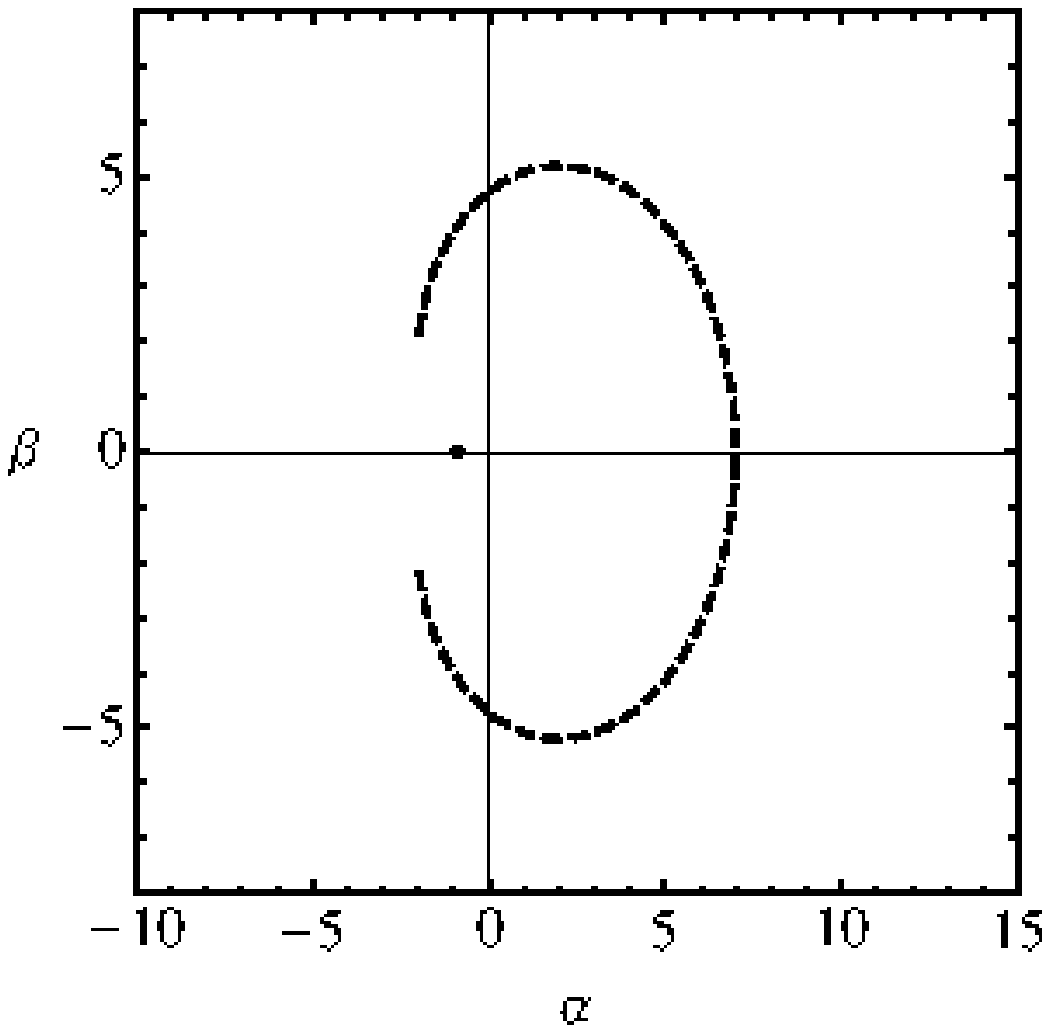}&&
	\end{tabular}
	\caption{\label{fig9}Silhouette of Kerr superspinars and naked singularities with identical spin is given for representative values of the spin $a=1.001,\, 1.5,\, 2.0,\, 4.0,\, 6.0$ (from bottom right to top left). In the twins, the superspinar(naked singularity) is at left (right).
      Solid line represents the surface of superspinar at $R=0.1$ (region of photons escaping to $-\infty$ in naked singularity case). The dashed line represents the arc corresponding to unstable spherical photon orbits. The distant observer is assumed at latitude $\theta_o=85^\circ$.}
	\end{center}
	\end{figure}

\begin{figure}[h!]
	\begin{center}
	\begin{tabular}{cccc}
	\includegraphics[width=4cm]{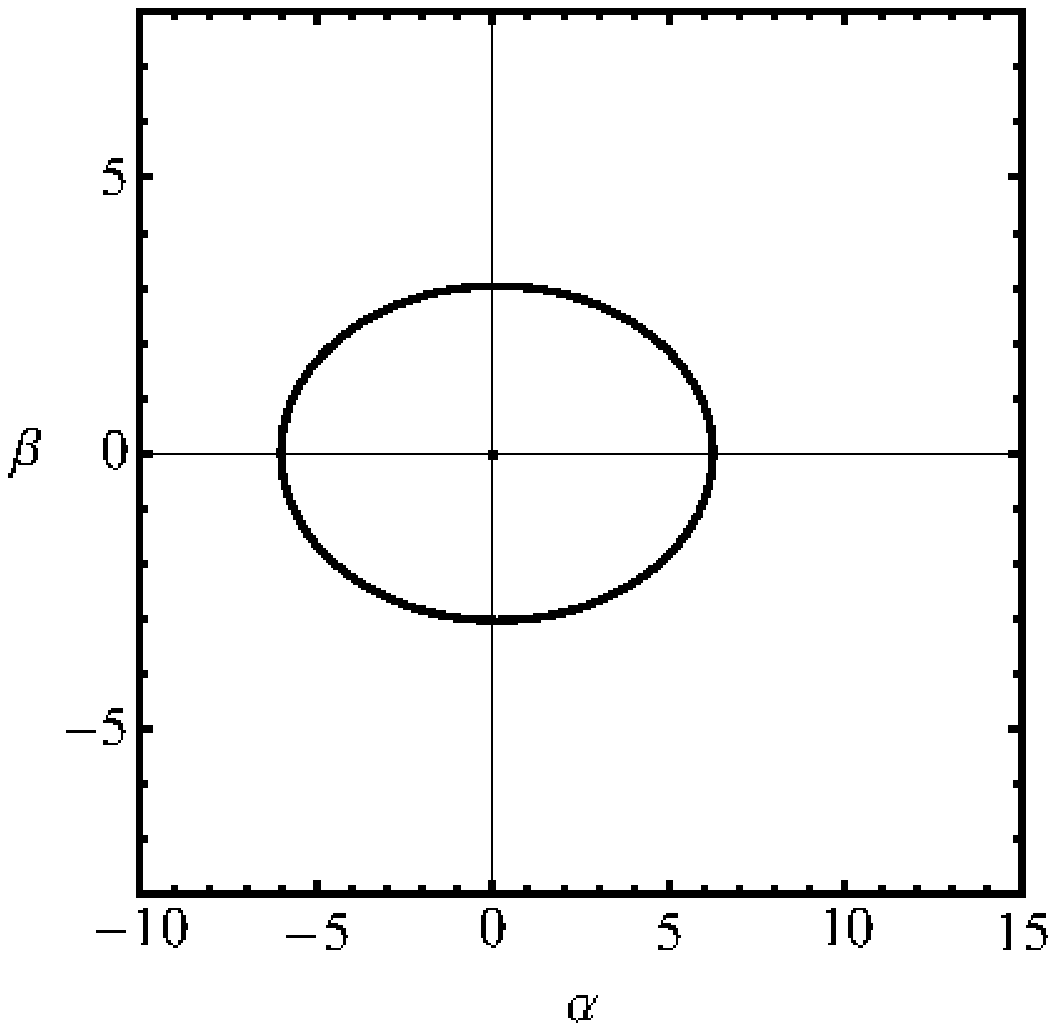}&\includegraphics[width=4cm]{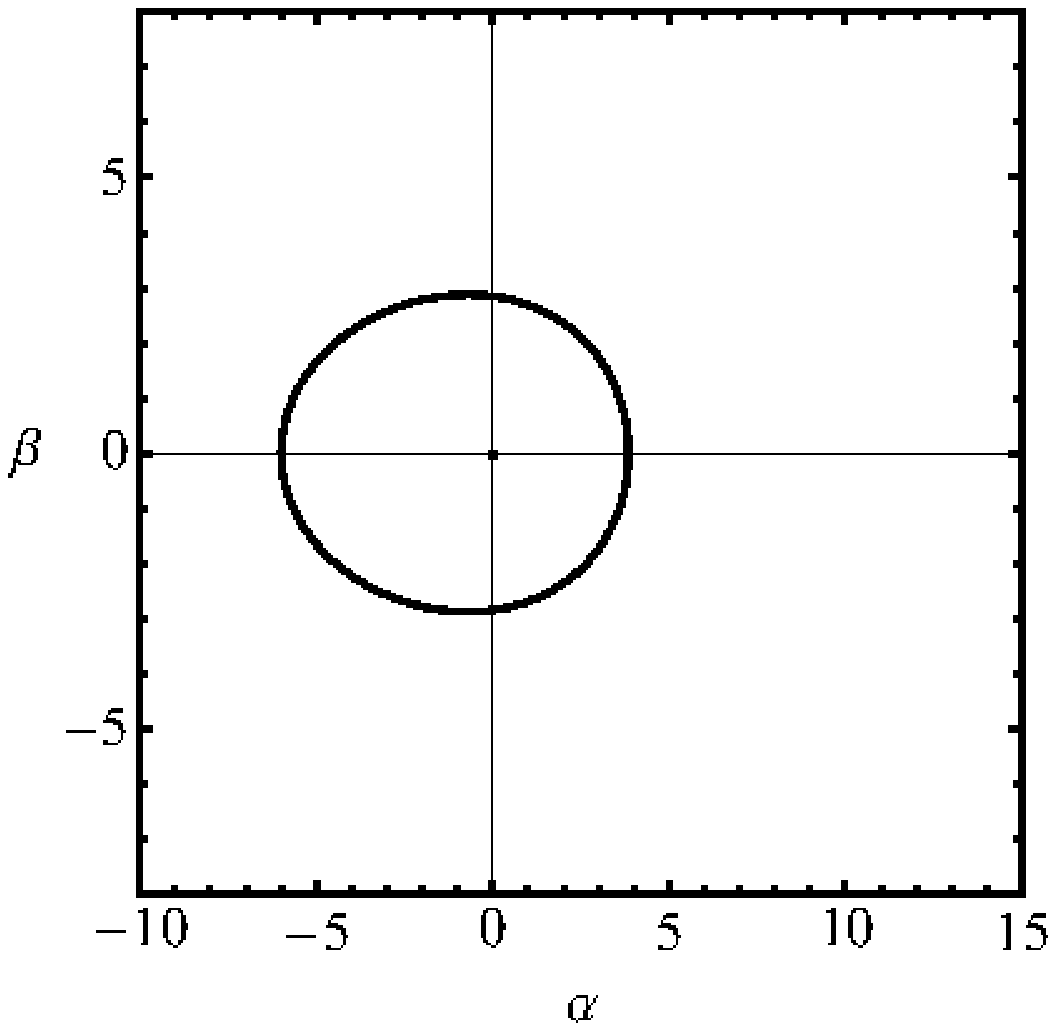}&
	\includegraphics[width=4cm]{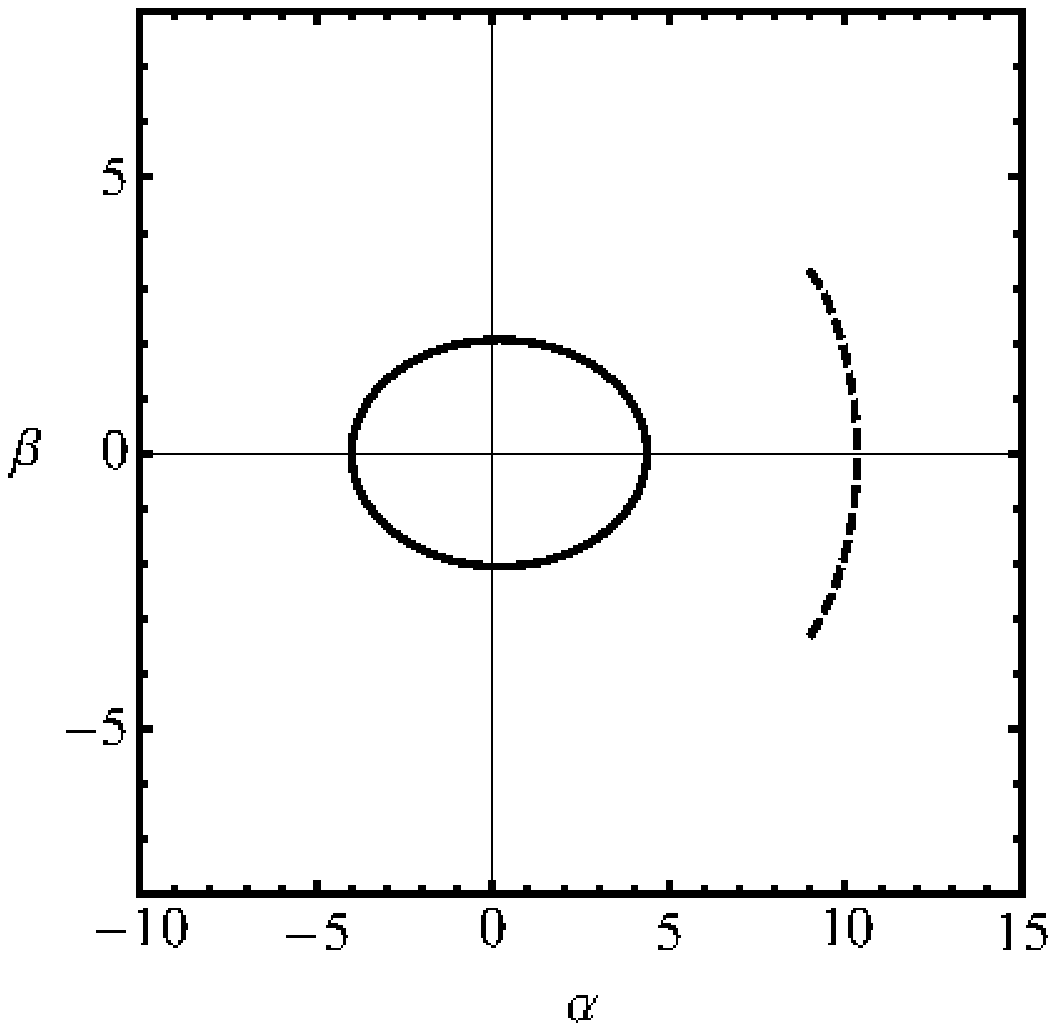}&\includegraphics[width=4cm]{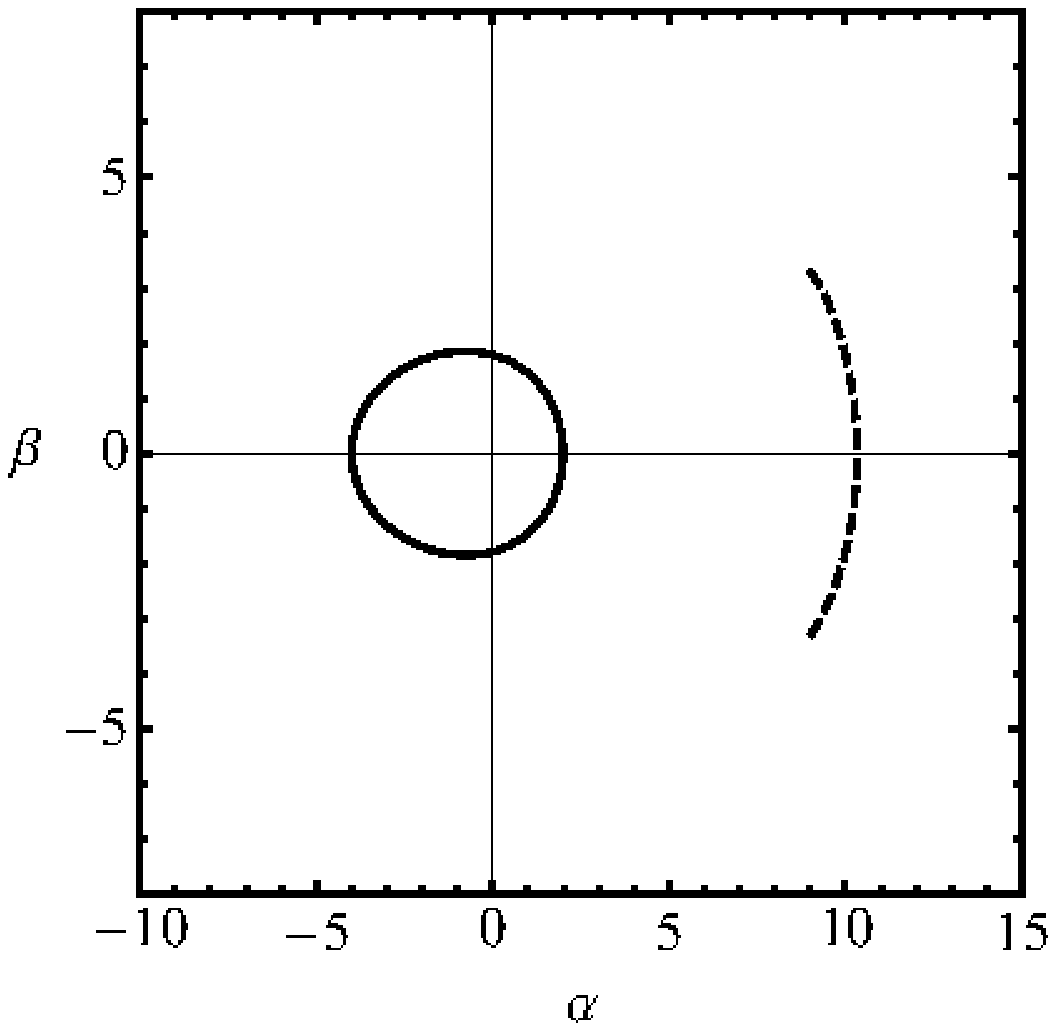}\\
	\includegraphics[width=4cm]{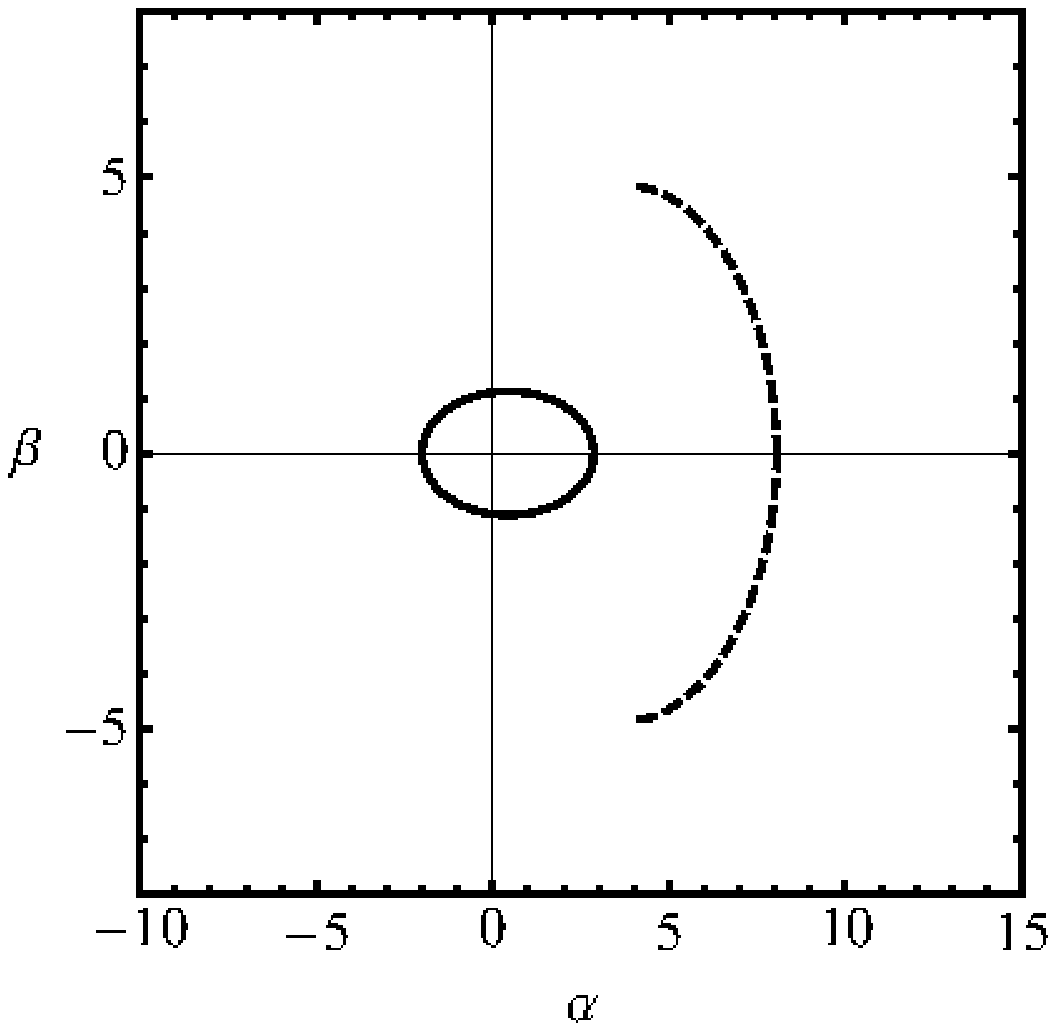}&\includegraphics[width=4cm]{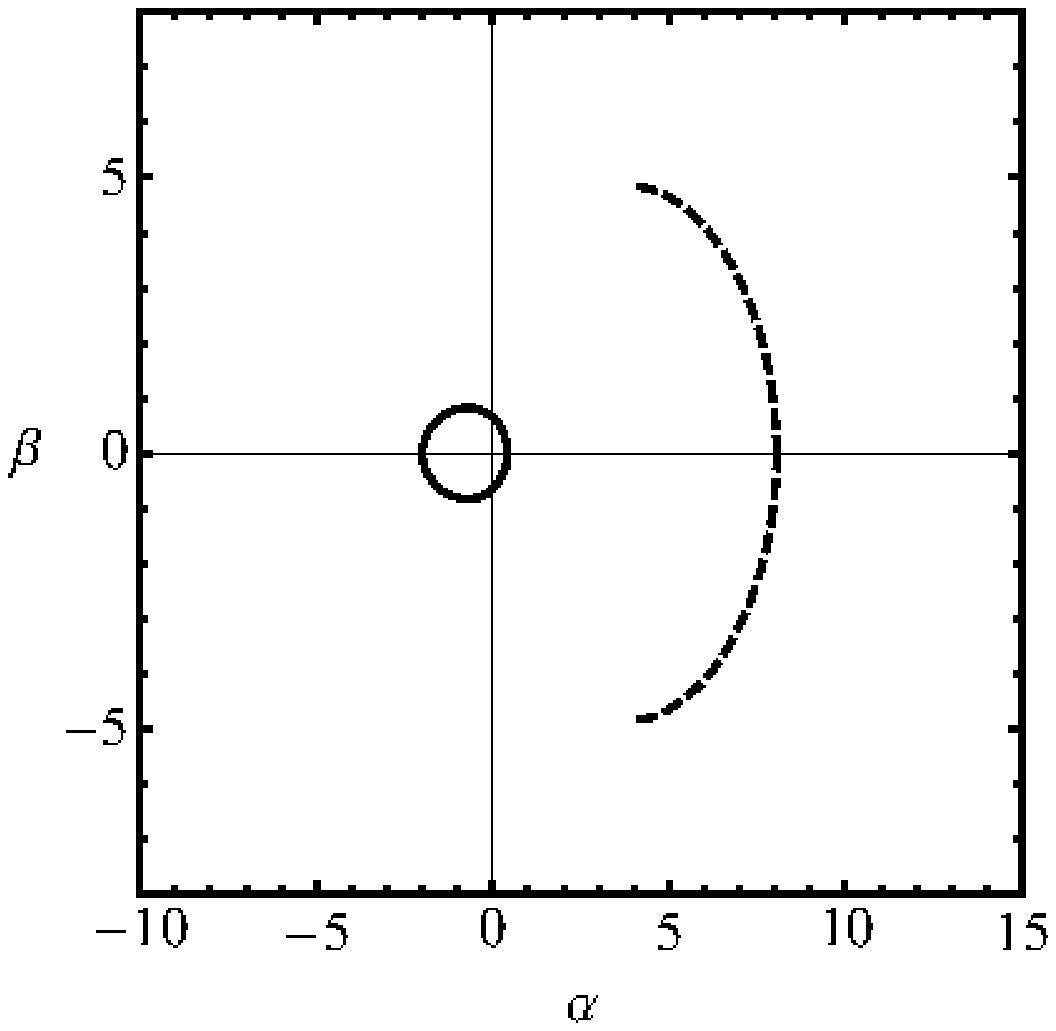}&
	\includegraphics[width=4cm]{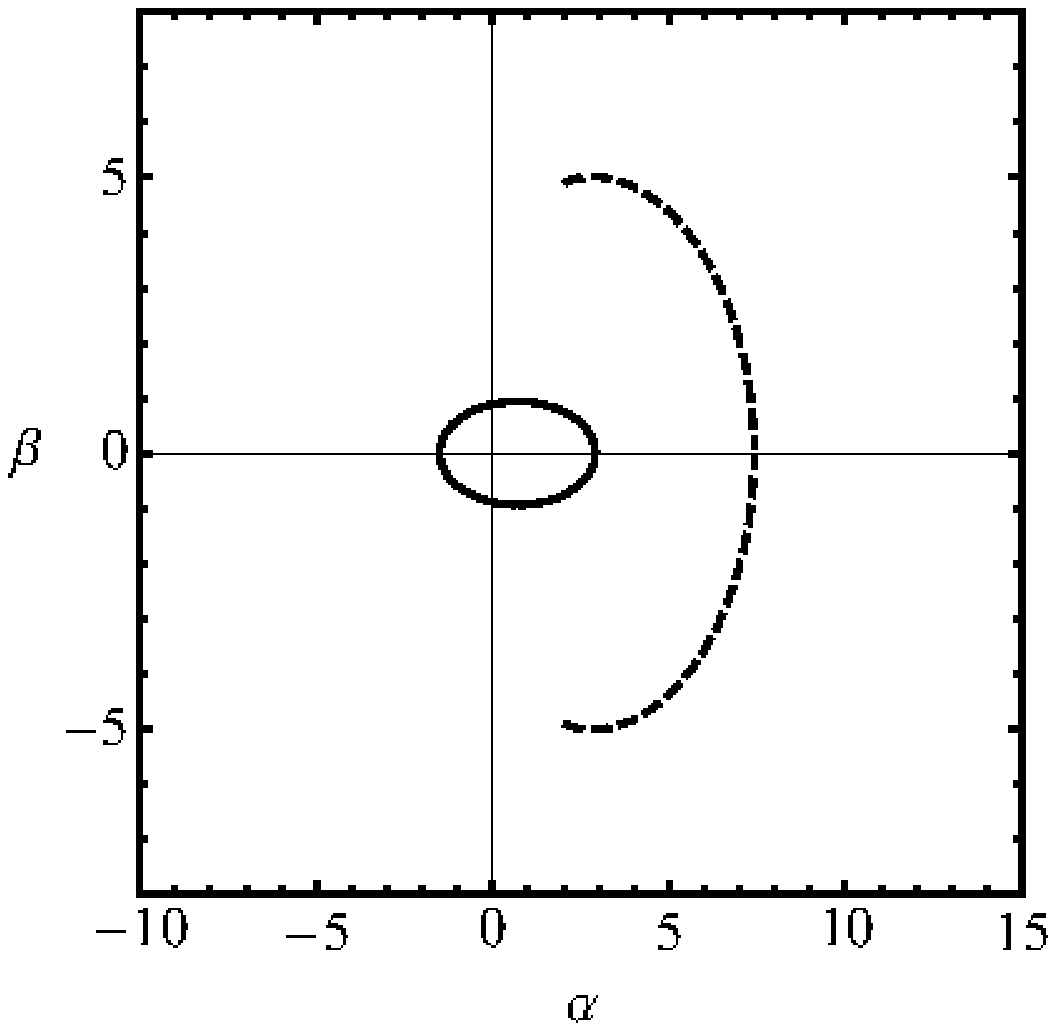}&\includegraphics[width=4cm]{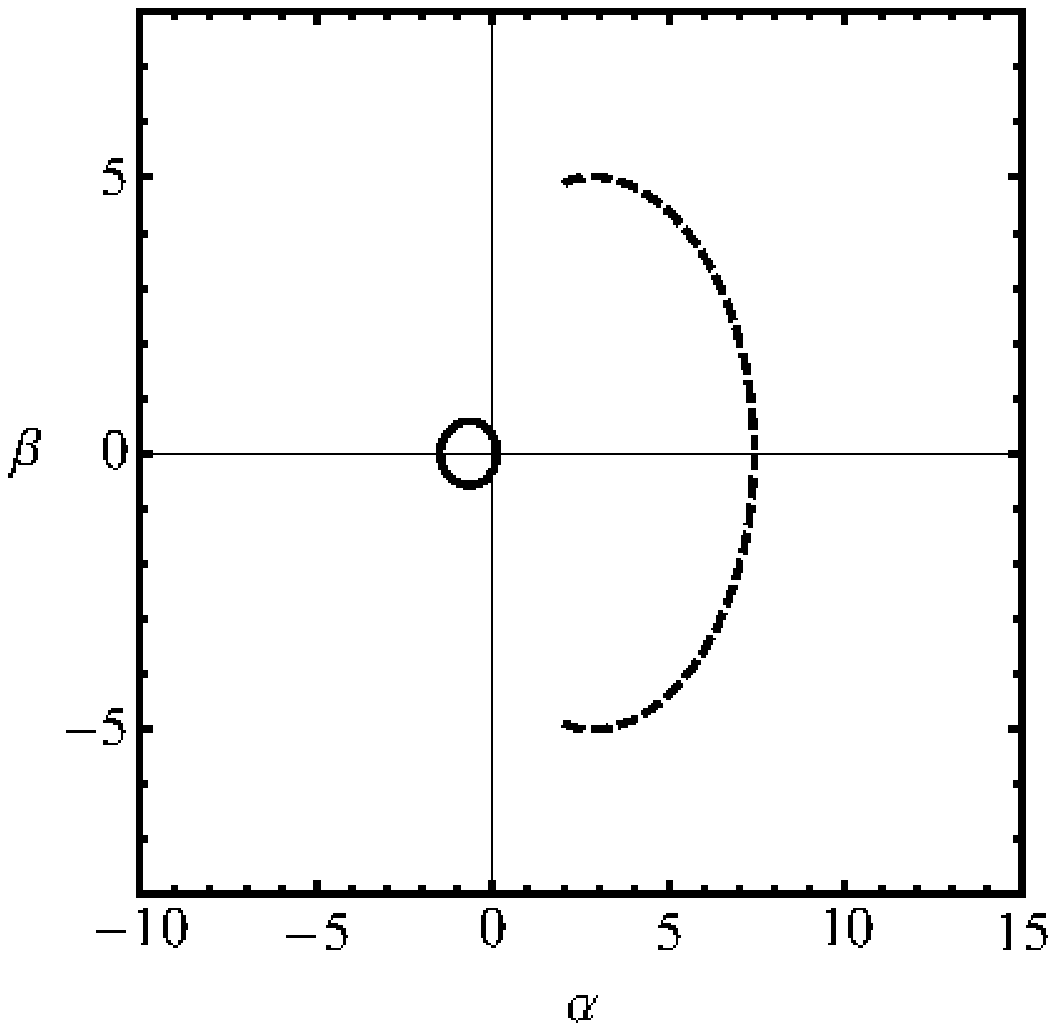}\\
	\includegraphics[width=4cm]{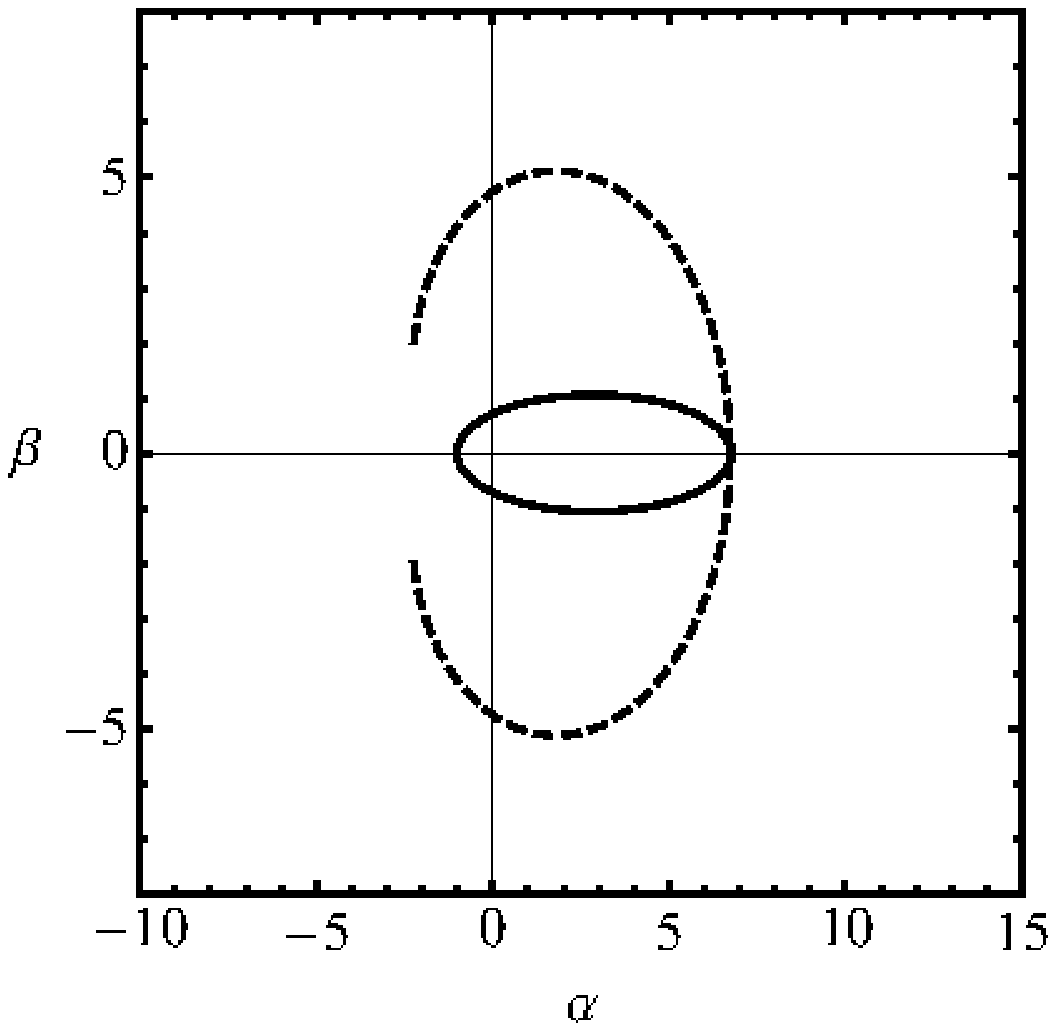}&\includegraphics[width=4cm]{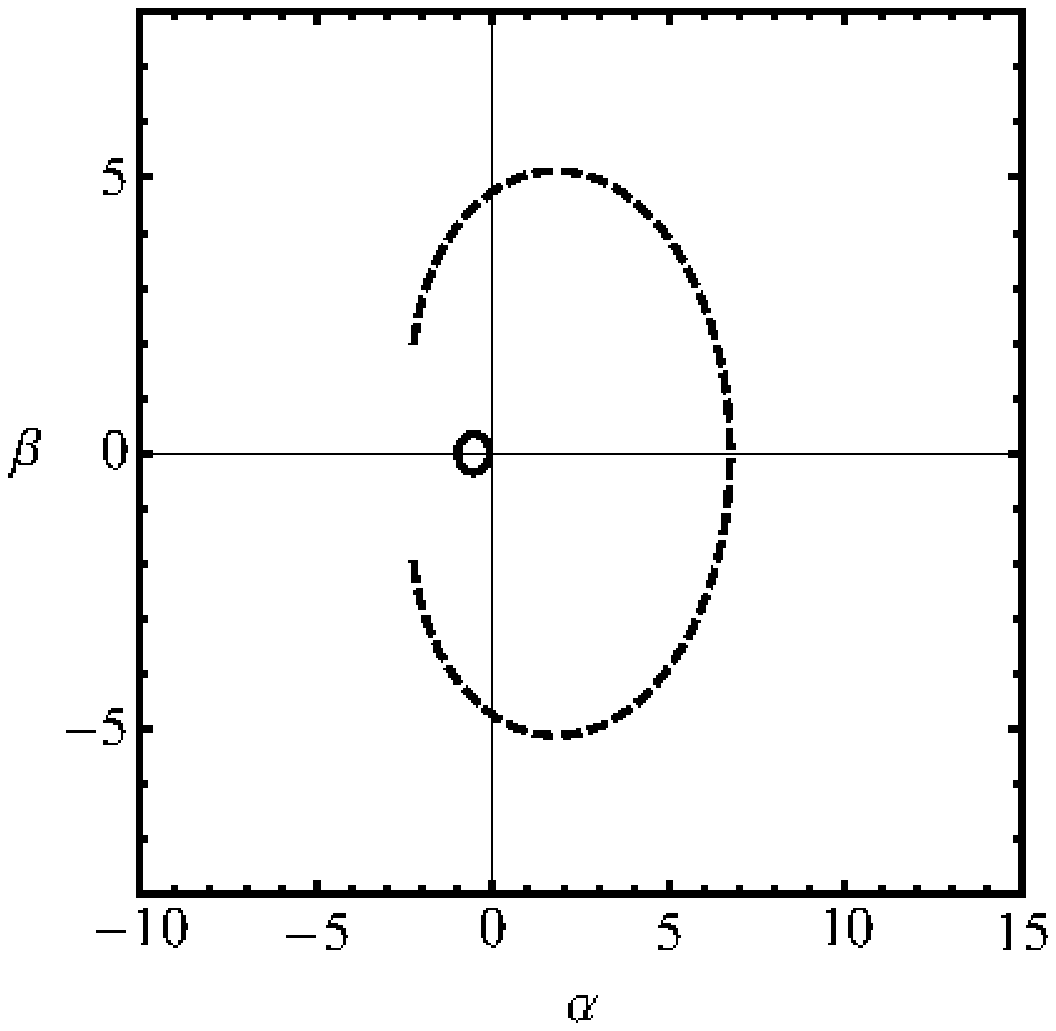}&&
	\end{tabular}
\caption{\label{fig10}Silhouette of Kerr superspinars and naked singularities with identical spin is given for representative values of the spin $a=1.001,\, 1.5,\, 2.0,\, 4.0,\, 6.0$ (from bottom right to top left). In the twins, the superspinar(naked singularity) is at left (right).
      Solid line represents the surface of superspinar at $R=0.1$ (region of photons escaping to $-\infty$ in naked singularity case). The dashed line represents the arc corresponding to unstable spherical photon orbits. The distant observer is assumed at latitude $\theta_o=60^\circ$.}
		\end{center}
	\end{figure}

We give an illustrative picture of the spin influence on the silhouette properties for two characteristic inclination angles $\theta_0=85^\circ$ (Fig.\ref{fig9}) and $\theta_0=60^\circ$ (Fig.\ref{fig10}) when the superspinar (naked singularity) rotational effects are very strong and mediate.

\begin{figure}
	\begin{center}
		\includegraphics[width=8cm]{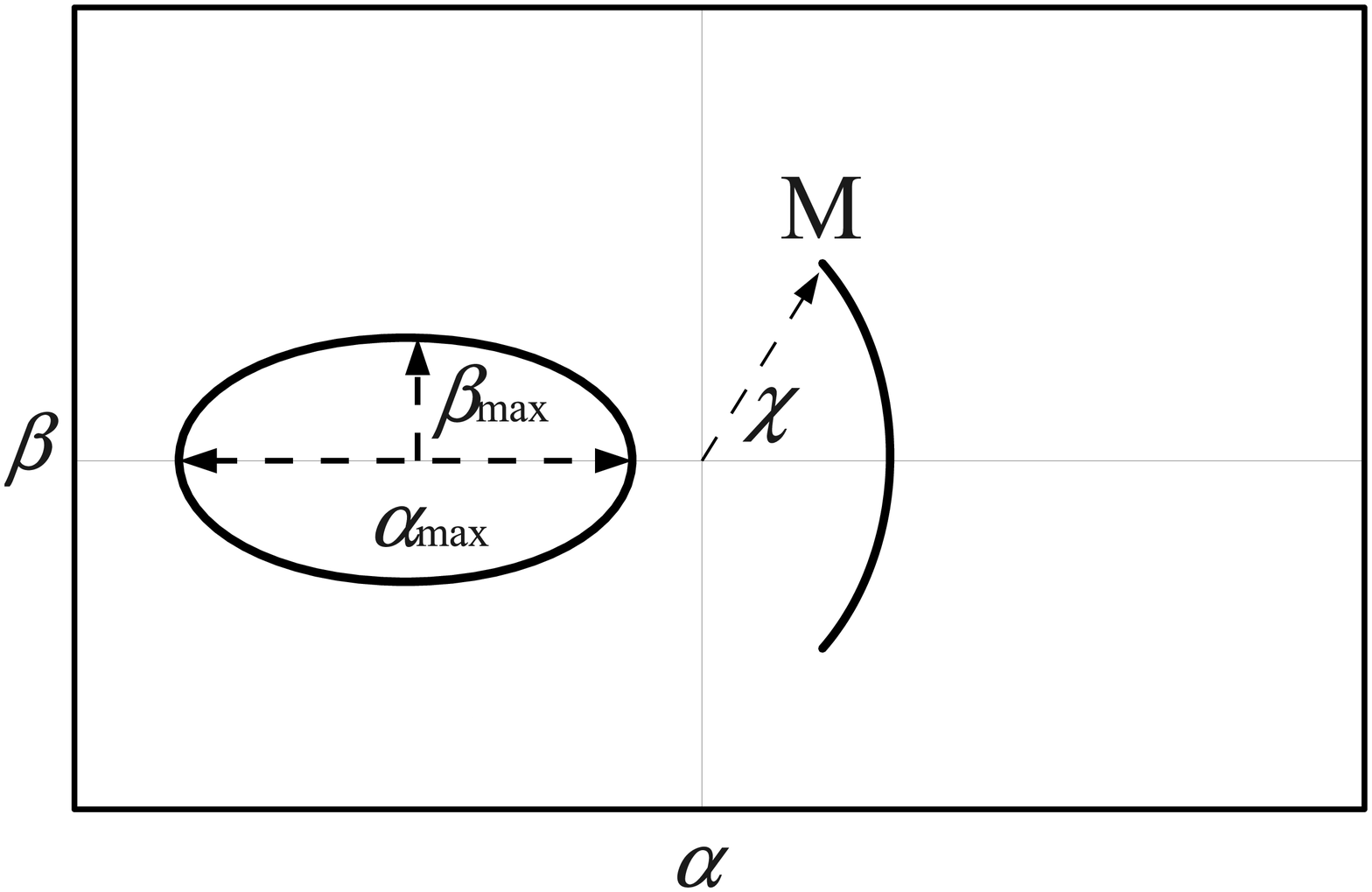}
		\caption{\label{fig11}Illustration of the $\chi$ and $\epsilon$ parameters definition.}
	\end{center}
\end{figure}

\begin{figure}
		\begin{center}
			\begin{tabular}{cc}
				\includegraphics[width=5.5cm]{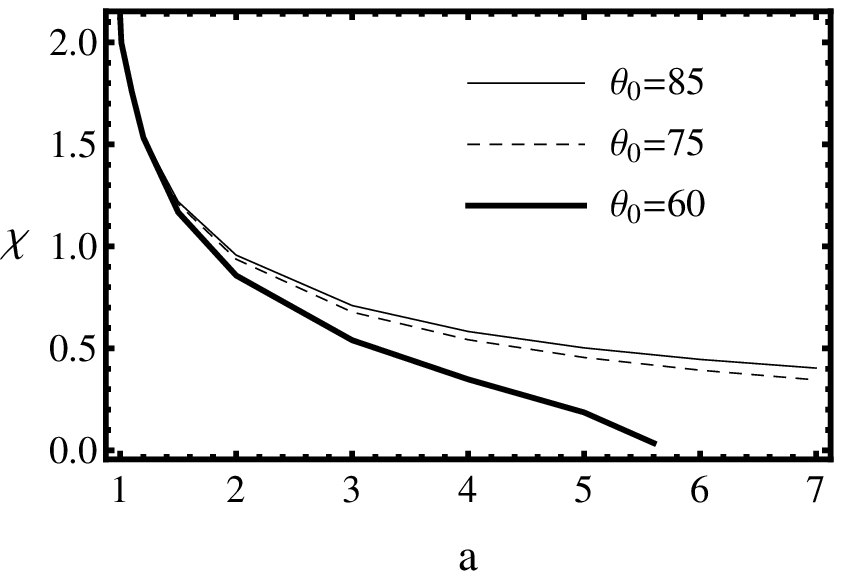}&\includegraphics[width=5.5cm]{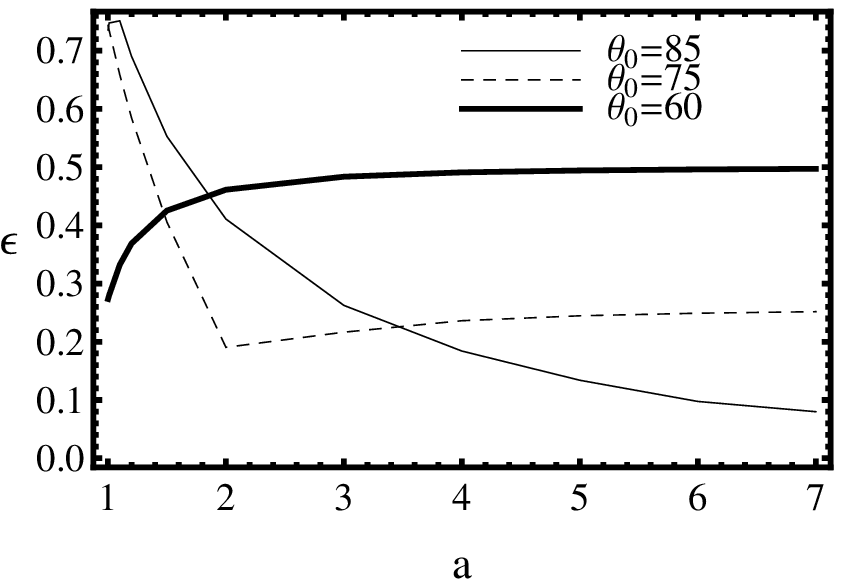}
			\end{tabular}
			\caption{\label{fig12}Left: plot of the $\chi=\chi(a)$ angle parameter as function of spin parameter $a$ for three representative values of observer's inclination $\theta_0=60^\circ.\, 75^\circ$ and $85^\circ$.  Right: plot of the $\epsilon=\epsilon(a)$ ellipticity parameter as a function of the spin parameter $a$ for three representative values of observer's inclination $\theta_0=60^\circ.\, 75^\circ$ and $85^\circ$. }
		\end{center}
\end{figure}

The silhouette shape can be efficiently used to determine the parameters of Kerr naked singularities \cite{Hio-Mae:2009:PhysRevD} and black holes \cite{Sch-Stu:2009:IJMPD,Hio-Mae:2009:PhysRevD}.  
In order to characterize the influence of the spin on the silhouette of a Kerr superspinar or naked singularity we define here two quantities in principle measurable by distant observers. The angle parameter $\chi$ of the silhouette arc   
\begin{equation}
 	\chi=\arctan\left(\frac{\beta_{M}}{\alpha_{M}}\right),\label{eqA}
\end{equation}
and the silhuette \emph{ellipticity} $\epsilon$
\begin{equation}
 	\epsilon=\frac{2\beta_{\mathrm{max}}}{\alpha_{\mathrm{max}}},\label{eqB}
\end{equation}
where $[\alpha_M, \beta_M]$ are coordinates of the edge point of the silhouette arc on the observers sky and $\alpha_{\mathrm{max}}$ and $\beta_{\mathrm{max}}$ are maximal width and height of the silhouette. The definition of the angle parameter $\chi$ and \emph{ellipticity} $\epsilon$ is illustrated in Fig.\ref{fig11}. (For an analogous definition of parameters characterizing black-hole silhouettes see \cite{Sch-Stu:2009:IJMPD,Hio-Mae:2009:PhysRevD,Tak-Tak:2010:arXiv:}.)
We calculated angle parameter $\chi$ and ellipticity $\epsilon$ of the Kerr naked singularity silhouette as a function of its spin taken from the interval $a=1.001 - 7.0$ (see Fig.\ref{fig12}).
These are quantities that could be measured and used for a Kerr naked singularity spin and the inclination angle of the observer estimates, if observational techniques could be developed to the level enabling the silhouette detailed measuring. Such a possibility seems to be realistic in near future for the case of the supermassive object (black hole or superspinar) predicted in the Galaxy Centre (Sgr $A^*$) \cite{Bro-Loe-Nar:2009:APJ,Sch-Stu:2009:IJMPD}. Since the ellipticity parameter is not giving unique predictions in dependence on the inclination angle $\theta_{0}$ (see Fig.\ref{fig12}), probably some more convenient silhouette parameter has to be introduced. 

Clearly, we have to introduce three properly chosen characteristics of the superspinar silhouette and its arc in order to determine three parameters related to observed superspinars - namely spin, inclination angle of the observer and superspinar surface radius. For Kerr naked singularities only the spin and inclination angle have to be determined using two characteristics of the silhouette and its arc as shown in \cite{Hio-Mae:2009:PhysRevD}. In the case of Kerr superspinars, an additional silhouette-characteristic parameter related to the superspinar surface radius $R$ has to be introduced. Such a parameter has to reflect relative positions of the silhouette and the arc, however, it is not clear yet if such a parameter could be uniquely related to the superspinar parameters and the observer inclination angle.

It is evident that the problem of defining the three superspinar silhouette and its arc parameters and relating them to the three parameters $(a, \theta_{0}, R)$ is much more complex in comparison with the problem of relating two Kerr naked singularity parameters and two silhouette and arc characteristics as presented in \cite{Hio-Mae:2009:PhysRevD}. It will be discussed in a future paper.

\clearpage

\section{\label{sec:DirAndIndirImages}Optical appearance of radiating Keplerian disc}

We focus our attention to images of radiating Keplerian discs orbiting a Kerr superspinar. Using the results of findings of the local escape cones, we can obtain the appearance of the discs as seen by distant observers in dependence on the superspinar spin and the viewing angle of the observer and then, at least in principle, obtain estimates on the astrophysically acceptable values of the spin $a$. 

\subsection{Direct and indirect images of isoradial geodesics}
	
Calculating images of an Keplerian accretion disc (ring) in the equatorial plane is the basic step in treating the optical phenomena in the field of a Kerr superspinar. Generally, one could obtain a direct and an indirect image (see Figs \ref{fig19} and \ref{fig20}), but in special cases the situation can be much more complicated due to complex character of the latitudinal and azimuthal photon motion in the Kerr backgrounds. We concentrate our attention to the direct and indirect images of isoradial geodesics. We consider situations with radius fixed to ISCO in the inner edge of the radiating disc and with radius fixed in terms of the superspinar mass ($20M$) in the outer edge of the disc.

In order to find all relevant positions of points forming the rotating ring on observer's sky, we have to find photon trajectories between the ring particles and the observer, i.e., we seek for such doubles of local observational angles $[\alpha_0,\beta_0]$ that satisfy the condition \cite{Rau-Bla:1994:ASTRJ2:}

\begin{equation}
 	I_U(\alpha_0,\beta_0;n_u,u_{sgn}) - I_M(\alpha_0,\beta_0;n,p,s)=0.\label{bvp}
\end{equation}
Here we introduced the modified radial coordinate $u=1/r$ and cosine of the latitudinal coordinate $\mu=\cos\theta$ \cite{Rau-Bla:1994:ASTRJ2:}. In the condition (\ref{bvp}), $n_u$ is the number of turning points in $u$ coordinate, $n$ is the number of turning points passed in $\mu$ coordinate, $p=mod(n,2)$, $s=(1-\mu_{sgn})/2$. In terms of $u$ and $\mu$ we define the functions $I_U$ and $I_M$ by

\begin{equation}
	I_U(\alpha_0,\beta_0;n_u,u_{sgn})\equiv\left\{\begin{array}{lcr}
					-u_{sgn}\left(\int^{u_0}_{u_t} +\int^{u_e}_{u_t}\right) & \textrm{for} & n_u=1\\
					u_{sgn}\int^{u_e}_{u_0} & \textrm{for} & n_u=0
					\end{array}\right.
\end{equation}
and
\begin{eqnarray}
	I_M(\alpha_0,\beta_0;n,p,s)&\equiv&\mu_{sgn}\left[\int^{\mu_+}_{\mu_0} + (-1)^{n+1}\int^{\mu_+}_{\mu_e}+\right.\\ \nonumber
&+&\left.(-1)^s[(1-p)n+p[(1-s)(n-1)+s(n+1)]]\int^{\mu_+}_{\mu_-} \right]
\end{eqnarray}
 with
\begin{eqnarray}
	\int^{u_2}_{u_1}&\equiv&\int^{u_2}_{u_1}\frac{\diff u}{\sqrt{U(u)}},\label{u_int}\\
	U(u)&=&1+(a^2-\lambda^2-q)u^2+2[(\lambda^2-a^2)^2+q]u^3 - \nonumber\\
	&-&[q a^2]u^4
\end{eqnarray}
and

\begin{eqnarray}
	\int^{\mu_2}_{\mu_1}&\equiv&\int^{\mu_2}_{\mu_1}\frac{\diff \mu}{\sqrt{M(\mu)}},\label{mu_int}\\
	M(\mu)&=&q+(a^2-\lambda^2-q)\mu^2-a^2\mu^4.
\end{eqnarray}

\subsection{Integration of photon trajectories}

We express the integrals (\ref{u_int}) and (\ref{mu_int}) in the form of the standard elliptic integrals of the first kind. Rauch and Blandford presented the tables of reductions of $u$-integrals and $\mu$-integrals for the case of photons in Kerr geometry \cite{Rau-Bla:1994:ASTRJ2:}. Here we use those reductions referring the reader to the original paper \cite{Rau-Bla:1994:ASTRJ2:} for details.

There are two cases we distinguish in the latitudinal integral. In the first case there is one positive, $M_+>0$, and one negative, $M_-<0$, root of $M(m^2)$ - it implies that there are two turning points located symmetrically about the equatorial plane given by $\pm\sqrt{M_+}$ and corresponding to the so called orbital motion \cite{Bic-Stu:1976:BULAI:,Fel-Cal:1972:NCimB}. In the second case there are two positive roots, $0<M_-<M_+$ of  $M(m^2)$, which implies that the latitudinal motion is constrained to the region above or below of the equatorial plane (so called vortical motion). Being concentrated on the equatorial Keplerian discs, we can restrict our attention only onto the orbital photon motion.  
\par
Using transformations published in \cite{Rau-Bla:1994:ASTRJ2:},  we can write the integrals (\ref{u_int}) and (\ref{mu_int})  in the form

\begin{equation}
 \int^{u}_{u_1}\frac{1}{\sqrt{U(\tilde{u})}}\diff \tilde{u} = c_1\mathcal{F}(\Psi;m)\label{ellint}
\end{equation}
and 
\begin{equation}
 \int^{\mu}_{\mu_1}\frac{1}{\sqrt{M(\tilde{\mu})}}\diff \tilde{\mu} = c_1\mathcal{F}(\Psi;m)\label{ellintM}
\end{equation}
where $\mathcal{F}$ is the elliptic integral of the first kind and $u_1$(resp $\mu_1$) depends on the case of root distribution of the quartic equation $U(u)=0$ (resp. $M(\mu)=0$) - see \cite{Rau-Bla:1994:ASTRJ2:} for details. 

 We consider two basic possibilities of trajectories, namely those corresponding to direct images when photons do not cross the equatorial plane and indirect images when photons cross once the equatorial plane.

\subsection{Disc images}

	It is very important and instructive to demonstrate the influence of the Kerr superspinar spin on the shape of images of rings in the equatorial plane representing parts of Keplerian accretion discs. Of course, as well known from the studies of the optical phenomena in the field of Kerr (and even Schwarzchild) black holes, the images strongly depend on the latitude of the observer. We calculate the direct and indirect images of flat discs for two representative values of viewing angle $\theta_0$ and appropriately chosen extension of radiating disc area. We present also the total images constructed from both direct and indirect parts assuming that the disc is opaque, capturing thus the corresponding part of the indirect image.


	We include the effect of frequency shift into the calculated images of the Keplerian discs assumed to be radiating at a given fixed frequency corresponding, e.g., to some emission line. The frequency shift is characterized by the $g$-factor that is determined by the ratio of observed ($E_0$) to emitted ($E_e$) photon energy
\begin{equation}
 g=\frac{E_0}{E_e}=\frac{k_{0\mu} u_0^\mu}{k_{e\mu} u_e^\mu},
\end{equation}
where $u_0^\mu$($u_e^\mu$) are components of the observer (emitter) 4-velocity and $k_{0\mu}(k_{e\mu})$ are components of the photon 4-momentum taken at the moment of emission (observation). For distant observers $u^\mu_0=(1,0,0,0)$. The emitter follows an equatorial circular geodesics at $r=r_e$, $\theta_e=\pi/2$. Therefore, $u_e^\mu=(u^t_e,0,0,u_e^\varphi)$, with components given by

\begin{eqnarray}
 u_e^t &=& \left[1-\frac{2}{r_e}(1-a\Omega )^2-(r_e^2+a^2)\Omega^2 \right],\\
 u_e^\varphi &=& \Omega u_e^t,
\end{eqnarray}
where $\Omega=\diff\varphi/\diff t$ is the Keplerian angular velocity of the emitter related to distant observers, given by equation (\ref{ang_vel_gf}).

The frequency shift including all relativistic effects is then given by

\begin{equation}
 g=\frac{\left[1-\frac{2}{r_e}(1-a\Omega)^2-(r_e^2+a^2)\Omega^2\right ]^{1/2}}{1-\lambda\Omega}
\end{equation}
where $\lambda\equiv-k_\varphi/k_t$ is the impact parameter of the photon being a motion constant for an individual photon radiated at a specific position of the radiating disc; notice that $g$ is independent of the second photon motion constant  (impact parameter) $q$. Of course, depending on the position of the emitter along the circular orbit, the impact parameters
$\lambda$, $q$ of photons reaching a fixed distant observer will vary periodically (see, e.g., \cite{Bao-Stu:1992:ASTRJ2:,Stu-Bao:1992:GENRG2:}). For each position of the emitter the impact parameters are determined by the procedure of integration of photon trajectories. 

The appearance of the Keplerian disc including the influence of the frequency shift on the disc images is demonstrated in Figs \ref{fig13} and \ref{fig14}. To demonstrate the frequency shift influence in the clearest form, we assume a disc radiating at a specific frequency (energy) corresponding to the Fe emission line and the radiating part of the Keplerian disc limited by the inner radius at ISCO $r_{ms}(a)$ and the outer radius at $r=20M$. The role of the spin is illustrated both for small ($\theta_0=30^\circ$) and high ($\theta_0=80^\circ$) inclination angles. 


\begin{figure}
	\begin{center}
	\begin{tabular}{ccc}
		\parbox[c]{6cm}{\includegraphics[width=6cm]{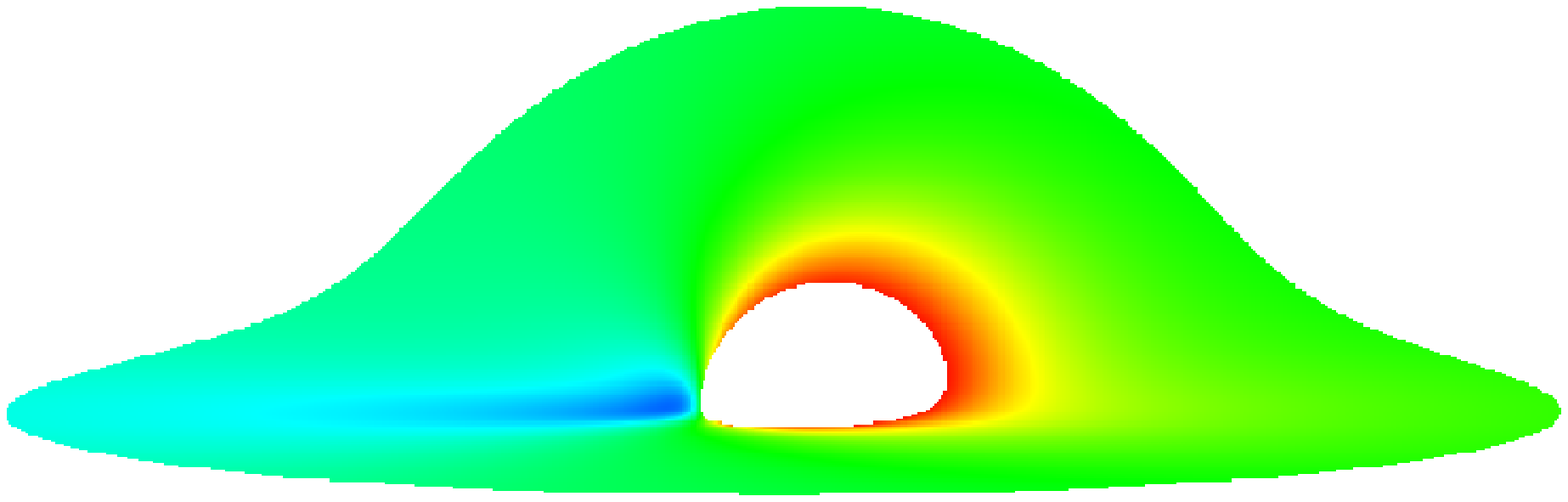}}&\parbox[c]{3cm}{\includegraphics[width=3cm]{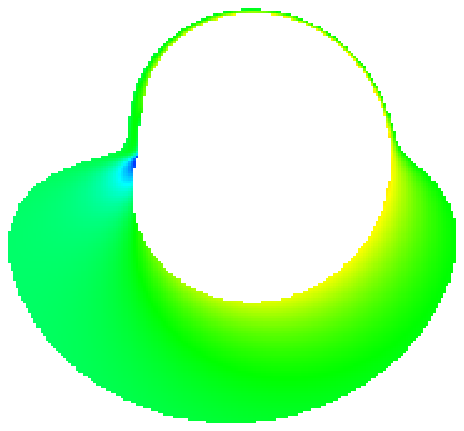}}&\parbox[c]{6cm}{\includegraphics[width=6cm]{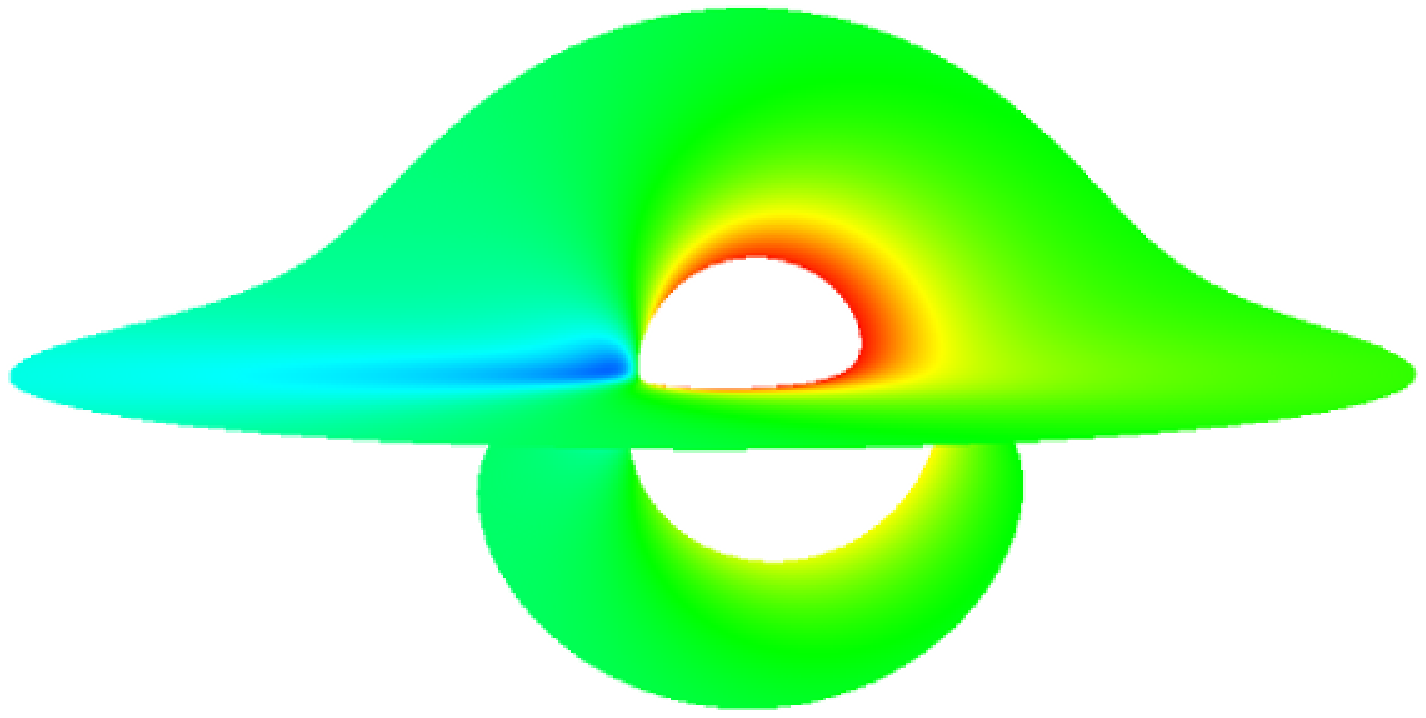}}\\
		\parbox[c]{6cm}{\includegraphics[width=6cm]{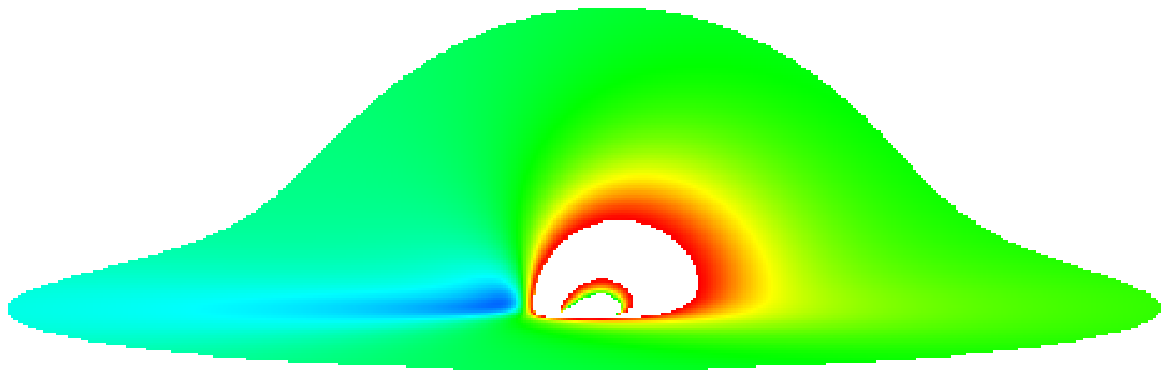}}&\parbox[c]{3cm}{\includegraphics[width=3cm]{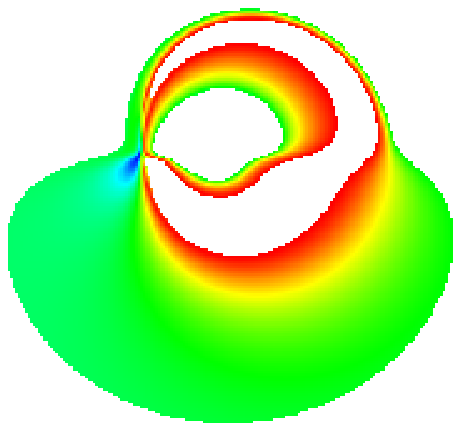}}&\parbox[c]{6cm}{\includegraphics[width=6cm]{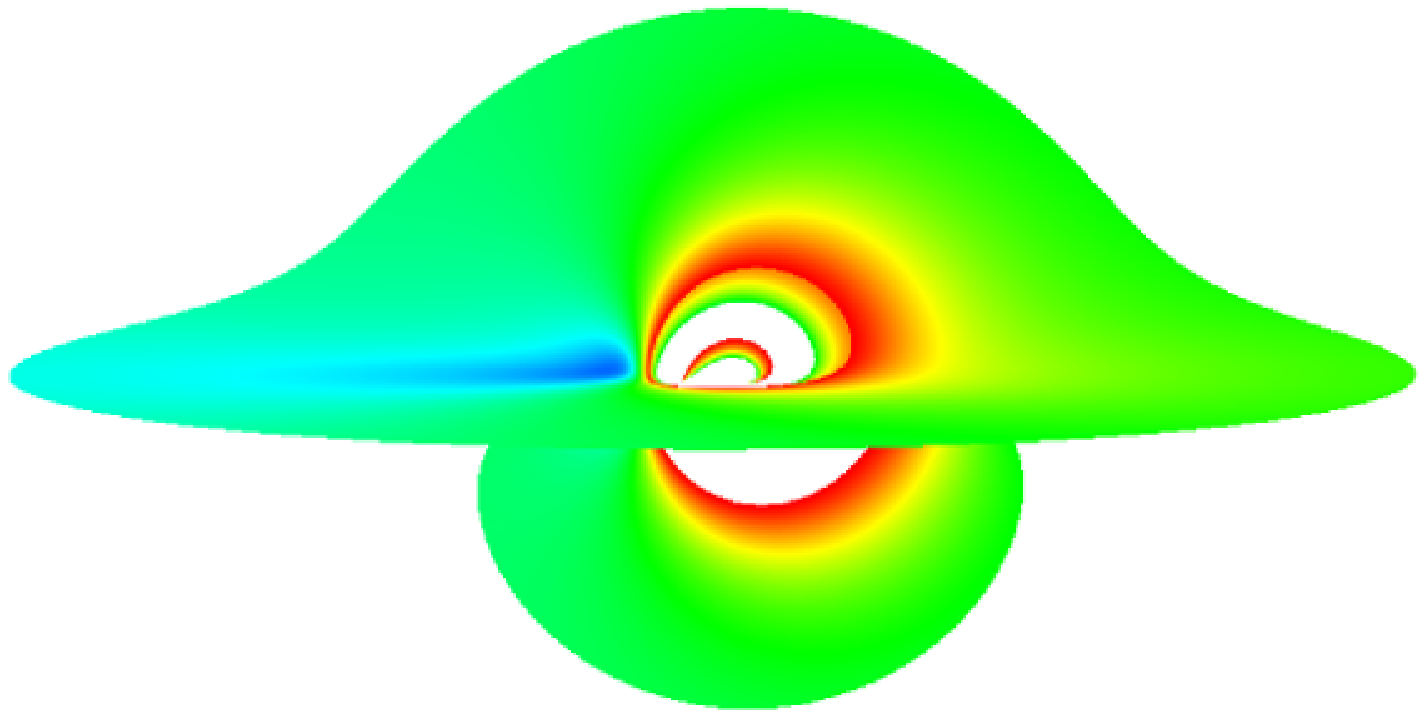}}\\
		\parbox[c]{6cm}{\includegraphics[width=6cm]{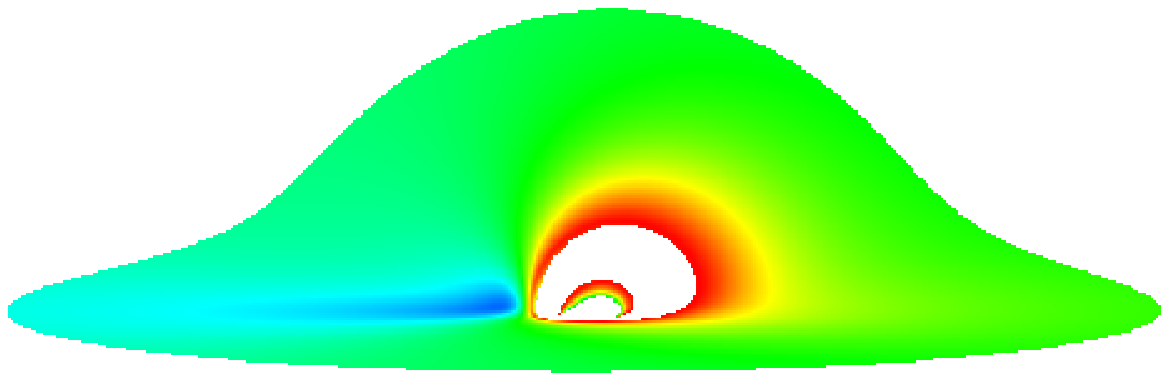}}&\parbox[c]{3cm}{\includegraphics[width=3cm]{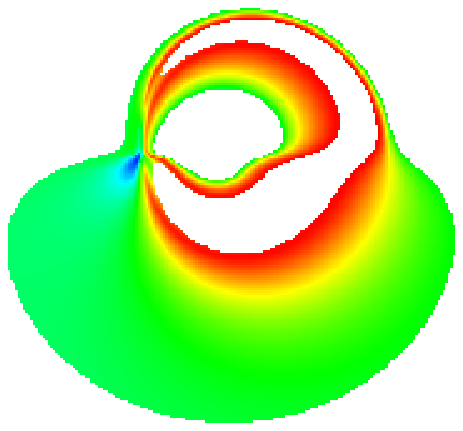}}&\parbox[c]{6cm}{\includegraphics[width=6cm]{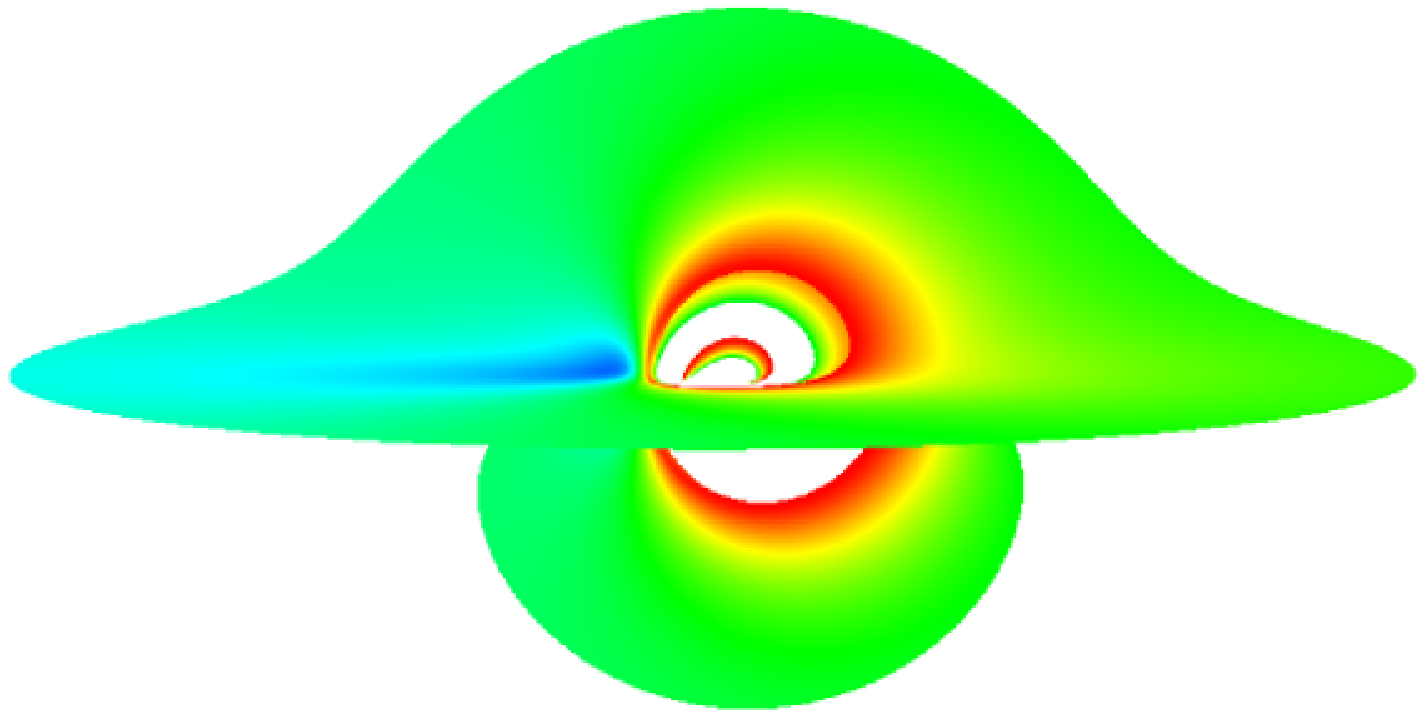}}\\
		\parbox[c]{6cm}{\includegraphics[width=6cm]{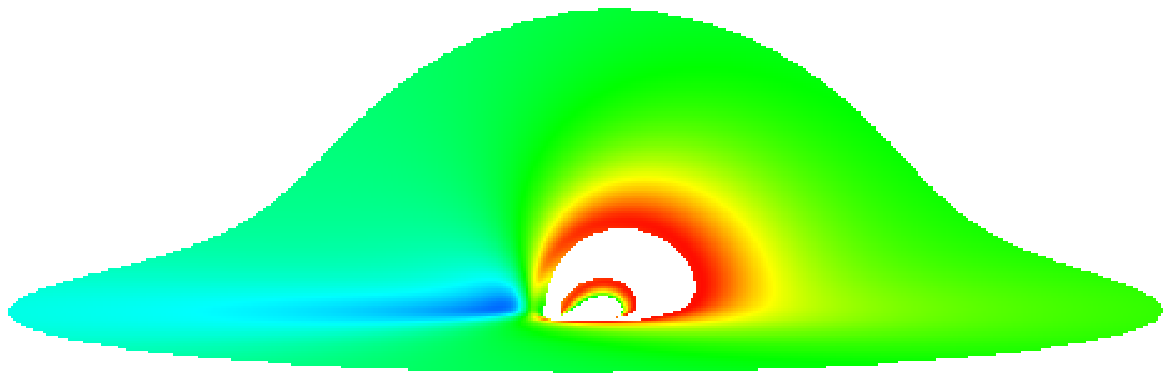}}&\parbox[c]{3cm}{\includegraphics[width=3cm]{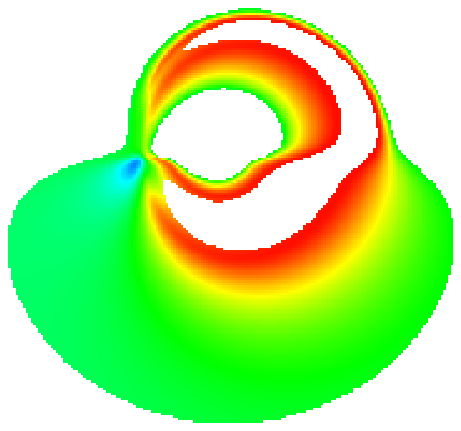}}&\parbox[c]{6cm}{\includegraphics[width=6cm]{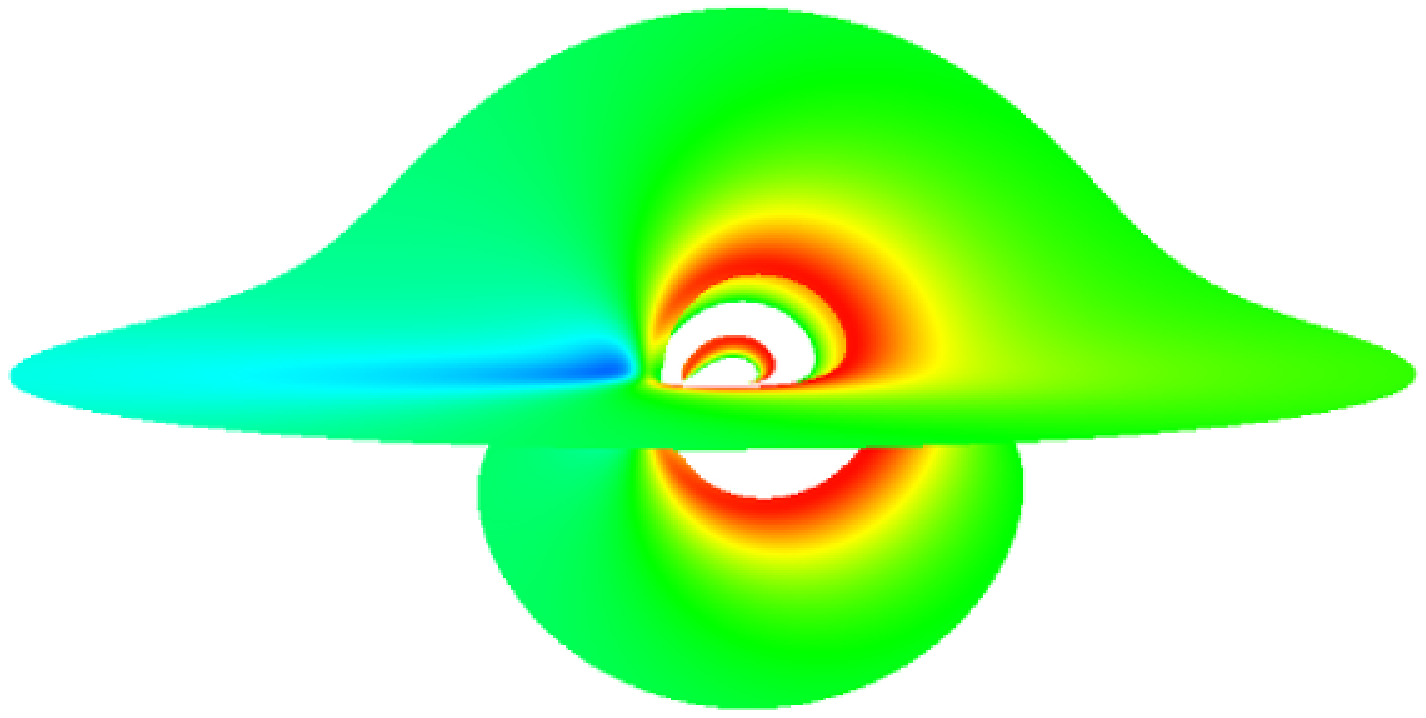}}
	\end{tabular}
	\caption{\label{fig13}Direct(left) and indirect(middle) images of Keplerian discs around Kerr black hole and naked singularities are plotted in false colors representing frequency shift parameter $g$ of radiation emitted from the locally isotropically emitting monochromatic sources forming the disc for representative values of spin parameter $a=0.9981$, $1.0001$, $1.001$ and $1.01$ (from top to bottom). The outer edge of the disc is at $r=20\mathrm{M}$ and the inned edge is at $r_{\mathrm{ms}}=r_{\mathrm{ms}}(a)$. The observer inclination angle is $\theta_o=85^\circ$. The total images (right) are contructed under assumption of opaque discs.}
	\end{center}
\end{figure}

\begin{figure}\ContinuedFloat
	\begin{center}
	\begin{tabular}{ccc}
		\parbox[c]{5.5cm}{\includegraphics[width=5.5cm]{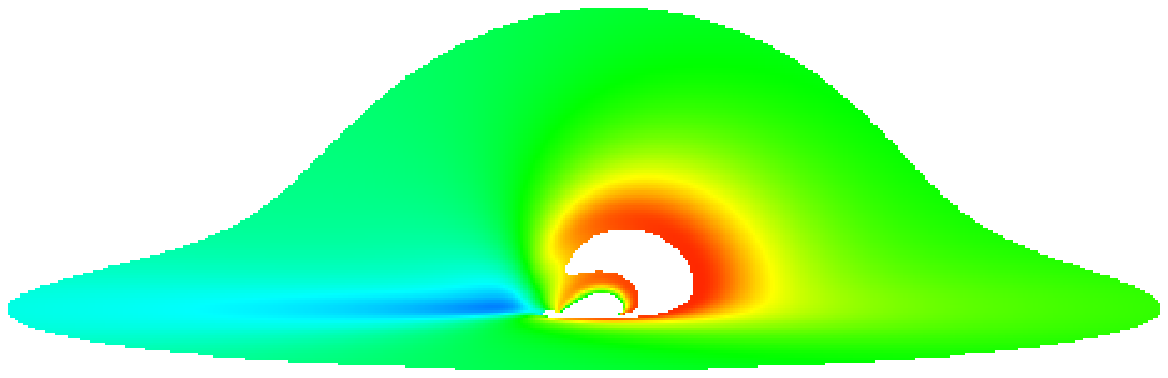}}&\parbox[c]{3cm}{\includegraphics[width=3cm]{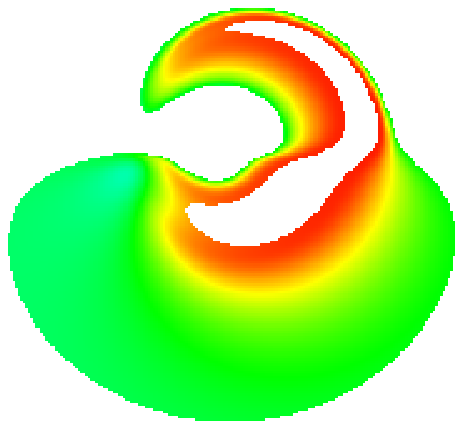}}&\parbox[c]{5.5cm}{\includegraphics[width=6cm]{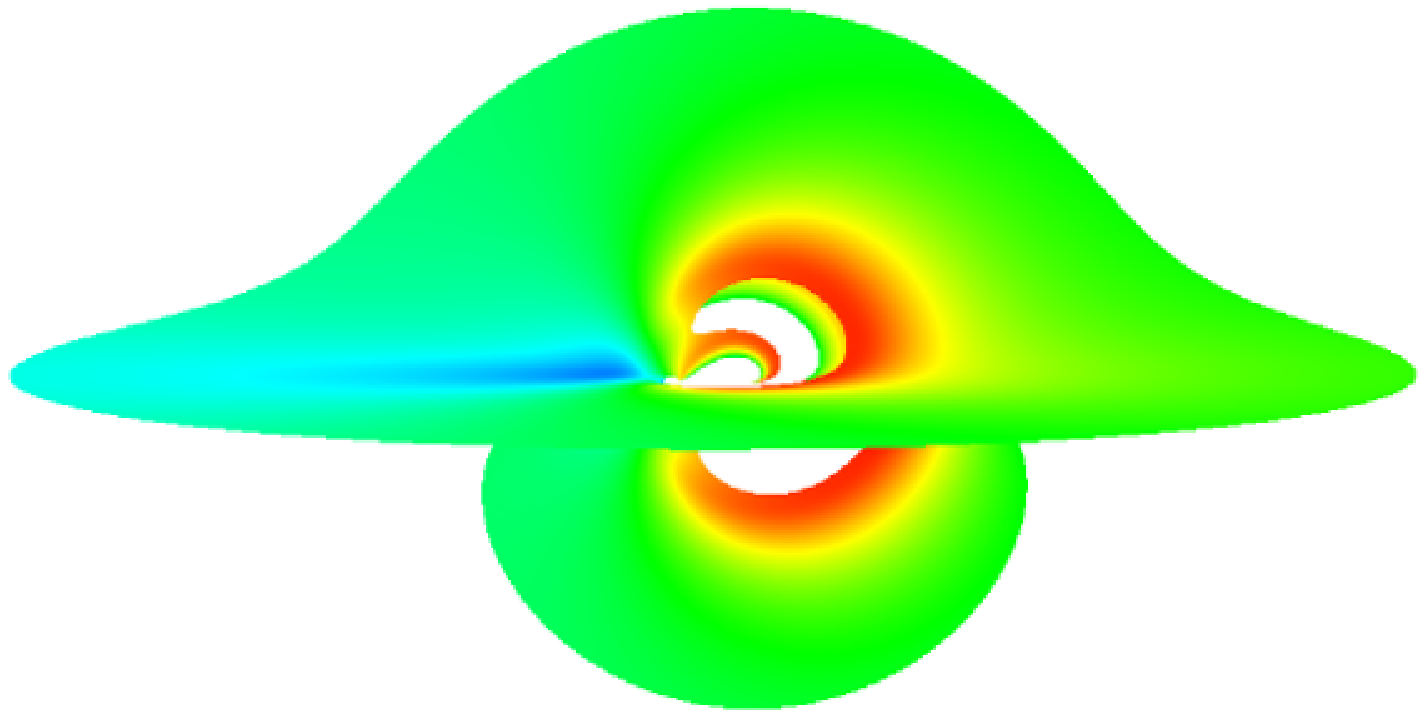}}\\
		\parbox[c]{5.5cm}{\includegraphics[width=5.5cm]{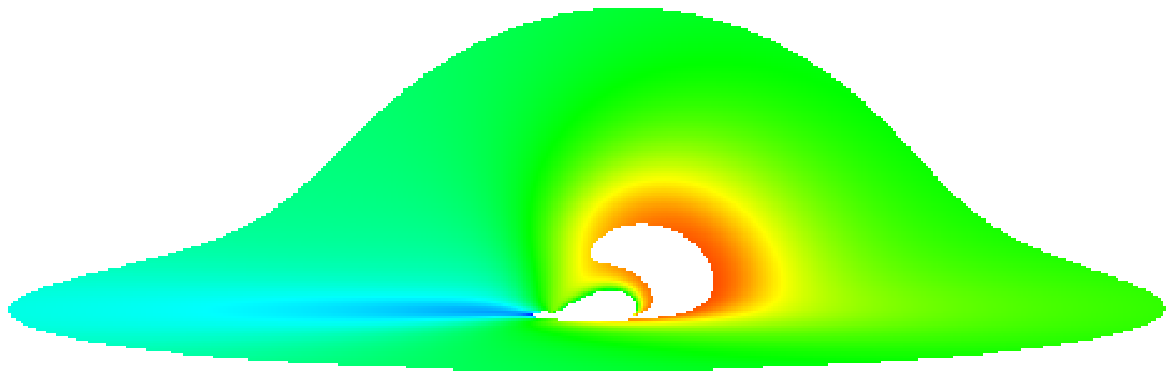}}&\parbox[c]{3cm}{\includegraphics[width=3cm]{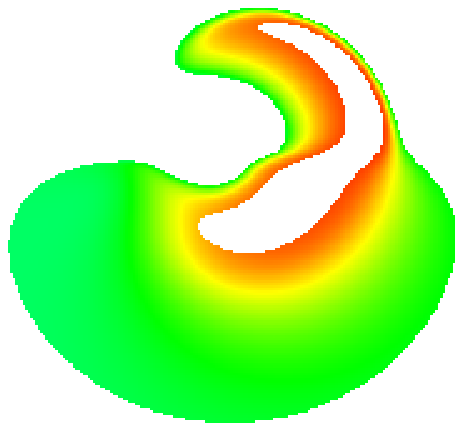}}&\parbox[c]{5.5cm}{\includegraphics[width=6cm]{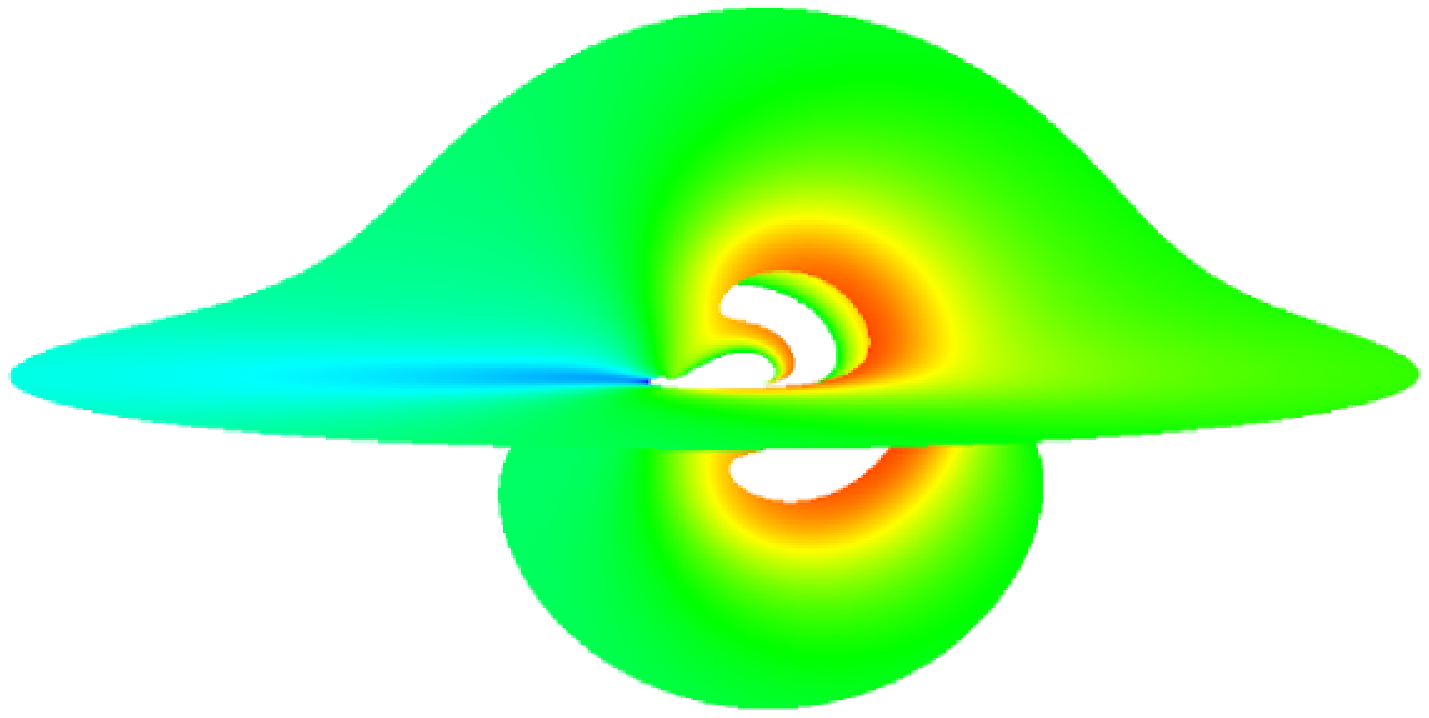}}\\
		\parbox[c]{5.5cm}{\includegraphics[width=5.5cm]{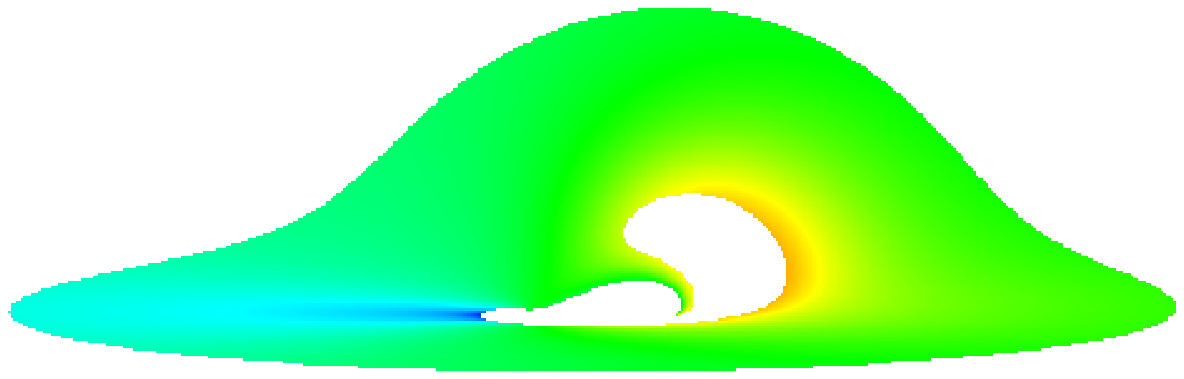}}&\parbox[c]{3cm}{\includegraphics[width=3cm]{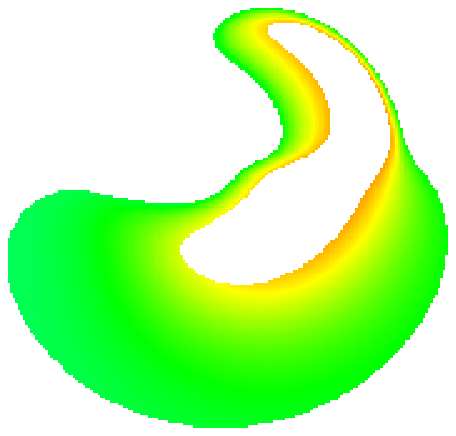}}&\parbox[c]{5.5cm}{\includegraphics[width=6cm]{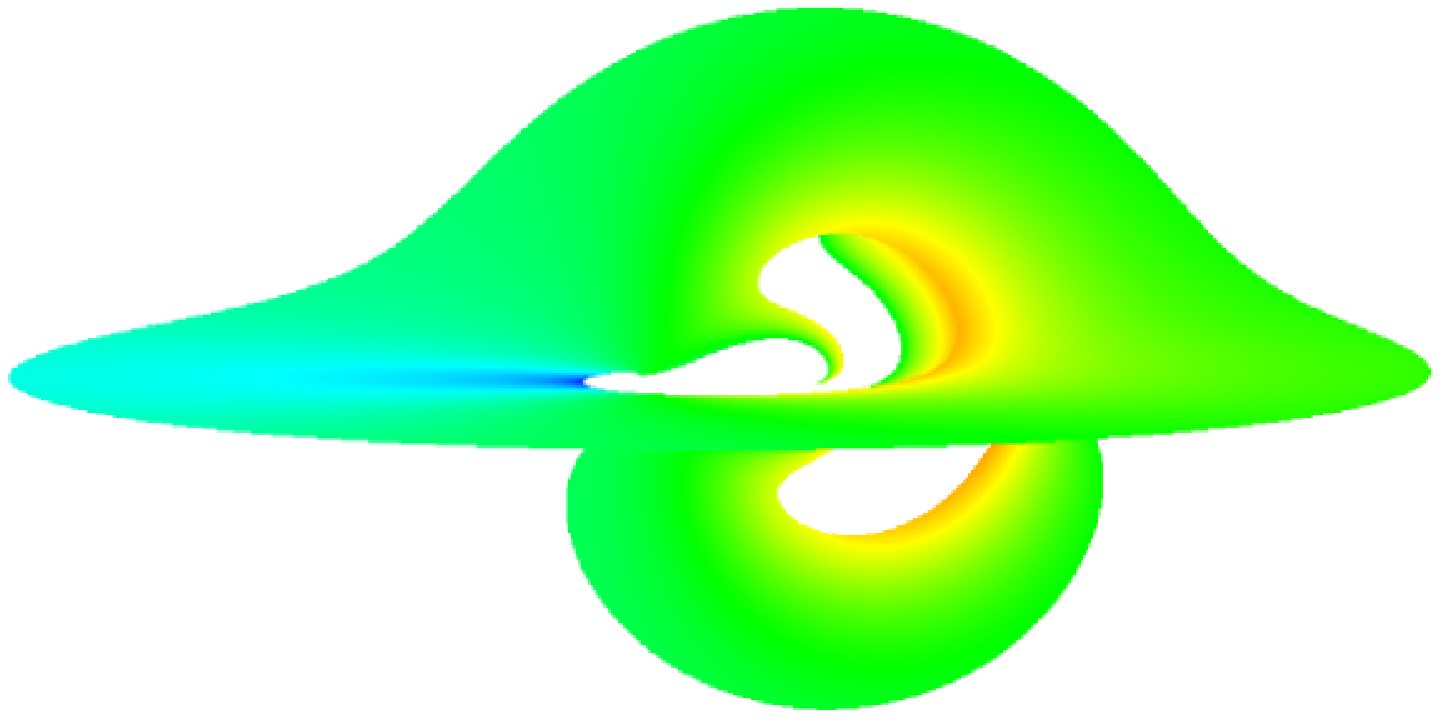}}\\
		\parbox[c]{5.5cm}{\includegraphics[width=5.5cm]{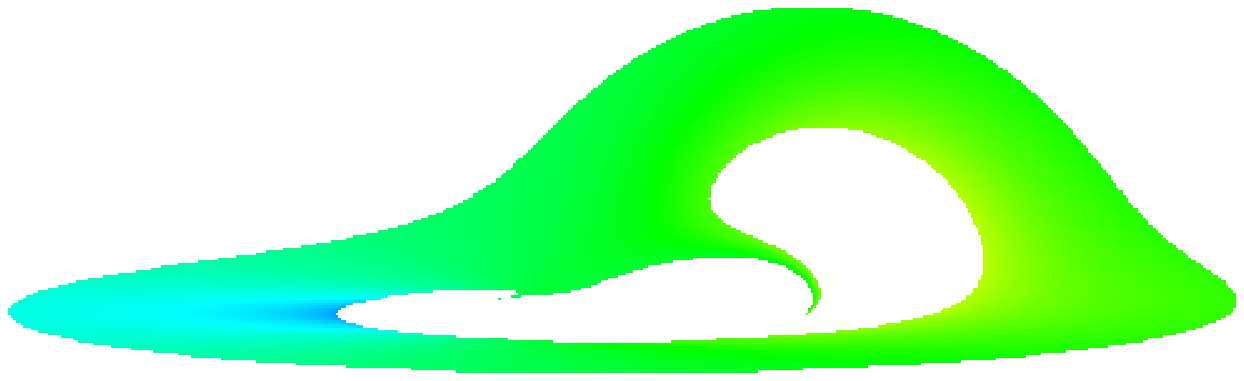}}&\parbox[c]{3cm}{ \includegraphics[width=3cm]{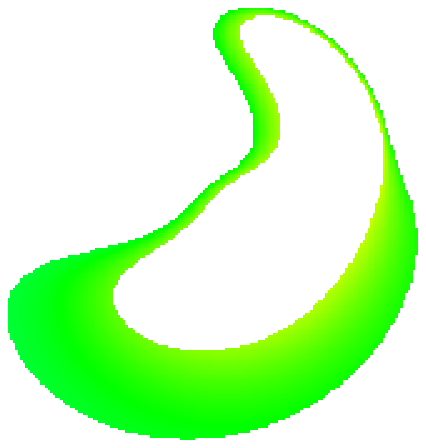}}&\parbox[c]{5.5cm}{\includegraphics[width=6cm]{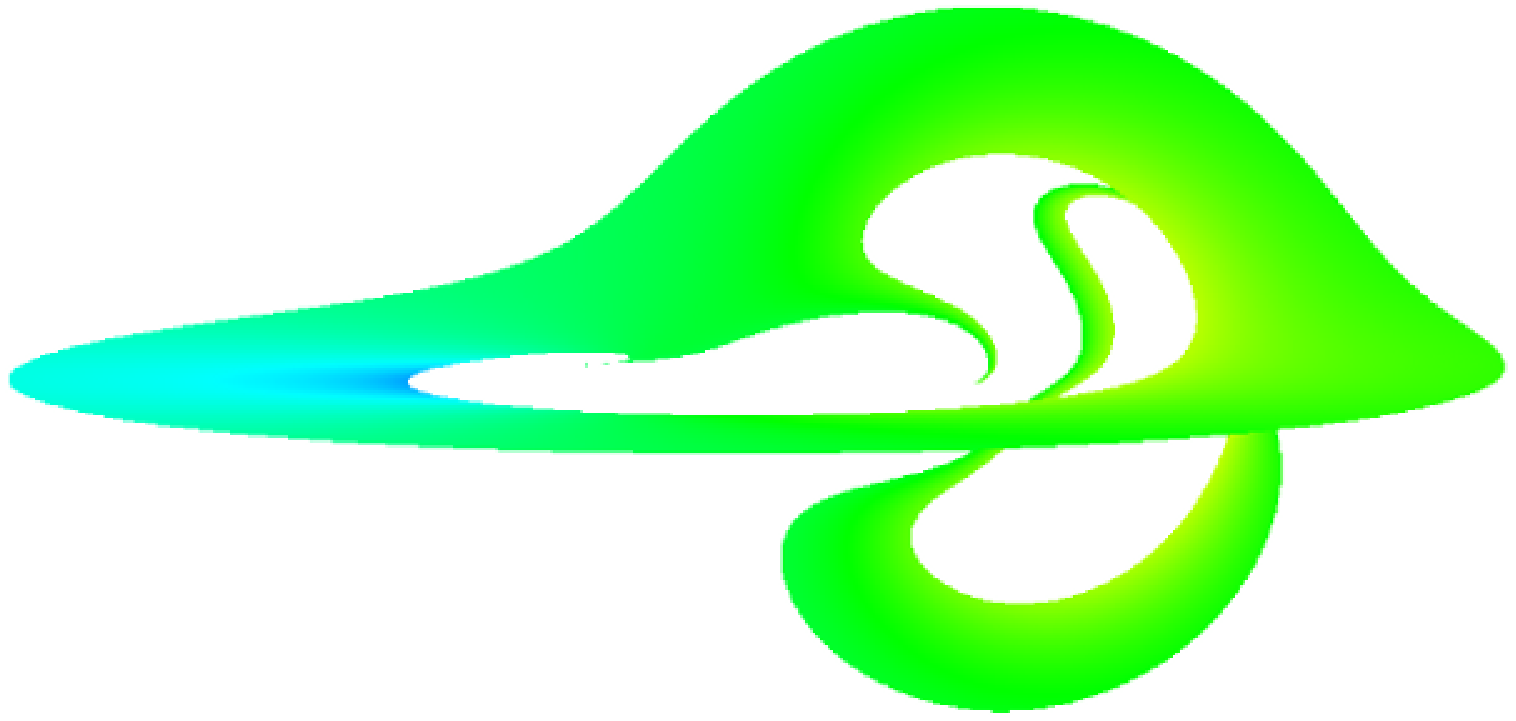}}
	\end{tabular}
	\captionsetup{labelformat=empty}	
	\caption{Fig.\ref{fig13} continuing with $a=1.1$, $1.5$, $3.0$ and $7.0$ (from top to bottom).}
	\end{center}
\end{figure}

\begin{figure}
	\begin{center}
	\begin{tabular}{ccc}
		\includegraphics[width=5.5cm]{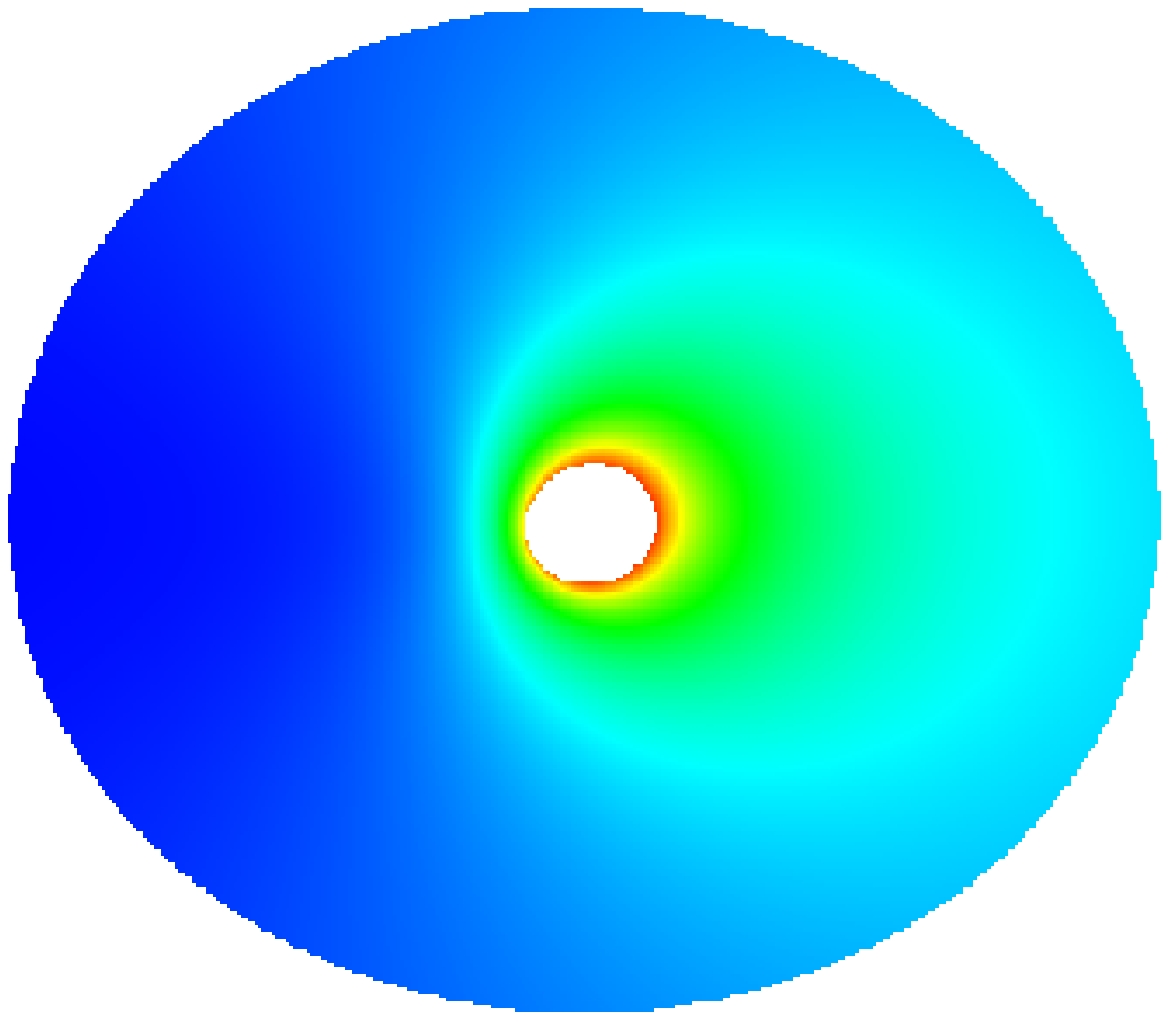}&\includegraphics[width=4cm]{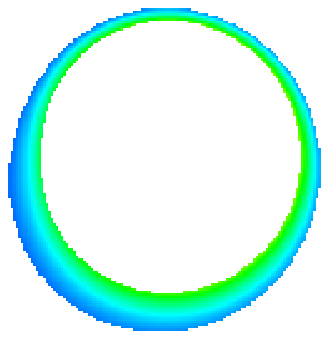}&\includegraphics[width=5cm]{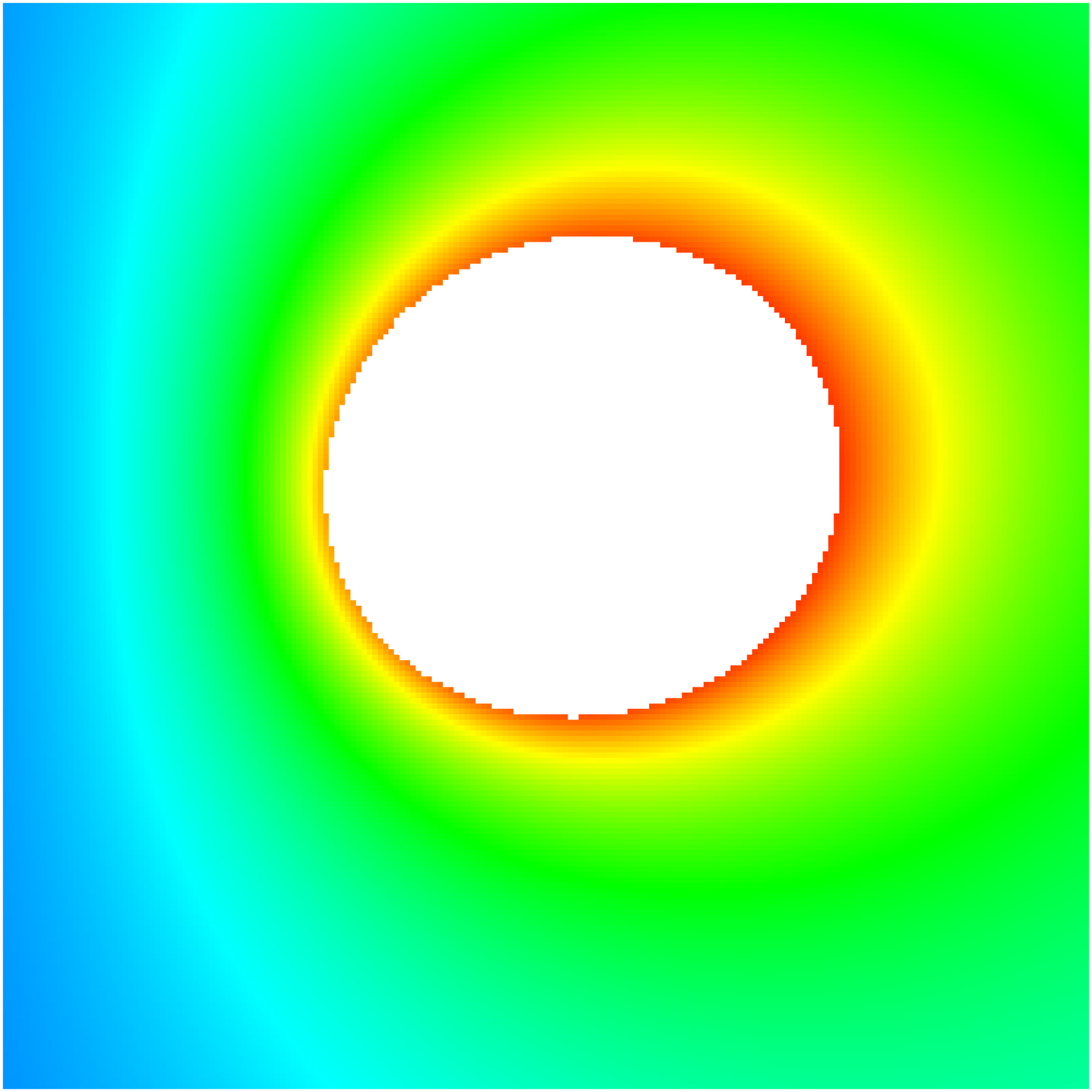}\\
		\includegraphics[width=5.5cm]{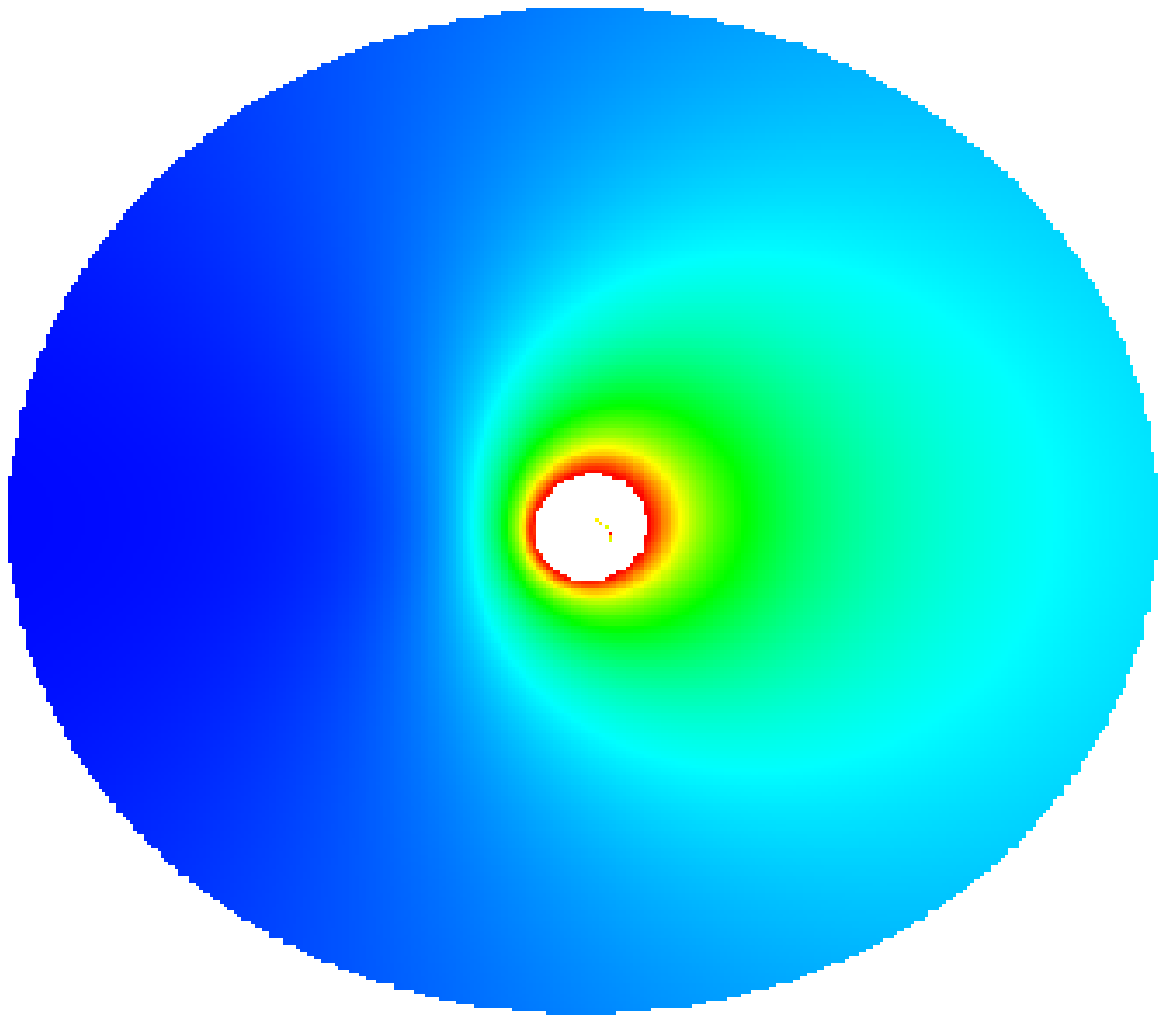}&\includegraphics[width=4cm]{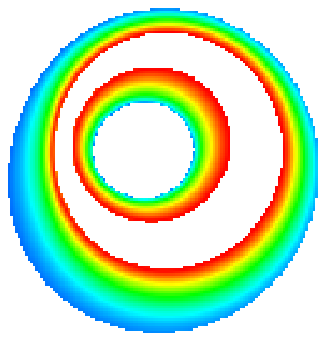}&\includegraphics[width=5cm]{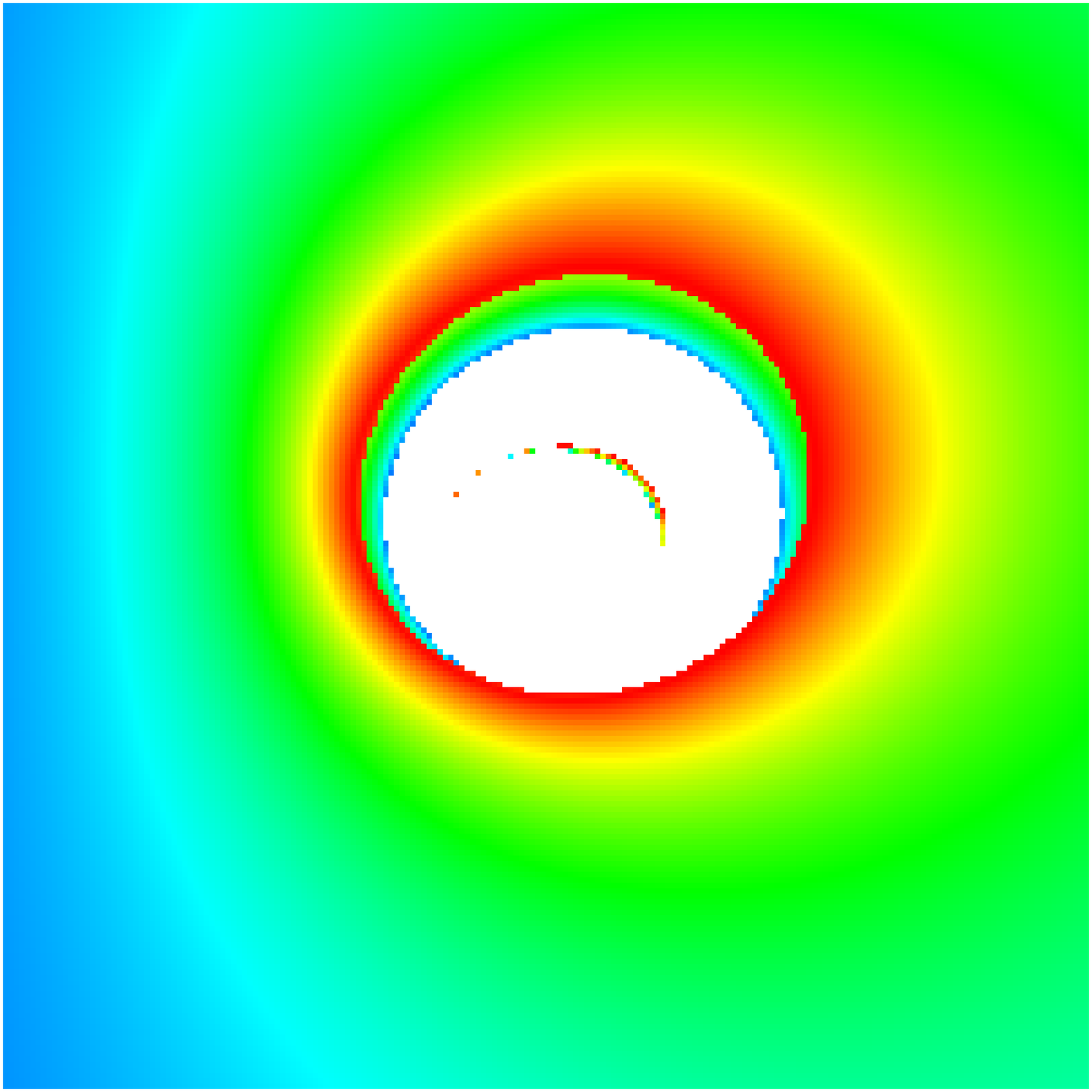}
	\end{tabular}
	\caption{\label{fig14}Direct(left) and indirect(middle) images of Keplerian discs around Kerr black hole and naked singularities are plotted in false colors representing frequency shift parameter $g$ of radiation emitted from the locally  isotropically emitting monochromatic sources forming the disc for representative values of spin parameter $a=0.9981$, $1.0001$ (from top to bottom). The outer edge of the disc is at $r=20\mathrm{M}$ and the inned edge is at $r_{\mathrm{ms}}=r_{\mathrm{ms}}(a)$. The observer inclination angle is $\theta_o=30^\circ$. The total images (right) are constructed under assumption of opaque discs.}
	\end{center}
\end{figure}

\begin{figure}\ContinuedFloat
	\begin{center}
	\begin{tabular}{ccc}
		\includegraphics[width=5.5cm]{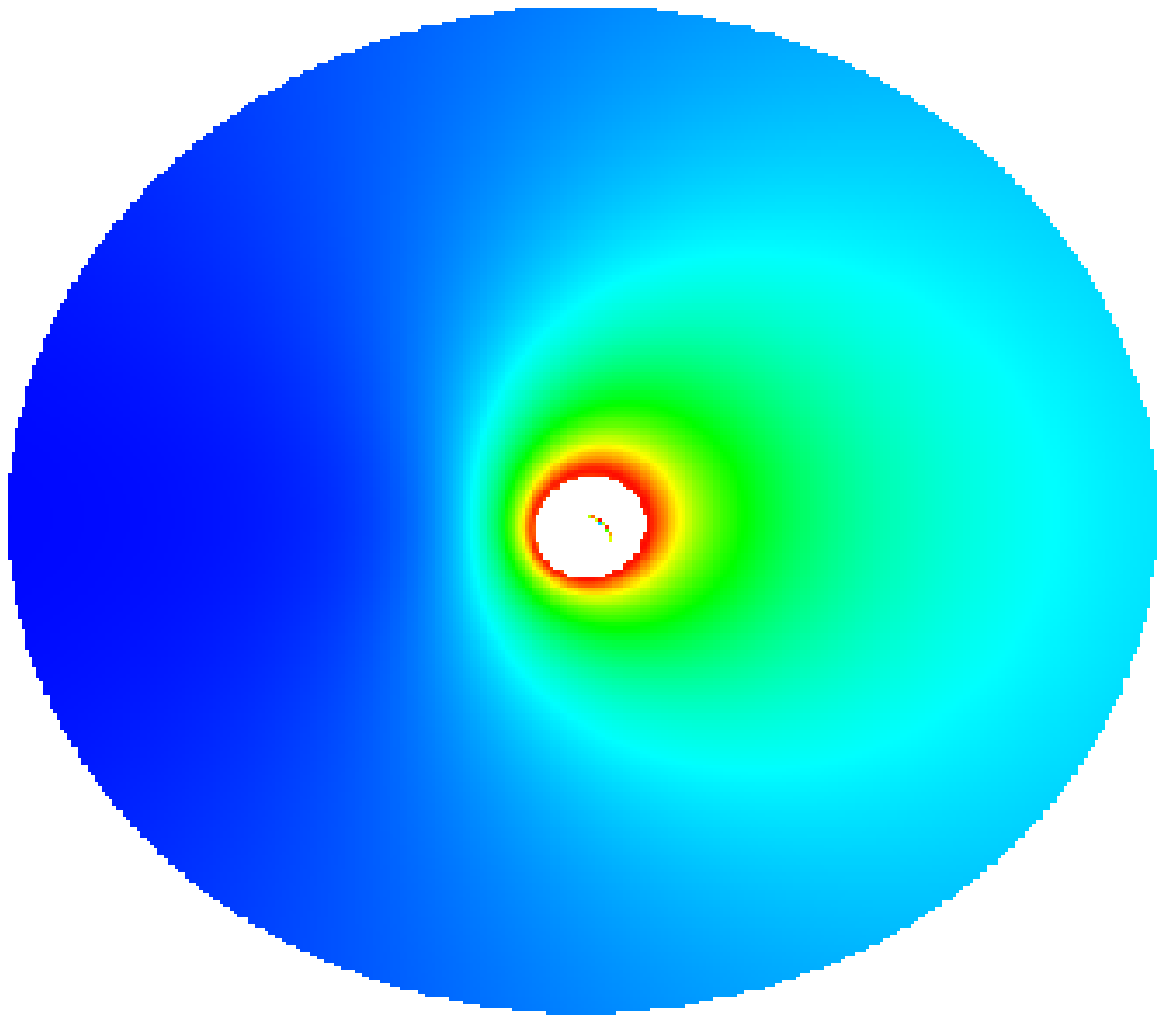}&\includegraphics[width=4cm]{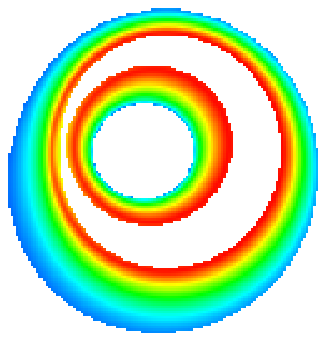}&\includegraphics[width=5cm]{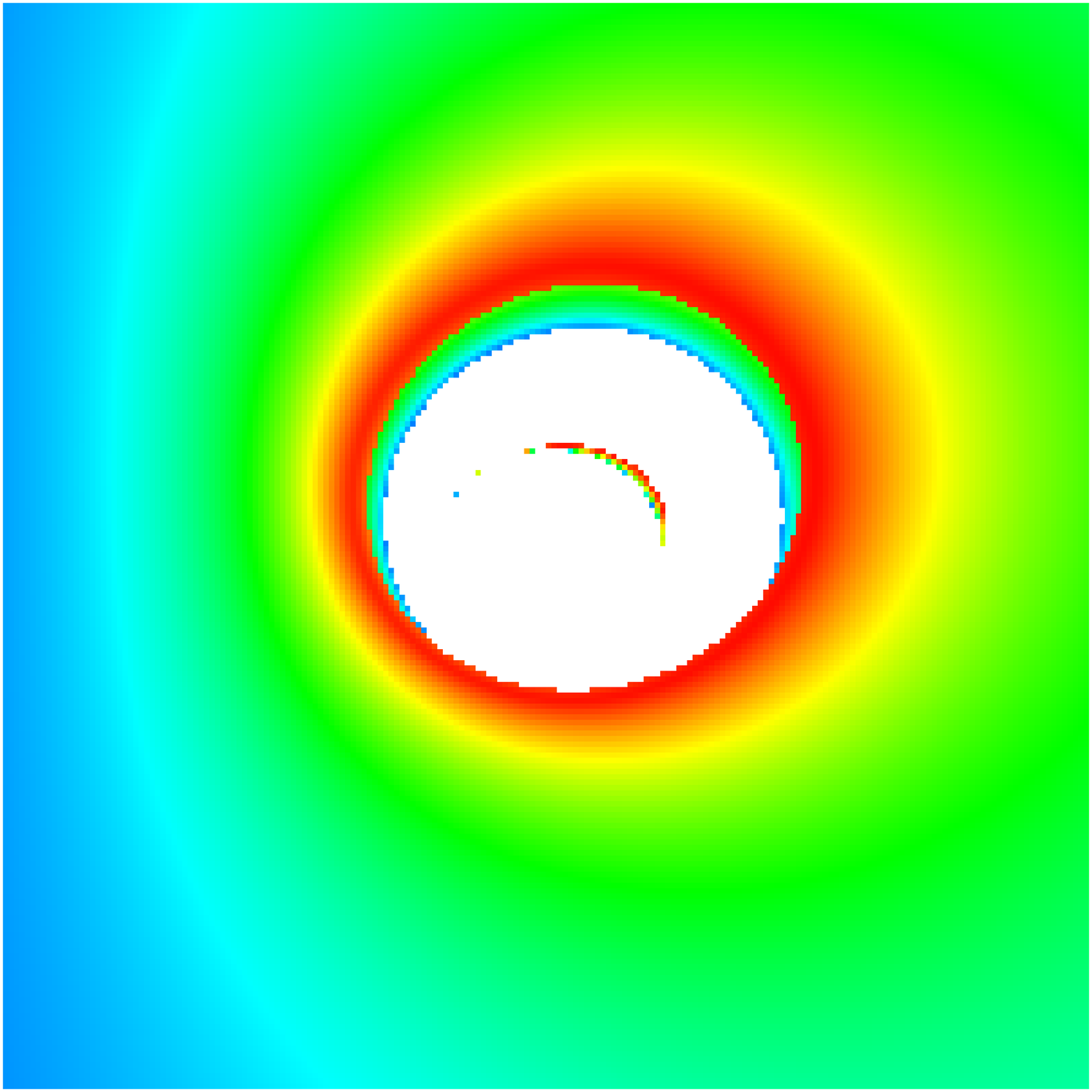}\\
		\includegraphics[width=5.5cm]{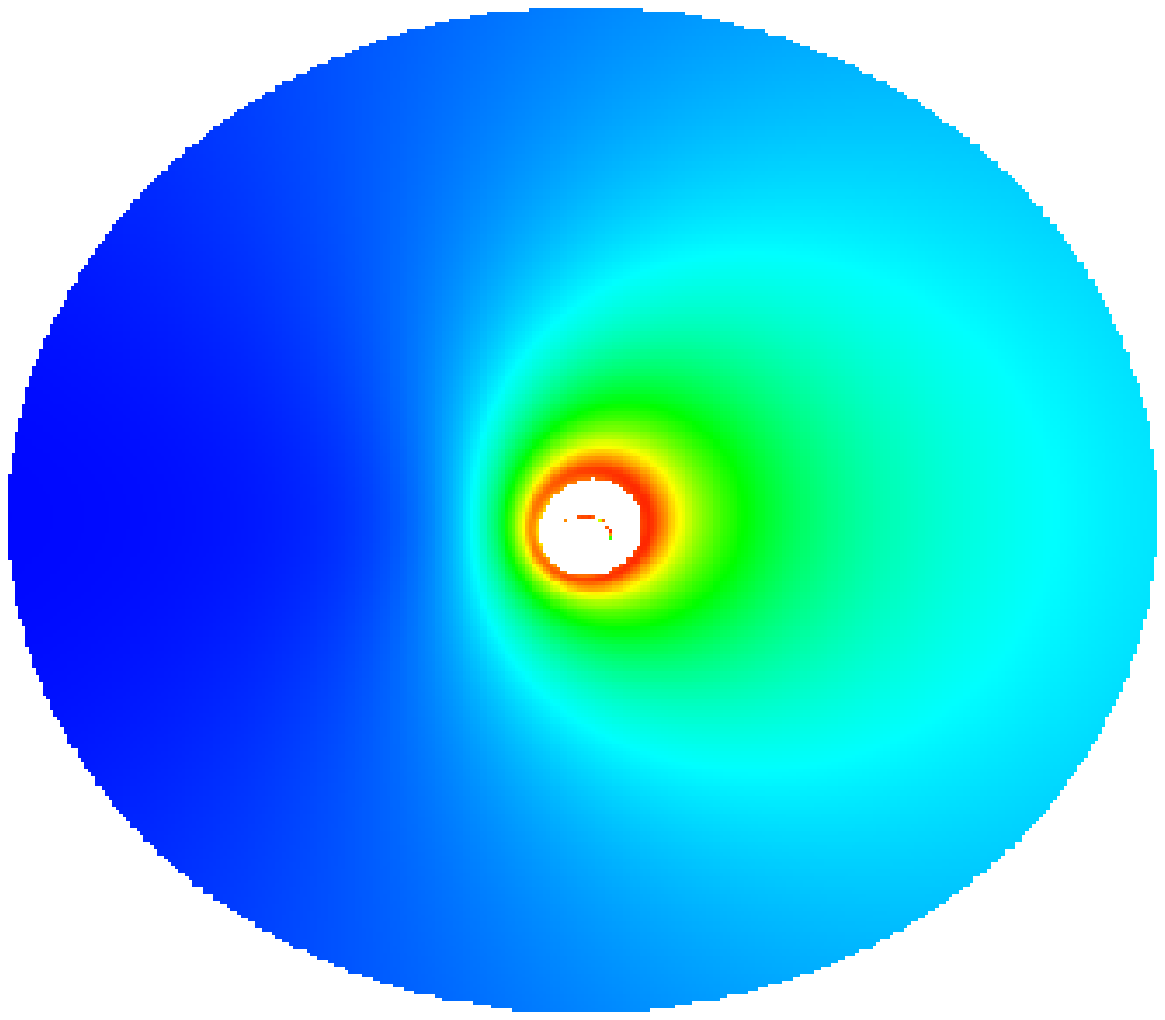}&\includegraphics[width=4cm]{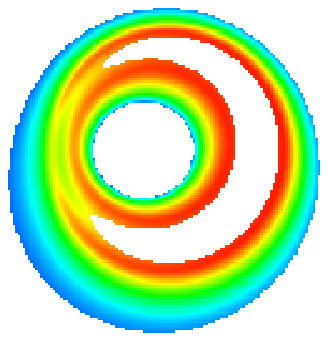}&\includegraphics[width=5cm]{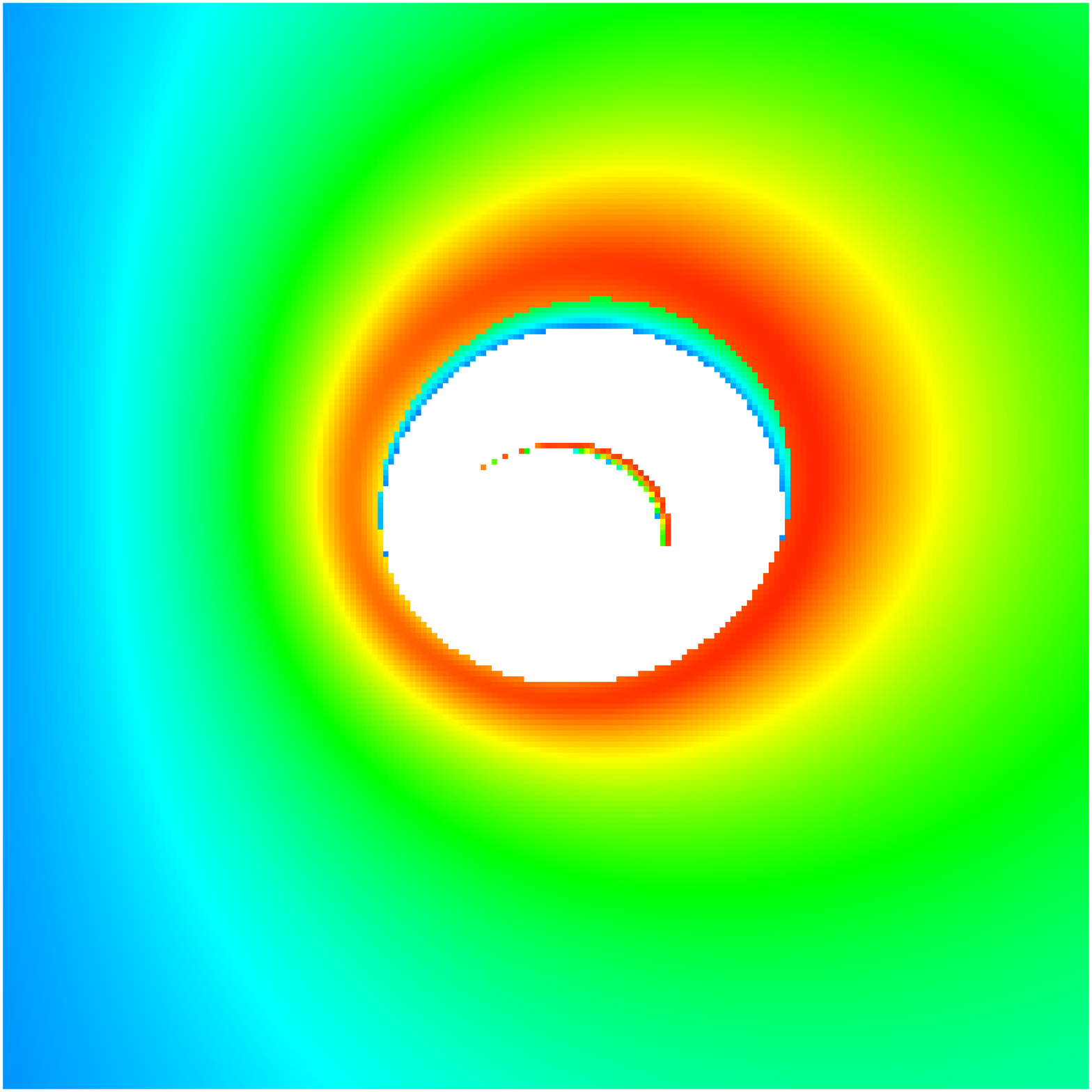}
	\end{tabular}
	\captionsetup{labelformat=empty}
	\caption{Fig.\ref{fig14} continuing with $a=1.001$, $1.01$ (from top to bottom).}
	\end{center}
\end{figure}

\begin{figure}\ContinuedFloat
	\begin{center}
	\begin{tabular}{ccc}
		\includegraphics[width=5.5cm]{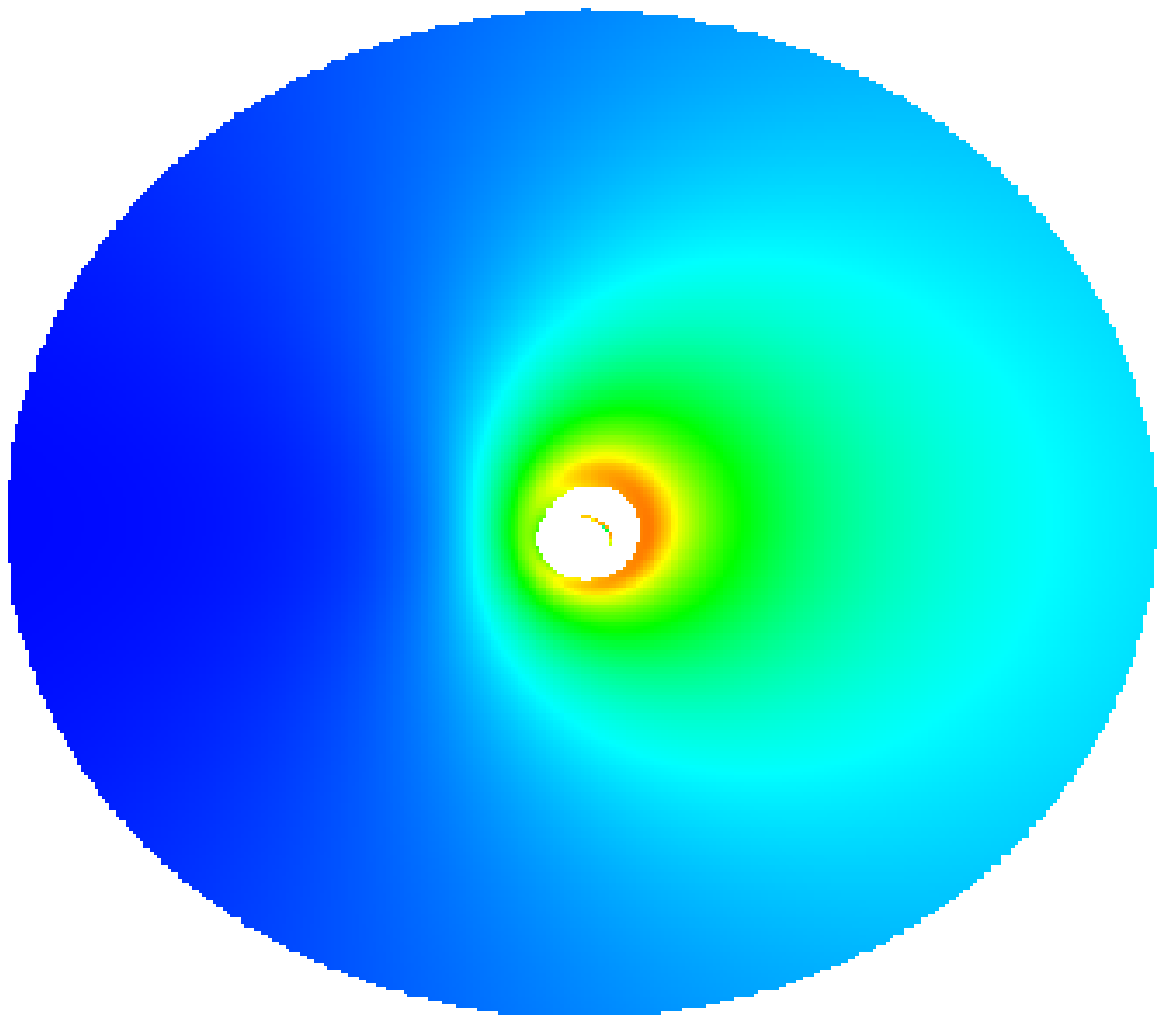}&\includegraphics[width=4cm]{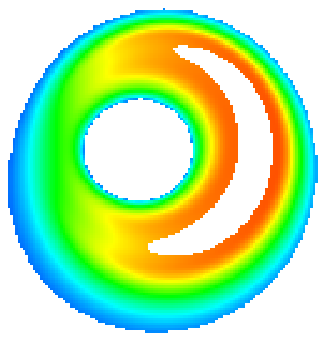}&\includegraphics[width=5cm]{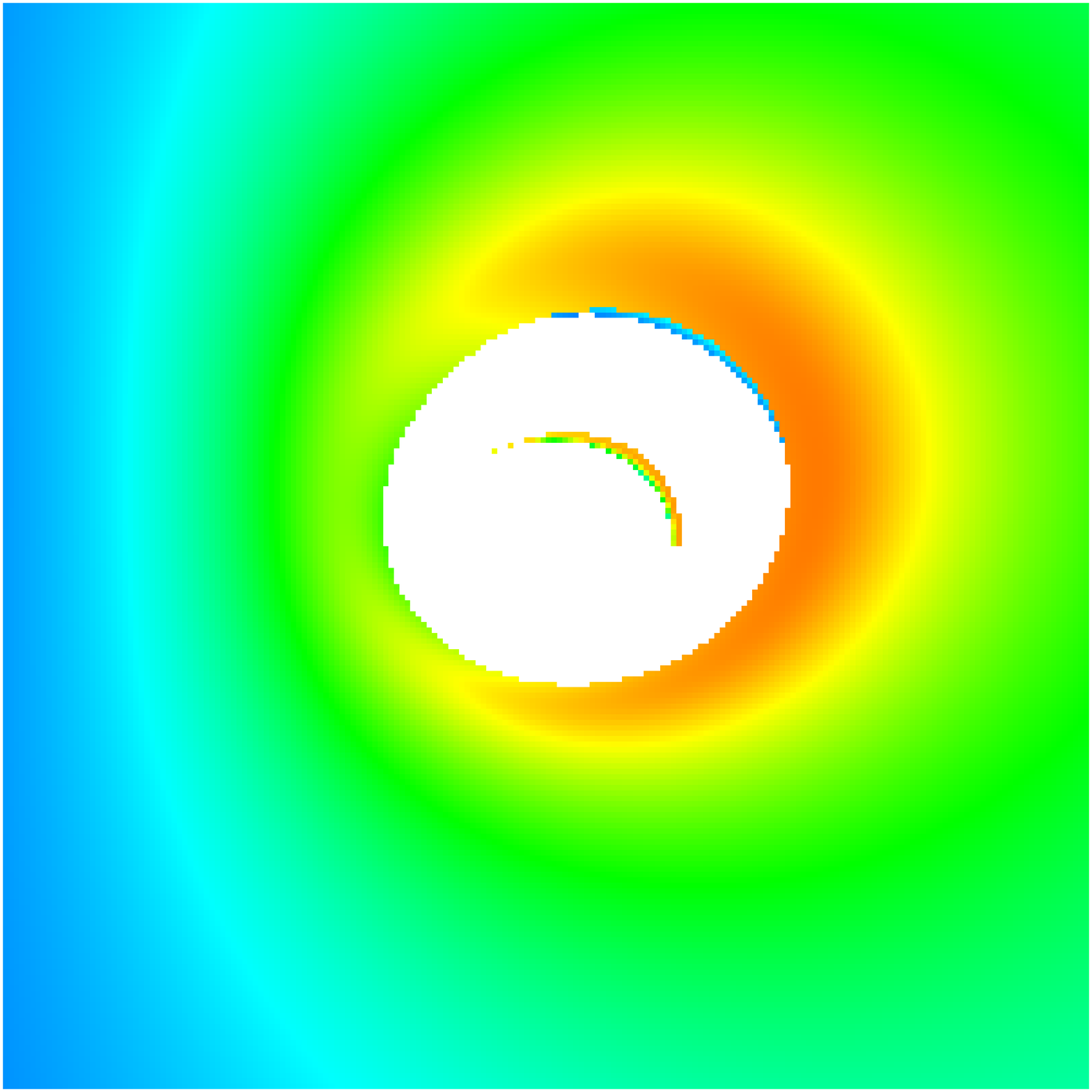}\\
		\includegraphics[width=5.5cm]{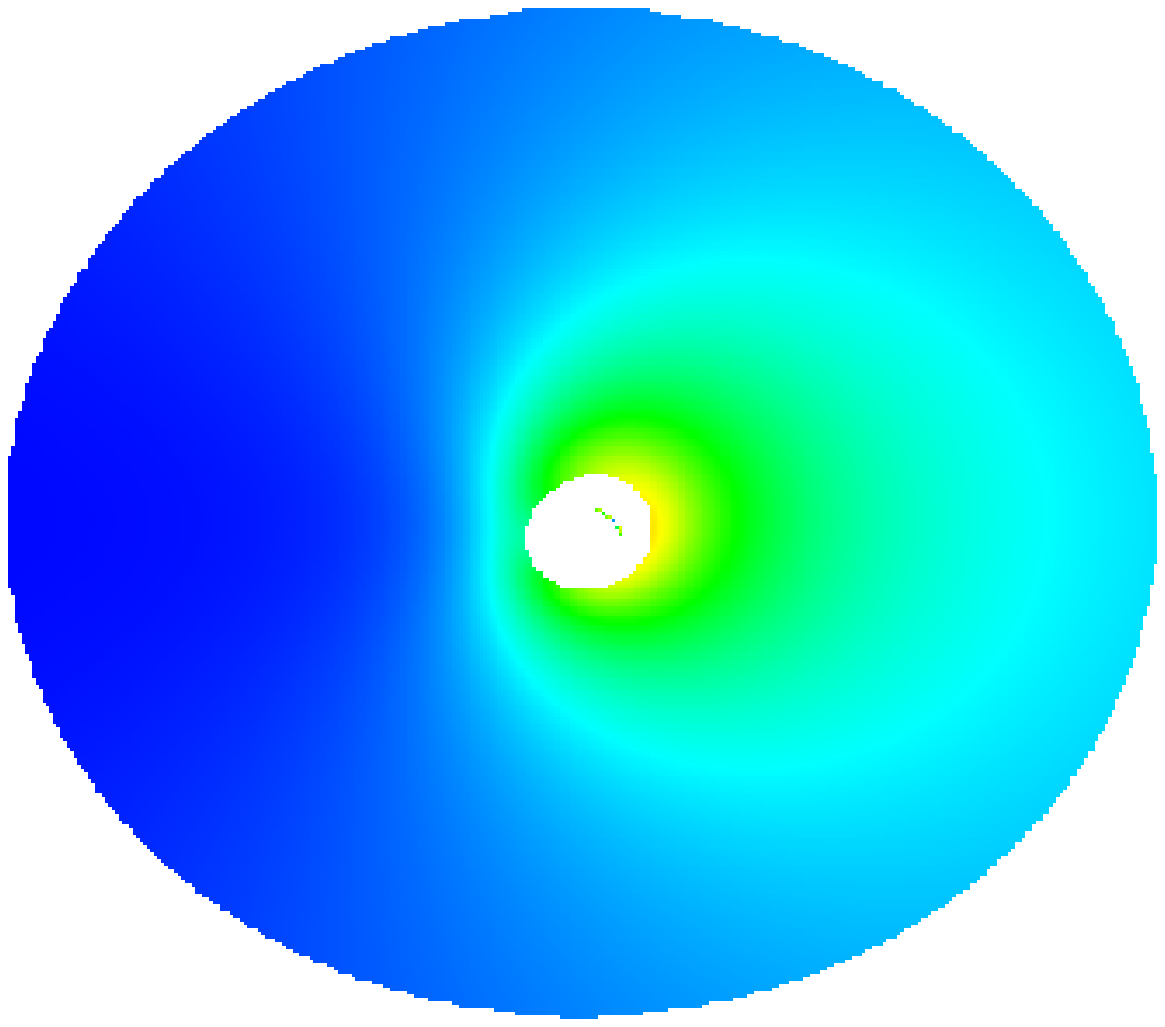}&\includegraphics[width=4cm]{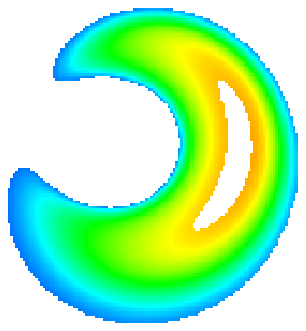}&\includegraphics[width=5cm]{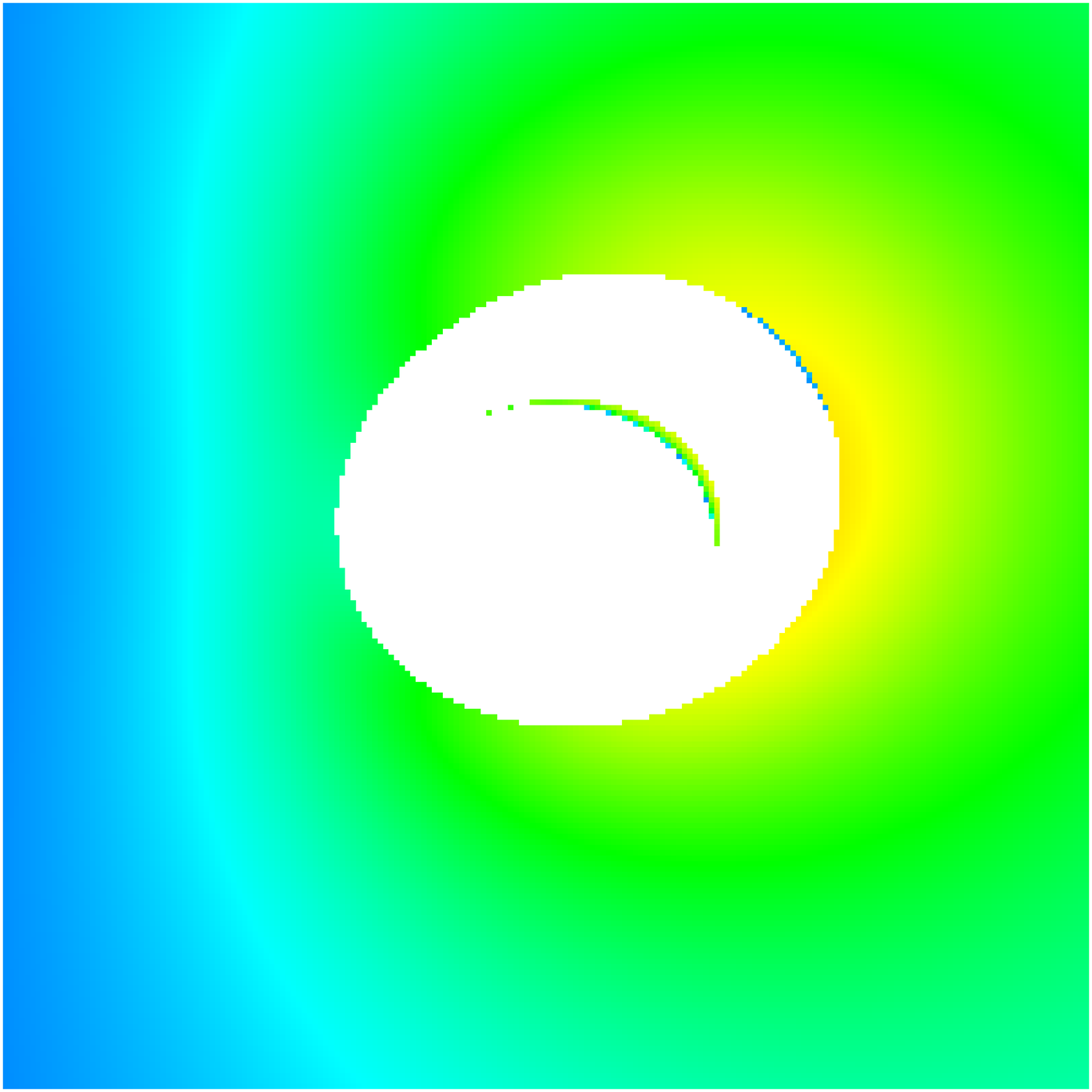}
	\end{tabular}
	\captionsetup{labelformat=empty}
	\caption{Fig.\ref{fig14} continuing with $a=1.1$, $1.5$ (from top to bottom).}
	\end{center}
\end{figure}

\begin{figure}\ContinuedFloat
	\begin{center}
	\begin{tabular}{ccc}
		\includegraphics[width=5.5cm]{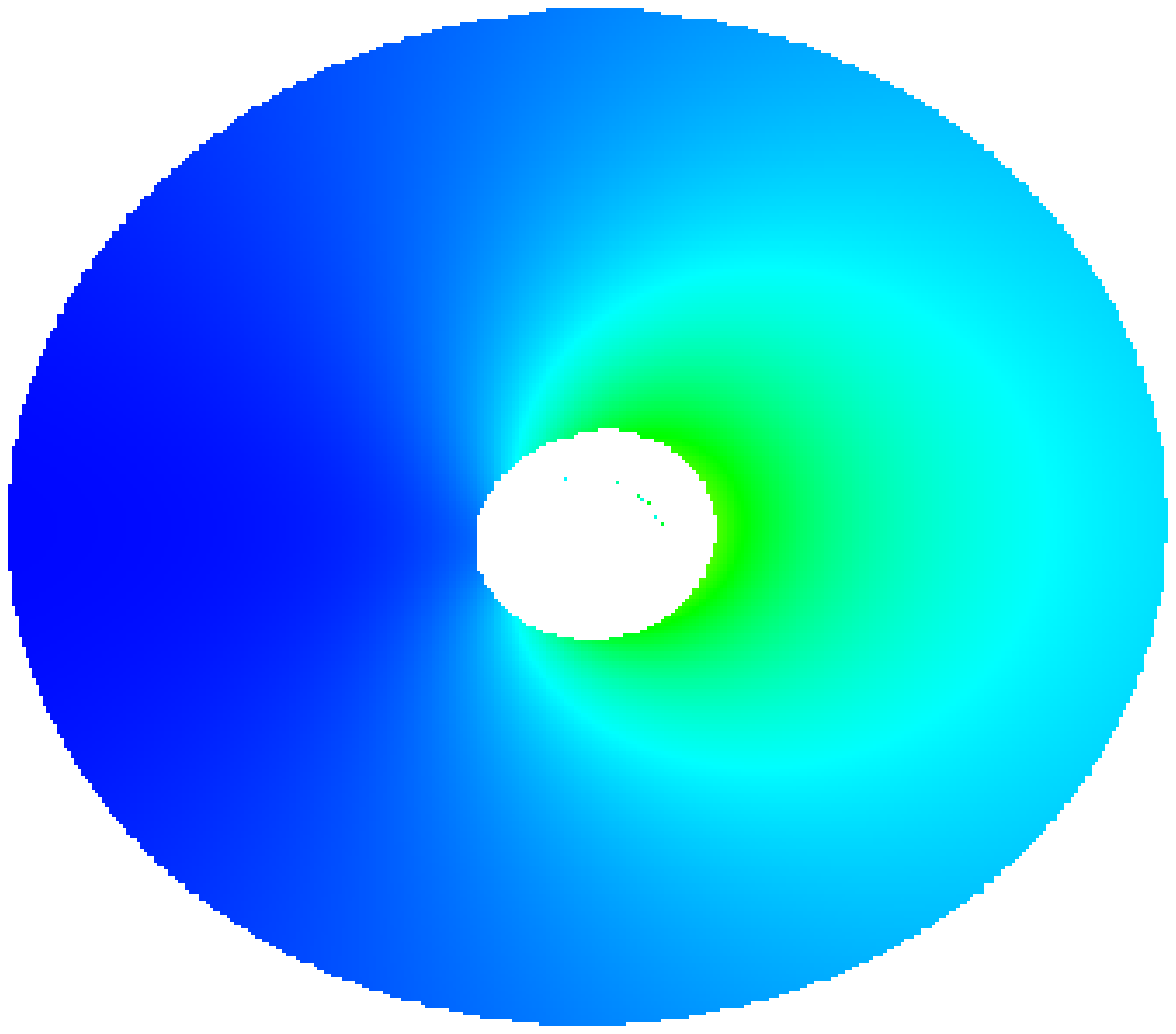}&\includegraphics[width=4cm]{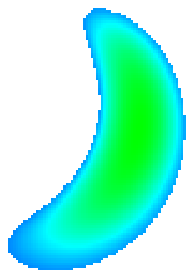}&\includegraphics[width=5cm]{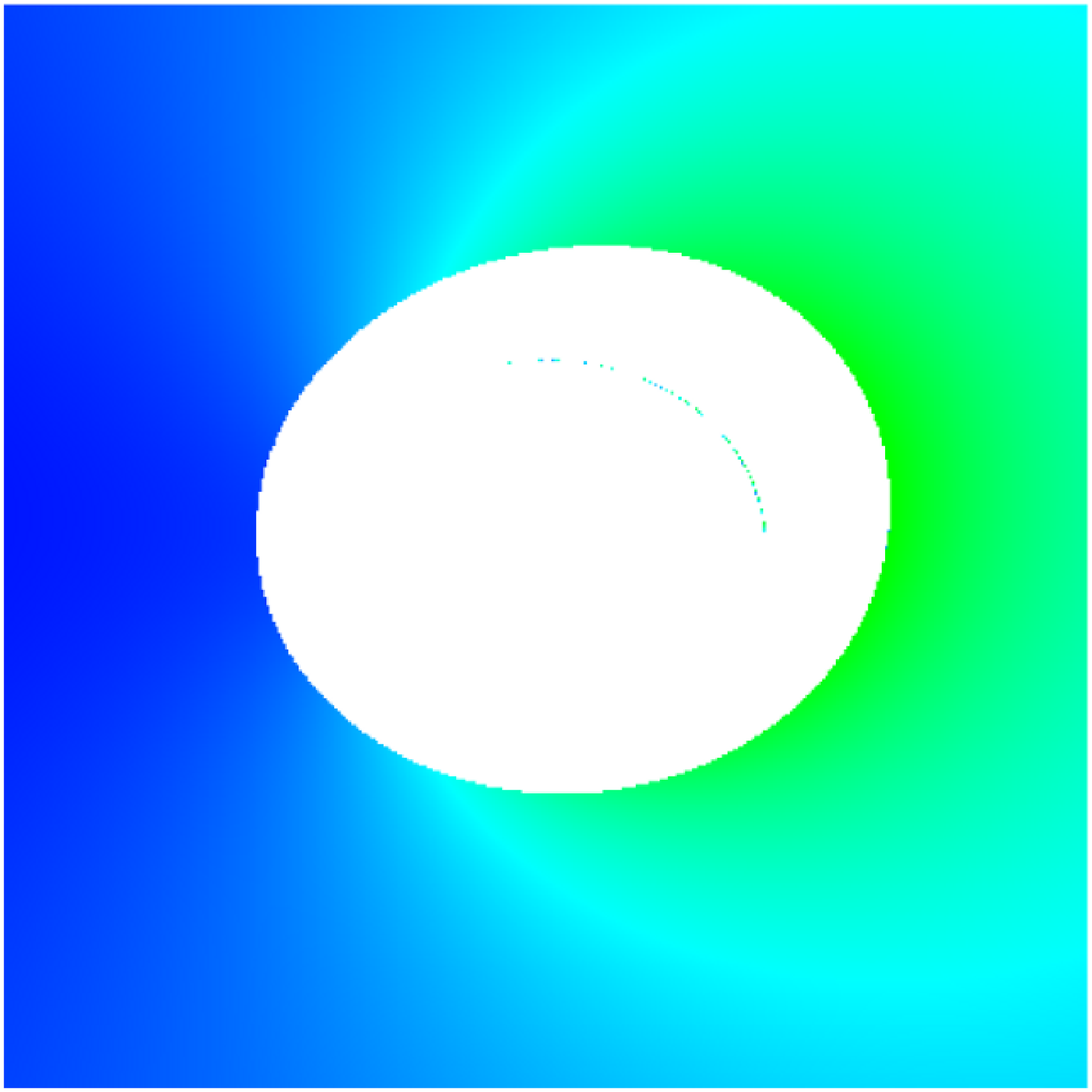}\\
		\parbox[c]{5.5cm}{\includegraphics[width=5.5cm]{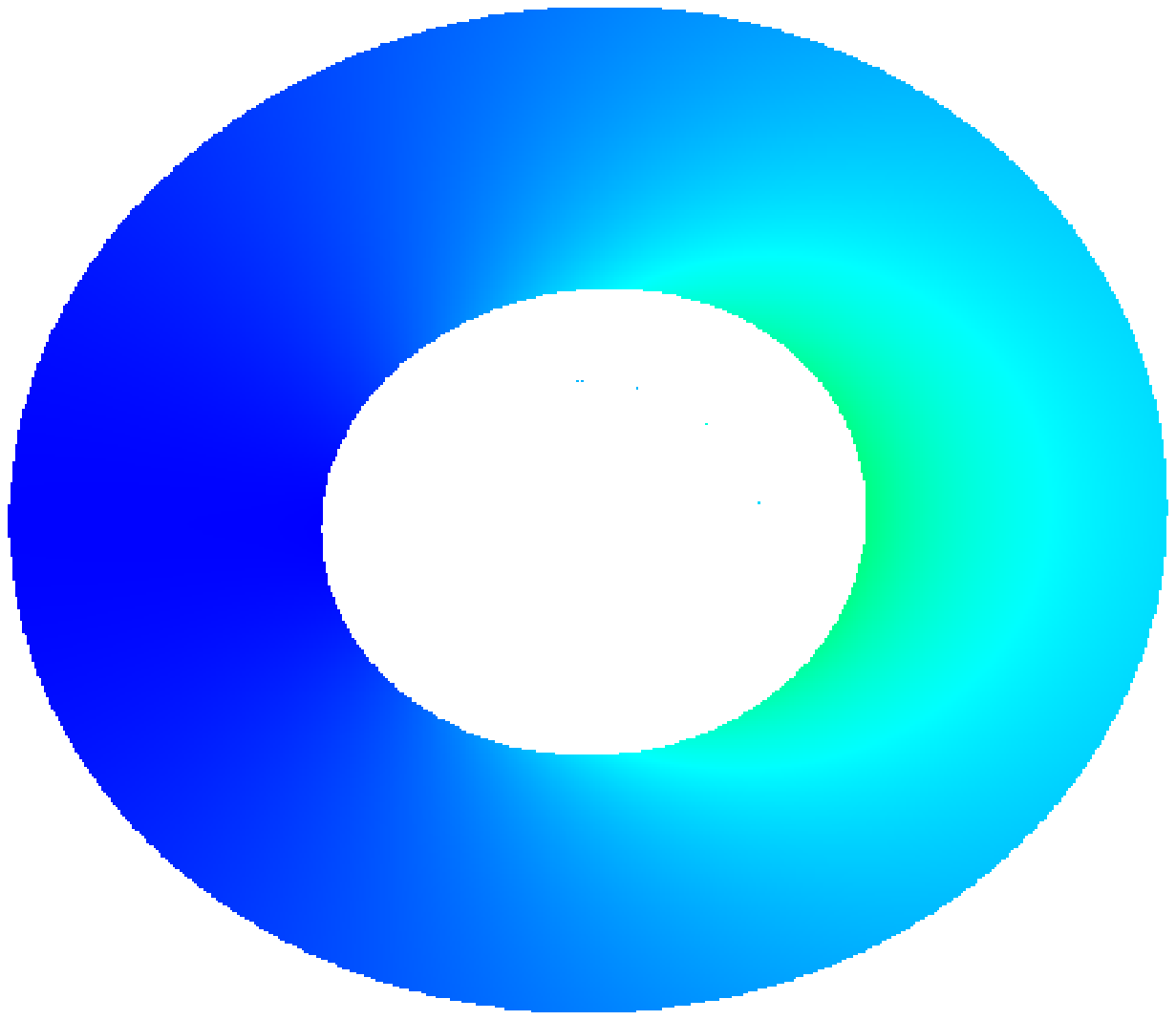}}& \parbox[c]{4cm}{\includegraphics[width=4cm]{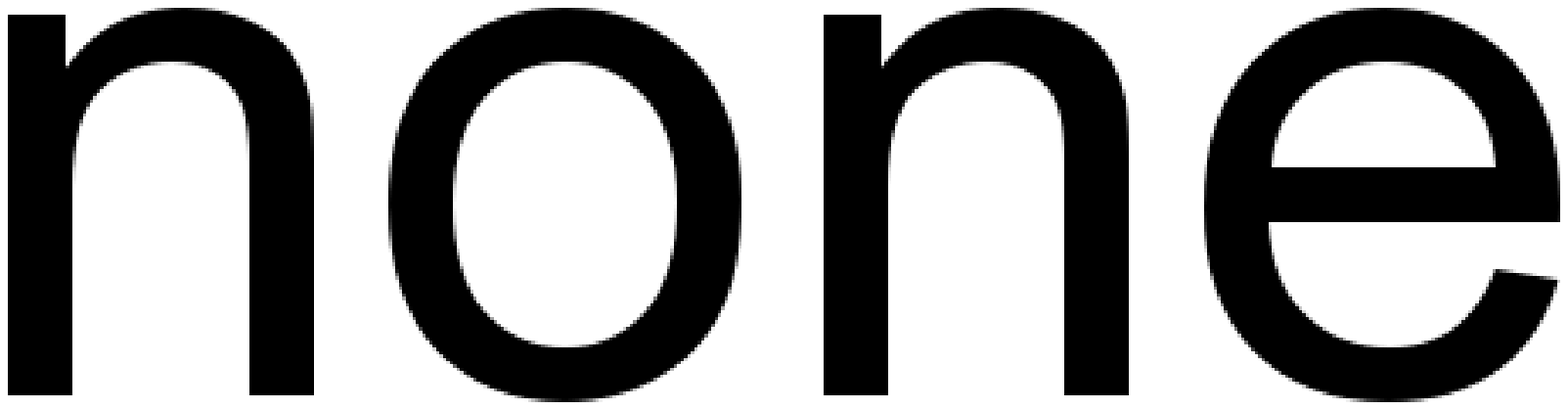}} & \parbox[c]{4cm}{\includegraphics[width=4cm]{fig14w}}
	\end{tabular}
	\captionsetup{labelformat=empty}
	\caption{Fig.\ref{fig14} continuing with $a=3.0$ and $7.0$ (from top to bottom).}
	\end{center}
\end{figure}

\begin{figure}
	\begin{center}
	\includegraphics[width=7.5cm]{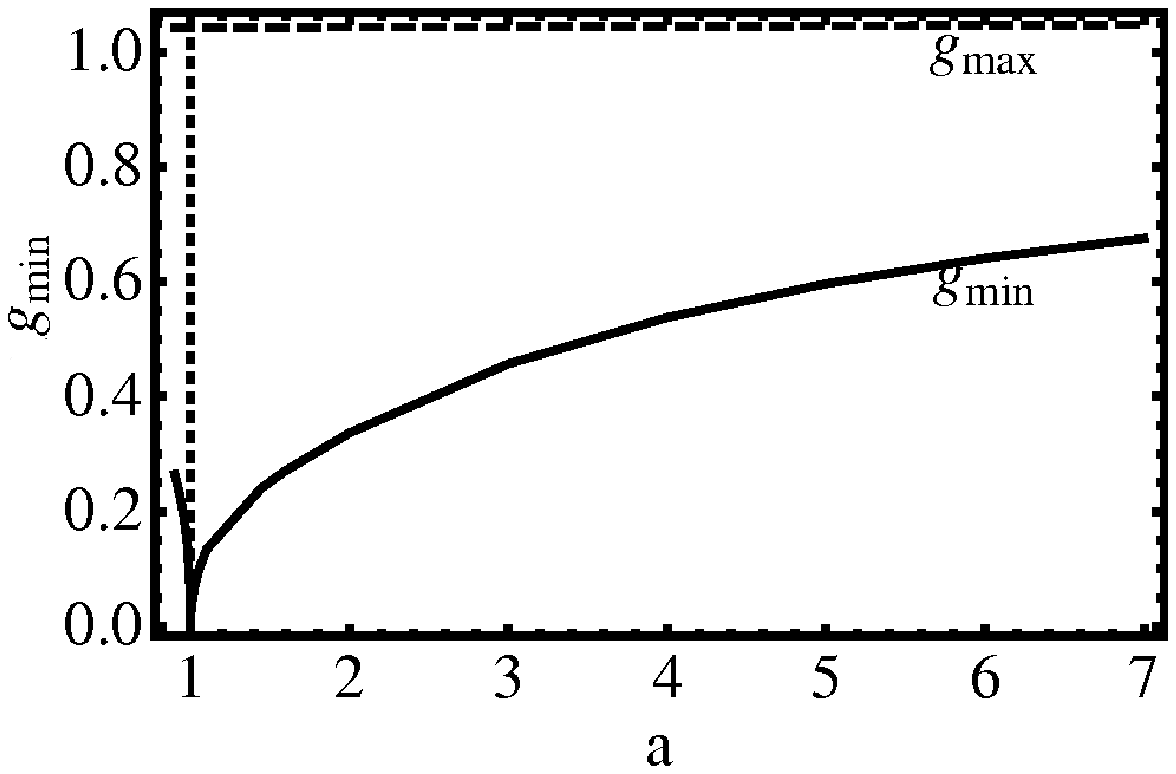}
	\begin{tabular}{ccc}
		\includegraphics[width=6.5cm]{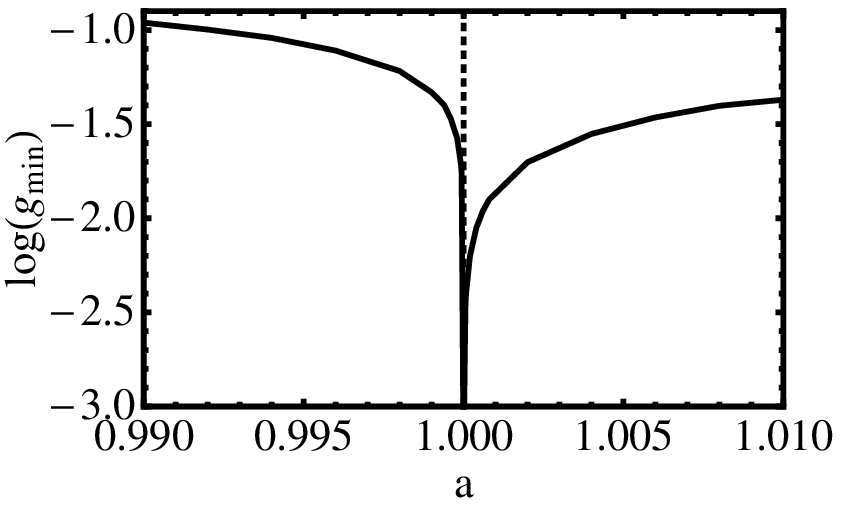}&\includegraphics[width=6.5cm]{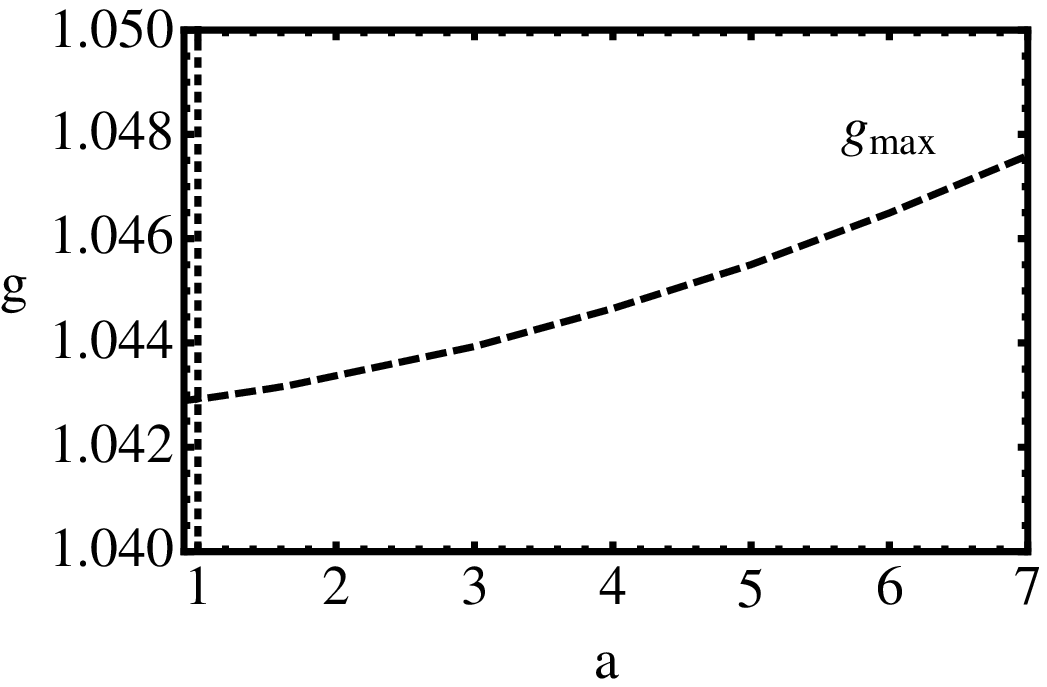}
	\end{tabular}
	\caption{\label{fig15}Top: The extension of the frequency shift of the Keplerian discs direct image represented by the g-factor. Bottom left: The enlarged region of $g_{\mathrm{min}}$ in the vicinity of $a=1$. Bottom right: The enlarged region of $g_{\mathrm{max}}$. The inclination of the observer $\theta_o=30^\circ$.}
	\end{center}
\end{figure}

\begin{figure}
	\begin{center}
	\begin{tabular}{cc}
		\includegraphics[width=6.5cm]{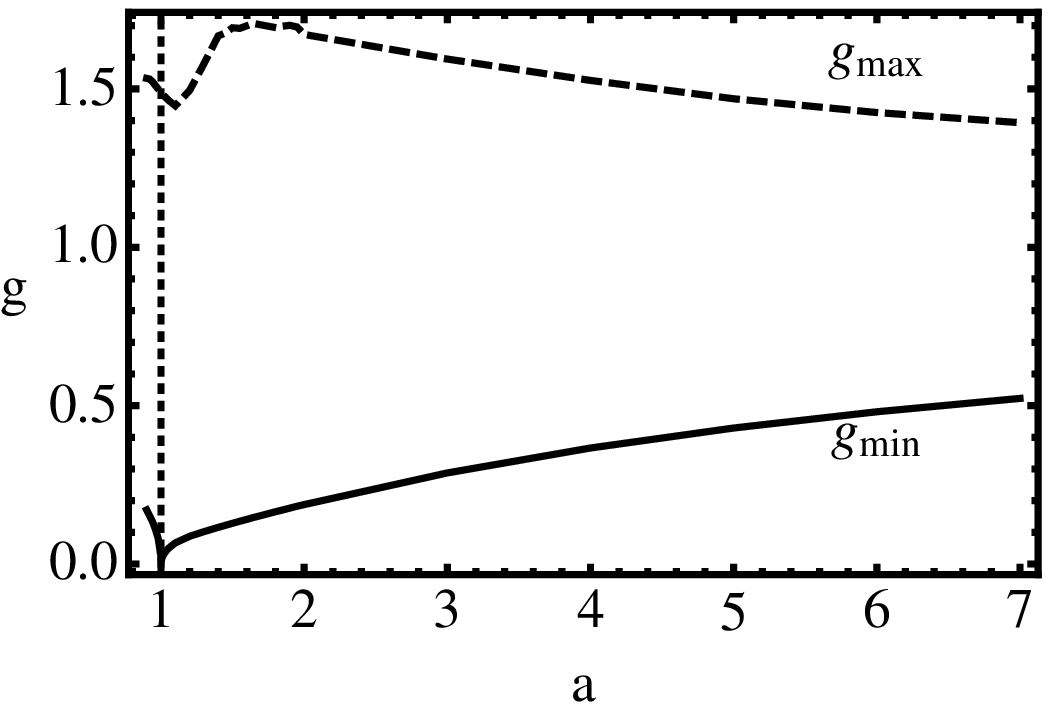}&\includegraphics[width=6.5cm]{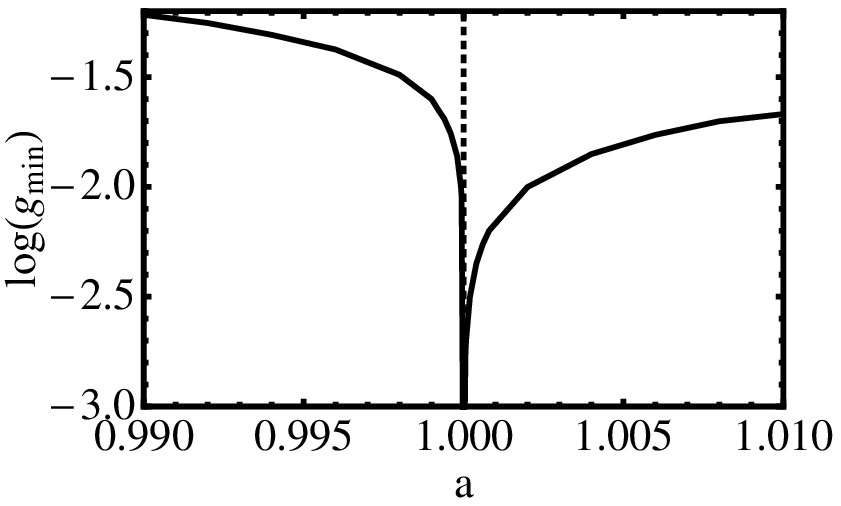}
	\end{tabular}
	\caption{\label{fig16}Left: The extension of the frequency shift of the Keplerian discs direct image represented by the g-factor. Right: the enlarged region of $g_{\mathrm{min}}$ in the vicinity of $a=1$. The inclination of the observer $\theta_o=85^\circ$.}
	\end{center}
\end{figure}

\begin{figure}
  \begin{center}
  \begin{tabular}{cc}
      \includegraphics[width=6.5cm]{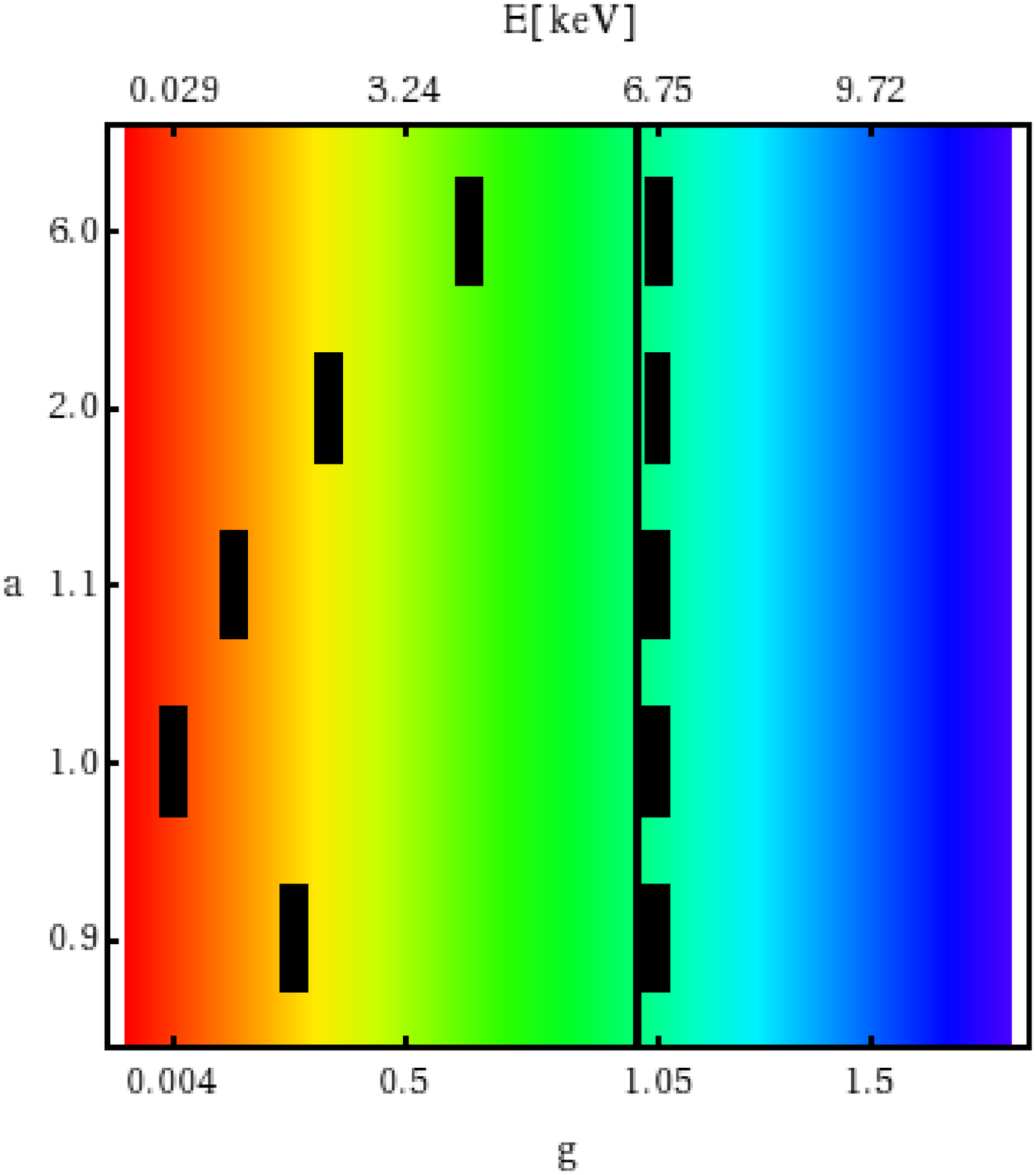}&\includegraphics[width=6.5cm]{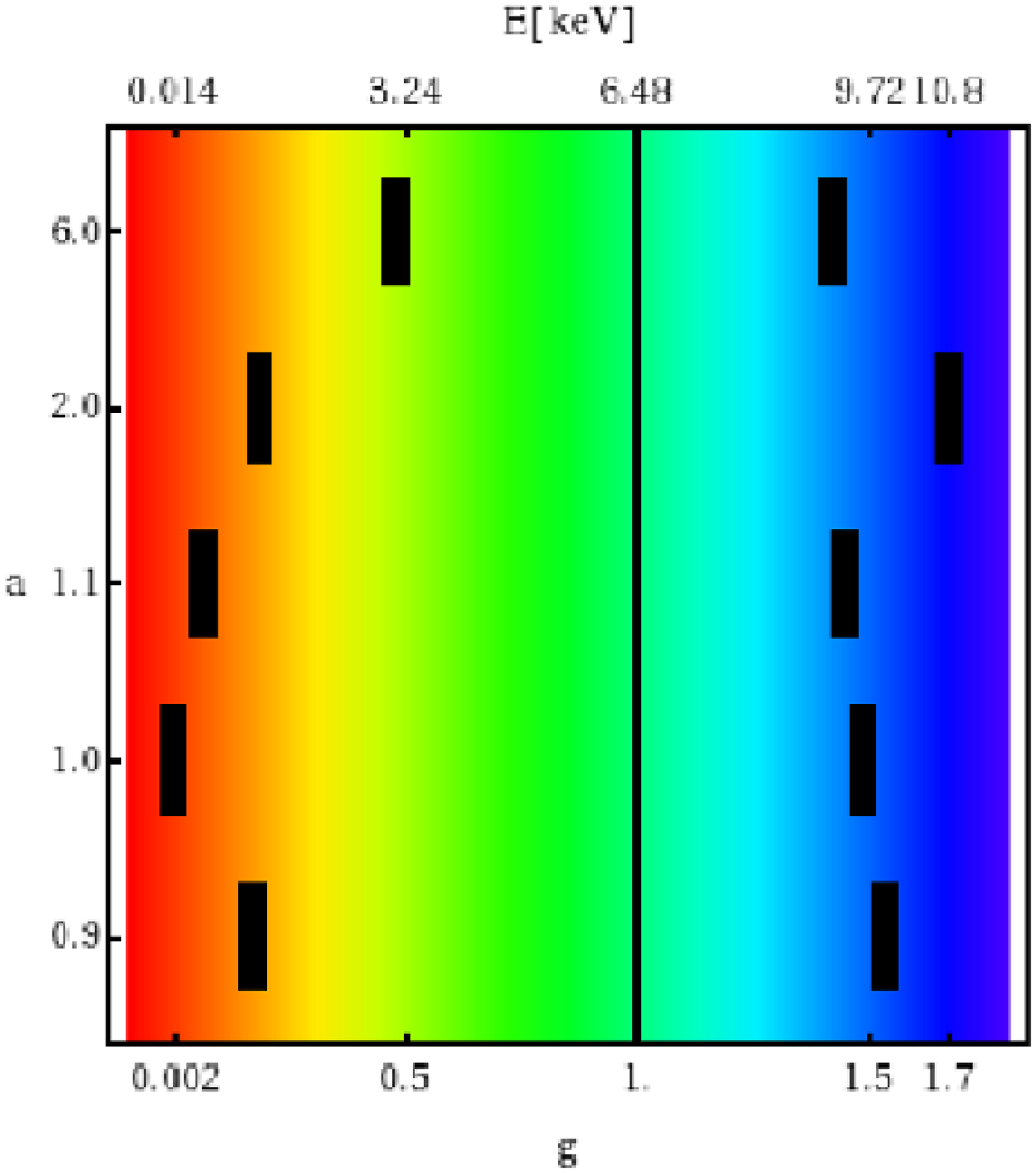}
  \end{tabular}
  \caption{\label{fig17}The extension of the observed photon energy $E$ and the frequency the g-factor (observed energy $E$ related to the rest frame energy $E_0=6.48$keV) are given for Keplerian discs direct image for representative values of spin parameter $a=0.9$, $1.0$, $1.1$, $2.0$ and $6.0$. The left image is plotted for observer inclination $\theta_o=30^\circ$ and the right image for $\theta_o=85^\circ$.}
  \end{center}
\end{figure}

In order to map the frequency shift $g$-factor into colour palette we define modified frequency shift factor $\bar{g}=(g-g_{min})/(g_{max}-g_{min})$ where $g_{min}$ ($g_{max}$) is the minimal (maximal) value of frequency shift, which is fixed in a particular set of images.

We can see from Figs \ref{fig13} and \ref{fig14} that growing spin has the tendency to enlarge and asymmetrize the disc images. The shape of the non-radiating area located inside the ISCO is a relevant signature of the superspinar spin due to the strong dependence of the ISCO radius on the spin \cite{Stu:1980:BULAI:}. Moreover, inside the internal area some remnants corresponding to multiple images appear and their shape and frequency shift strongly depend on the spin. 

The frequency scatter of the radiating part of the disc significantly depends on the superspinar spin. It should be stressed that both the strong gravitational redshift and Doppler shift generated by extremely fast motion of radiating matter of the disc combine to give the high frequency scatter. The frequency scatter range shrinks with growing spin and it is larger for the direct images as compared to those of the indirect images. We demonstrate the frequency-shift-range dependence on the spin in detail, as it could represent an impressive signature of the Kerr superspinars. We give the observed frequency scatter range for some typical values of superspinar spin, including the range of observed photon energy (assuming the rest photon energy corresponding to the Fe line with $E_0=6.48keV$), in Fig.\ref{fig17}. Observed frequency shift g-factor range $g_{min} - g_{maqx}$ is shown as a function of the superspinar spin for direct images and inclination angle $\theta_0 = 30^\circ$ in Fig.\ref{fig17}(left) and for inclination angle $\theta_0 = 85^\circ$ in Fig.\ref{fig17}(right). For comparison, we also include the range of the $g$ factor for near-extreme Kerr black holes to demonstrate clearly the possibility to distinguish the Kerr black holes and Kerr superspinars by treating simultaneously the frequency $g$-factor range and the shape of the internal part of the disc image corresponding to the innermost edge of the disc near the ISCO. 

For small inclination angles the maximal blueshift of the disc radiation $g_{max}$ is only slightly overcoming $1$ and is only slightly growing with the spin (see Fig.\ref{fig15}), but for large inclination angles it can be much stronger and its spin dependence is more complex - with spin growing up to $a \sim 2$, the maximal factor $g_{max}$ grows, but it descends for $a>2$ (see Fig.\ref{fig16}).

On the other hand, at the redshift end of the disc radiation the behavior of the frequency shift factor $g_{min}$ is of the same character independently of the inclination angle (see Figs \ref{fig15} and \ref{fig16}). The value of $g_{min}$ (corresponding to photons coming from the ISCO) is descending with spin $a$ descending to $a=1$; the same statement holds when $a$ grows to $a=1$ for discs orbiting Kerr black holes. It should be stressed that the frequency factor $g_{min} \rightarrow 0$ for the spin approaching $a=1$ from both the black hole and superspinar side; the limit of the ISCO radius $r_{ms}(a) \rightarrow 1$ and the formula (89) immediately imply this conclusion in such a situation.

We can see from Figs \ref{fig15} and \ref{fig16} that knowing the inclination angle, even the frequency shift from the radiating disc could be used to determine the Kerr superspinar spin because of the specific character of the g-factor dependence on the inclination angle when at the red end of the frequency shift ($g_{min}$) it is quite different than at he blue end ($g_{max}$).

\begin{figure}
	\begin{center}
	\begin{tabular}{cc}
		\includegraphics[width=6.0cm]{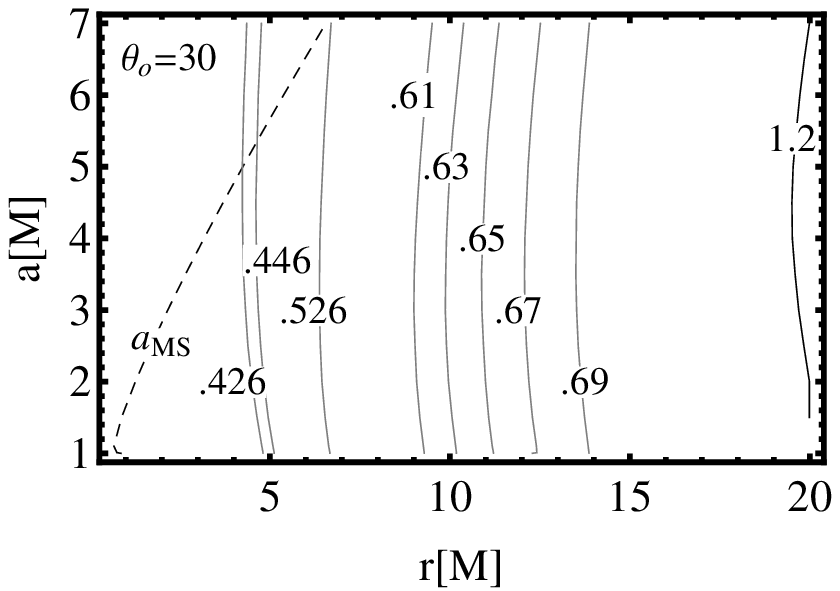}&\includegraphics[width=6.0cm]{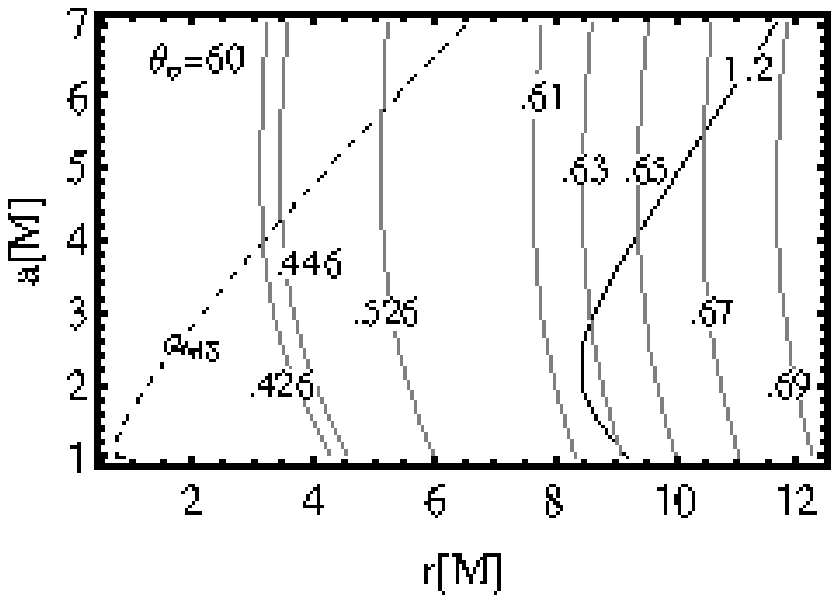}
	\end{tabular}
	\caption{Plots of spin $a$ versus hot spot radial coordinate  $r$. The solid gray(black) curves are plotted for fixed minimal(maximal) value of frequency shift $g=g_{min}$ (resp. $g_{max}$) of the hot spot integrated profiled line. Each of mentioned curves is marked by the value of $g_{min}$(resp. $g_{max}$). By fixing the edge ($g_{min}$ or $g_{max}$) of the profiled line spectra one gets , by varying the radii $r$, various values of spin parameter $a$.  Points on the curve marked with $a_{MS}$ correspond to the marginally stable orbits.  The left(right) image is plotted for observer inclination $\theta_o=30^{\circ}(60^\circ)$.\label{murphy} }
	\end{center}
\end{figure}

Our results can be applied also to observed frequency shift range $g_{min} -- g_{max}$ related to small hot spots radiating at a fixed local energy $E_0$ (corresponding, e.g., to some of the Fe K line profile created during at least one full orbit of the hot spot) from Keplerian discs orbiting Kerr superspinars (Kerr black holes). The full range of observed extremal values of the frequency shift generated by an infinitesimally small spot enables determination of the Kerr black hole spin $a$ and the orbital hot-spot radius $r_e$ for known inclination angle of the observer $\theta_o$, as demonstrated by \cite{Mur-etal:2009:}. We show that this method could be applied also in the case of Kerr superspinars. For an easy comparison, we use the same inclination angles $\theta_o = 30^o, 60^o$ as in \cite{Mur-etal:2009:}, fixing $g_{max}$ and varying $g_{min}$. Our results are given in Fig \ref{murphy}. We can immediately see important qualitative differences between the case of Kerr superspinars and Kerr black holes as presented in Fig.7 of \cite{Mur-etal:2009:}:

\begin{itemize}
\item generally, the spin and the radius are not given uniquely for fixed values of $g_{min}$ and $g_{max}$,
\item usually the relevant points are located above the radius of marginally stable orbit.
\end{itemize}
Clearly, when the method predicts the Kerr superspinar spin in a non-unique way, some other method (e.g., based on the measurements of quasiperiodic oscillations of high frequency \cite{Tor-Stu:2005:AA}) has to be applied to find the spin.

\section{Discussion}
The construction of photon escape cones of local emitters (observers) presented in this paper enables us to determine in an efficient way both the appearance of accretion discs and the silhouette of the Kerr superspinars. 

The silhouette of the superspinar can be defined as a complementary to the escape cone for distant static observers. We have constructed the silhouette shape for the case of Kerr naked singularities when photons are allowed to escape to the region of negative values of the radial coordinate \cite{Car:1973:blho,Stu:1981:BULAI:} and compared them with the case when a superspinar with surface located at $R=0.1M$ can swallow but not emit photons. It is known that in the simple case of the Kerr naked singularity spacetimes the spin and inclination angle of the observer relative to the symmetry axis can be determined by measurements of some characteristic parameters of the naked singularity silhouette shape \cite{Hio-Mae:2009:PhysRevD}. It could be quite interesting to look for an analogous way to determine the superspinar parameters, when an additional parameter giving its radius must be taken into account. We have demonstrated that the situation is more complex for two reasons. First, we have to consider three dimensional space of parameters and the choice of the parameters characterizing the superspinar silhouette shape is not straightforward. Second, we have shown that the results obtained for our choice of the parameters characterizing the silhouette shape are not unique, since the characteristic curves cross each other generally (see Fig.\ref{fig12}). undoubtedly, this problem is worth further detailed studies which we postpone for future work.

We have further considered appearance of the Keplerian, thin accretion discs with quasigeodesic character of the disc-matter motion, but our method can be straightforwardly generalized to any kind of accretion discs. Our results clearly demonstrate the strong dependence of the disc appearance on the superspinar spin both for the shape of the disc and the frequency shift of its radiation. The differences are quite evident for both the direct and indirect images of the disc and can represent a strong tool for astrophysical tests of the superspinar existence.

In the studies of the disc appearance we have shown how in the both direct and indirect images the frequency shift of different parts of the disc is reflected, assuming for simplicity that locally the disc radiates at a particular frequency, fixed across all the radiating part of the Keplerian disc. The results obtained for superspinars having a selected typical values of the spin are confronted to analogous results obtained in the case of the near-extreme Kerr black hole with spin $a=0.9981$. The frequency shift g-factor represents an important signature of the superspinar spin - it is quite interesting that the frequency shift differs significantly for the direct and indirect images. In comparison with the black hole case, the g-factor in the discs orbiting Kerr superspinars has larger range and substantially higher shift to red end of the spectrum when the same distance of the spin from the extreme state $a=1$ is assumed. 

In both the direct and indirect images a dark central region appears resembling the silhouette of the superspinar surface, however, the dark region is doubled. With spin growing, the extension of the disc and the central dark region grows. Clearly, the dark regions correspond to both photons that are captured by the superspinar surface and those that are trapped in the vicinity of the surface. The doubled internal structure in the radiating disc appears due to complex photon motion near the superspinar. The internal part of the disc image shows a complex structure that gets more complicated with spin descending to the extreme black hole state (see Figs \ref{fig13} and \ref{fig14}). We have demonstrated strong effects of the Kerr superspinar spin on the disc appearance for both large and small inclination angles of the distant observer.

\begin{figure}[ht]
	\begin{center}
		\includegraphics[width=7cm]{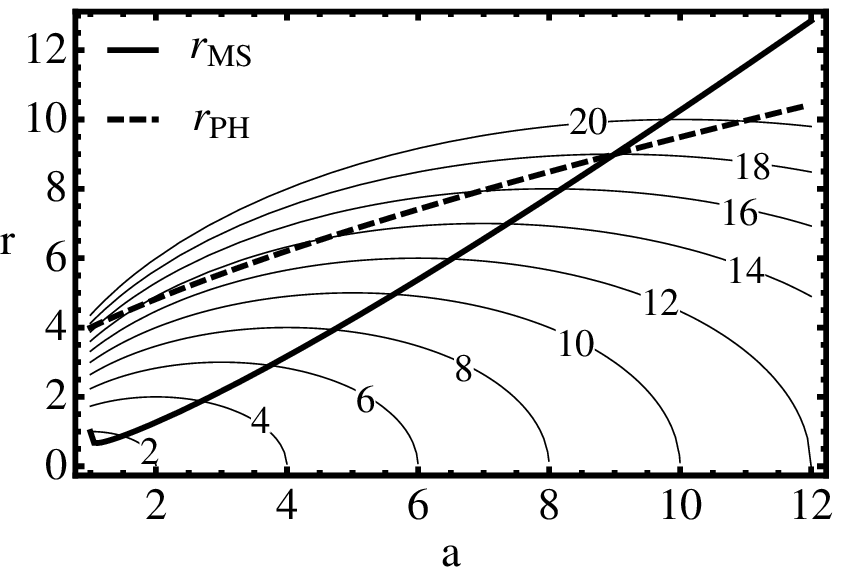}
		\caption{\label{fig18}The dependence of the radii of the unstable circular photon orbit $r_{ph}$ and ISCO radius $r_{ms}$ on the superspinar spin parameter $a$. The thin solid lines represent radii of the outermost unstable photon spherical orbit for fixed value of impact parameter $\lambda$ as functions of the superspinar spin parameter $a$.}
	\end{center}
 \end{figure}

\begin{figure}
	\begin{center}
	\begin{tabular}{ccc}
		\includegraphics[width=5cm]{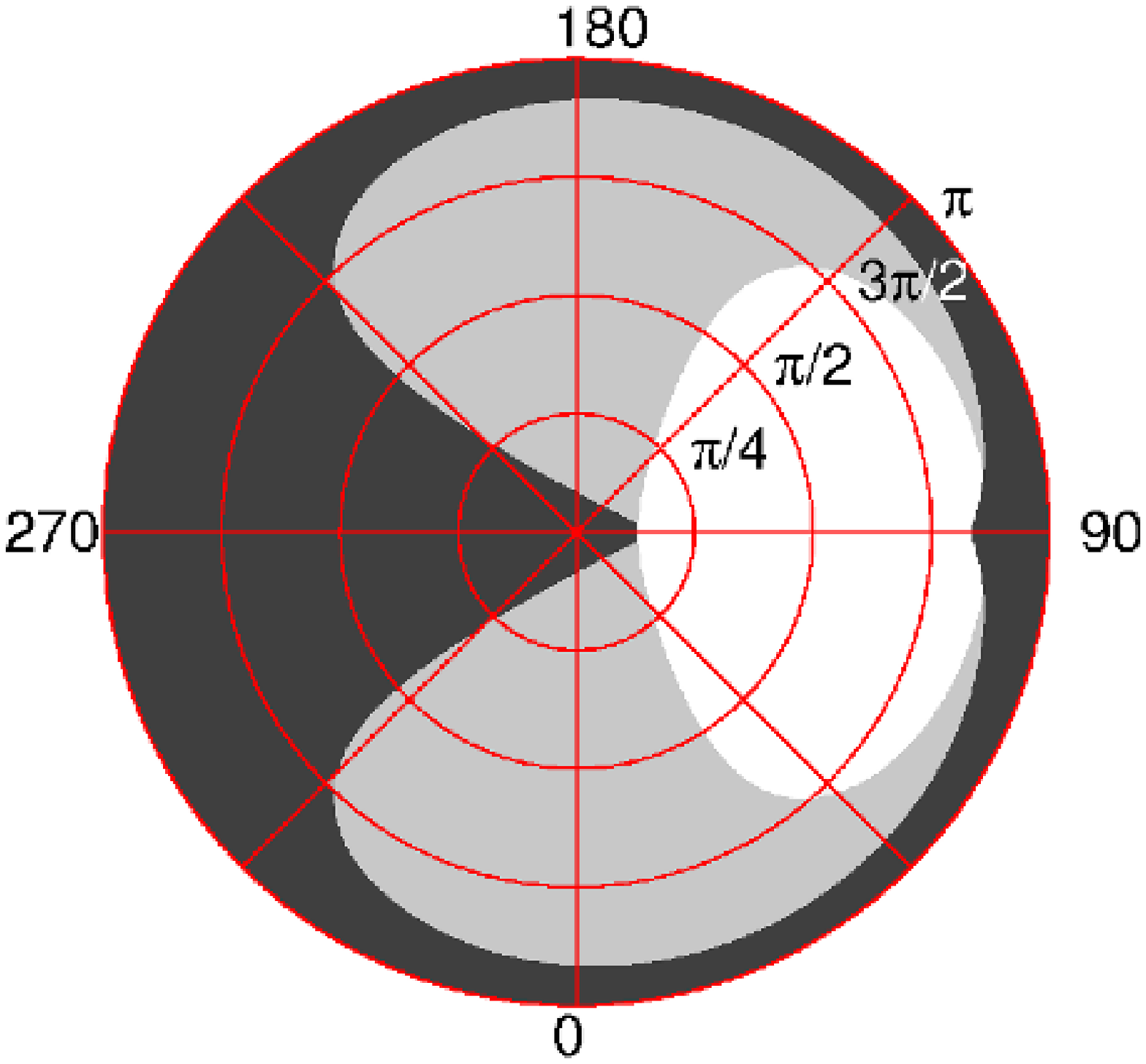}&\includegraphics[width=5cm]{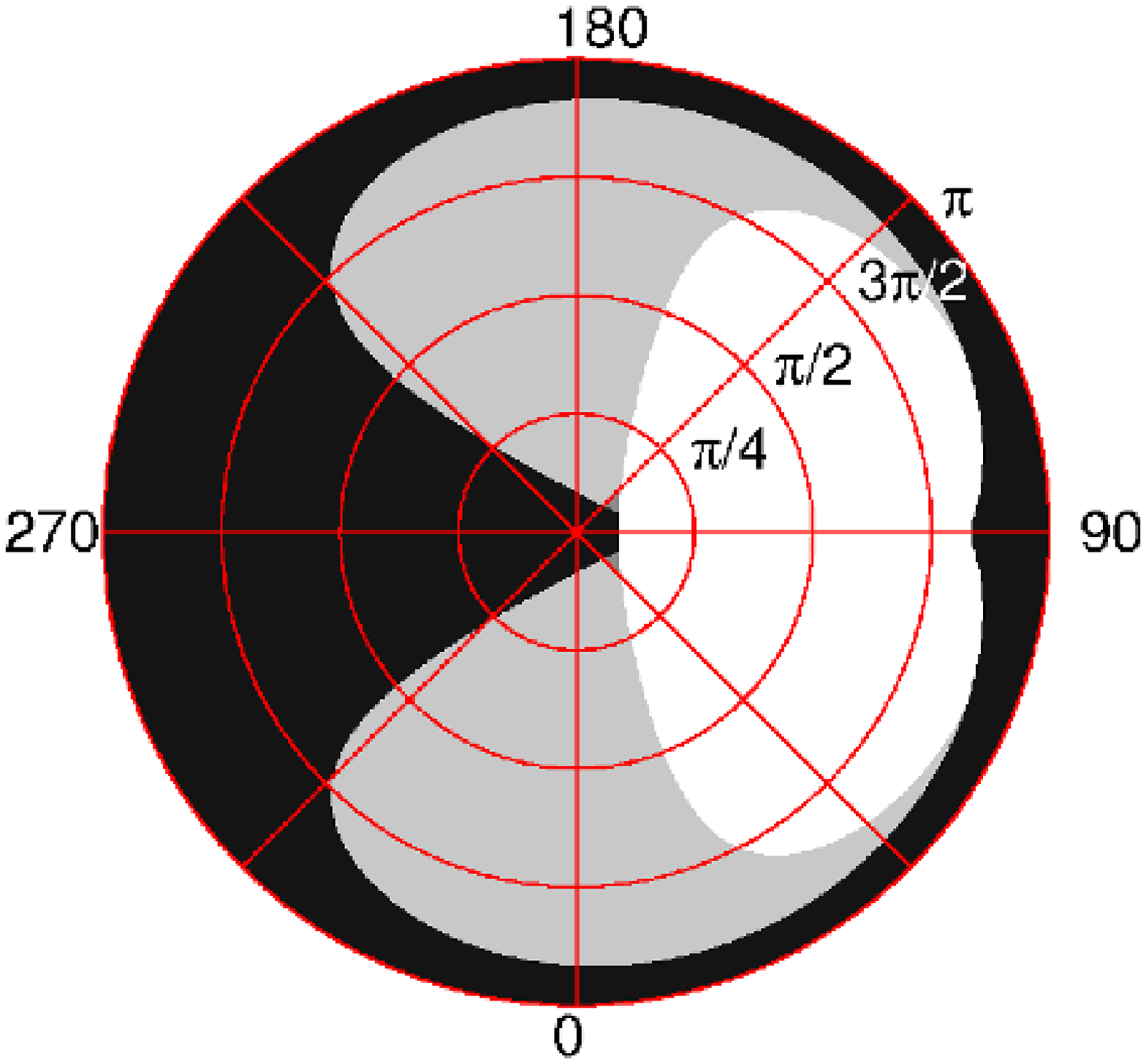}&\includegraphics[width=5cm]{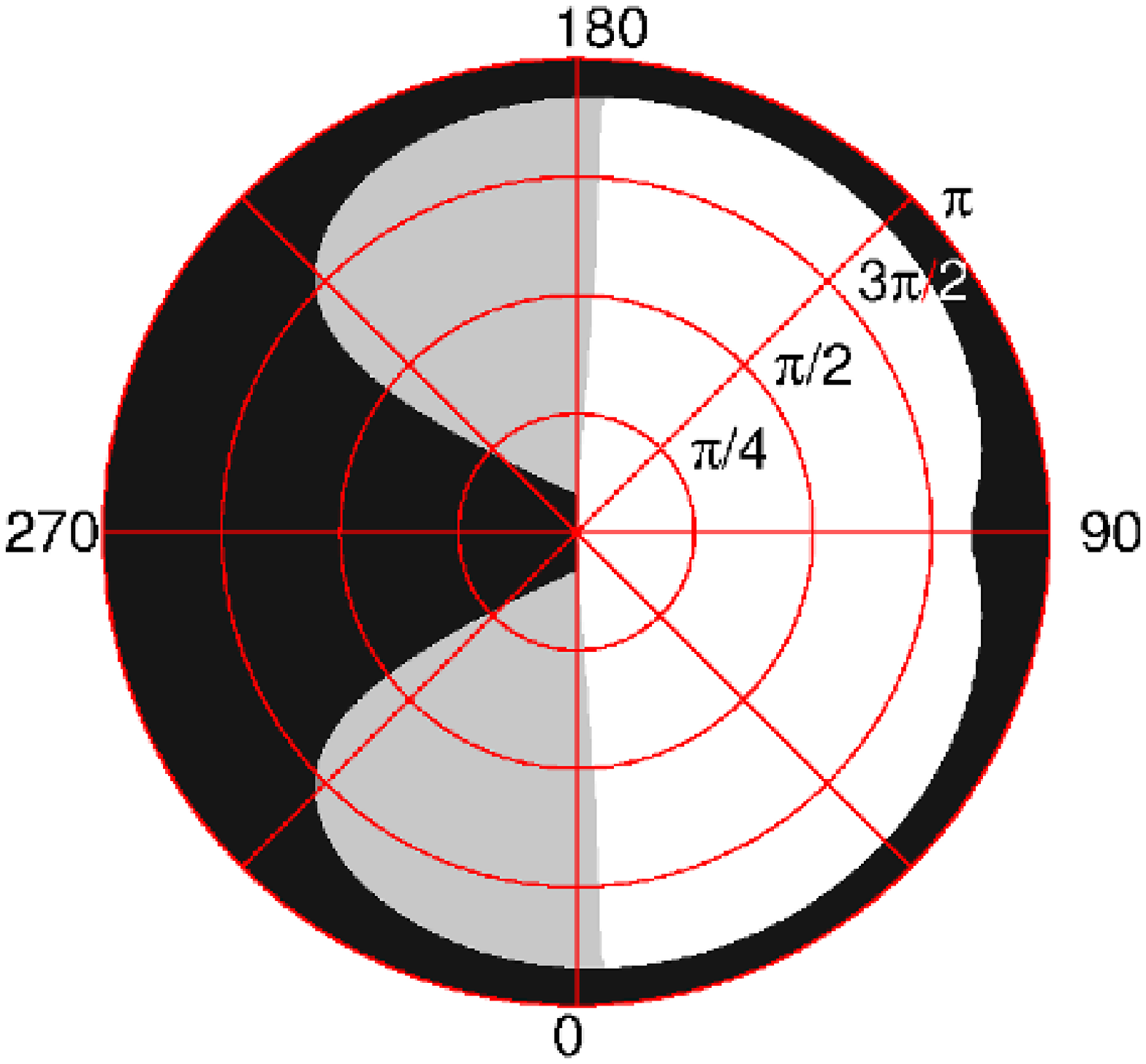}\\
\includegraphics[width=5cm]{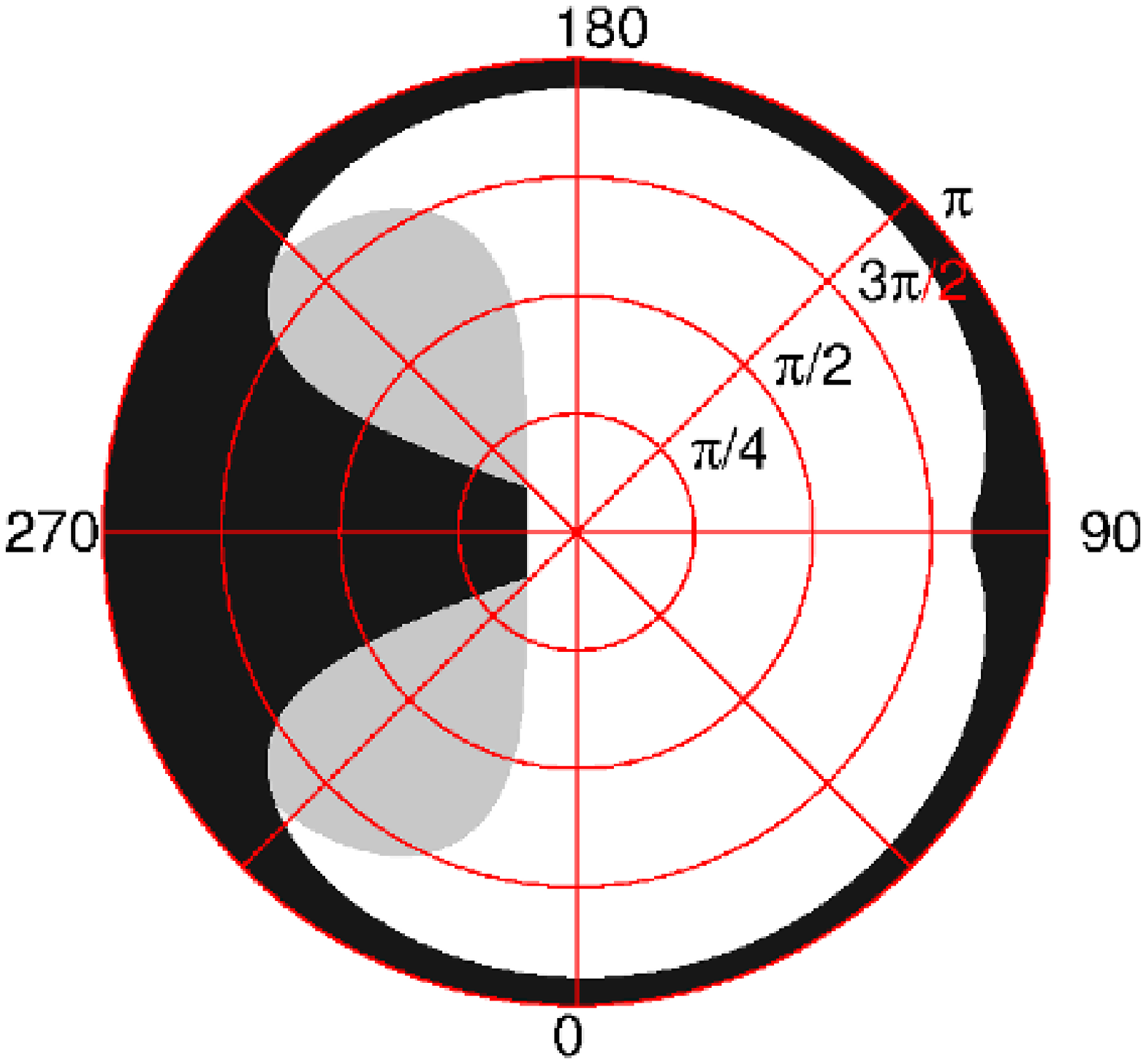}&\includegraphics[width=5cm]{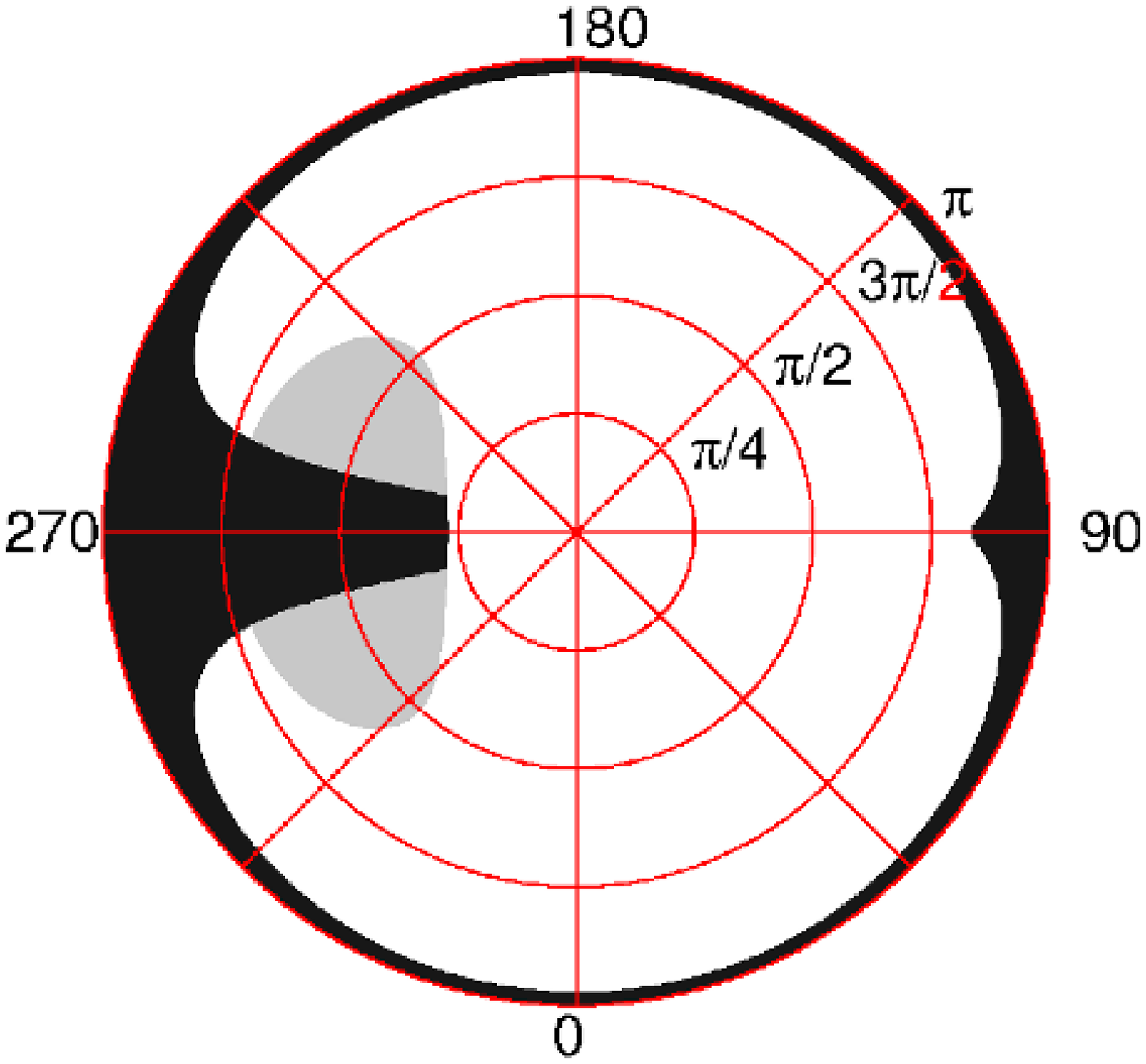}&\includegraphics[width=5cm]{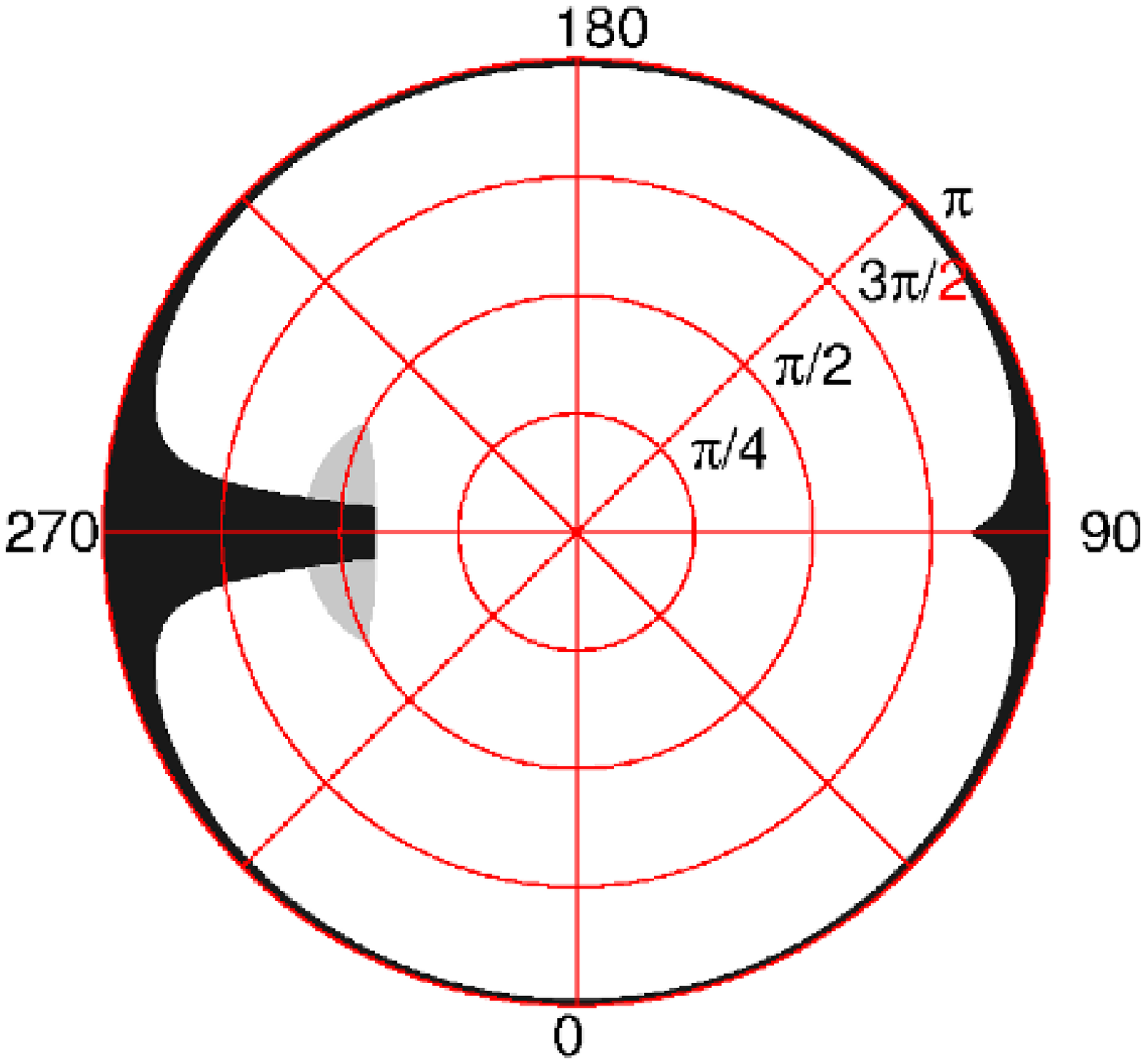}		
	\end{tabular}
	\caption{\label{fig19}GF+ local cones for trapped (black), captured (gray) and escaping (white) photons. The cones are plotted for six representative values of spin parameter $a=1.001$, $1.01$, $1.1$, $1.5$, $3.0$ and $6.0$ (from top left to bottom right). The emitter is moving along ISCO with radial coordinate $r_{\mathrm{ms}}=r_{\mathrm{ms}}(a)$.}
	\end{center}
\end{figure} 

\begin{figure}
	\begin{center}
	 \begin{tabular}{ccc}
			\includegraphics[width=5.5cm]{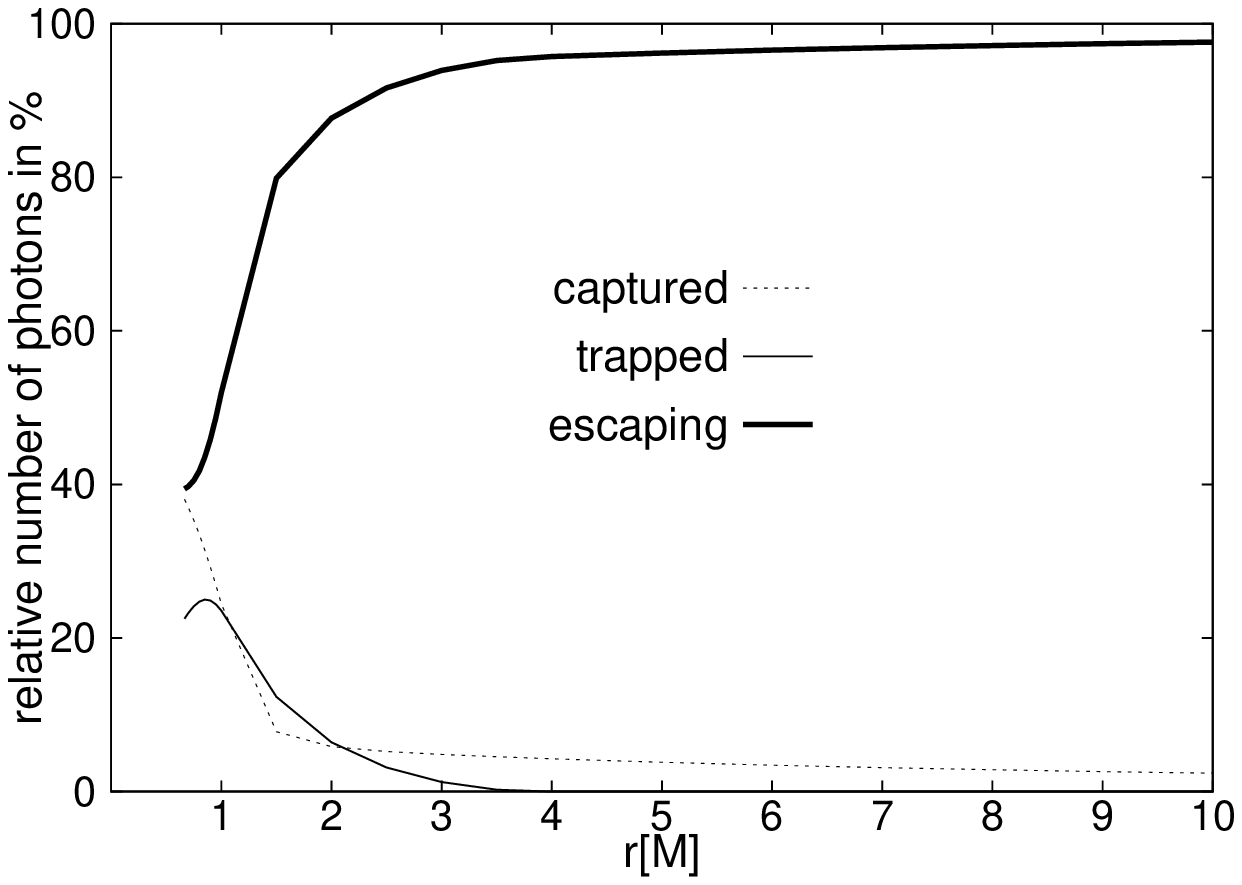}&\includegraphics[width=5.5cm]{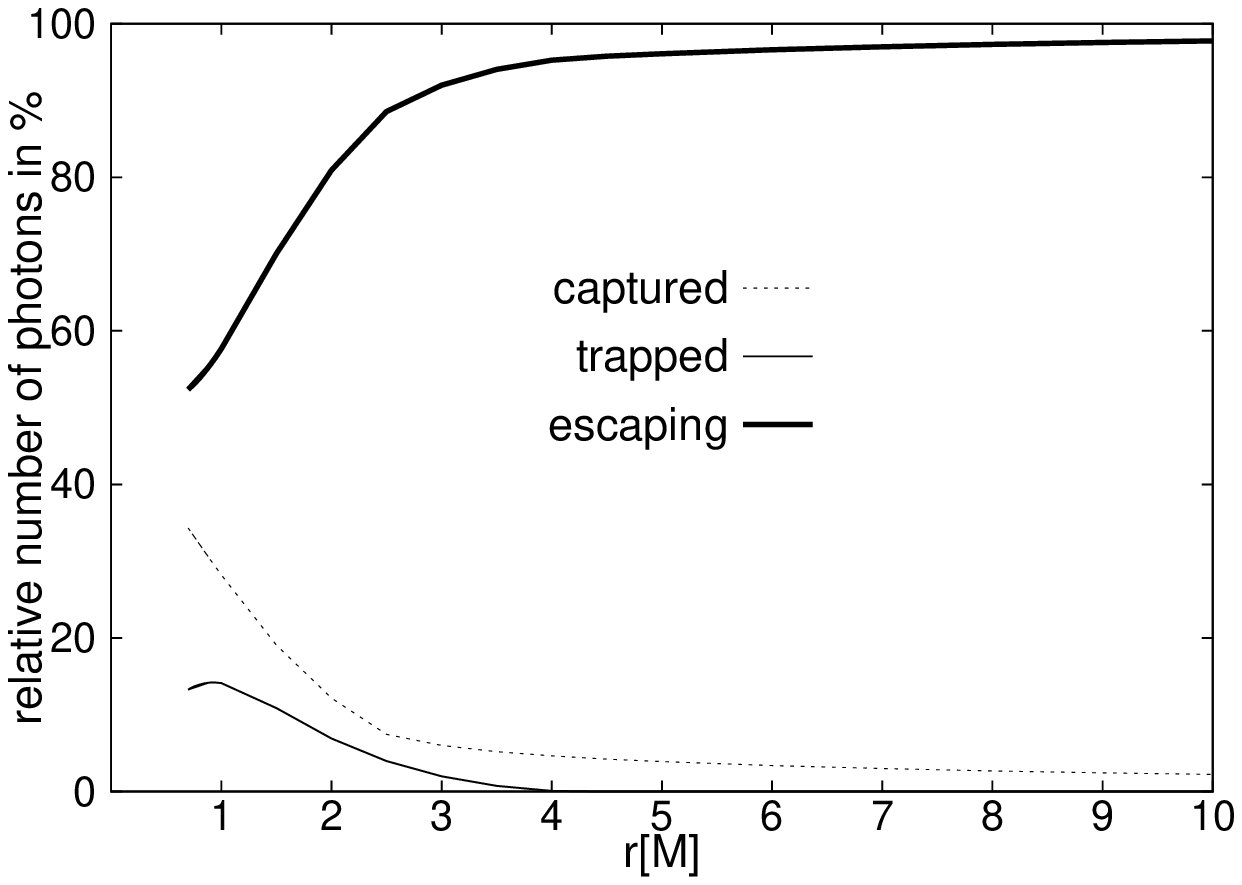}&\includegraphics[width=5.5cm]{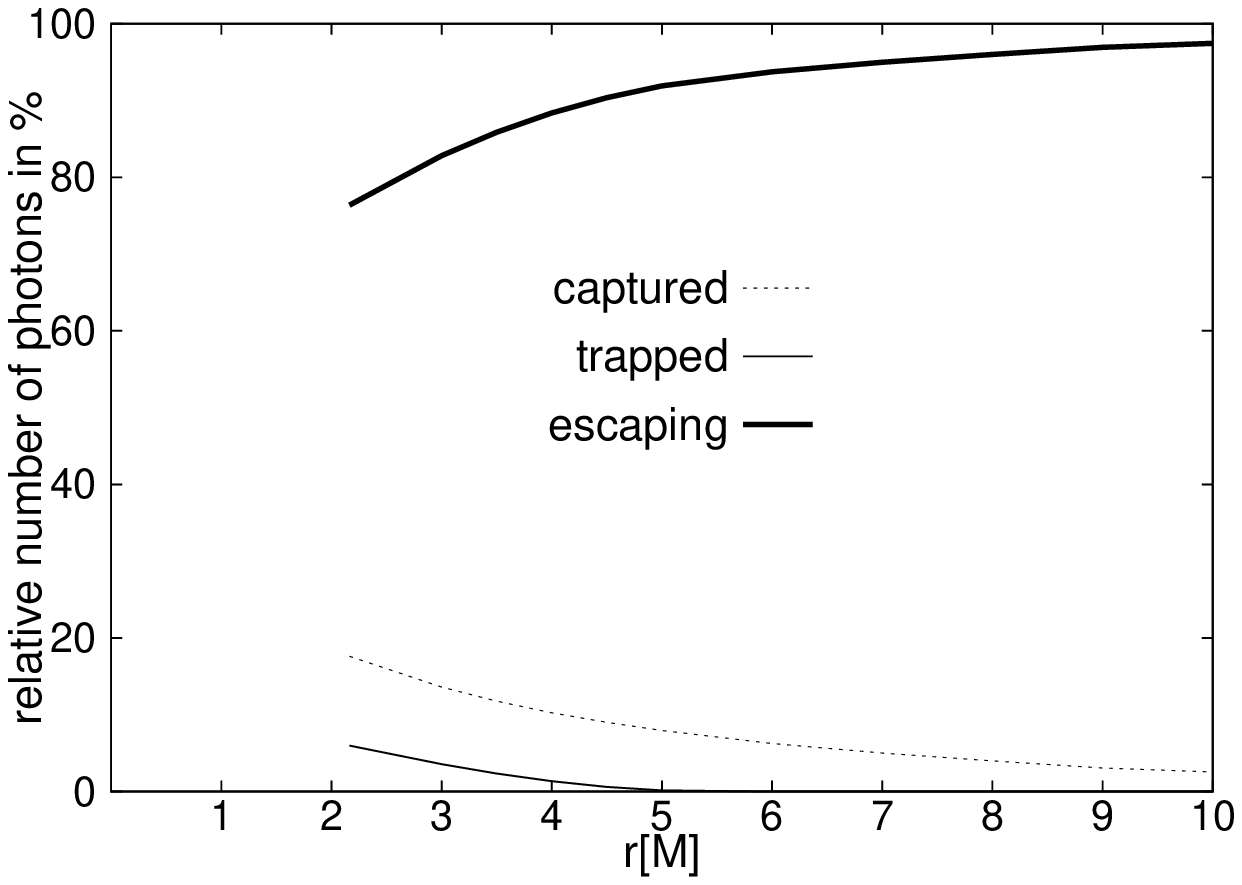}
	  \end{tabular}
			\caption{\label{fig20}The relative number of photons that are captured, trapped and escaping (given by the ratio of areas corresponding to escaped, captured and trapped photons) as functions of radial position of the emitter $r_e$ for superspinar with spin $a=1.1$, $1.5$ and $3.0$ (from left to right).}
	\end{center}
\end{figure}



There is some new aspect of the photon motion in the field of Kerr superspinars, crucial for the character of the accretion discs radiation and even for the disc structure, that is not present in the case of Keplerian discs orbiting Kerr black holes. Namely, there is a family of photons that are trapped by the strong gravitational field of the superspinars without being captured by them. Such trapped photons can serve as an efficient source of reflection component of the discs radiation and, moreover, they could influence structure of the accretion discs, potentially causing significant deviations from the standard Keplerian character even for subcritical accretion rates. 

The trapped photon orbits are centered around spherical photon orbits that are stable with respect to radial perturbations (see Figs \ref{fig3}a and \ref{fig3}f). Such trapped photons could cause a strong self-illumination effect even in the Keplerian accretion discs, influencing thus strongly their radiation. For Kerr superspinars with spin close to the extreme value ($a<1.1$), trapped photons emitted in the outer regions allowing existence of trapped photons can be reflected from the innermost parts of the discs near the ISCO, with strong increase of their energy (frequency) due to the motion in the extremely steep gravitational potential of such superspinars. On the other hand, the trapped photons emitted near the ISCO could influence substantially the external parts of the trapped-photon region. It should be stressed that the self-illumination effect can in principle change strongly even the character and physical properties of the accretion discs since a substantial amount of heat generated by the friction phenomena will be captured by the discs enabling possibility of a subsequent change of an originally thin equatorial accretion discs to a toroidal structure. The self-illumination and self-reflection effect in optical phenomena and the discs structure will be considered in future papers. Here we restrict our attention to some basic estimates of the relevance of the trapped photons and the self-illumination effect.

The influence of trapped photons could be crucial for the accretion discs orbiting Kerr superspinars with their edge entering the region of the trapped photon orbits. The trapped orbits are surely located under the photon circular orbits with radius $r_{ph}$ given by Eq.(\ref{eq25}). Therefore, the Keplerian accretion discs will be influenced by the trapped photons when the condition $r_{ms} < r_{ph}$ will be satisfied. Since the ISCO is determined by the relations (\ref{isco}), we can conclude that the self-illumination effect has to be relevant for Keplerian discs orbiting Kerr superspinars with spin $a < a_{si}=9$ - see Fig.\ref{fig18}. 

\begin{figure}
	\begin{center}
		\includegraphics[width=8cm]{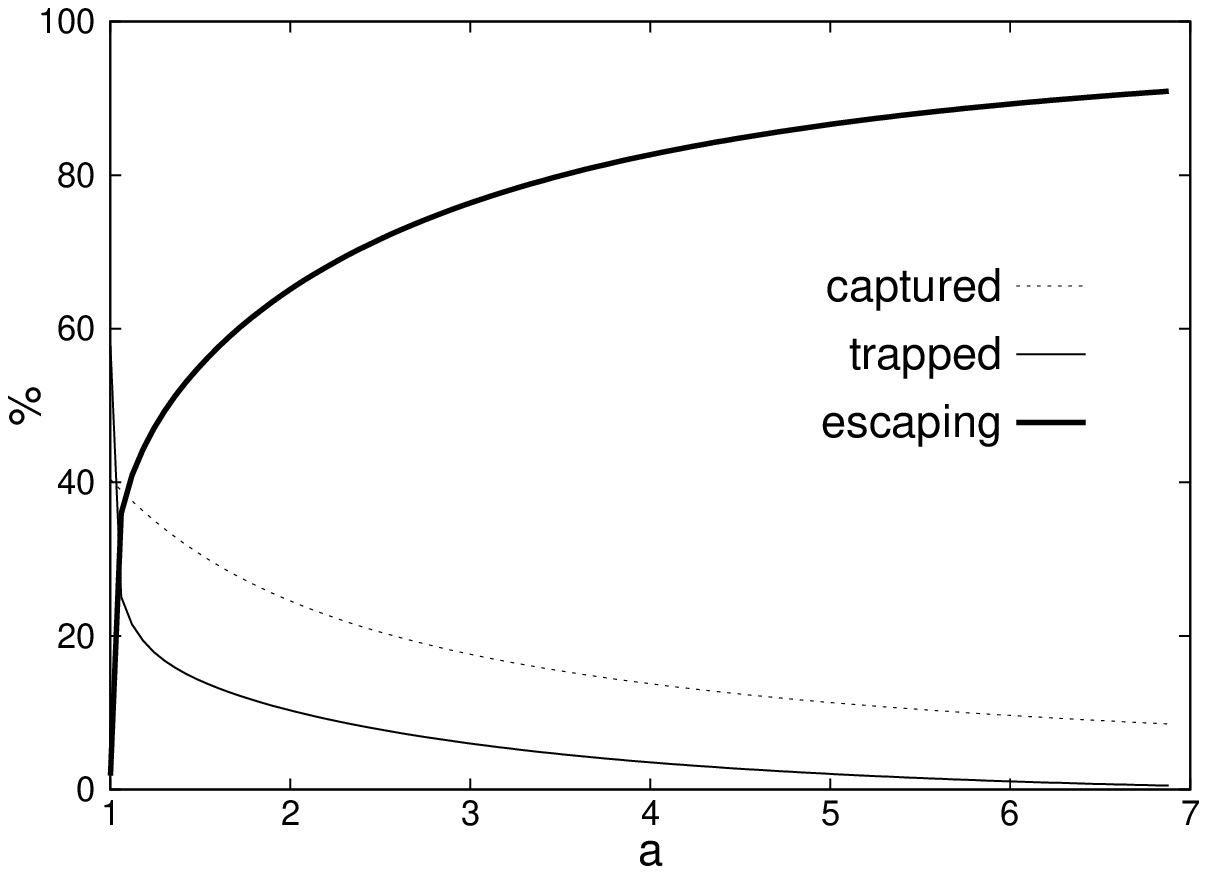}
		\caption{\label{fig21}Relative number of photons radiated by a source moving at ISCO which are escaping to infinity (captured - hit the superspinar surface, trapped - orbiting the superspinar) as a function of the spin parameter $a$.}
	\end{center}
\end{figure}

Further, we briefly discuss relevance of the self-illumination effect by constructing detailed structure of the complementary parts of the local escape cones, separating them into parts corresponding to the captured and trapped photons. The method used is quite analogical to the construction of escape cones, but the relevant notion here is the presence of the stable photon spherical orbits, contrary to the case of escape cones, when the unstable photon spherical orbits play the crucial role. The results are illustrated for the case of emitters orbiting the superspinar along the ISCO orbit for some typical values of its spin (see Fig.\ref{fig19}.). The efficiency of capturing and trapping photons along the Keplerian disc is demonstrated in Fig.\ref{fig20} for some typical values of spin. The relative efficiency of these processes at the ISCO is given as a function of the spin in Fig.\ref{fig21}.
The results clearly demonstrate that the effect of trapping photons in the accretion dics strongly grows with descending spin of the superspinar. The amount of trapped photons overcomes those of the captured photons for Kerr superspinars with $a<1.1$, while it is substantially smaller for $a>2$. In the liming case of $a \rightarrow 1$ the radiation at ISCO is separated between the trapped and captured photons in the ratio of $60 : 40$. Clearly, for superspinars close to the extreme state $a=1$ the self-illumination effect has to be quite relevant for both appearance and structure of accretion discs. Undoubtedly, future detailed studies are necessary in order to clear up all aspects of the trapped-photons phenomenon. 

\clearpage
\newpage

\section{Conclusions}

Studying the optical appearance of the Keplerian accretion discs orbiting Kerr superspinars we have demonstrated that there exists a plenty of observational phenomena enabling clear distinguishing of Kerr superspinars and Kerr black holes. This is an important result, especially due to the recent results demonstrating that considering the spectral profiles of Keplerian discs as observed at infinity, it is not possible to decide if the spectrum is generated by a Kerr superspinar (Kerr naked singularity) or a Kerr black hole due to one to one correspondence of their spin that determines location of the ISCO orbit  relevant in spin determination \cite{Tak-Har:2010:arXiv}.

Our results concerning the appearance of the Keplerian discs demonstrated a potential possibility to obtain relatively good estimates of the superspinar spin due to the character of the dependence of the innermost part of the radiating disc near the ISCO in its shape and the frequency shift represented by the g-factor. In fact, the shape is very sensitive on large values of the spin ($a>2$), while, quite inversely, the g-factor is very sensitive on the spin for its small values 
($a<2$). The combination of both measurements of the disc shape at the most internal part and the g-factor thus enables relatively good estimates of the superspinar spin in the whole considered range of $1<a<10$ as these measurements serve in a complementary way. Moreover, independent estimates of the inclination angle could substantially improve the superspinar spin estimates.

It is important that the Kerr superspinar silhouette resulting by capture of photons radiated by an extended source located behind the superspinar can also enable to clearly distinguish Kerr superspinars, Kerr naked singularities violating the Penrose cosmic censorship hypothesis, and Kerr black holes \cite{Hio-Mae:2009:PhysRevD,Bam-Fre:2009:PhysRevD}. Nevertheless, we have shown that in the case of superspinars endowed with a characteristic parameter representing radius of their surface, the relations between the spacetime parameters (and inclination angle of the observer) and the characteristics representing properties of the superspinar silhouette and its arc are much more complex in comparison with the case of Kerr naked singularities having no characteristic radius to be determined and deserves further attention.

Considering the superspinar surface properties, we assume a strong similarity with the black hole horizon properties, namely, we assume one-way membrane property of this surface. Therefore, all photons are captured, no photons are able to escape from the Kerr superspinar surface. If the superspinar surface could radiate and make a backreaction on the accreting matter, the optical phenomena will be modified in comparison to our results, but such modifications have to be related to the innermost parts of the images constructed here. We are not able to address this problem at the present state of our knowledge. Clearly, it depends on the detailed internal solutions of the superspinar structure that has to be matched to the external Kerr field and would be given by the string (or some other multidimensional) theory. Then the important problem of the superspinar stability could be addressed too. Here, stability (or metastability on long time scales) of the Kerr superspinars is assumed, similarly to a number of related studies of the superspinars.

We have shown that photons can be trapped by strong gravitational field in vicinity of the surface of Kerr superspinars. No such effect exists outside the event horizon of black holes where all photons must escape to infinity, or be captured by the black hole. The existence of the trapped photons strongly influences a variety of optical effects in the field of Kerr superspinars both directly and indirectly through its influence on the structure of accretion discs in close vicinity of Kerr superspinars. Trapped photons are potentially able to affect structure of accretion discs in the field of Kerr superspinars with spin $a < 9$ when the ISCO radius reaches the region of trapped photons - detailed understanding of all the related problems remains a task for future studies. Here, we restricted our attention to the problems related to the potential effectiveness of the photon trapping on the appearance and structure of the Keplerian accretion discs. We have demonstrated that the role of the self-illumination effects caused by the trapped photons can be significant for Kerr superspinars with $a<1.1$ becoming very strong for near-extreme superspinars when both the disc appearance and the even its structure could be substantially shifted from the standard model.

We can conclude that the Kerr superspinars (naked singularities) could influence the optical phenomena in their vicinity in quite different way as compared with those given by the presence of Kerr black holes. Moreover, there are strong indications that photons trapped by the strong gravitational field of superspinars, without being captured by them, can strongly affect even the structure of the accretion discs. Therefore, there is a plenty of astrophysical phenomena that could principally lead to a confirmation or rejection of the hypothetical existence of Kerr superspinars as predicted by the string theories and deserve further extensive attention in both theoretical models of the optical phenomena and their relation to the observational data obtained from some exotic sources. 

\section*{Acknowledgements}

This work was supported by the Czech grant MSM~4781305903, LC06014, GACR 202/09/0772. One of the authors (ZS) would like to express his gratitude to the Committee for Collaboration of Czech Republic with CERN for support and the Theory Division of CERN for perfect hospitality. He would also like to thank Prof. P. Ho\v{r}ava for stimulating discussions.


\end{document}